\documentclass[11pt,twocolumn]{article}
\usepackage{amssymb,epsfig,wrapfig,graphics,aas_macros,appendix,url,multirow}
\usepackage{longtable}
\usepackage[dvips]{color}
\usepackage[footnotesize]{caption}
\def\comment#1{{}}
\vspace{-0.5in}
\newlength{\cvindent}
\newlength{\cvhang}

\newlength{\refindent}
\def\lsim{\lower0.6ex\vbox{\hbox{$ \buildrel{\textstyle <}\over{\sim}\ $}}}
\def\gsim{\lower0.6ex\vbox{\hbox{$ \buildrel{\textstyle >}\over{\sim}\ $}}}

\def\beq{\begin{equation}}
\def\eeq{\end{equation}}

\def\lsim{\mathrel{\hbox{\rlap{\hbox{\lower4pt\hbox{$\sim$}}}\hbox{$<$}}}}
\def\gsim{\mathrel{\hbox{\rlap{\hbox{\lower4pt\hbox{$\sim$}}}\hbox{$>$}}}}

\begin{document}
\onecolumn

\widowpenalty=10000
\clubpenalty=10000
\title{The Status and future of ground-based TeV gamma-ray astronomy\\
A White Paper prepared for the \\ Division of Astrophysics of the American Physical Society}

\author{Organizers and working-group chairs: \\
J. Buckley, Washington University, St. Louis,\\
K. Byrum, Argonne National Laboratory,\\
B. Dingus, Los Alamos National Laboratory,\\
A. Falcone, Pennsylvania State University,\\
P. Kaaret, University of Iowa,\\
H. Krawzcynski, Washington University, St. Louis,\\
M. Pohl, Iowa State University,\\
V. Vassiliev, University of California, Los Angeles,\\
D.A. Williams, University of California, Santa Cruz}
\date{}
\maketitle

\tableofcontents
\newpage
\twocolumn
\setcounter{section}{0}
\pagenumbering{arabic}



\section{Summary and Overview}
\label{summary-subsec}
\subsection[Executive Summary]{Executive Summary}

High-energy $\gamma$-ray astrophysics studies the most energetic
processes in the Universe. It
explores cosmic objects such as supermassive black holes and exploding
stars which produce
extreme conditions that cannot be created in experiments on Earth.
The latest generation of $\gamma$-ray instruments has discovered objects that
emit the bulk of their power in the form of high-energy $\gamma$-rays.
High-speed imaging technology,
with gigahertz frame rates allow us to detect individual cosmic
$\gamma$-ray photons produced
under the most violent and extreme conditions. Gamma-ray astronomy has
unique capabilities
to reveal the nature of the elusive dark matter that dominates the
matter contents of the Universe.

Motivated by the recent advances of TeV $\gamma$-ray astronomy, the
Division of Astrophysics of the
American Physical Society (APS) charged the editorial board of this
White Paper to summarize
the status and future of ground-based $\gamma$-ray astronomy.
The APS requested a review of the science accomplishments and
potential of the field.
Furthermore, the charge called for a description of a clear path
beyond the immediate future to
assure the continued success of this field.
The editorial board solicited input from all sectors of the
astroparticle physics community
through six open working groups, targeted international meetings, and
emails distributed through the
APS and the High-Energy Astrophysics Division of the American
Astronomical Society.
The board also enlisted senior advisers that represent ground-based
and satellite-based $\gamma$-ray
astronomy, particle physics, and the international community of
astroparticle physicists.

This section summarizes the findings and recommendations of the White
Paper team. It also
gives a brief introduction to the science topics that can be addressed
with TeV $\gamma$-ray astronomy.
The interested reader is referred to the detailed discussions in the
reports of the working groups and to
excellent review papers about the status and accomplishments of TeV
$\gamma$-ray astronomy
by Hinton\footnote{Hinton, J., 2007, http://arxiv.org/abs/0712.3352}, 
and by Aharonian et al.\footnote{Aharonian, F., Buckley, J., Kifune, T., 
Sinnis, G., 2008, Reports on Progress in Physics, 71, 9,
http://stacks.iop.org/0034-4885/71/096901}.
Appendix A lists the authors of the White Paper, the make-up of the
working groups, and the meetings
organized by the White Paper team. Appendix B reproduces the APS charge.

\subsubsection[Summary of findings]{Summary of Findings}
{\bf (F1)} The current generation of instruments has demonstrated that
TeV $\gamma$-ray astronomy is a rich field of research.
TeV $\gamma$-ray astronomy saw its first major success in the year 1989
with the firm detection of a cosmic source of TeV $\gamma$-ray
emission, the Crab Nebula, with the Whipple 10-m Cherenkov telescope.
To date, advances in instrumentation and analysis techniques have established
TeV $\gamma$-ray astronomy as one of the most exciting new windows
into the Universe.
The H.E.S.S., MAGIC, and VERITAS experiments have shown us a glimpse of the
discovery potential of this new type of astrophysics. The Milagro and
Tibet experiments have explored
an alternative experimental technique that permits a full survey of the sky.
Due to the increased sensitivity of these instruments, the number of known TeV
$\gamma$-ray sources has increased by an order of magnitude (from
$\sim$10 to $\sim$100) in the
past 3 years. Known source classes include the remnants of supernova
explosions, neutron stars,
supermassive black holes, and possibly groups of massive stars.
Many new TeV sources and source classes have been discovered.
Several of them have
heretofore unobserved substructures resolved by TeV $\gamma$-ray telescopes,
indicating isolated regions in which energy is transferred to
high-energy particles.
Several sources show brightness variations clearly resolved
in the TeV data; the previously known sources Mrk 421 and
PKS 2155-304, both of which involve a supermassive black hole,
have been observed to vary on timescales as short as 2 minutes,
indicating that the emission regions may be comparable in size to the
event horizon
of the parent black hole, and revealing the inner workings of these
powerful systems.
The very short variability timescales are
also facilitating studies of quantum gravity through the search for
violations of Lorentz invariance. Among the most important
discoveries is the fact that there are many mysterious unidentified TeV
objects that have no currently known counterpart at any other wavelength.\\

{\bf (F2)} Primary scientific drivers of ground-based $\gamma$-ray astronomy:
\begin{itemize}
\item High-energy particles are an ubiquitous but insufficiently
studied component of
cosmic plasmas. TeV $\gamma$-ray astronomy makes it possible to study
the acceleration
and propagation of these high-energy particles in a wide range of different
environments, from the remnants of exploding stars to the formation
of the largest gravitationally bound structures in the Universe.
\item The combination of $\gamma$-ray observations, accelerator
experiments, and direct detection experiments
may lead to one of the most spectacular discoveries of the 21$^{\rm
st}$ century: The
unambiguous identification of the mysterious dark matter that holds
together the cosmic
entities in which we live: Galaxies and galaxy clusters.
\item Supermassive black holes 
reside at the centers of most galaxies.
A black hole that is well-fed with infalling gas will produce
collimated outflows, or jets, of
gigantic proportions, reaching out beyond the bounds of its host galaxy.
TeV $\gamma$-ray astronomy offers the possibility to study the formation of jets
and to obtain key insights into how black holes grow,
thus revealing the cosmic history of supermassive black holes and
their influence on the
cosmological evolution. The strong beams of $\gamma$-rays from these
sources can be used to probe
the extragalactic infrared background radiation and thus to constrain
the star-formation history
of the Universe.
\end{itemize}

{\bf (F3)} A large-scale ground-based $\gamma$-ray observatory would
substantially increase the scientific
return from a number of present and future ground-based and
space-based 
observatories including (but
not limited to) the Low-Frequency Array (LOFAR) and the
Square-Kilometer Array (SKA)
at radio wavelengths, the Large Synoptic Survey
Telescope (LSST), the Fermi Gamma-Ray Space Telescope, the
neutrino experiments IceCube and ANITA, and
the gravitational wave experiments LIGO and LISA.\\

{\bf (F4)} The experimental techniques for the ground-based detection
of $\gamma$-rays,
Imaging Atmospheric Cherenkov Telescopes (IACTs) and Water Cherenkov
Arrays (WCAs),
were pioneered in the United States and have achieved
a state of high maturity. The U.S.-led Fermi satellite was
successfully launched in June 2008, but no follow-up
experiment is on the horizon. In view of the long lead times of new
initiatives, it is mandatory
to now start designing and constructing
a major ground-based $\gamma$-ray observatory.\\

{\bf (F5)} TeV $\gamma$-ray experiments are very broad in the scientific
problems that they can address, and therefore a substantial increase in their
sensitivity will provide answers to many different questions.
A next-generation experiment could improve on the sensitivity
by a factor of 5-10, and could make measurements of the $\gamma$-ray
sky in unprecedented detail.
Technical reasons make sensitive measurements at very low and very
large energies difficult,
i.e. expensive to achieve, and therefore the 30~GeV to 100~TeV energy range
appears the optimal energy band in which the advances in sensitivity
should be accomplished.
The Fermi satellite experiment will conduct very sensitive studies in
the energy band below 50~GeV,
and while overlap with Fermi is desirable, there is diminishing return in making
GeV $\gamma$-ray measurements with TeV $\gamma$-ray experimental techniques.\\

{\bf (F6)} The IACT and WCA techniques complement each other.
IACTs achieve unprecedented instantaneous sensitivity over a small
field of view, as well as excellent angular and energy resolution for
detailed studies of
cosmic objects. With a large field of view, WCAs can alert the
IACTs about the brightest transient phenomena. Furthermore, WCA arrays
achieve a high sensitivity for steady extended sources and at $>$10
TeV energies.\\

{\bf (F7)} VHE $\gamma$-ray astronomy was pioneered in the U.S., as was
the imaging technique
which has lead to the success of the current suite of experiments.
Due to a lack of sufficient
funding and an aggressive effort in other nations, this leadership
position is being challenged.
However, novel ideas and unique expertise still reside within the U.S.
(wide field-of-view optical systems, novel low-cost low-weight mirror 
technologies, advanced camera 
design and electronics, and intelligent array triggers) and with
sufficient funding the U.S. can regain its leadership position in this
area of research. In addition, the U.S. is still the leader in the 
WCA technique; however, other nations are now beginning to invest 
in this area and we must provide sufficient funding here to 
retain our leadership position.
\subsubsection{Recommendations}
{\bf (R1)} The IACT and WCA techniques have achieved a high state of maturity
that allows high-fidelity extrapolations in cost and performance. A
next-generation experiment
at an installation cost of \$120M could achieve a factor of 5-10
better sensitivity than
current experiments. This level of investment is warranted by the
guaranteed rich astrophysics
return and the exciting potential for more fundamental discoveries in
a number of key areas.
\\

{\bf (R2)} While U.S. groups pioneered ground-based $\gamma$-ray
astronomy, in the last few years
the position of the U.S. has been challenged as European funding agencies
were quicker to
recognize the potential of the field. To maintain a worldwide
leadership role, it is imperative that
appropriately funded R\&D and design studies for the next-generation
experiment start immediately. \\

{\bf (R3)} The space born $\gamma$-ray observatory Fermi is poised to
revolutionize the field of $\gamma$-ray astronomy.
However, owing to Fermi's limited angular and energy resolutions and
rather small collection area, many results will need
follow-up observations with a ground based experiment. The Large
Hadron Collider (LHC) might find the first evidence for
dark matter particles. The design of the next-generation $\gamma$-ray
experiment (especially the energy band for which
it will be optimized) will depend on the science results from Fermi
and the LHC.
Therefore, the decision on the final design of the experiment should 
be made two to three years from now.
The construction of the full experiment should start 
4 to 5 years from now.\\

{\bf (R4)} In parallel to work on technology R\&D, the U.S. groups
should work on
establishing a site on which a large-scale experiment can be built
during the coming decade. The site should allow for 
step-wise
enlargement; therefore, 
sufficient space and a long-term lease agreement
are mandatory. Procuring a site should be pursued as early as possible
to avoid the delays that affected VERITAS and the Heinrich-Hertz
SMT on Mt. Graham.
To maximize the science return of the experiment, a site should be
chosen that allows one 
to observe the Galactic Center.
\\

{\bf (R5)} The next generation ground based gamma-ray instrument should 
be an international project. The U.S. groups should continue and intensify the
collaboration with the European and Japanese groups. The U.S. groups have already 
formed joint scientific and technical working groups. This process should be continued.
The merits of distributed and largely independent experiments with telescopes deployed 
at two or three sites should carefully be compared with the merits of building a single
large experiment supported by a world wide collaboration.
\\

{\bf (R6)} To maximize the return of the investments, broader impact
strategies need to be considered along
with the development of the 
scientific and technical aspects of the
next-generation experiment.
In particular, the U.S. groups should take the lead in efforts to
incorporate broader impacts from
the beginning of the development phase. These efforts should include:
\begin{itemize}
\item Developing observatory and data access policies that encourage
full participation of the
astroparticle, astronomy, and particle physics communities. One
component of the experiment
should be a vigorous guest-investigator program and strong
multi-wavelength partnerships.
In contrast to existing P.I.-type instruments like VERITAS, Milagro,
or H.E.S.S., a next-generation detector should be
an observatory, so any scientist can apply for observing time and
receive support for analyzing the data.
The instrument teams should be charged with, and a budget allocated by
the funding agencies for,
the development and maintenance of the appropriate tools and support
systems for researchers outside
of the experiment collaboration.
\item Many of the most difficult challenges facing our nation in the
areas of nuclear non-proliferation,
nuclear terrorism, and the identification and reaction to conventional
terrorists attacks require technological
advances in the areas of ultra-fast low-light imaging systems and
event classification and response in real time
in the presence of an enormous data volume.  These needs are common to
the next generation of VHE instruments.
The community should work with the appropriate government agencies to
ensure that the technology developed
can be utilized to find solutions for these critical national needs.
\item Building the future generation of scientists and engineers
through involving
undergraduate and potentially high-aptitude high-school students
in all phases of development.
\item Partnering with science centers and planetaria to engage the
public in the exciting science
opened up by TeV $\gamma$-ray astronomy, and also to raise the
level of science appreciation within the general population.
\end{itemize}

\subsection{Ground based $\gamma$-ray astronomy - historical milestones}
Our atmosphere absorbs energetic $\gamma$-rays. However, at sufficiently high
$\gamma$-ray energies, it becomes possible to detect radiation from secondary particles 
produced by the primary $\gamma$-rays in the atmosphere with detectors stationed on the ground. 
In the following, a summary of the major discoveries made with ground based $\gamma$-ray 
experiments is given.

\smallskip\noindent
{\bf 1987:} Detection of the first cosmic source of TeV $\gamma$-rays, the Crab
Nebula (Whipple 10m).\\
The technique of imaging air showers produced by $\gamma$-rays in
Earth's atmosphere with fast pixelated cameras
permitted the classification of $\gamma$-ray like and cosmic ray-like
events. The suppression of cosmic ray initiated air 
showers allowed the Whipple collaboration to detect the Crab Nebula
in TeV $\gamma$ rays.
Powered by the strongly magnetized wind of the Crab pulsar, the Crab
Nebula now serves as the standard
candle of TeV astronomy.
Sixty-five cosmic sources of TeV $\gamma$-rays have
been observed to date.

\smallskip\noindent
{\bf 1992:} Detection of the first extragalactic source of TeV
$\gamma$-rays, the
active galactic nucleus Mrk 421 (Whipple 10m).\\
Active galactic nuclei harbor black holes of about a billion solar
masses that eject plasma
at approximately the speed of light. In these collimated plasma
outflows, called jets, particles are efficiently accelerated
and radiate a significant fraction of their energy in the form of
$\gamma$-rays.  Relativistic aberrations make active galactic nuclei that have a jet pointed towards the observer
prominent $\gamma$-ray sources.

\smallskip\noindent
{\bf 1992:} First constraints on the intensity of the Diffuse Extragalactic
Infrared Background based on energy spectrum of Mrk 421
(Whipple 10m).\\
TeV $\gamma$-rays produce electron-positron pairs in collisions with
extragalactic infrared and optical light. Observed as energy-dependent
extinction, this effect
permits a measurement of the intensity of the Extragalactic
Infrared Background which traces the star-formation history of the Universe.

\smallskip\noindent
{\bf 1996:} Discovery of extremely fast $\gamma$-ray flux variability from the
active galactic nucleus Mrk 421 (Whipple 10m).\\
Significant brightness fluctuations can only be produced in an
emission region of limited extent. Although the jets of 
active galactic nuclei are often larger than their host galaxies, the variable $\gamma$-ray
emission comes from a portion of the jet not bigger than the solar system.

\smallskip\noindent
{\bf 1996:} Discovery of Mrk 421 X-ray/TeV-gamma-ray flux correlation (ASTRO-E,
Whipple 10 m).\\
The correlations indicate that the processes leading to X-ray and TeV
$\gamma$-ray emission are
related. They give important clues to the nature of the radiating
particles and the radiation processes.

\smallskip\noindent
{\bf 1997:} Discovery of extreme flares of the active galactic nucleus
Mrk 501 with X-ray
emission up to 100 keV (BeppoSAX) and TeV $\gamma$-ray emission up to
16 TeV (HEGRA).\\
Active galactic nuclei not only change their brightness, but also their
spectrum. Correlated measurements
of these variations in X-rays and TeV $\gamma$-rays allow us to
observe the acceleration of the radiating
particles nearly in realtime.

\smallskip\noindent
{\bf 2003/2006:} Detection of $\gamma$-rays from the Radio Galaxy M87
(HEGRA), and
discovery of day-scale flux variability (H.E.S.S.).\\
M 87 is a relatively nearby active galactic nucleus whose jet is not
directed to us.
Measuring TeV $\gamma$-rays and fast $\gamma$-ray brightness
fluctuations give fundamental clues on
particle acceleration and radiation processes in an object that can be
spatially resolved in many wavebands,
thus revealing the geometrical structure, and for which relativistic
corrections are not as severe and difficult to estimate as in system, that have a jet pointed toward us.

\smallskip\noindent
{\bf 2003:} Discovery of the first unidentified TeV $\gamma$-ray
source TeV~J2032+4130 (HEGRA).\\
This source of TeV $\gamma$-rays is the first dark accelerator, of which a few
dozens have been found to date. The question of which objects are bright in 
high-energy $\gamma$-rays, but relatively dim in all other wavebands, 
is a fascinating one.

\smallskip\noindent
{\bf 2004:} Discovery of $\gamma$-rays from the Galactic Center
(CANGAROO, Whipple
10m, H.E.S.S.).\\
The Galactic Center is a region of particular interest because
$\gamma$-rays can be used to probe
the three-million solar-mass black hole residing there, the elusive
dark matter presumed to be
concentrated in that region, and many other systems. Spectral
measurements suggest that the bright
emission seen is not produced by dark matter.

\smallskip\noindent
{\bf 2004:} First spatially and spectroscopically resolved TeV
$\gamma$-ray  image of the Supernova remnant
RX J1713.7-3946 (H.E.S.S.).\\
The remnants of supernova explosions have long been suspected to
be the main sites of particle acceleration in the Galaxy. Advances in imaging now permit us
to measure TeV $\gamma$-ray energy spectra and their
changes across the remnants, thus deciphering the distribution of
radiating particles with unprecedented detail.

\smallskip\noindent
{\bf 2005:} Scan of the inner region of the Galactic plane reveals a large
population of sources, including Pulsar Wind Nebulae and a
considerable number of unidentified sources (H.E.S.S.).\\
A survey of the inner Galaxy proved what researchers have suspected
for many years: the Galaxy is filled with a variety of objects that accelerate particles to very high
energies and shine prominently
in TeV  $\gamma$-rays. The Galaxy is much more than the stars it contains.

\smallskip\noindent
{\bf 2005:} Discovery of the periodic emission from the X-ray binary LS 5039
(H.E.S.S.).\\
A compact companion, perhaps a black hole, is exposed to the strong
stellar 'wind' and the
intense light radiated by a massive blue star.
The interaction of the compact object with the stellar wind
accelerates particles. However, the star light can absorb the 
$\gamma$-rays produced by the high-energy particles, thus leading 
to a complex modulation pattern of the $\gamma$-ray emission.
This discovery opens the way to a better understanding of the dynamics
of such binary systems.

\smallskip\noindent
{\bf 2005:} First detection of super-TeV $\gamma$-rays from the
Galactic Plane (Milagro).\\
The Galaxy is permeated with energetic particles called cosmic rays,
whose origin is one
of the fundamental problems in modern physics. Measurements of a diffuse glow
of $\gamma$-rays that cosmic rays produce at very high energies give
invaluable insight
into the properties of cosmic rays far from the solar system.

\smallskip\noindent
{\bf 2006:} Discovery of diffuse TeV $\gamma$-ray emission from the
Galactic Center region (H.E.S.S.).\\
The bright diffuse emission is evidence for episodic states of high
activity in the Galactic Center region
during which extreme amounts of energy are transferred to energetic
particles, possibly by the massive black hole
that resides at the Galactic center. The separation of this diffuse
emission from compact sources of $\gamma$-rays
gives testimony of the imaging capabilities of modern Atmospheric
Cherenkov telescopes.

\smallskip\noindent
{\bf 2007/2008:} Discovery of 3 min flux variability from $10^9$ solar mass
black hole systems Mrk 501 (MAGIC) and PKS 2155-304 (H.E.S.S.).\\
The extremely short variability timescale indicates that the
$\gamma$-rays come from a region
not bigger than the black hole that is the central part of an active
galactic nucleus.

\smallskip\noindent
{\bf 2007:} Discovery of degree-size unidentified sources of super-TeV
$\gamma$-rays in the Galactic Plane (Milagro).\\
The $\gamma$-ray sky at energies above 10 TeV shows bright extended
features that are likely associated
with localized sources. Those features are too large to be caused by
the wind of a pulsar, and too powerful
to be identified with the remnant of a Supernova explosion, suggesting
they may be sites in which many Supernovae
exploded and pulsars were born.

\smallskip\noindent
{\bf 2008:} Discovery that TeV $\gamma$-rays from M 87 most likely
come from the compact
core rather than the outer jet (VERITAS).\\
Correlations between X-ray and TeV $\gamma$-ray emission from the
nearby active galactic nucleus M~87 indicate
that the powerful TeV $\gamma$-ray emission comes from the central
core around the supermassive black hole rather
than from various active regions further out in the jet. M~87 is one
of the very few active galactic nuclei
for which the jet can be resolved and directly imaged in high-energy emission.

\smallskip\noindent
{\bf 2008:} Discovery of pulsed $\gamma$-ray emission above 20 GeV
from the Crab Nebula (MAGIC).\\
40 years after their discovery, pulsars and their radiation mechanisms
are a very active field of research.
Precision measurements of the high-energy end of the spectrum of
pulsed $\gamma$-ray emission provide
fundamental insights into the structure of pulsar magnetospheres and
the main energy transfer processes at work.

\subsection{Scientific overview}
\subsubsection[Unveiling an important component of our Universe:
high-energy particles]
{High-Energy Particles}
The Universe is filled with energetic particles, electrons and
fully-ionized atoms, traveling
through space very close to the speed of light. Their origin is one of
the fundamental
unsolved problems in modern astrophysics. We know
that our Galaxy contains astrophysical systems capable of accelerating
particles to
energies beyond the reach of any accelerator built by humans.
Candidates for particle accelerators
are shocks formed in cosmic plasmas when a star explodes, when a rapidly
spinning neutron star expels electromagnetic energy, or when a black hole
spews out matter at nearly the speed of light.
What drives these accelerators is
a major question in physics and understanding these accelerators has
broad implications.

The charged energetic particles constitute a very tenuous medium; each
particle carries extreme energy, but the particles themselves are very
few. However, in our Galaxy high-energy particles carry on average as much energy per unit volume
as the gas, the magnetic field between stars, and as star light. The processes that determine their
energy and spatial distribution
are different from those that shape ordinary gases on Earth, because they rely
almost entirely on electric and magnetic fields.
Most gases on Earth are ``thermal'': the energy of the gas is
distributed approximately equally among
the atoms or molecules of the gas. In contrast, matter in the Universe
is often far from equilibrium.
In the dilute plasmas that fill most interstellar and intergalactic
space, nature chooses to endow a small
number of particles with an extreme amount of energy. We witness a
fundamental self-organization that,
through interactions between particles and electromagnetic fields,
arranges the atoms and available energy
in three components: a cool or warm gas that carries the bulk of the
mass, energetic particles with a wide range of
energies, and the turbulent electromagnetic fields that link the two.

Why does nature produce energetic particles? What is the fate of
the turbulent magnetic field? Do interactions of high-energy particles
generate the magnetic field
that permeates large structures in the Universe such as clusters of galaxies?

To address those questions, we need to measure the properties of
energetic particles in detail. The energetic particles can be observed
through high-energy $\gamma$-rays,
electromagnetic radiation with very high-energy.
Because $\gamma$-rays carry so much energy,
they must be produced by particles with even more energy, and in particular
TeV-$\gamma$-ray radiation is a key diagnostic of highly energetic particles.
By emitting TeV $\gamma$-ray emission,
the high-energy particles give us information about some of the most
extreme physical environments
and the most violent processes in the Universe that cannot be obtained
in any other way.

\subsubsection{Radiation processes and the sky in high-energy $\gamma$-rays}
The $\gamma$-ray sky is very different from the sky that we see with
our own eyes: the Sun is dark,
and a bright glow of $\gamma$-rays produced by cosmic rays (as energetic
particles in the Milky Way Galaxy are called) fills a large fraction
of the visible sky. Some structures associated
with the remnants of exploded stars cover a few square degrees of the night sky;
others may be the annihilation sites of the most mysterious particles
that make up dark matter.
Embedded in that extended, diffuse emission are compact or point-like sources,
some of which can be surprisingly variable: a large fraction of
compact TeV $\gamma$-ray sources
show bright flares on time scales of minutes to days. In most cases
these flares are powered by matter
falling into black holes.

\begin{figure*}[tb]
\centerline{\psfig{file=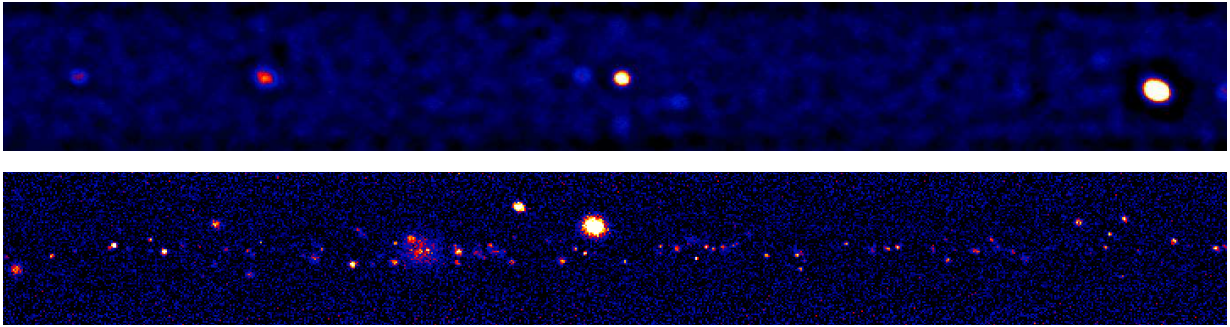,width=6.4in,clip=} }
\caption{Simulation of a sky survey conducted with a future
atmospheric Cherenkov telescope. The top
panel gives a map of the inner Galaxy as actually measured with
H.E.S.S. during its sky survey. The bottom panel
shows a simulated sky map that would be observed with a future
atmospheric Cherenkov telescope
ten times as sensitive as are H.E.S.S., MAGIC, or VERITAS.} \label{simu}
\end{figure*}

Energetic particles such as cosmic rays are ionized, comprising, as
individual particles,
atomic nuclei and the electrons that we would usually find in the
shells of atoms.
They belong to different families of particles: electrons, together
with the elusive neutrinos, are leptons, whereas the protons and
neutrons in atomic nuclei are hadrons.
The interactions and, in particular, the radiation processes of electrons
are different from those of nuclei because of their mass difference. 
Electrons radiate efficiently when they accelerate or decelerate quickly.
Interacting with dense plasmas, electrons emit Bremsstrahlung radiation; 
synchrotron radiation is emitted by
electrons that spiral around magnetic field lines; electrons can
transfer a large fraction of their
energy to photons in ``inverse Compton'' processes, when they scatter
off low energy photons.
We understand the characteristics of these radiation processes, and so
by measuring the radiation spectrum from a source we can infer the
energy distribution
of the radiating particles, provided the emission process is known.

From most astronomical objects synchrotron radiation of electrons is
typically observed from the radio band
up to X-rays. Inverse Compton emission generally extends from the
X-ray band up to very high-energy $\gamma$-rays.
Correlated measurements of X-ray and TeV $\gamma$-rays thus provide us
with two views of the same
radiating electrons, differing only by the process through which the
electrons radiate, thus providing a
measure of the number of radiating electrons and the strength of the
magnetic field and the radiation environment. A detector of
high-energy $\gamma$-rays should, therefore, have
the capability to measure the radiation spectrum with fine energy
resolution over a wide range of wavelengths.

Even though
the emission processes described above are inefficient for atomic
nuclei on account of their large mass, nuclei
can radiate through collisions with ordinary gas by creating
unstable particles, about a third of which are neutral pions ($\pi^0$)
that directly decay into two high-energy gamma
rays. The intensity of emission is proportional to the abundance of
interaction partners,
gas in the case of the radiating nuclei and infrared or other
radiation in the case of electrons
that undergo inverse Compton scattering. Very often we know the
distribution of gas from independent measurements,
or we know the strength of the ambient radiation because it is
dominated by the cosmological microwave background,
and so the spatial distribution of TeV $\gamma$-ray emission holds
clues to the nature of the radiating particles.
An excellent angular resolution is thus a key ingredient for a future
$\gamma$-ray experiment.

Many sources in the Universe emit at a constant level, but not all. If a source
significantly changes its brightness within a certain time,
the size of the emission region cannot exceed the distance traveled
by light in that time (A correction
factor of known character must be applied to this relation, if the
emission region moves with a
velocity close to the speed of light). Suppose a source flares 
for a minute: One would estimate it is smaller than about
20 million kilometers across, or about
one eighth of the distance between sun and Earth, yet the objects may
outshine an entire galaxy.
The current generation of TeV $\gamma$-ray
telescopes has detected flux variability from active galactic nuclei
down to timescales as short as about 2 minutes, and there has been no
lower limit
in the distribution of variability timescales.  Future telescopes
with 10-fold improved sensitivity will be capable of probing timescales less
than 10 seconds.  In studying such systems we are investigating the
most extreme and violent
conditions and processes in the Universe.

Brightness variations within a short time require that the radiating
particles be produced rapidly and that
they lose their energy swiftly, so the source can fade again. In most
cases the dominant means
of energy loss of energetic particles is radiation, which leads to a
characteristic time evolution
of the emission spectrum.
Variability, in particular when observed at different wavelengths,
therefore carries
important information about the size of and the physical conditions
in astronomical sources. A detector of high-energy $\gamma$-
rays should therefore have the sensitivity to detect sources within a
short time, so their brightness
variations can be followed.

Both the spectrum and the variability of high-energy emission
are shaped by the efficacy and the energy
dependence of the processes that accelerate the radiating electrons or
nucleons to high energies. These
particles are constantly deflected by fluctuating magnetic fields, but
magnetic fields alone can't
change their energy. What is required are fluctuating magnetic fields
or plasmas, dilute ionized gases
in which the magnetic field is embedded, that move relative to each
other. More than fifty years ago
the eminent Italian physicist Enrico Fermi classified processes of
that nature: If the motions of the
magnetic-field irregularities are random, we speak of 2nd-order Fermi,
or stochastic, acceleration. If the
motions of the magnetic-field irregularities change systematically and
abruptly, as in a shock front, the
associated acceleration process is called 1st-order Fermi, or shock,
acceleration.

Historically, scientists have often been too conservative in their
predictions of what nature can do.
A substantial increase in measurement capability has often led to the
discovery of new phenomena
or new classes of sources. Likewise for known variable sources, the
active phases have been
notoriously difficult to predict. To find the unexpected, or to seize
the opportunities offered by a
source awakening from dormancy, a detector should have a rather large
(5$^{\circ}$-10$^{\circ}$ diameter)
field of view.
\subsubsection{Diffuse emission and the nature and distribution of dark matter}
The study of diffuse Galactic $\gamma$-ray emission is important for a
number of reasons.
It provides direct information on the energy distribution of cosmic rays in various
locations in the Galaxy, which is needed to understand the origin and
interactions
of cosmic rays, in particular to separate the characteristics of the
production of cosmic rays
from those of their propagation.
Also, this emission must be understood to properly analyze extended
$\gamma$-ray sources
and to derive self-consistent limits on the amount of dark matter in the Galaxy.

Physicists have assumed for a long time that the matter surrounding
us, atoms made of
nuclei and electrons, is representative of all the matter in the Universe.
In the 1930's, Fritz Zwicky found that additional dark matter
beyond the luminous matter is required to explain the existence of galaxies.
In the 1970's, Vera Rubin measured the motion of spiral galaxies like our
Milky Way and also found strong evidence that non-luminous
matter holds the galaxies together. The picture that has emerged is
that we live
in an highly non-representative concentrate of ordinary matter that accumulated
at the center of massive structures of dark matter, so-called dark-matter halos.
Although the Universe contains
five times as much dark matter as normal matter, we do not yet know
what dark matter is.

Ground based $\gamma$-ray observations promise to lead to the
detection of annihilation $\gamma$-rays from
accumulations of dark matter particles at the center of dwarf galaxies,
at the center of the Milky Way, or in so-called mini-halos that
populate the Milky Way.
Recent results of particle physics suggest that the very early Universe
was filled with massive particles, only the lightest of which survived
to constitute dark matter.
Those particles would interact with other particles only very rarely,
but on those rare occasions
would produce high-energy $\gamma$-rays that can be observed with the
next generation of $\gamma$-ray
detectors. Depending on the mass of the dark matter particles and the
particulars of their decay or annihilation, 
a $\gamma$-ray signal is expected in either the GeV band, to be
observed with the Fermi telescope, or beyond 30~GeV,
where the next generation of ground-based $\gamma$-ray instruments can
detect them.
The $\gamma$-ray measurements could reveal the total mass of 
dark matter and 
its distribution in the
Milky Way and other galaxies, and thus give unique information about
the nature of
dark matter that is complementary to the results from laboratory experiments.

\subsubsection[Powerful particle accelerators in our Milky Way Galaxy:
supernova remnants, pulsars, and
stellar mass black holes]{Powerful particle accelerators in our Milky
Way Galaxy:\\supernova remnants,
pulsars, and stellar-mass black holes}
It appears that efficient acceleration of cosmic rays proceeds in
systems with outflow phenomena,
in which a fraction of the 
energy can be transferred to cosmic
rays. Some of those
systems are shell-type supernova remnants (SNR), in which material
from the exploded star slams
into the ambient gas,
forming a shock front. In fact, SNRs have
long been suspected
as production sites of Galactic cosmic rays because of their 
total energy output and 
rate.
SNRs in the Galaxy change slowly, so we can compare their appearance in
high-energy $\gamma$-rays to that in other wavebands such as X-rays,
which allows us to determine
their spatial structure. The remnants are large enough that they can
be resolved in
gamma rays, so we have an opportunity to perform spatially-resolved
studies in systems
with known geometry. The question of cosmic-ray acceleration in SNRs
includes aspects of
the generation, interaction, and damping of magnetic turbulence in
non-equilibrium plasmas.
The physics of the coupled system of turbulence, energetic particles,
and colliding plasma flows
can best be studied in young SNRs, for which X-ray and $\gamma$-ray
observations indicate very efficient
particle acceleration and the existence of a strong turbulent magnetic field.
The amplification of magnetic fields in shocks is of particular
interest because it may play an
important role in the generation of magnetic fields in the Universe.

Compact objects in the Galaxy can also accelerate particles to very
high energies.
Among the remnants of massive stars are pulsars, highly magnetized
remnants of stars crushed
to densities greater than atomic nuclei. Pulsar masses are
approximately equal to that of the Sun,
but their diameters are typically on the order of 10 miles. They are
rapidly spinning and
produce a beam of radiation like a lighthouse. At the same time, they
emit a wind of
relativistic particles that moves almost at the speed of light. Gamma
rays can be very
efficiently produced in those winds, and the entire system is
generally referred to as
a pulsar-wind nebula (PWN). They provide unique laboratories for the study of
relativistic shocks because the properties of the pulsar wind are
constrained by our
knowledge of the pulsar and
because the details of the interaction of the relativistic wind can be
imaged in the X-ray,
optical, and radio bands. Relativistic shock acceleration may be key to many
astrophysical sources, such as active galactic nuclei and Gamma-Ray Bursts.
PWNe are, perhaps, the best laboratory to understand the detailed
dynamics of such shocks.

When a stellar mass black hole or a pulsar is bound to a companion
star, it seems to be able to form
tightly collimated plasma beams or jets; TeV $\gamma$-rays have been
observed from
those jets. In fact, TeV emission provides a unique probe of the
highest-energy particles
in a jet, allowing us to address key questions: Are jets made up of
normal ions, or a mixture of
matter and antimatter? What is the total energy carried by jets? What
accelerates particles in jets?
Energetic particles often dominate the energy budget of the jet and
the accurate measurement of their
spectrum and acceleration time is essential for addressing these
questions, which, in turn, are
fundamental to our understanding of the physics of jets and their formation.

\subsubsection[Extragalactic sources of TeV
$\gamma$-rays]{Extragalactic sources of\\TeV $\gamma$-rays}

The Big Bang resulted in a remarkably homogeneous Universe.
For the last 13.7 billion years, the history of the Universe has
been one of matter clumping together under the influence of gravity.
We now think that small clumps formed first and made stars.
Later, larger clumps formed, resulting first in galaxies made of
billions of stars,
and subsequently in galaxy clusters, comprising up to several thousand
galaxies.
As these large structures grow and draw in matter, large shocks form in which
incoming material is heated and cosmic rays are accelerated.
Large structures like galaxy clusters thus consist of several components:
a dark matter halo that holds the cluster together, the individual
galaxies visible
with optical telescopes, hot plasma that radiates X-rays, and finally
cosmic rays that are expected
to carry a substantial fraction of the available energy.
This energetically important component has evaded detection so far.
A next-generation, ground-based $\gamma$-ray observatory has an excellent
chance to
discover $\gamma$-ray emission from this component and to deliver
detailed information
about its spatial and spectral properties.
Such an experiment could thus make a substantial
contribution to our understanding of the energy and pressure composition
of the plasma in the largest structures in the Universe.

High-energy particles, or cosmic rays, 
also play an important role in
other galaxies.
Massive stars produce strong plasma winds and, towards the end of their life,
spectacular supernova explosions that violently expel plasma into space.
The stellar winds and the
supernova outflows are both thought to be efficient cosmic ray accelerators.
A next-generation $\gamma$-ray observatory will allow us to detect
$\gamma$-rays
from these cosmic rays in a large number of nearby galaxies, and
to study the relationship between star formation and cosmic ray
acceleration in very different galactic environments.
The $\gamma$-ray studies will thus revolutionize our understanding
of the role of stellar feedback in the formation and evolution of
stars and plasma in galaxies.

The deaths of some stars are thought to be responsible for some of the
most violent explosions in
the Universe, Gamma Ray Bursts. These events may lead to the
acceleration of the highest-energy
cosmic rays through multiple shocks driven at highly relativistic
speeds in their collimated plasma outflows (jets),
and they are thought to produce significant TeV $\gamma$-ray emission,
which has eluded detection thus far. By detecting this emission from
gamma-ray bursts and measuring its properties, we would make
great strides towards understanding the extreme nature and environments
of $\gamma$-ray bursts, particularly the local opacity and the bulk
Lorentz factor. It
could also contribute to our understanding of the decades-old problem
of the origin of
ultra-high-energy cosmic rays, as well as permit Lorentz-invariance
violation studies, $\gamma$-ray burst progenitor studies, and thus star
formation history studies. An observation of
violations of Lorentz invariance would be a major step towards a
quantum theory of gravity, which is
the only fundamental interaction in nature for which a quantum
description has not been successfully
formulated to date.

We know that other galaxies harbor supermassive black holes with a
mass between a million and a
few billion solar masses that are part of what astronomers call active
galactic nuclei (AGN).
These black holes offer physicists a unique opportunity to test
Einstein's theory of the
nature of space, time, and the gravitational force. Recent radio and
X-ray observations indicate
that black holes may play an important role in galaxies and galaxy
clusters by regulating the rate
of star formation. TeV $\gamma$-ray astronomy affords the possibility
to study the environment
of supermassive black holes, and the processes by which the black
holes grow.\\ 

The jets from supermassive black holes are laboratories to study turbulence and
particle acceleration in the most extreme setting; the flow velocity
is much higher
than in supernova remnants, and the energy content vastly exceeds that of
galactic compact objects. The main questions astrophysicists ask are
similar to those relevant for jets of
galactic solar-mass black holes, but the parameters are different.
Studying the same issue with
galactic black holes and with AGN can be likened to probing the same
physical behavior with two
laboratory experiments that use different techniques, thus offering
complementary views that give
a better and more complete picture. Better measurements of the
spectrum of the highest-energy particles
in a jet and its rapid changes, using more refined TeV $\gamma$-ray
observations and X-ray studies in parallel,
are by far the best approach to addressing
how nature organizes energy and matter in these most violent conditions.

TeV $\gamma$-rays from extragalactic sources also carry information
about the infrared light between
galaxies: High-energy $\gamma$-rays can be absorbed upon collision
with infrared and optical light,
thus modifying the $\gamma$-ray spectra of sources in a way that has a
characteristic dependence
on their distance. The absorption also depends on the total intensity
of optical and infrared light
ever emitted by stars. In this way, TeV $\gamma$-ray observations
constrain the early history
of star and galaxy formation in the Universe.

\subsection{Technology and the path toward a future observatory}
High-energy $\gamma$-rays can be observed from the ground by either
imaging the Cherenkov
light produced by the secondary particles
once the $\gamma$-ray interacts high in the atmosphere or, using
Extended Air Shower (EAS) arrays, by directly detecting
the shower particles (electrons, muons and photons) that reach the ground.
The former method employs imaging atmospheric Cherenkov telescopes (IACTs).
Modern IACT experiments like VERITAS, MAGIC, and H.E.S.S. detect point
sources with a
TeV $\gamma$-ray flux of 1\% of the flux from the Crab Nebula.
EAS arrays such as Milagro have complementary capabilities to IACTs.
While their instantaneous sensitivity is currently a factor of
$\sim$150 lower than that
of IACTs, their field of view is over 200 times larger and their duty
factor is close to 100\% as compared to 10\%
for IACTs. EAS observatories are, therefore,  suited
to performing unbiased surveys to search for new or transient sources.

It is possible to improve the sensitivity of both techniques by
another order of magnitude
at a total cost only one order of magnitude higher than that of the
present instruments; that is,
installation costs of the order of \$~100M.
At the core of designing a next-generation ground-based $\gamma$-ray
detector is the requirement to improve the integral flux sensitivity
in the 50 GeV to 50 TeV
regime where the techniques are proven to give excellent performance.
At lower energies (below 50 GeV) and at much higher energies (50-200
TeV) there is
great discovery potential, but new technical approaches must be explored.
For particle-detector (EAS) arrays, the technical roadmap is relatively
well-defined. Simulations
indicate that by moving to a higher altitude, enlarging the detection
area, and optically
isolating the detector modules, the proposed High Altitude Water
Cherenkov (HAWC)
experiment can achieve a sensitivity factor of 10-15 better than
that of Milagro.
The joint US-Mexico collaboration estimates the total cost of HAWC below \$10M.

In considering the design of future IACT arrays, the development is
likely to follow several different
(although complementary) branches, with
the aim of covering a broad energy range from 10~GeV up to 100~TeV.
Achieving an order of magnitude sensitivity improvement in the 200 GeV
to 10 TeV regime
will require an experiment with an effective area of $\sim$1~km$^2$ and
a large number ($\sim$50) of telescopes. The design, construction, and
operation of a
large-scale ground based $\gamma$-ray experiment also brings
new challenges: Efficiently mass-producing telescopes,
simplifying the process of checking out and calibrating the telescopes, and
minimizing the maintenance required to keep the
telescopes fully operational. The design of a future IACT array can be
based on the
well-studied performance of the existing VERITAS, MAGIC, and H.E.S.S.\
instruments, for which preliminary
studies indicate that the sensitivity improves faster than the square
root of the number of
telescopes, as would be expected for a large number of telescopes
operated as independent detectors.
Various technological advances substantially reduce the cost
per telescope; e.g., the availability of high-quantum-efficiency
photodetectors, the development
of fast, integrated, application-specific integrated circuits (ASICs)
for the front-end electronics, the use of optimized mechanical 
and optical designs, and the development of novel mirror technology. 
Also, the costs of the current experiments were largely driven by the 
one-time engineering that will form a lower percentage of the project cost
for a larger experiment.

Such a next-generation IACT instrument could be designed and built on
a time scale of $\sim$5 years.
We would recommend a 3-year R\&D program to provide a better
understanding of the design options, cost uncertainties and reliability. 
This time scale also makes it possible to adapt the design of the experiment 
to new science opportunities opened up by discoveries of the Fermi and LHC experiments. 
We also encourage the community to
pursue technologies which could have a large impact on cost, operation,
and scientific capability.  The R\&D program should include the
following design studies:
\begin{itemize}
\item Based on Monte Carlo simulations, the strengths and weaknesses
of different array configurations
need to be fully explored. Two specific issues that should be studied
further are (i)
the impact of the pixel size on the energy threshold, background-rejection
capability and angular resolution of the telescopes, and (ii)
the impact of the altitude at which the instrument is operated.
\item The development and evaluation of different camera options, with
emphasis on
achieving a higher quantum efficiency of the photodetectors, and a
modular design of the
camera to reduce assembly and maintenance costs.
\item The development of ASIC-based front-end-electronics to further
minimize the power
and price of the readout per pixel.
\item A next-generation experiment should offer the flexibility to
operate in different configurations, so that specific
telescope combinations can be used to achieve certain science
objectives. Such a system requires the development of
a flexible trigger system. Furthermore, the R\&D should explore
possibilities to combine the trigger signals of
closely spaced telescopes to synthesize a single telescope of larger aperture.
\item The telescope design has to be optimized to allow for mass
production and to minimize the maintenance costs.
\item The telescopes should largely run in robotic operation mode to
enable a small crew to operate
the entire system.
\end{itemize}
The R\&D should coincide with the establishment of a suitable
experimental site and the build-up of basic
infrastructure.
Ideally, the site should offer an easily accessible area exceeding 1 km$^2$.
For an IACT array, an altitude between 2 km and 3.5 km will give the
best tradeoff between low energy thresholds,
excellent high-energy sensitivity, and ease of construction and
operation. EAS arrays should be located at higher
altitudes to facilitate the direct detection of shower particles.

Beyond the immediate future, alternative optical designs should be
explored in greater detail.
Such designs have the potential
to combine excellent off-axis point-spread functions, large
field-of-views, and isochronicity
with significantly reduced camera size. Key issues that need to be
addressed are the cost and
reliability of suitable mirror elements, the procedure of adjusting the mirrors,
and the price increase arising from the required mechanical precision
and stability of the
support structure, the more complex mirror assembly, and
primary/secondary obscuration. The reduced camera size
would permit using integrated photodetectors such as multi-channel plates and
Geiger-mode Si detectors, that are independently developed by the
industry. The superior performance, low price,
and extreme reliability of both alternative optics and integrated
photodetectors must be demonstrated
in the next years, before these technologies can form the design
baseline of a future IACT array.

The U.S. teams have pioneered the field of ground based $\gamma$-ray
astronomy during the last 50 years. The U.S. community has formed the
AGIS collaboration
(Advanced Gamma ray Imaging System) to  optimize the design of a
future $\gamma$-ray detector.
A similar effort is currently under consideration in Europe by the
CTA (Cherenkov Telescope Array) group, and the
Japanese/Australian groups building CANGAROO are also exploring
avenues for future progress.
Given the scope of a next-generation experiment, the close
collaboration of the US teams with the European and
Japanese/Australian groups should be continued and intensified. If
funded appropriately, the US teams are in
an excellent position to lead the field to new heights.

\subsection{Synergies with other wavebands and particle astronomy missions}

TeV $\gamma$-rays are the high energy cousins to photons at lower
energies. The closest in energy
are  GeV photons  and are detected with satellite telescopes such as
Fermi and AGILE. Together,
the ground-based and satellite detectors span 6 orders of magnitude in
energy  (0.1~GeV to 100~TeV)
for probing particle acceleration and emission processes in cosmic
accelerators, allowing one to apply
the most rigorous tests of theoretical models .

Another relative in the family of fundamental particles is the
neutrino.  Measurable neutrino fluxes
are expected  to accompany
gamma-ray emission when observing astrophysical objects that harbor an
acceleration site of
cosmic-ray protons and nuclei. The IceCube detector at the south pole
and soon the Antares experiment in the
Mediterranean sea will provide information about the hadronic
components in cosmic accelerators. Neutrino telescopes,
together with the ground-based TeV $\gamma$-ray telescopes, could
trace, identify and carefully inspect potential sites
where the highest energy cosmic rays (UHE short for Ultra High Energy)
have their origin. For example, AGNs were
identified as TeV $\gamma$-ray emitters, providing a possible
connection to UHE cosmic ray acceleration.

The direct identification of the sources of cosmic rays is being
pursued with the AUGER experiment, which is the
largest air shower array capable of detecting weak fluxes of UHE
cosmic rays.  In fact, evidence for the correlation
of the arrival direction of UHE cosmic rays with AGNs was reported
recently.  These findings give even more urgency
to searching for these cosmic Zevatrons, identifying their nature and
understanding their production mechanism.  The use
of different messenger particles such as Neutrinos (IceCube), GeV to
TeV $\gamma$-rays (satellite and ground-based $\gamma$-ray
detectors) and UHE cosmic rays  (AUGER) are indispensable in
understanding the origin of the highest energy radiations
in the Universe.

The leptonic component is becoming visible in photons via synchrotron
radiation and inverse Compton scattering and is also
a big contributor to high energy radiation, sometimes considered an
unwelcome background for understanding
cosmic-ray sources. Identifying and understanding its role in
different types of
cosmic accelerators  requires the collaboration of radio, optical,
X-ray  telescopes and $\gamma$-ray telescopes.
This is also essential in separating out the leptonic and  hadronic
cosmic ray production in astrophysical objects.

TeV $\gamma$-ray instruments have a key role as they provide an
important link between X-ray telescopes (Chandra, Swift, RXTE,
BeppoSAX, Suzaku, etc.) and cosmic-ray and neutrino telescopes. TeV
telescopes bridge the energy gap between the lower
energy photon emissions and the highest energy cosmic rays, and are
sensitive to radiation  of leptonic
and hadronic origin, thus holding a key to understanding the energy
budget in different types of cosmic accelerators.

\clearpage

\section{Galactic diffuse emission, supernova remnants, and the origin of cosmic rays}
\label{GDE-subsec}
Group membership:\\ \\
A. Abdo, A. Atoyan, M. Baring, R. Blandford, Y. Butt, D. Ellison, S. Funk, F. Halzen, E. Hays, B. Humensky, T. Jones, P. Kaaret,
D.Kieda, S. LeBohec, P. Meszaros, I. Moskalenko, M. Pohl,
P. Slane, A. Strong, S. Wakely\\ \\ Independent Reviewers: \\ \\John Beacom, Andrei Bykov



\subsection{Why are they important?}
The origin of Galactic cosmic rays and the mechanisms of their acceleration are
among the most challenging problems in astroparticle physics and also
among the oldest. Cosmic rays are energetically important in our
understanding of the interstellar medium (ISM) because they contain at
least as much energy as the other phases of the ISM.  They also provide, along with
interstellar dust, the only sample of ordinary matter from outside the heliosphere. Yet, the origin
of cosmic rays in the Galaxy remains uncertain more than 90 years after their discovery by
Victor Hess in 1912 (for a recent review, see \cite{hillas05}). 
Improving our knowledge of the interaction between highly
energetic particles and the other elements of the ISM could help
understand other systems, such as active galactic
nuclei (AGN) that produce
strong outflows with highly energetic particles.

High-energy gamma rays are a unique probe of cosmic rays. Observations in the TeV band 
are a sensitive probe of the highest energy physical processes occurring in a 
variety of astronomical objects, and they allow us to measure the properties of energetic 
particles anywhere in the Universe, such as their number, composition, and 
spectrum. From such measurements we know already
that our Galaxy contains astrophysical systems capable of accelerating particles to
energies beyond the reach of any accelerator built by humans. What drives these
accelerators is a major question in physics and understanding these accelerators has broad 
implications, but more sensitive gamma-ray detectors are needed to address these questions.
Among the many types of Galactic
gamma-ray sources, observations of high-energy emission from shell-type 
supernova remnants (SNR) are particularly beneficial because:

\begin{itemize}
\item The acceleration of relativistic charged particles is one of the main unsolved,
yet fundamental, problems in modern astrophysics. Only in the case of SNRs do we have an 
opportunity to perform
spatially resolved studies in systems with known geometry, and the plasma physics 
deduced from these observations will help us to understand other systems where rapid particle 
acceleration is believed to occur and where observations as detailed as those of SNRs are 
not possible.

\item The acceleration of particles relies on interactions between
energetic particles and magnetic turbulence, so the question of cosmic-ray acceleration is, in fact,
one of the generation, interaction, and damping of turbulence in a non-equilibrium plasma. 
The physics of the coupled system of turbulence, energetic particles, and colliding plasma flows 
can be ideally studied in young SNRs, for which observations in X-rays 
\cite{Koyama95} and TeV-scale gamma rays \cite{aha04} indicate a
very efficient particle acceleration to at least 100 TeV and the existence of a
turbulent magnetic field that is much stronger than a typical shock-compressed
interstellar magnetic field. The amplification of magnetic fields by streaming
energetic particles is of particular
interest because it may play an important role in the generation of cosmological magnetic fields. 
\item SNR are the most likely candidate for the sources of cosmic rays, either
as isolated systems or acting collectively in groups in so-called superbubbles, although to date
we do not have conclusive evidence that they produce cosmic-ray ions in addition to
electrons. An understanding of
particle acceleration in SNR may solve the century-old question of the
origin of cosmic rays.
\item SNR are a major source of heat and turbulence in the interstellar medium of galaxies, 
and thus have an impact on the evolution of the galactic ecosystems. 
In particular, when new insights are 
extended to shocks from other sources, e.g. the winds of massive stars, they will help in
advancing our understanding of the energy balance and evolution of the interstellar medium in
galaxies. 
\item The evolution and interaction of turbulence and cosmic rays determines how the cosmic rays
will eventually be released by the SNR, which has an impact on the amplitude and frequency
of variations of the cosmic-ray flux near Earth and at other locations in the Galaxy
\cite{pe98}.  
\end{itemize}

The study of diffuse Galactic gamma-ray emission is important for a number of reasons.

\begin{itemize}
\item It provides direct information on the cosmic-ray spectrum in various 
locations in the Galaxy, which is needed to understand the origin of cosmic rays 
near and beyond the knee.

\item It must be understood to properly analyze extended gamma-ray sources,
in particular in terms of possible spatial variations of its
spectrum resulting from non-stationary cosmic ray transport. 

\item It will enable us to analyze the gamma-ray spectra of supernova remnants
self-consistently in the 
light of their function as possible sources of Galactic cosmic rays.

\item It allows us to derive self-consistent limits on the amount of
dark matter in the Galaxy by determining both the cosmic ray propagation and 
radiation properties
and the gamma-ray emissivities of dark matter for a variety of spatial distributions.

\end{itemize}

\subsection{What do we know already?}

\medskip
\subsubsection{Supernova remnants}

Cosmic rays consist of both electrons and hadrons. However, the hadrons
dominate the energy budget, and the acceleration of hadrons is the key
issue in understanding the origin of cosmic rays. The energy density in
local cosmic rays, when extrapolated to the whole Galaxy, implies the
existence of powerful accelerators in the Galaxy.  Supernova remnants
(SNRs) have long been thought to be those accelerators
\cite{ginzburg64}, but there is no definitive proof that hadrons are
accelerated in SNRs. The classical argument that shocks in shell-type
SNRs accelerate cosmic rays is that supernova explosions are
one of the few Galactic phenomena capable of satisfying the energy
budget of cosmic-ray nuclei, although even supernovae must have a high
efficiency ($\sim$10\%-20\%) for converting the kinetic energy of the
SNR explosions to particles \cite{drury89}. However, these arguments are indirect. 
Other source
classes may exist that have not been considered to date, and one may ask
what role is played by the many sources seen in the TeV-band that do not have 
an obvious counterpart in other wavebands \cite{hess-survey}. In any case,
observations of TeV photons from SNRs offer the most promising direct way to confirm 
whether or not SNRs are, in fact, the main sources of CR ions below
$10^{15}$~eV. 

\begin{figure*}[tb]
\centerline{\psfig{file=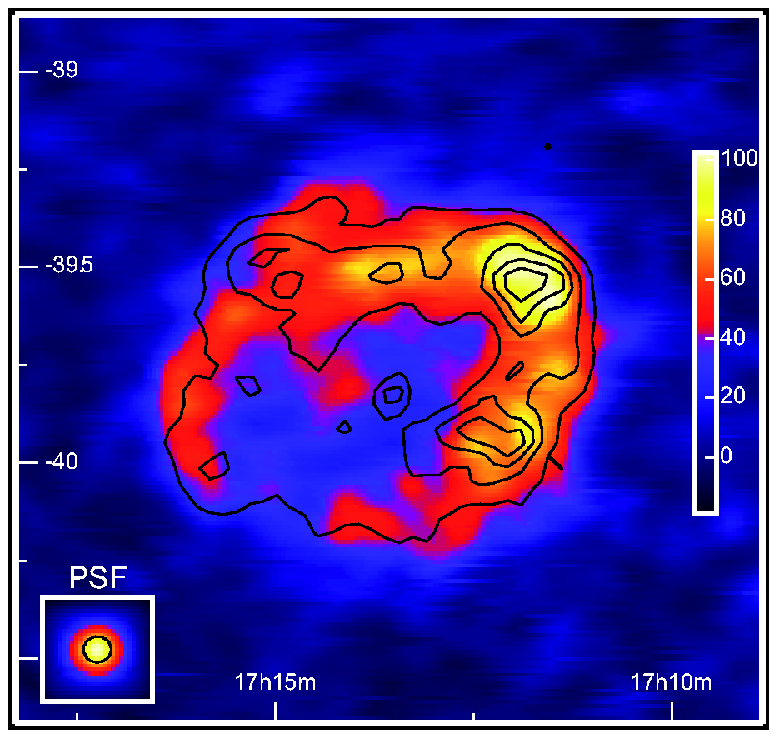,width=2.8in,clip=} \hspace{0.2in}
\psfig{file=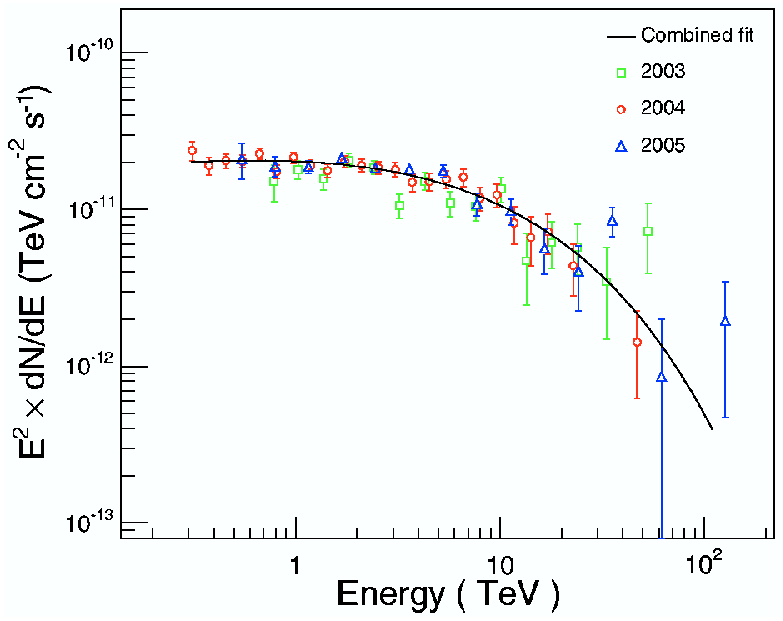,width=3.3in,clip=}} 
\caption{The left panel shows an image of the acceptance-corrected
gamma-ray excess rate in the TeV band as observed with H.E.S.S. from the SNR
RX~J1713-3946 \cite{hess-rx}.
The insert labeled PSF indicates how a point source would appear in this image. Overlaid are
black contour lines that indicate the X-ray intensity at 1-3 keV. Note the similarity between the
X-ray and TeV-band images. The right panel shows the TeV-band spectrum for the entire remnant
broken down for three different observing seasons.} \label{ic443} 
\end{figure*}

Even though very early measurements showed that the fluxes of TeV emission from SNRs
are lower than originally predicted if SNRs really do accelerate the
bulk of Galactic cosmic rays \cite{buckley98}, later observations with H.E.S.S.
established shell-type SNRs such as RX~J1713-3946 \cite{hess-rx} and
RX~J0852.0-4622 \cite{hess-vj} as TeV-band gamma-ray sources. The maturity of high-energy
gamma-ray astrophysics is best illustrated by the ability of current atmospheric Cherenkov detectors
such as H.E.S.S., MAGIC, and VERITAS
to resolve sources and to map the brightness distribution in TeV-band gamma rays. 
Figure~\ref{ic443} shows such a gamma-ray map and the TeV-band spectrum of RX~J1713-3946.
The interpretation of these TeV observations is complicated because two competing 
radiation processes, pion-decay photons from ion-ion interactions and 
Inverse-Compton
(IC) emission from TeV electrons scattering off the cosmic microwave background and 
the ambient galactic radiation, 
can produce similar fluxes in the GeV-TeV energy range. 
In the hadronic scenario neutrinos
would be produced through the decay of charged pions. If even a few
neutrinos are detected from a source at high enough energies, where the atmospheric neutrino background is minimal, then this alone
could decisively indicate the hadronic mechanism \cite{kis+bea}. 

\begin{figure*}[tb] \centerline{\psfig{file=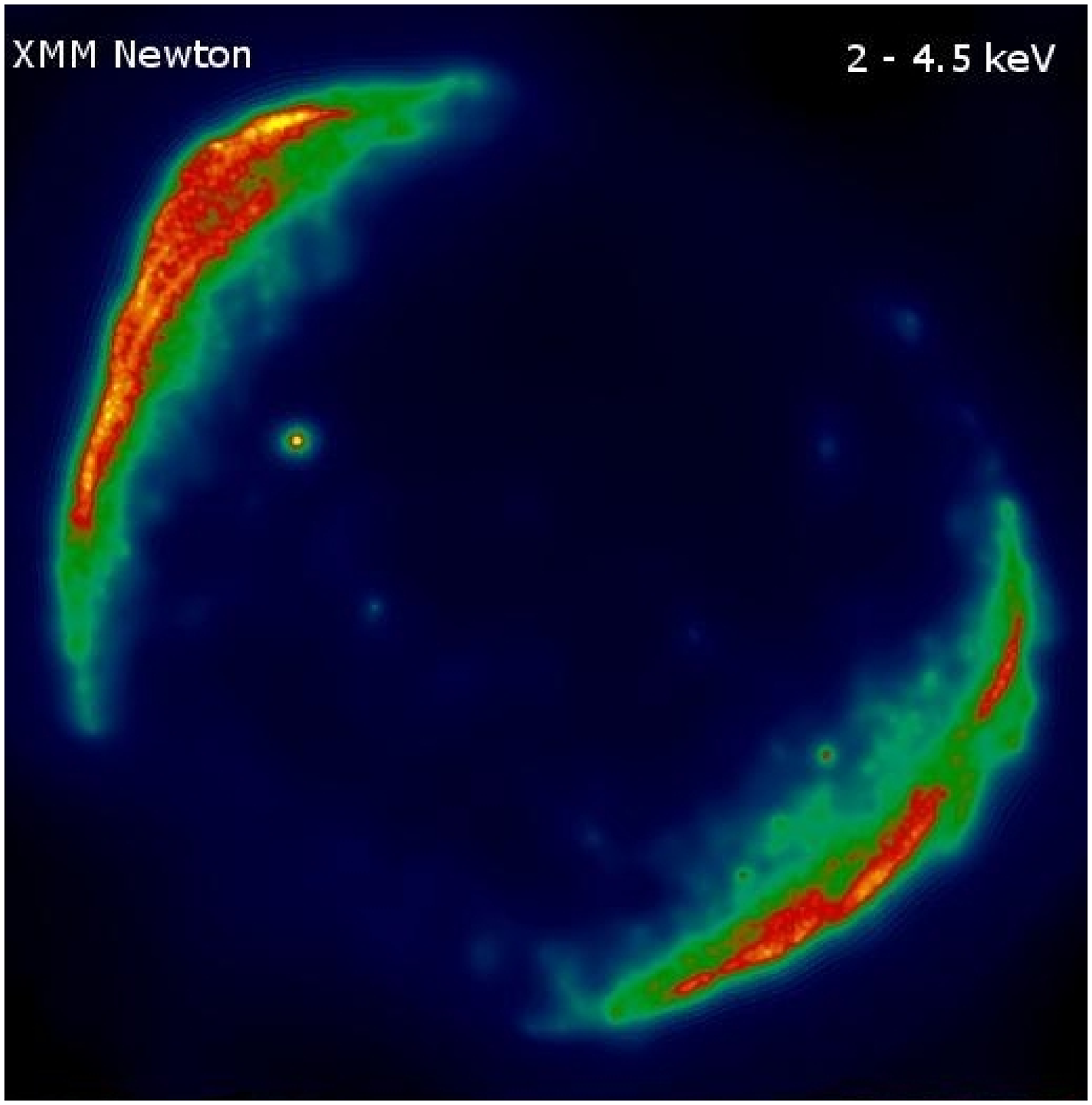,width=3.0in,clip=} ~
\psfig{file=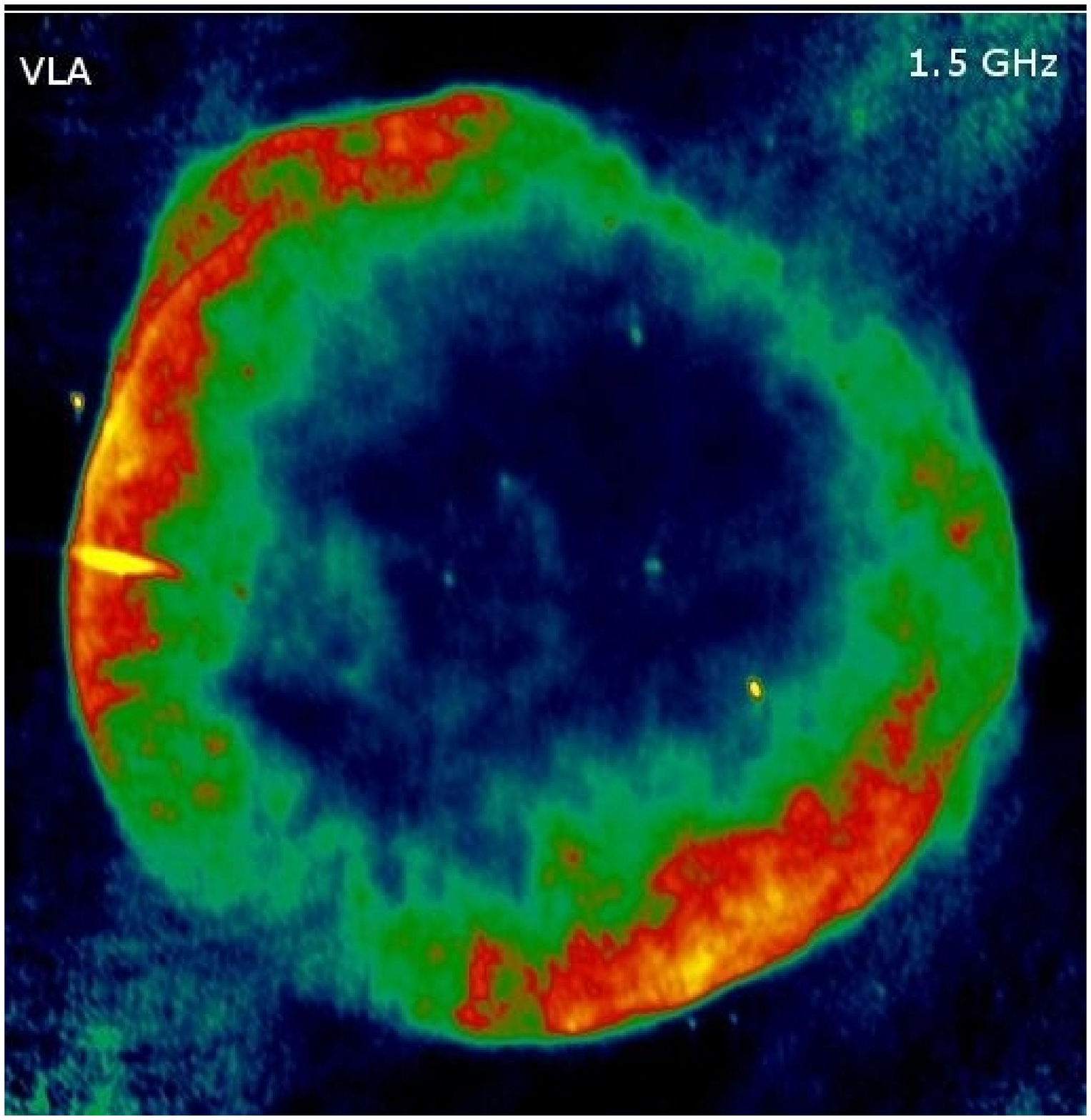,width=3.in,clip=}} \caption{X-ray and radio
images of SN 1006 \cite{rothenflug04}. Hard X-rays (left) are mainly
produced by very high-energy electrons ($\sim 100$~TeV) emitting
synchrotron radiation.  Radio emission (right) is produced by electrons
with energies in the GeV range emitting synchrotron radiation.  Imaging
TeV observations will enable us to map the inverse-Compton emission
from high-energy electrons and make a measurement of the magnetic field
strength in the vicinity of the shock.  Such mapping is also essential
for distinguishing TeV photons produced by electronic versus hadronic
cosmic rays.  The angular size of the image is 35~arcmin.  (Image
courtesy of CEA/DSM/DAPNIA/SAp and ESA.)} \label{sn1006} \end{figure*}

SNRs do accelerate {\it electrons}. As has long been known from radio observations, 
GeV-scale electrons are accelerated in SNRs, and now compelling evidence
for acceleration of electrons at the forward shocks of SNRs comes from observations of
non-thermal X-rays from several shell-type SNRs.  The X-ray
emission is synchrotron radiation from electrons accelerated to TeV
energies. In the case of SN 1006, the electrons must have energies of
at least 100~TeV \cite{Koyama95}, see Fig.~\ref{sn1006}.  These
electrons must be accelerated in situ because such energetic electrons
cannot travel far from their origin before they are attenuated by energy losses
due to synchrotron radiation.  The same electrons should produce TeV emission via
inverse-Compton scattering. The intensity and spectrum of the emission
are determined by the electron density, maximum electron energy, and
local magnetic field. Combining radio, X-ray, and TeV data can provide a
measurement of the magnetic field strength in the vicinity of the
shock.  This important parameter is not provided by X-ray
spectroscopy alone, because the photon cut-off energy is insensitive to the
magnetic field strength if it is generated by the competition of
strong synchrotron cooling and gyroresonant acceleration of
electrons. 

An important clue to the nature of the parent particles comes
from correlation studies with X-rays in the 2--10~keV band. For the
two prominent SNRs, RX\,J1713.7-3946 and
RX\,J0852.0-4622, one finds a spatial correlation down to angular scales of
$\sim 0.1^\circ$, between the
X-ray emission and the TeV-band gamma-ray emission, with correlation factors
in excess of 70\%. This correlation suggests a common emission origin. The non-thermal 
X-ray emission is known to have structure on scales $\lesssim 0.01^\circ$, 
and it is the limited angular resolution and sensitivity of the current TeV observatories
that prevents a correlation analysis on the physically more relevant small scales.
Nevertheless, if the TeV gamma-ray
emission was of leptonic origin as suggested by the spatial
correlation, the spectra in X-rays and 
gamma rays should also be similar.
As hadronic gamma-ray production requires interaction
of the cosmic-ray nucleons with target nuclei, this emission will be
stronger for those SNRs located near or interacting with dense gas,
such as molecular clouds.  The TeV emission should be brightest in
those regions of the SNRs where the target density is highest.
 
In situ observations in the heliosphere show that collisionless shocks can accelerate particles.
The process of particle acceleration at SNR shocks is intrinsically efficient
\cite{kj05}. Thus, the shocks should be strongly modified,
because the energetic particles have a smaller adiabatic index and a much larger mean free path 
for scattering than does the quasi-thermal plasma. 
In addition, the particles at the highest 
energy escape, thus making energy losses significant and 
increasing the shock compression ratio \cite{be87}.
A fundamental consequence of particle acceleration at cosmic-ray modified shocks
is that the particle spectrum is no longer a power law, but a concave spectrum, 
as hard as $N(p)\propto p^{-1.5}$ at high momenta
\cite{ab06,vla06}. Gamma-ray observations in the GeV-TeV band appear to be the 
best means to measure the particle spectra and thus probe the
acceleration processes in detail.

Particle confinement near the shock is supported by self-generated magnetic 
turbulence ahead of and behind the shock that quasi-elastically scatters 
the energetic charged particles and thus makes their propagation diffusive. 
The amplitude of the turbulence 
determines the scattering frequency, and thus the acceleration 
rate \cite{drury83}. The instabilities by which cosmic rays drive turbulence 
in the upstream region were long thought to be weak enough so that
quasilinear approximations were realistic, i.e. $\delta B/B < 1$,
but recent research suggests that
the process by which streaming cosmic rays excite MHD
turbulence is different from that usually supposed, if the cosmic-ray 
acceleration is efficient.
The amplitude of the turbulent magnetic field may actually
exceed that of the homogeneous, large-scale field \cite{bl,lb}. More recent studies 
\cite{bell04} suggest
that ahead of the shock non-resonant, nearly purely growing modes of 
short wavelength may be more efficiently excited than resonant plasma waves.

The observation of narrow synchrotron X-ray filaments indicates 
that the magnetic field must be very strong at the particle
acceleration sites \cite{bamba,vink}, thus supporting the notion of magnetic-field 
amplification by cosmic rays.
Those strong magnetic fields will decay as the plasma convects
away from the forward shocks of SNRs, and it is an open question 
how far the regions of high magnetic field strength extend \cite{ply,cc07}.
The magnetic-field generation in shocks is also a 
candidate process for the creation of primordial magnetic fields in the cosmological context 
\cite{wid02,ss03,med06}. 
TeV-band gamma-ray observations, together with high-resolution X-ray studies, are the key to
understanding the generation of magnetic fields by energetic particles.  

\begin{figure*}[tb] \centerline{\psfig{file=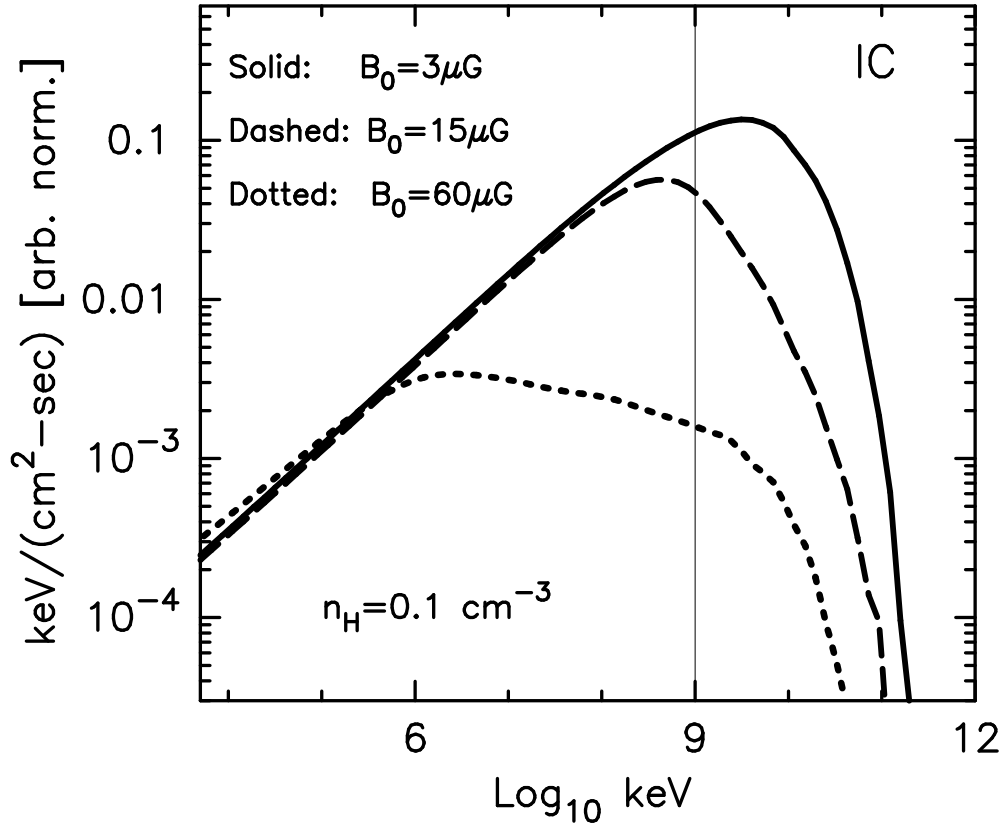,width=5.2in,clip=}} 
\caption{Expected GeV-TeV band gamma-ray emission from Inverse-Compton scattering
of the microwave background on highly relativistic electrons, according to recent
model calculations \cite{ell07}. Shown are three spectra for different values of the
magnetic field strength upstream of the SNR forward shock. For a high field strength
strong radiative losses and evolution make the IC spectrum significantly softer
above about 10~GeV, so it becomes similar to the expected gamma-ray spectrum
produced by energetic hadrons. The thin vertical line marks 1~TeV photon energy.} 
\label{ell-f12} \end{figure*}

If the acceleration efficiency is kept constant, a strong magnetic field would 
reduce the TeV-band gamma-ray emission arising
from IC scattering of energetic electrons relative to their synchrotron
X-ray emission, thus arguing against an IC origin of observable TeV-band emission.
Yet it would make the expected IC spectrum similar
to that of the hadronic pion-decay gamma rays \cite{voelk,ell07}, because strong energy
losses and evolution would produce a spectral change in the electron spectrum, 
as shown in Fig.~\ref{ell-f12}, and would also tie the spatial distribution of the 
gamma-ray emissivities even closer to that of the synchrotron X-rays. It is therefore mandatory
to combine sensitive spectral gamma-ray
measurements with a better angular resolution, so as to avoid confusion and
to effect discrimination between the hadronic and leptonic origin of the
gamma rays.

On the other hand, the pion-decay spectra in the GeV-TeV region, predicted by nonlinear particle 
acceleration models (e.g., \cite{bkv}), depend on 
uncertain parameters such as the ambient density and also somewhat on the strength of the 
interstellar magnetic field. RX~J1713-3946, the brightest shell-type SNR in the TeV band, 
harbors very little gas \cite{cassam}, thus making less likely a pion-decay origin of the
observed TeV-band emission. A robust discriminator, however, 
is the maximum photon energy. Since large magnetic fields produce severe radiation 
losses for electrons, there is a strong correlation between the ratio of maximum 
energy from ion-ion collisions to the maximum energy from IC and the magnetic field strength. 
The shape of the gamma-ray spectrum above $\sim$100~GeV also contains clues to the 
efficiency of the underlying acceleration process, and some SNRs, e.g., 
RX~J1713-3946 (see Fig.~\ref{ic443}), clearly show gamma-ray spectra too soft to be the result of 
efficient acceleration of cosmic-ray nucleons to the knee at 3~PeV \cite{huang}, where the spectrum
of Galactic cosmic rays starts to deviate from a simple power-law form. 

Since the massive stars of type O and B that explode as supernovae are 
predominantly formed in so-called OB associations, most SNRs \cite{hl80, sp90, hl06}
reside in superbubbles \cite{bru80, par04}, giant structures formed by the collective effect
of stellar winds and supernovae. Cosmic rays accelerated in superbubbles may achieve a higher particle
energy than those produced in isolated SNRs, possibly on account of stochastic acceleration processes
in the magnetic turbulence induced by the powerful multiple interacting supersonic stellar winds
\cite{cesarsky83,bykov01}. The winds from superbubbles, therefore, are a 
possible alternative cosmic-ray source class, and some aspects of the isotopic composition
of Galactic cosmic rays support their origin in superbubbles \cite{binns},
although the composition of the bulk of
the cosmic rays is that of the well-mixed interstellar medium \cite{meyer}, which somewhat limits
the role superbubbles can play as the main sources of Galactic cosmic rays. Even though a stellar cluster
may have already been seen in TeV-band gamma rays \cite{reimer}, it is very difficult to arrive at firm
theoretical estimates and interpretations for superbubbles because of their generally poorly known geometry and history, even though some of them are associated with
shell-like structures of atomic hydrogen \cite{mccg02}. There is the possibility of detecting the presence of high-energy heavy nuclei through their interaction with
the intense stellar radiation in clusters of massive stars. Nuclei with energies of a few PeV 
($10^{15}$~eV) will disintegrate upon collision with the starlight, and the subsequent 
de-excitation of the nuclear fragments gives rise to characteristic gamma-ray emission
with a distinct peak in power at about 10~TeV gamma-ray energy \cite{anch}.
 
Superbubbles in the Galaxy are typically 
several degrees in size (eg. the X-ray emitting Cygnus superbubble is \~15~$^{\circ}$ across), 
and therefore low surface brightness, confusion, and varying absorption
complicates their analysis in the radio, optical, and X-ray bands, so 
Galactic superbubbles may be incompletely cataloged \cite{tw05}. The Large Magellanic Cloud (LMC) 
is likely a better location to study superbubbles on account of its distance
(roughly 6 times the distance to the Galactic Center) and low foreground absorption. Numerous 
superbubbles have been found in the LMC \cite{dunne}, which are typically 10 arcminutes 
in apparent size, similar to Galactic SNRs. Nonthermal X-rays were observed from the outer
shell of the superbubble 30 Dor C \cite{bamba04} with a spectrum very similar to the 
nonthermal X-ray seen from Galactic SNRs like SN~1006, but with a luminosity 
about a factor of ten higher and an angular size of only a few minutes of arc, which 
partially compensates for the larger distance. The overall appearance
of superbubbles can, therefore, be likened to that of young SNRs, with one exception: 
the superbubbles are probably much older, so they can maintain efficient particle 
acceleration for $\sim 10^5$ years, in contrast to the $\sim 10^3$ years after which 
shell-type SNRs turn into the decelerating (Sedov) phase and gradually lose their ability
to efficiently accelerate particles. 

The TeV-band to keV-band nonthermal flux ratio of SNRs varies from object to object; 
that ratio is at least a factor of 20 lower for SN~1006 and Cas~A than for RX~J1713-3946. 
However, for superbubbles the flux ratio may actually be significantly higher than 
for isolated SNRs on account of the typical age of the objects, because 
the X-ray emitting electrons are severely loss-limited, whereas for gamma-ray-emitting ions 
that may not be the case. A relatively conservative TeV-band flux estimate can 
be made by taking the measured flux of nonthermal X-rays and
the flux ratio, as for RX~J1713-3946. With this, one would expect TeV-band gamma-ray emission 
from 30 Dor C at a level of a few milliCrab (1 Crab refers to the flux measured 
from the Crab Nebula: the standard candle in high-energy astrophysics), which is a factor 5-10 below the 
sensitivity threshold of present-generation imaging Cherenkov telescopes. Galactic
superbubbles may be much brighter, with about 1 Crab, but likely a few degrees in size, thus rendering their
detection and physical analysis equally difficult.

A future sensitive gamma-ray instrument is needed to
perform studies on a whole class of SNRs to 
finally understand the acceleration and interactions of both
energetic nucleons and electrons. It would also investigate in detail the new and exciting topic
of magnetic field amplification. Therefore, 
an advanced gamma-ray facility can, in conjunction
with current X-ray telescopes, provide detailed information on the
division of the energy budget in shocked SNR environs; namely, how
the global energetics is apportioned between cosmic ray electrons,
ions and magnetic field turbulence.  This is a principal goal that
will elucidate our understanding of plasma shocks, generation of
magnetic turbulence and cosmic ray acceleration in the cosmos.

\medskip
\subsubsection{Diffuse galactic emission}

\begin{figure*}[tb]
\centerline{\psfig{file=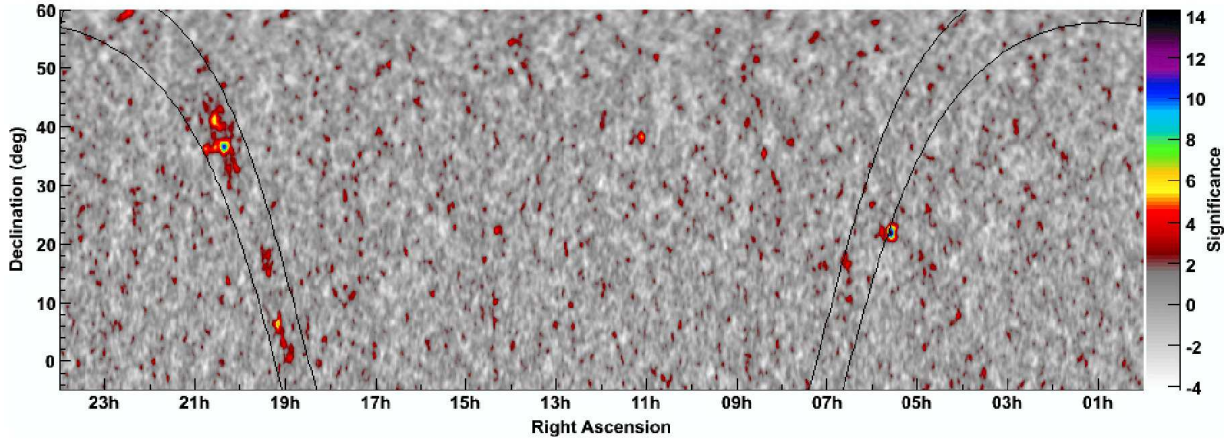,width=6.4in,clip=}} \caption{
A sky map from 5 years of Milagro data 
taking \cite{abdo}.  Clearly detected in this plot are the Crab Nebula and the 
Galactic ridge. The brightest portion of the inner Galaxy is the Cygnus region and we 
have strong evidence for an extended source embedded within the larger 
diffuse emission region.  The additional structure observed at lower 
Galactic latitudes has not yet been analyzed in detail and we cannot 
comment on the significance of any apparent features.} \label{milagro} \end{figure*}

In contrast to the case of SNRs, most of our knowledge of diffuse galactic gamma-ray emission
was obtained with survey-type instrumentation that combines a very large field-of-view with moderate
angular resolution of about one degree. Detectors like the Milagro instrument, that used the 
water-Cherenkov technique to measure gamma rays around 10~TeV energy, or the satellite experiment EGRET,
a pair-production instrument sensitive to GeV gamma rays that operated in the Nineties, 
fall into this category. Survey instruments
can provide good sensitivity to large-scale structures, but often suffer from confusion because the 
small-scale distribution of the signal cannot be determined, and so point sources and extended
emission cannot be reliably separated. On the other hand, atmospheric Cherenkov telescopes such as
H.E.S.S., MAGIC, and VERITAS offer a high angular resolution, so the angular structure of
compact sources can be properly determined; but they generally have a small field-of-view and a reduced
sensitivity for structures larger than a few degrees. The different characteristics of survey
instruments and high-resolution cameras are evident in the scientific results of existing experiments.

EGRET has produced an all-sky map of the gamma-ray sky up to 10 GeV; at these
energies inverse Compton (IC) scattering is still a major component of diffuse emission,
possibly even dominant. NASA's next-generation experiment, Fermi, will
clarify the nature of excess emission seen with EGRET 
at a few GeV, dubbed the GeV excess \cite{hunter}, 
produce an allsky map of GeV-scale diffuse gamma-ray emission, 
and also extend the coverage to 100 GeV. While at 1~GeV the statistical accuracy will be very high, with more 
than a hundred detected gamma rays per year and angular resolution element, the angular resolution
as measured through angle around the true photon direction for 68\% containment still exceeds 0.5~degrees,
so confusion will be an issue.
At higher energies, around 30~GeV, the angular resolution is better than $0.1^\circ$, but we can expect only
about one detected photon per $0.1^\circ$ resolution element through the 5-year mission, so
the angular resolution cannot be fully exploited. Fermi will provide
invaluable spectral information on the diffuse Galactic gamma-ray emission in the GeV band with 
degree-scale angular resolution, but TeV-band measurements will produce complementary images and spectra 
with very high angular resolution for selected regions of the sky that will be particularly useful 
where imaging with Fermi suffers from confusion. GeV- and TeV-band observations can be combined to extract
the information required to understand the propagation of energetic particles in the Galaxy.

The IC contribution to the diffuse Galactic gamma-ray emission
can be large and not easy to separate from that of pion-decay. The
separation of the diffuse gamma-ray signal into the contributions of cosmic-ray ions and those of
electrons is desirable, because the propagation properties of the two particle populations is different.
Also, measurements of the isotopic composition of cosmic rays near earth with
appropriate particle detectors such as, e.g., PAMELA \cite{malvezzi} 
allows us to additionally constrain the propagation history of cosmic-ray ions, although it appears
very difficult to both fit the EGRET data and the locally measured spectra of cosmic-ray ions
and electrons \cite{andy04}.

For gamma rays with energy above 10 TeV, the electron energies have to be at least a few 
tens of TeV, but in view of the rapid energy losses it is probable that the electrons do
not have the time to propagate away from their
acceleration sites; hence, IC is of much less importance for the diffuse emission at those very high energies.
An all-sky map above 10 TeV would provide a 'clean' view of the distribution 
and spectrum of cosmic-ray hadrons over the whole Galaxy. Such a skymap could 
provide the key to the origin of cosmic-ray hadrons, in particular when it could be combined with 
information on the intensity of neutrinos. 

A measurement of the intensity of diffuse emission at TeV energies would be extremely valuable, 
provided one is able to separate truly diffuse
emission from individual sources such as pulsar-wind nebulae. To date, 
the Milagro collaboration
reports evidence of TeV-scale gamma-ray emission from the Galactic plane, in particular the 
Cygnus region (see Figure~\ref{milagro}). 
The intensity measured with Milagro at 12~TeV is 70~Crab/sr (about 0.02 Crab/deg$^2$)
and thus extremely sensitive to the point source content. For comparison, the intensity of 
the 14 new sources detected during the H.E.S.S. survey of the inner Galaxy
\cite{hess-survey}, if they were unresolved, would
be 17~Crab/sr above 200~GeV. Assuming the measured spectrum extends to 12~TeV, the equivalent
intensity of the 14 H.E.S.S. sources at 12~TeV, which is more relevant for a comparison 
with the Milagro result, would be 140~Crab/sr, i.e. twice the intensity observed with Milagro. 
The source density in the region observed with Milagro is probably smaller, but there are also sources
not seen with H.E.S.S. or known prior to the survey, and therefore
a significant fraction of the Milagro result will be due to unresolved sources, and confusion is
a substantial problem.

\begin{figure*}[tb]
\centerline{\psfig{file=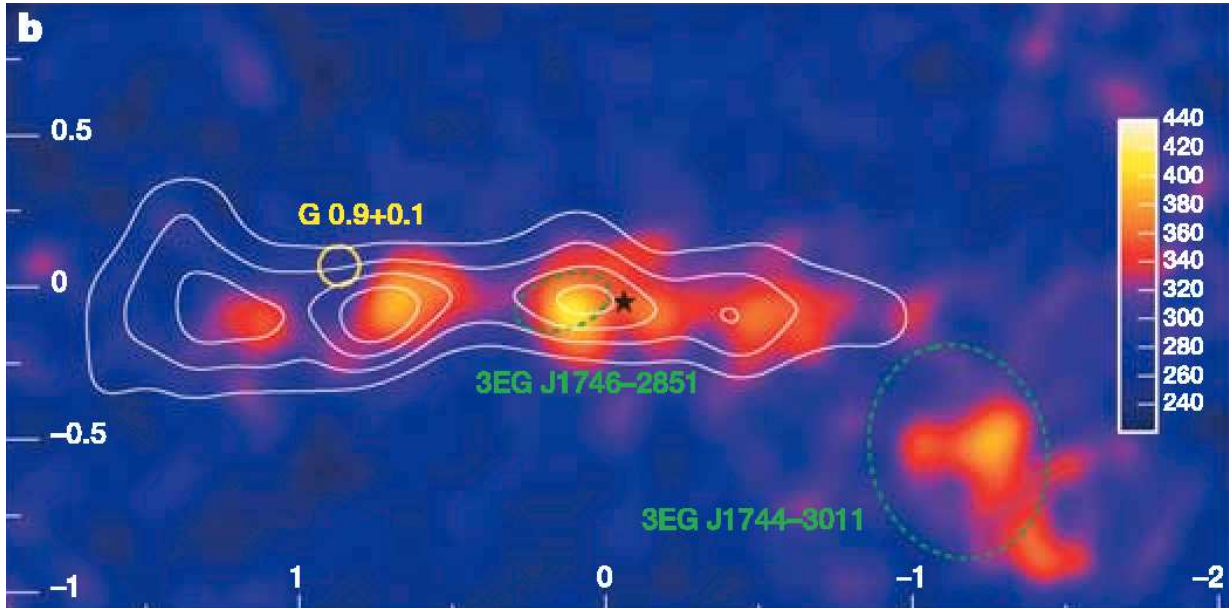,width=6.4in,clip=}} \caption{
H.E.S.S. gamma-ray count map after subtraction of two
bright point sources. The white contour lines indicate the column density of 
molecular gas traced by CS line emission.} \label{hess} \end{figure*}

The H.E.S.S. collaboration has published a map and the spectrum of diffuse emission from the 
inner degree of the Galaxy, after subtracting two dominating point sources (see Figure~\ref{hess}). 
In a fit of the observed spectrum as $dN/dE\propto E^{-\gamma}$, the spectral index is 
$\gamma=2.3\pm 0.08$, much harder than expected and measured anywhere else, as cosmic ray
ions with a spectrum as directly measured at earth should produce a gamma-ray spectrum with 
$\gamma\simeq 2.7$. The measured intensity 
corresponds to 590 Crab/sr at a TeV and is about the highest one may expect anywhere 
in the Galaxy based on the intensity distribution of GeV-band gamma rays.

\subsection{What measurements are needed?}

\subsubsection{Supernova remnants}

With multi-waveband data, it
is possible to provide quantitative constraints on the particle acceleration
mechanism.  Because the maximum IC power output from these objects is
expected to be in the TeV region, TeV observations provide information
unavailable via any other means.  High-resolution maps and accurate
spectra of the TeV emission, when compared with data from other wavebands,
will permit estimation of the magnetic field and the maximum energy of
the accelerated particles.  Comparison of the maps from various
wavelengths will increase our understanding of the diffusion and
lifetimes of the highly energetic electrons. 

We stress the importance of GeV-band
gamma-ray data that will be shortly provided by Fermi. However, the sensitivity of
Fermi at 10--100~GeV is limited: if we extrapolate the TeV-band spectrum of RX~J1713-3946 
(see Fig.~\ref{ic443}) to lower energies as $dN/dE\propto E^{-2}$, or a flat line in the figure, then
we expect Fermi to detect two photons per year with energy above 100~GeV from the entire SNR. 
Above 10~GeV, Fermi would find 20 photons per year, so even after five years, a Fermi gamma-ray excess map
would have much lower statistical accuracy than the existing H.E.S.S. map; the single-photon resolution
is also worse. At energies below 10~GeV, the number of Fermi-detected photons increases, 
but the angular resolution deteriorates. Fermi will perform important
studies of shell-type SNRs, but TeV-band measurements will provide complementary and, in many cases, 
richer images and spectra.
 
A key for any future VHE observatory will be to unambiguously
disentangle the emission from electronic versus hadronic cosmic rays.
Spectral studies may help arriving at a discrimination between 
gamma rays from electrons and those produced by hadrons, but they are not sufficient. 
TeV gamma rays from IC scattering of the microwave background should have a spectral shape
that reflects that of synchrotron X-rays below approximately 1 keV, where 
the discrimination of synchrotron emission and thermal radiation of ordinary hot gas
is often difficult and requires a very good angular resolution.

\begin{figure*}[tb]
\centerline{\psfig{file=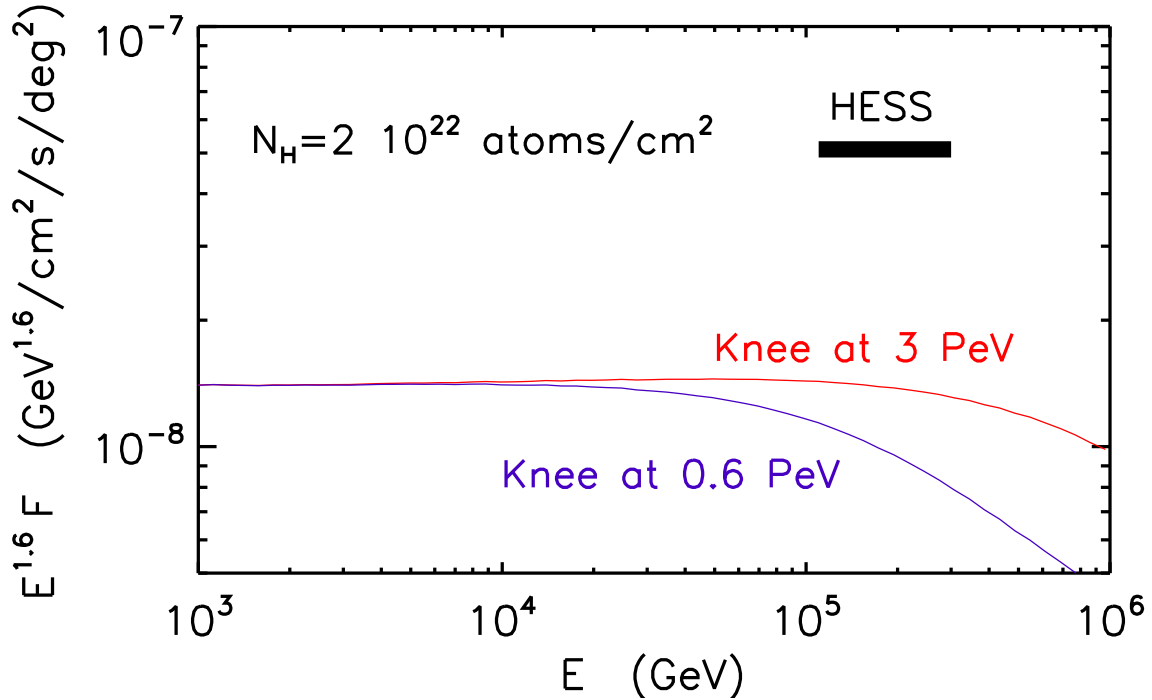,width=6.0in,clip=}} \caption{
Shown as red line is the intensity of diffuse Galactic gamma rays, multiplied with $E^{1.6}$, 
for a standard cosmic-ray spectrum with the knee at 3~PeV and one of the Orion molecular clouds. 
If near some molecular gas complex the
knee was at 0.6~PeV, the spectrum of diffuse gamma rays from that region would follow the blue line. 
Observing a location dependence of the knee energy would provide important clues on the nature of the knee,
as do similar measurement for individual sources of cosmic rays (e.g. \cite{huang}).
The black bar indicates 
an estimate of the current H.E.S.S. sensitivity in the 100--300~TeV band, based on published
spectra of RX~J1713-3946. An increase by a factor 10 in sensitivity around 200~TeV would be needed to
discriminate the blue and the red curve.} \label{knee} \end{figure*}

On the other hand, TeV gamma rays of hadronic origin reflect the spectrum of
energetic nuclei at about 1--100~TeV energy. If the SNR in question 
accelerates hadronic cosmic rays to energies beyond the knee at 3~PeV, then we should see 
a continuation of the gamma-ray emission up to 100 TeV and beyond, which would be a good
indication of a hadronic origin of gamma rays (see also Fig.~\ref{knee}). 
It will therefore be important to maintain
a sensitivity up to and beyond 100~TeV. We also note an obvious relation to neutrino astrophysics:
all gamma-ray sources at the Crab flux level that do not cut off below 100~TeV energy should be observable
with neutrino detectors \cite{kis+bea}, if the gamma-ray emission arises from interactions of 
energetic nuclei.
In the near term, only IceCube at the South Pole will be large enough
to observe Galactic neutrino sources such as SNRs.  Since it must look through
Earth, it is very important that it be paired with sensitive gamma-ray
instruments in the Northern hemisphere.  The H.E.S.S. experiment and its possible successor, CTA, 
planned for the same site in Namibia, can observe many Southern gamma-ray sources,
but they won't be paired with an adequately sensitive km$^3$-scale Mediterranean neutrino detector until
much later. The US-led VERITAS experiment currently observes sources in the Northern sky.
A more sensitive successor could optimally exploit the scientific opportunities that lie in
the synergies with IceCube. An additional future survey-type instrument could search 
for very extended sources and serve as a pathfinder for high-angular-resolution observations with
the atmospheric Cherenkov telescopes.
 
High-resolution imaging of the TeV emission, combined with good spectral information
is, therefore, required.
TeV emission from hadronic interactions should
trace the distribution of target material, while
TeV emission from electrons should be well correlated with non-thermal
X-ray emission (see Figs.~\ref{ic443}). 
Because a significant fraction of the non-thermal X-ray intensity is organized in thin
filaments, arcminute-scale resolution in the TeV band combined with the
appropriate sensitivity would permit a clear separation of hadronic and leptonic emission,
and would allow a direct measurement of the magnetic field strength at the forward shock of 
SNR, and hence a clean assessment of the efficacy of magnetic-field amplification by energetic 
particles. The required angular resolution is, therefore, a factor 3--5 better than 
what is currently achieved with H.E.S.S. and VERITAS. The sensitivity needed to derive 
well-defined spectra with angular resolution below $0.1^\circ$ is about a factor of 10 higher 
than that afforded by the current generation of atmospheric \v Cherenkov telescopes. This
is also the sensitivity likely needed to detect TeV-band gamma rays from superbubbles. 

Non-detection of hadronically produced gamma rays
would require either a very steep source spectrum, inconsistent with
that needed to produce the local spectrum, or a greatly reduced
cosmic-ray intensity, inconsistent with the energy budget for cosmic
rays.  Either of these possibilities would lead to
serious revisions in our understanding of the origin of cosmic rays. 
Detection of TeV photons from hadronic cosmic rays would immediately
constrain the spectrum and total energy budget of the cosmic rays, and
would provide invaluable constraints on the relative acceleration efficiency
of electrons and protons or other ions in shocks. This may help resolve  
the hundred-year-old question of the origin of cosmic rays, and will yield important 
information on shock physics that can be used in other shock systems. If hadronic
cosmic rays are accelerated in shocks produced by the winds from OB
associations, the TeV photons produced by those cosmic rays should,
again, trace the distribution of target material. The angular
resolution requirements are similar to those discussed for supernova
remnants.

\subsubsection{Diffuse galactic emission}

Currently most pressing questions are the following:
Are cosmic rays above the knee at 3~PeV, where the spectrum of local cosmic rays
considerably steepens, really Galactic in origin? What is the origin of the
knee? Is 
the knee a source property, in which case we should see a corresponding
spectral feature in the gamma-ray spectra of cosmic-ray sources, or the result of propagation, so we
should observe a knee that is potentially dependent on location, because the propagation properties
depend on position in the Galaxy? 

Another series of questions concerns cosmic-ray electrons, whose source power is significant, but whose
spectrum above 1 TeV is essentially unknown. What is the distribution of cosmic-ray electrons at 
energies beyond 1 TeV? Measuring electron spectra inside and outside their sources
carries direct information on the particle acceleration rate, and thus on the nature of the 
acceleration process, as well as on the propagation properties of cosmic rays up to the knee.

Via their gamma-ray emission, we would therefore wish to independently measure

\begin{itemize}
\item the spectrum and flux of cosmic nuclei, 
which are expected to produce a gamma-ray signal that
largely correlates with the density of interstellar gas. The expected intensity on a half-degree scale
in, e.g., the Cygnus region is 60 Crab/sr (0.02 Crab/deg$^2$) in the 100~GeV--1~TeV band, and
40 Crab/sr (0.013 Crab/deg$^2$) above 1~TeV. 
Note that gamma rays with energy higher than about 100 TeV map 
the cosmic-ray nuclei spectrum around the knee, so an increase by at least
a factor 10 in sensitivity and a
good energy resolution up to and beyond
100~TeV is required to potentially prove a location dependence of the knee (see also Fig.~\ref{knee}).
At these energies, pair production with ambient radiation can attenuate the gamma-ray signals as they
travel across the Galaxy, but observations of relatively nearby complexes of molecular clouds
would ensure that absorption in the Galaxy is 
negligible and that the intensity measurement can be made by integration over typically a 
square-degree in solid angle.
One should note that a high angular resolution is nevertheless needed 
for those measurements, both to account for point-source contributions and to verify the spatial correlation
of the signal with the distribution of atomic
molecular gas, which is known on scales $\lesssim 0.1^\circ$. 

\item the spectrum and flux of cosmic electrons, which will produce a patchy and spectrally variable
gamma-ray signal that does not correlate with the gas density but may have structure on a 
1--3~$^{\circ}$ scale. The intensity is impossible to estimate without insight into the nature of 
the EGRET GeV excess, but may be stronger than the hadronic emission in the 100~GeV--1~TeV band.

\item the point-source content of the gamma-ray signal to properly separate sources from truly 
diffuse emission.
\end{itemize}

None of these measurements requires a very low energy threshold, though one would wish to 
not have a large gap to the energy range accessible with Fermi, which will make reliable 
measurements up to about 50~GeV gamma-ray energy. The measurement of hadronic and, in particular, 
leptonic gamma rays chiefly requires advances in both the effective area and, in particular,
the background rejection of future observatories. The field-of-view, in which sensitive 
gamma-ray observations are taken, is of minor importance,
as long at it is at least 4~$^{\circ}$ in diameter. This requirement arises
from the necessity to independently determine a reliable gamma-ray zero level, which is best done
by having the gamma-ray-sensitive field-of-view
cover both the source, e.g. an SNR or a molecular cloud complex, and empty regions surrounding it 
for background measurements. 

As a byproduct, one would also be interested in measuring the direct \v Cerenkov light of local cosmic rays, 
which provides unique information on the cosmic-ray composition in the PeV energy range. 
Although cosmic rays form the primary background for ground-based gamma-ray 
detectors, this background can be used to make a 
unique measurement of cosmic ray composition.  Recently, the H.E.S.S. 
collaboration has measured the 
direct \v Cherenkov light of local cosmic rays \cite{hess-direct}, which provides unique 
information on the cosmic-ray composition in the PeV energy range \cite{kieda2001}. 
This method is more direct than that 
used in extensive air shower experiments because it avoids the 
dependence on hadronic simulations in identifying the primary particle 
type. Also, the main air showers display strong 
statistical fluctuations in their evolution, thus making the observation of 
direct \v Cherenkov light a much more precise measurement. 
The current atmospheric \v Cherenkov arrays like H.E.S.S. or VERITAS lack the 
angular and timing resolution to fully exploit this method, and
a future instrument with 0.01-degree image pixel 
resolution and nanosecond time resolution 
could further improve the determination of the cosmic-ray composition at 
high energies. 

\subsection{What is the required instrument
performance?}

For the study of SNR 
the two key instrument parameters are angular resolution and a
sufficiently high count rate to effectively exploit the angular resolution. 
The
required angular resolution can be estimated from the known angular sizes of the
non-thermal X-ray filaments and of the dense molecular clouds and shock
regions.  For the closer SNRs, the typical angular resolution required
is about one arcminute. A key point is that a sufficient number of gamma rays 
must be detected to make effective use of the angular resolution.  The
detection rate can be increased relative to current instruments either
by increasing the effective area or by reducing the energy threshold.
The goal should be to image several SNRs with arcminute-level resolution with 
a minimum of 150 events in each image bin,
so reliable spectra can be reconstructed.

To maximize the scientific return for Galactic sources, a future
instrument should be located at sufficiently southern latitude to give good
coverage of a large fraction of the Galactic plane extending to the
inner Galaxy. At the same time it is desirable that a large overlap is maintained with the
coverage of neutrino experiments such as IceCube, which makes a Southern location less advantageous. 

To achieve scientifically significant observations of the diffuse Galactic gamma-ray emission with
the next-generation instrument, the angular resolution is important, but does not need to be as good as for
observations of SNR. Mainly one needs to model and subtract individual sources. A good
angular resolution is also needed to find intensity that correlates with the gas distribution. 

It is mandatory to achieve a very high sensitivity for extended emission.
Given that a number of emission components have to be fitted in parallel
and that, at least for the diffuse emission, the data are background-dominated, a
strongly improved background rejection is required to achieve the desired sensitivity. 
A large aperture alone appears insufficient, as it is necessary to achieve both large event 
numbers and a very low background contamination level.

The instrument requirements can thus be summarized as follows:

\begin{itemize}
\item an angular resolution of $\le 0.02^\circ$ at 1 TeV.
\item for the bright parts of SNRs at least 150 gamma rays in each image bin for a 
reasonable observing time.
\item a sensitivity for extended emission that is significantly better than 10 Crab/sr above a 
TeV and better than 15 Crab/sr below a TeV.
\item maintain a high sensitivity up to and possibly beyond 100~TeV.
\item a good energy resolution of $\delta E/E\lesssim 15\%$ at all energies.
\item a gamma-ray sensitive field-of-view of at least 4~$^{\circ}$ in diameter. Bigger is better, 
but tens of degrees are not needed.
\end{itemize}

%
%



\clearpage

\section{Galactic compact objects}
\label{GCO-subsec}
Group membership: \\ \\ 
P. Kaaret, A. A. Abdo, J. Arons, M. Baring,
W. Cui, B. Dingus,
J. Finley, S. Funk, S. Heinz, B. Gaensler, A. Harding, E. Hays, J. Holder, D. Kieda, A. Konopelko, S. LeBohec, A. Levinson,
I. Moskalenko, R. Mukherjee, R. Ong, M. Pohl, K. Ragan, 
P. Slane, A. Smith, D. Torres

\subsection{Introduction}
Our Galaxy contains astrophysical systems capable of accelerating
particles to energies in excess of several tens of TeV, energies beyond
the reach of any accelerator built by humans.  What drives these
accelerators is a major question in astrophysics and understanding these
accelerators has broad implications.  TeV emission is a key diagnostic
of highly energetic particles.  Simply put, emission of a TeV photon
requires a charged particle at an energy of a TeV or greater. 
Observations in the TeV band are a sensitive probe of the highest energy
physical processes occurring in a variety of Galactic objects.  Galactic
TeV emitters also represent the sources for which we can obtain the most
detailed information on the acceleration and diffusion of high-energy
particles and are, thus, our best laboratories for understanding the
mechanisms of astrophysical ultra-relativistic accelerators.

\begin{figure*}[tb] \centerline{\psfig{file=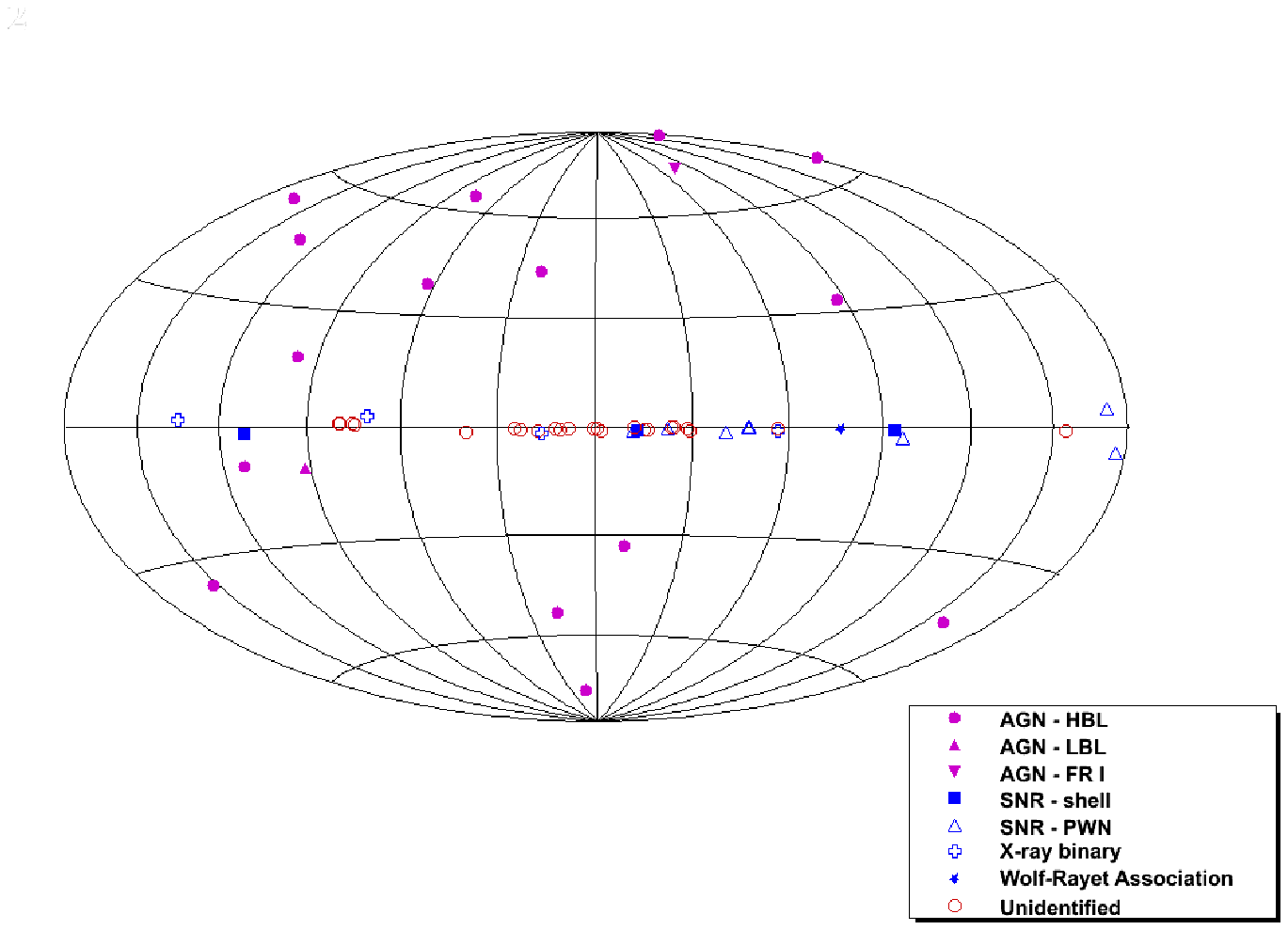,width=6.5in}}
\caption{Known TeV emitting objects plotted in Galactic coordinates. 
The center of the Milky Way is at the center of the ellipse.  The
Galactic plane is the horizontal midplane.  The symbols and colors
indicate the source type.  Figure courtesy of Dr.\ E.\ Hays.}
\label{fig:tevmap} \end{figure*}

Recent results from the new generation of TeV observations, primarily
H.E.S.S., have revealed a large population of Galactic sources; see
Fig.~\ref{fig:tevmap} which shows the known TeV sources in Galactic
coordinates.  Galactic sources now comprise a majority of the known TeV
emitters with object classes ranging from supernova remnants to X-ray
binaries to stellar associations to the unknown.  Future TeV
observations with a more sensitive telescope array will lead to the
discovery of many more TeV emitting objects and significantly advance
our understanding of the acceleration of the highest energy particles in
the Galaxy.

\subsection{Pulsar wind nebulae}

Pulsar wind nebulae (PWNe) are powered by relativistic particles
accelerated in the termination shock of the relativistic wind from a
rotation-powered pulsar.  The basic physical picture is that the
rotating magnetic field of the pulsar drives a relativistic wind.  A
termination shock forms where the internal pressure of the nebula
balances the wind ram pressure.  At the shock, particles are thermalized
and re-accelerated to Lorentz factors exceeding $10^{6}$.  The energy in
the Poynting flux is transferred, in part, to particles.  The high
energy particles then diffuse through the nebula, partially confined by
nebular magnetic fields, and cool as they age due to synchrotron losses,
producing radio to X-ray emission, and inverse-Compton losses, producing
gamma-ray emission.

Studies of PWNe address several central questions in high-energy
astrophysics, the most important of which is the mechanism of particle
acceleration in relativistic shocks.  PWNe provide a unique laboratory
for the study of relativistic shocks because the properties of the
pulsar wind are constrained by our knowledge of the pulsar and because
the details of the interaction of the relativistic wind can be imaged in
the X-ray, optical, and radio bands.  Relativistic shock acceleration is
key to many astrophysical TeV sources, and PWNe are, perhaps, the best
laboratory to understand the detailed dynamics of such shocks.  Studies
of pulsar-powered nebulae also target a number of crucial areas of
pulsar astrophysics, including the precise mechanism by which the pulsar
spin-down energy is dissipated, the ratio of magnetic to particle energy
in the pulsar wind, the electrodynamics of the magnetosphere, and the
distribution of young pulsars within the Milky Way.

Observations of TeV emission are essential to resolve these questions.
Measurement of the spectrum from the keV into the TeV range allows one
to constrain the maximum particle energy, the particle injection rate,
and the strength of the nebular magnetic field.  Observation of TeV
emission from a significant set of pulsar-powered nebulae would allow us
to study how the pulsar wind varies with pulsar properties such as
spin-down power and age.  Detection and identification of new nebulae
may also lead to the discovery of new young pulsars, particularly those
lying in dense or obscured parts of the Galaxy where radio searches are
ineffective because of dispersion.

PWNe have proven to be prolific TeV emitters.  The Crab nebula was the
first TeV source to be discovered.  H.E.S.S. has recently detected a number of other Galactic
sources, several of which are confirmed to be, and many more thought to
be, PWNe \cite{DeJager06}.  Significantly, H.E.S.S. has discovered new PWNe
that were not previously detected at other  wavelengths.  Furthermore,
the high resolution capabilities of H.E.S.S. have allowed  imaging of the
first TeV jet in the PWN of PSR1509-58 \cite{Aharonian05}, which is 
also the first astrophysical jet resolved at gamma-ray energies. 
Comparison of the  gamma-ray jet with the one detected by Chandra in
X-rays, which is less extended and  has a flatter spectral index, shows
that the evolution of emitting particles in the jet is  consistent with
synchrotron cooling.  In addition, TeV imaging has provided a clearer
picture of PWNe such as PSR B1823-13 and Vela X \cite{Aharonian06} that
are offset  from the position of the pulsar, an effect which may be due
to the pressure of the reverse  shock \cite{Blondin01}.

\subsubsection{Measurements needed}

\paragraph{Broadband modeling of PWNe}

The broadband spectrum of a PWN provides constraints on the integrated
energy injected by the pulsar as well as the effects of adiabatic
expansion and the evolution of the magnetic field.  The spectrum
consists of two components: 1) synchrotron emission extending from the
radio into the X-ray and, in some cases, the MeV band, and 2)
inverse-Compton emission producing GeV and TeV photons.  Emission in the
TeV band originates primarily from inverse-Compton scattering of ambient
soft photons with energetic electrons in the nebula.  The ratio of TeV
luminosity to pulsar spin-down power varies strongly between different
PWN and understanding the cause of this effect will advance our
understanding of the physics of PWNe.

All PWNe show spectra that are steeper in the X-ray band than in the
radio band, but the nature of the spectral changes between these bands
is not well understood.  Synchrotron losses result in a spectral break
at a frequency that depends on the age and magnetic field strength,
while other spectral features can be produced by a significant change in
the pulsar input at some epoch, by spectral features inherent in the
injection spectrum, and by interactions of the PWN with the reverse
shock from its associated supernova remnant.  TeV observations provide
an independent means to probe the electron energy distribution. 
Addition of TeV data breaks many of the degeneracies present in analysis
of synchrotron emission alone and allows independent estimates of the
electron energy distribution and the nebular magnetic field.  TeV
observations are essential to understand the PWN electron energy
distribution and its evolution.
\begin{figure}[tb] \centerline{\psfig{file=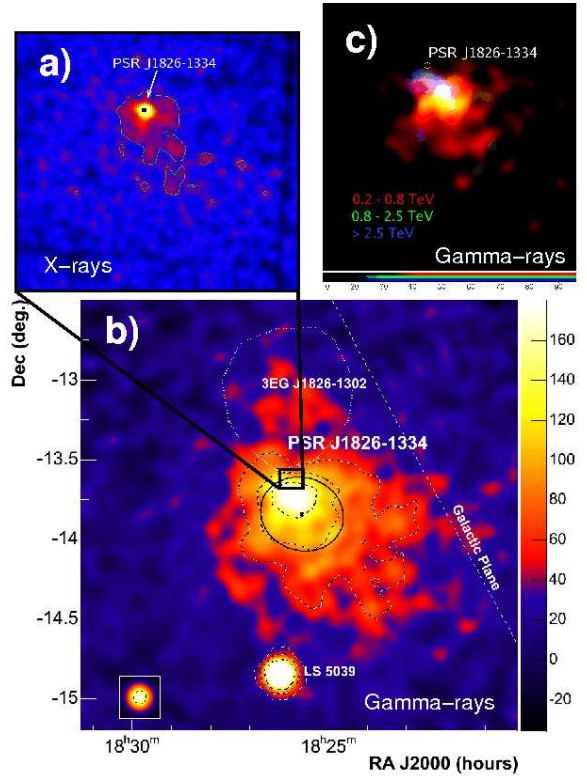,width=3.0in}}
\caption{H.E.S.S. map of TeV emission from H.E.S.S. J1825-137 (b), X-ray image
of the central part of the field showing the PWN G18.0--0.7 (a), three
color image of the TeV emission showing that the nebula is the most
compact at the highest energies (c).  From \cite{Funk06}.}
\label{fig:hess1825} \end{figure}
\paragraph{Highly Extended PWNe}
Several of the recently-discovered H.E.S.S. sources appear to be PWNe, due
to the presence of young radio pulsars nearby, but have unexpected
morphologies.  Examples include H.E.S.S. J1804-216 \cite{hess1804}, H.E.S.S.
J1825-137 \cite{hess1825}, and H.E.S.S. J1718-385 \cite{hess1718}.  There
are two issues that require considerable further study for these
sources.  First, the young pulsars suggested as the engines for these
nebulae are distinctly separated from the TeV centroids.  The most
common explanation is that the supernova remnants in which these PWNe
formed (most of which are not observed, to date) are sufficiently
evolved that the reverse shocks have disturbed the PWNe, as appears to
be the case in Vela X, which is also observed as an extended TeV source
offset from its pulsar \cite{hessVelax}.  This requires an asymmetric
interaction with the reverse shock, which can occur if the SNR expands
into a highly non-uniform medium, and there are suggestions that these
systems may indeed be evolving in the vicinity of molecular clouds.  In
this scenario, the reverse shock encounters one side of the PWNe first,
and the disruption leaves a relic nebula of particles that is
concentrated primarily on one side of the pulsar.  More sensitive TeV
observations are required to produce higher fidelity maps of these
nebulae, and to search for evidence of a steepening of the spectrum with
distance from the pulsar.

A second and more vexing question centers on the very large sizes of
these PWNe.  These sources are observed to be extended on scales as
large as $1^{\circ}$ \cite{hessVelax}, significantly larger than their
extent in X-rays.  One possible explanation for this is that the extent
of the synchrotron radiation observed in the X-ray band is confined to
the region inside the magnetic bubble of particles that is sweeping up
the ambient ejecta, while the IC emission is produced wherever energetic
particles encounter ambient photons. If the diffusion length of these
energetic particles is extremely large, they can escape the
synchrotron-emitting volume, but still produce TeV gamma rays.  Because
these sources are relatively faint, high-quality maps of this extended
emission do not yet exist. Higher sensitivity, along with somewhat
improved angular resolution, are crucial for probing more deeply into
the structure of these nebulae.
\paragraph{Jets/Magnetization}
X-ray observations with Chandra and XMM-Newton have revealed jet
structures in a large number of PWNe. Models for the formation of these
jets indicate that some fraction of the equatorial wind from the pulsar
can be redirected from its radial outflow and collimated by hoop
stresses from the inner magnetic field. The formation of these jets is
highly dependent upon the ratio of the Poynting flux to the particle
energy density in the  wind. H.E.S.S. observations of PSR B1509-58 reveal an
extended TeV jet aligned with the known X-ray jet.  New TeV observations
of similar jets should provide insight into the Poynting fraction and
the physics of jet formation.
\paragraph{Discovery Space}
The recently-discovered H.E.S.S. sources that appear to be previously
unknown PWNe highlight the potential for uncovering a large number of
PWNe in TeV surveys. For cases where the nebula magnetic field is low,
thus reducing the synchrotron emissivity, the IC emission could be the
primary observable signature. An increase in sensitivity will be
important to enhance the discovery space, and cameras with a large field
of view would enable large surveys to be conducted. Given that some of
the H.E.S.S. sources in this class are extended, improved angular resolution
also holds promise both for identifying the sources as PWNe and for
investigating the structure of these systems.

\subsection{Pulsed emission from neutron stars}

The electrodynamics of pulsars can be probed more directly via
observation of their pulsed emission.  The high timing accuracy
achievable with pulsars has led to Nobel prize winning discoveries, but
the mechanism which produces the pulsed emission, from the radio to
gamma rays, is not well understood.  TeV observations may provide key
insights.

A key question that has pervaded pulsar paradigms over the last two
decades is where is the locale of the high-energy non-thermal
magnetospheric emission?  Two competing models have been put forward for
gamma-ray pulsars: (1) polar cap scenarios, where the particle
acceleration occurs near the neutron star surface, and (2) the outer gap
picture, where this acceleration arises out near the light cylinder.
Data have not yet discriminated between these scenarios, and our
understanding of pulsar magnetopheres has stalled because of this. For
energetic young pulsars like the Crab and Vela, TeV telescopes/arrays
would offer the greatest impact if the outer gap model is operable.  For
millisecond pulsars, TeV telescopes should provide valuable insight
regardless of the emission locale.  Indeed, the answer to the question
may differ according to which subset of pulsars is examined.

Knowing the location of their radiative dissipation will permit the
identification of the pertinent physical processes involved and open up
the possibility for probing the acceleration mechanism.  This could then
enable refinement of pulsar electrodynamics studies, a difficult field
that is currently predominantly tackled via MHD and plasma simulations. 
Should polar cap environs prevail as the site for acceleration, then
there is a distinct possibility that pulsar observations could provide
the first tests of quantum electrodynamics in strong magnetic fields. 
An additional issue is to determine whether there are profound
differences in emission locales between normal pulsars and their
millisecond counterparts.  High-energy gamma-ray observations are
central to distinguishing between these competing models and accordingly
propelling various aspects of our knowledge of pulsar electrodynamics.

Detection of pulsed emission at TeV energies has so far been elusive.
The observation of high-energy cutoffs below 10 GeV in the pulsed
emission spectra of several normal pulsars with high magnetic fields by
EGRET \cite{Thompson04} has made the prospects of detecting emission at
energies above 100 GeV very unlikely. Indeed, such cutoffs are predicted
from magnetic pair production in polar cap models \cite{Daugherty96} and
from radiation reaction limits in outer gap models \cite{Romani96}.
However, outer gap models predict that a separate component produced by
inverse Compton scattering should be detectable at TeV energies, while
polar cap models do not expect such a contribution.  This provides a key
opportunity for distinguishing between these competing pictures.  Yet, the outer
gap scenario has suffered through a sequence of non-detections (e.g. 
see \cite{Lessard00} for Whipple limits on the Crab's pulsed signal)  in
focused observations by TeV telescopes, progressively pushing the pulsed
flux predictions down.  In a recent addition to this litany, MAGIC has
obtained constraining flux limits at 70 GeV and above to PSR B1951+32
\cite{magic_psr1951}, implying turnovers below around 35 GeV in the
curvature/synchrotron component, thereby mandating a revision of the
latest outer gap predictions of inverse Compton TeV fluxes
\cite{Hirotani07}.

This result highlights the importance of lowering the threshold of
ground-based ACT arrays.  Such saliency is even more palpable for the
study of millisecond pulsars (MSPs).  Polar cap model predictions can
give turnovers in the 30-70 GeV range for MSPs \cite{Harding05}, though
outer gap turnovers for MSPs are actually at lower energies due to
significant primary electron cooling by curvature radiation reaction. 
While possessing much lower magnetic fields than normal energetic young
pulsars, millisecond pulsars can be expected to be as luminous in some portion
of the gamma-ray band because their rapid periods imply large spin-down power. 
Hence, future sub-TeV observations of MSPs should significantly advance our
understanding of these objects.

\begin{figure*} \centerline{\psfig{file=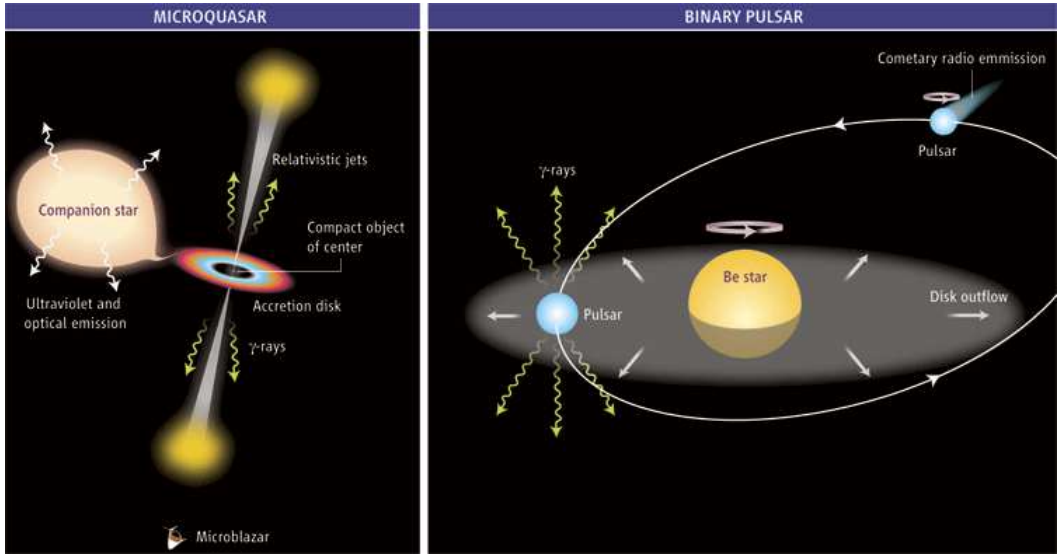,width=5.5in}}
\caption{ The two types of binaries systems producing TeV emission.  The
left image shows a microquasar powered by accretion onto a compact
object, neutron star or black hole.  The right image shows a
rotation-powered pulsar (neutron star) in a binary where the
relativistic wind from the pulsar leads to the production of TeV
photons.  From \cite{Mirabel06}.} \label{fig:binaries} \end{figure*}

\subsubsection{Measurements needed}

What is clearly needed to advance the pulsar field is a lower detection
threshold and better flux sensitivity in the sub-TeV band.  The goal of
lower thresholds is obviously to tap the potential of large fluxes from
the curvature/synchrotron component.  At the same time, greater
sensitivity can provide count rates that enable pulse-profile
determination at the EGRET level or better, which can then probe
emission region geometry. Pulse-phase spectroscopy is a necessary and
realizable goal that will enable both model discrimination and
subsequent refinement. Since the current generation ACTs cannot quite
reach thresholds below 70~GeV, and since the model predictions are very
dependent on emission and viewing geometry, it seems that detection of
very high-energy emission from millisecond and young pulsars will be
unlikely for the current instruments and will require new telescopes.
Hence, goals in the field are to both lower the threshold to the 30-50 GeV
band, and improve the flux sensitivity by a factor of ten.

\subsection[Relativistic jets from binaries]{Relativistic Jets from \\Binaries}

One of the most exciting recent discoveries in high-energy astrophysics
is the detection of TeV emission from binaries systems containing a
compact object, either a neutron star or black hole (see
Fig.~\ref{fig:binaries}).  TeV emission requires particles at TeV or
higher energies and promises to give unique insights into the
acceleration of ultrarelativistic particles in X-ray binaries.  The TeV
emission is found to be strongly time varying.  Hence, multiwavelength
(TeV, GeV, X-ray, optical, and radio) light curves will strongly
constrain models of high-energy particle acceleration and interaction
within these systems.

Key questions that will be addressed by TeV observations include:

$\bullet$~ What is the composition of ultra-relativistic jets?  Even though
ultra-relativistic jets are ubiquitous features of compact objects,
occurring in systems ranging from supermassive black holes to neutron
stars, the basic question of whether the jets are electron-positron or
have a significant hadronic component remains unanswered for almost all
objects.  The only case with a clear signature of the composition is SS
433, in which X-ray line emission reveals the presence of iron nuclei. 
However, even for SS 433, the matter may be entrained from the companion
star wind.  This question is fundamental in understanding the physics of
jet production.  Measurement of the time variation of the TeV/GeV/X-ray
spectrum from TeV emitting binaries has the potential to resolve this
question.

$\bullet$~ What is the total energy carried by jets?  TeV emission provides
a unique probe of the highest energy particles in a jet.  These
particles often dominate the total energy of the jet and their accurate
measurement is essential in understanding the energetics of jets.

$\bullet$~ What accelerates particles in jets?  Measuring the
acceleration time and the spectrum of the highest energy particles in a
jet is critical for addressing this question.

\subsubsection[Current status]{Current Status}

The first evidence that binary systems containing stellar-mass compact
objects could accelerate particles to TeV energies came from
observations of X-ray synchrotron radiation from the large-scale jets of
XTE J1550-564 \cite{Corbel02,Kaaret03}.  The detection of deceleration
in these jets suggests that the high-energy particles are accelerated by
shocks formed by the collision of the jet with the interstellar medium. 
The acceleration is likely powered by the bulk motion of the jets. More
recently, three TeV-emitting compact-object binaries have been found at
high confidence.  One, PSR B1259-63 contains a young, rotation-powered
pulsar \cite{hess1259}.  The nature of the other two systems, LS 5039
and LS I 61 303 \cite{magic_ls61} is less clear.  A lower significance
signal (3.2$\sigma$ after trials) has been reported from the black hole
X-ray binary Cyg X-1 \cite{magic_cygx1}.

PSR B1259-63 consists of a young, highly energetic pulsar in a highly
eccentric, 4.3 year orbit around a luminous Be star.  At periastron the
pulsar passes within about 1 A.U.\ of its companion star.  Radio and
hard X-ray emission, interpreted as synchrotron radiation, from the
source suggest that electrons are accelerated to relativistic energies,
mostly likely by shocks produced by interaction of the pulsar wind with
the outflow from the Be star \cite{Tavani96}.  However, the electron
energy and magnetic field strength cannot be determined independently
from the X-ray and radio data and alternative interpretations of the
X-ray emission are possible.  H.E.S.S. detected TeV emission from PSR
B1259-63 \cite{hess1259}.  TeV emission was detected over observations
within about 80 days of periastron passage and provides unambiguous
evidence for the acceleration of particles to TeV energies. 

LS I +61 303, a high mass X-ray binary system located at $\sim$2 kpc
distance which has been a source of interest for many years due to its
periodic outbursts in radio and X-ray correlated with the $\sim$26.5 day
orbital cycle and its coincidence with a COS-B and EGRET GeV gamma-ray
source \cite{Casares05,Leahy97,Taylor96}.  MAGIC found variable TeV
emission from this source \cite{magic_ls61}.  The nature of the compact
object in LS I +61 303 is not well established.   The identification of
LS I +61 303 as a microquasar occurred in 2001 \cite{Massi01} when what
appeared to be relativistic, precessing radio jets were discovered
extending roughly 200 AU from the center of the source.  However,
recent repeated VLBI imaging of the binary shows what appears to be the
cometary tail of a pulsar wind interacting with the wind from the
companion star.  This suggests that the binary is really a pairing of a
neutron star and a Be main sequence star \cite{Dhawan06}.  The (much)
shorter orbital period of LS I +61 303, as compared to PSR B1259-63,
makes the system much more accessible for observations.  Also, the
detection of LS I +61 303 at GeV energies will enable constraints on the
modeling which are not possible for PSR B1259-63.

H.E.S.S. has detected TeV emission from the high-mass X-ray binary LS 5039
\cite{hess_ls5039}.  The TeV spectral shape varies with orbital phase.
LS 5039/RX J1826.2-1450 is a high-mass X-ray binary. Radio jets from LS
5039 have been resolved using the Very Long Baseline Array
\cite{Paredes00,Bosch05}.  This suggests that the compact object is
accreting.  Optical measurement of the binary orbit also suggests a
black hole, although the measurements do not strongly exclude a neutron
star \cite{Casares05}.

\subsubsection{Measurements needed}

It should be possible to determine the correct emission mechanism for
the TeV emission in both neutron-star and black hole binaries via
simultaneous multiwavelength (radio, X-ray, GeV, TeV) observations of
the time variable emission.  Important in this regard will be measuring
how the various emission components vary with orbital phase. The key
here is adequate cadence, which requires good sensitivity even for short
observations.  Understanding the correct emission mechanism will place
the interpretation of the TeV observations on a firm footing and allow
one to use them to make strong inferences about the jet energetics and
the populations of relativistic particles in the jets.  If the TeV
emission from a given system can be shown to arise from
interactions of relativistic protons with a stellar wind, then this
would show that the jet contains hadrons.  This would provide a major
advance in our understanding of the physics of jets.

If the jets do have a significant hadronic component, then they are
potential neutrino sources.  The calculated neutrino flux levels,
assuming a hadronic origin for the observed TeV emission, are detectable
with neutrino observatories now coming on line, such as ICECUBE
\cite{Torres06}.  The detection of neutrinos from a compact object
binary would be very exciting in opening up the field of neutrino
astronomy and would be definitive proof of a hadronic jet.

Detailed light curves will also allow us to extract information about
the interaction of the pulsar wind or black hole jet with the  outflow
from the stellar companion.  This is a very exciting possibility which
will provide a direct confrontation of magnetohydrodynamical simulations
with observation and significantly advance our understanding of
time-dependent relativistic shocks.  The knowledge gained will be
important for essentially all aspects of high-energy astrophysics. If
the broad-band spectrum of PSR B1259-63 is modeled assuming that the TeV
photons are produced by inverse-Compton interactions of photons from the
companion star with the same population of accelerated electrons
producing the synchrotron emission, then the TeV data break the
degeneracy between electron energy and magnetic field and allow the
magnetic field to be estimated to be $\sim 1$~G.  This estimate is
similar to the values predicted by magnetohydrodynamical simulations of
the pulsar wind.  Future more sensitive observation would enable
measurement of the time evolution of the magnetic field.

The detection of TeV emission from a black hole binary, perhaps already
accomplished, would have important implications.  Acceleration of
particles to TeV energies is required to produce the TeV emission.  It
is unlikely that such acceleration occurs in the accretion disk or
corona; the particle acceleration likely occurs in the jet.  The same is
not true about X-ray or hard X-ray emission.  This is significant
ambiguity about whether any X-ray/hard X-ray spectral component can be
attributed to the jet, and the strong X-ray flux from the accretion
disk complicates isolation of any jet emission.  This makes TeV emission
a unique probe of the properties of jet and observation of TeV
gamma rays from the jets of accreting stellar-mass black holes should
lead to important information about the jet production mechanism.

There are two possible mechanisms for the generation of the TeV
emission.  Electrons accelerated to very high energies may
inverse-Compton scatter photons emitted from the O6.5V companion star.
However, the radiation density from the O star companion at the position
of the compact object is very high and the radiative time scale is $\sim
300$~s.  Very rapid acceleration would be required for the electrons to
reach the high energies required in the face of such rapid energy loss. 
Instead, the TeV emission may arise from the interactions of protons
accelerated in a jet with the stellar wind.  

Even with the ambiguity between an electron versus proton mechanism for
the TeV emission, the luminosity in the TeV band indicates an extremely
powerful outflow.  For very efficient, $\sim 10$\%, conversion of bulk
motion into VHE radiation, the jet power must be comparable to X-ray
luminosity.   For more typical acceleration efficiencies at the level
of a few percent, the energy in the outflow would be several times the
X-ray luminosity.  The result has major implications for our
understanding of accretion flows near black holes.  The balance between
accretion luminosity and jet power is currently a major question in the
study of microquasars, but estimation of the total jet kinetic energy
from the observed radio luminosity is uncertain \cite{Fender03}. 
Recently, a radio/optical ring was discovered around the long-known
black hole candidate Cyg X-1 \cite{Gallo05}.  The ring is powered by a
compact jet and acts as a calorimeter allowing the total jet kinetic
energy to be determined (the energy radiated by the jet is negligible). 
The jet power is between 7\% and 100\% of the X-ray luminosity of the
system.  This implies that the jet is a significant component of the
overall energy budget of the accretion flow.  It is remarkable that a
similar inference can be made directly from the observed TeV luminosity
of LS 5039.  This suggests that additional TeV sources of black hole
jet sources will be important in understanding the balance between
accretion luminosity and jet power and the fundamental role of jet
production in accretion dynamics.

A future TeV instrument with improved sensitivity would enable
observation of sources at lower luminosities than those currently
known.  An important current question in the study of Galactic black
hole is how the ratio of jet power to X-ray luminosity varies as a
function of accretion rate.  The observed relation between X-ray and
radio flux for black holes producing compact jets \cite{Corbel00} has
been interpreted as evidence that the jet dominates the accretion flow
at low accretion rates \cite{Fender03}.  Sensitive TeV observations
should enable us to directly probe this relation; the strategy would be
to observe a black hole transient in the X-ray and TeV bands as it decays
back to quiescent after an outburst.  This would provide important
information on the nature of the accretion flow at low luminosities
which would impact the question of whether the low quiescent
luminosities of black holes are valid evidence for the existence of
event horizons and also the effect of (nearly) quiescent supermassive
black holes (such as Sgr A*) on the nuclei of galaxies.

\subsection{Required instrument performance}

For the study of PWNe, the performance drivers are improved sensitivity,
angular resolution, and extension of the spectral coverage up to
100~TeV.  In order to detect large populations of fainter sources,
improved sensitivity in the band around 1~TeV is essential.  The
properties of PWNe and the resident pulsars vary significantly and a
large sample of sources is needed to fully understand these objects and it is
essential to use them as probes of pulsar astrophysics.  Improved
angular resolution, with sufficient counting statistics to make
effective use of the resolution, is needed to accurately map the TeV
emission.  Radio and X-ray maps are available with arcseconds precision
which cannot be matched in the TeV band.  However, angular resolution
sufficient to produce multiple pixel maps of the TeV emission is
adequate to map the distribution of high-energy particles as needed to
understand their diffusion within PWNe.  Extension of the spectral
coverage up to 100~TeV would enable us to measure the spectral break and
determine the highest energies to which particles are accelerated.  This
would provide fundamental information on the physics of the acceleration
process.

Since many pulsar spectra cut off below 10~GeV, extension of the energy
range down to the lowest energies possible is important for the study of
pulsed emission.  Detection of the pulsed emission from a significant
number of pulsars will likely require sensitivity below 50~GeV. 
However, a search for the inverse Compton component predicted in outer
gap models to lie at TeV energies will provide important constraints on
models.

For the study of binaries, both neutron star and black hole, sensitivity
is the main driver in order to detect additional sources and to study
known objects with high time resolution.  A factor of ten increase in
sensitivity in the `canonical' TeV band (0.2-5~TeV) should significantly
increase the number of binaries which are detected in the TeV band,
permitting studies of how TeV emission correlates with binary
properties; i.e., spin-down power, orbital separation, and companion
star type.  This will provide insights into the mechanism which produces
the TeV photons.  

Increased sensitivity is essential to study binaries at faster cadence. 
All of the binary sources are variable and the differing time
evolution at different wavebands will likely be the key to understanding
the dynamics of particle acceleration and TeV photon production in these
systems.  In addition, the ability to monitor a given source on a daily
basis for long periods is essential to allow studies of the dependence
of the TeV flux on orbital phase.  To search for jet emission from
quiescent black holes, a  flux sensitivity $10^{-14} \rm \, erg \,
s^{-1}$ in the 0.25-4~TeV band is required.

%



\clearpage

\section{Dark matter searches with a future VHE gamma-ray observatory}
\label{dms-subsec}
Group membership: \\ \\
J.H. Buckley, Edward Baltz, Gianfranco Bertone, Francesc Ferrer,
Paolo Gondolo, Savvas Koushiappas, Stefano Profumo,
Vladimir Vassiliev, Matthew Wood, Gabrijela Zaharijas
\normalsize


\subsection{Introduction}
In the last decade, a standard cosmological picture of the universe (the
$\Lambda$CDM cosmology) has emerged, including a detailed breakdown of the main
constituents of the energy-density of the universe.  This theoretical framework
is now on a firm empirical footing, given the remarkable agreement of a diverse
set of astrophysical data \cite{Spergel:2006hy,Percival:2006gt}.  In the
$\Lambda$CDM paradigm, the universe is spatially flat and its energy budget is
balanced with $\sim$4\% baryonic matter, $\sim$26\% cold dark matter (CDM) and
roughly 70\% dark energy.

\begin{figure*}[tb]
\label{fig:halos}
\begin{center}
\includegraphics[width=0.96\textwidth]{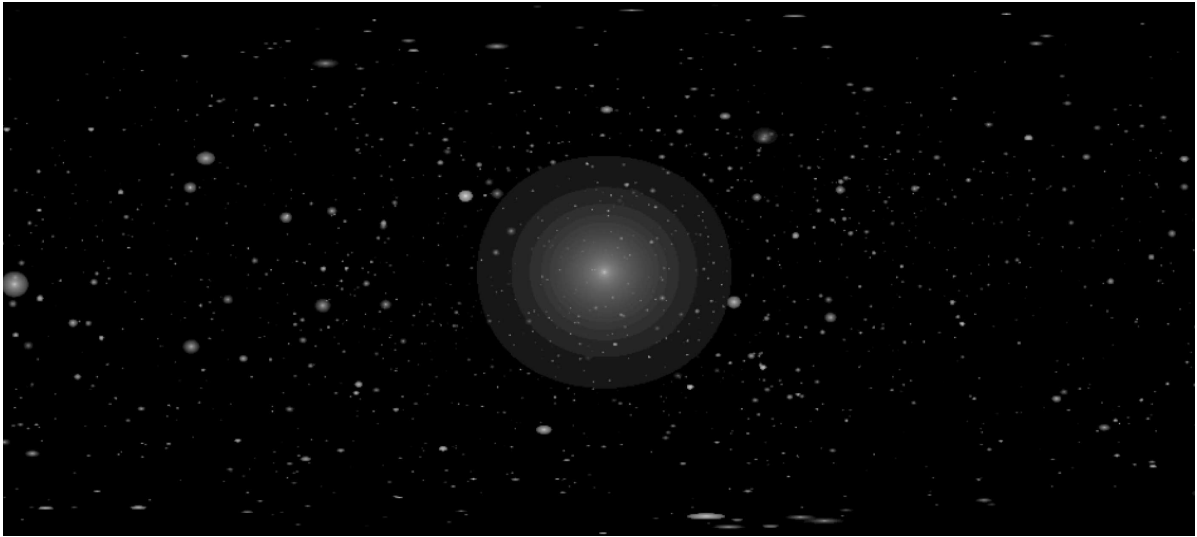}
\end{center}
\caption{Simulated appearance of the gamma-ray sky from neutralino annihilation
in the galactic halo plotted
as the intensity in galactic coordinates \cite{baltz_p5_06}.  The galactic
center appears as the bright object at the center of the field of view.
If the sensitivity of a future ACT experiment were high enough, a number of the
other galactic substructures visible in this figure could be detected with
a ground-based gamma-ray experiment.}

\end{figure*}

While the dark matter has not been directly detected in laboratory experiments,
the gravitational effects of dark matter have been observed in the Universe on
all spatial scales, ranging from the inner kiloparsecs of galaxies out to the
Hubble radius.  The Dark Matter (DM) paradigm was first introduced by Zwicky ~\cite{Zwicky1933} in
the 1930s to explain the anomalous velocity dispersion in
galaxy clusters.

In 1973, Cowsik and McClelland \cite{CowsikMcClelland(1973)} proposed that
weakly-interacting massive neutrinos could provide the missing dark matter
needed to explain the virial mass discrepancy in the Coma cluster. However,
since neutrinos would be relativistic at the time of decoupling, they would
have a large free-streaming length. While neutrino dark matter would provide an
explanation for structure on the scale of clusters, this idea could not explain
the early formation of compact halos that appear to have seeded the growth of
smaller structures, such as galaxies.

This observation motivated the concept of cold dark matter (CDM) consisting of
weakly interacting massive particles (WIMPs) with rest energy on the order of
100 GeV that were nonrelativistic (cold) at the time of decoupling.  CDM would
first form very small, dense structures that coalesced into progressively larger
objects (galactic substructure, galaxies, then galaxy clusters and
superclusters) in a bottom-up scenario known as hierarchical structure
formation.  A plethora of diverse observations suggests the presence of this
mysterious matter: gravitational lensing, the rotation curves of galaxies,
measurements of the cosmic microwave background (CMB), and maps of the
large-scale structure of galaxies.

Measurements of the CMB have been the key to pinning
down the cosmological parameters; the angular distribution of temperature
variations in the CMB depends on the power spectrum of fluctuations produced in
the inflationary epoch and subsequent acoustic oscillations that resulted from
the interplay of gravitational collapse and radiation pressure.  These acoustic
peaks contain information about the curvature and expansion history of the
universe, as well as the relative contributions of baryonic matter, dark matter
and dark energy. Combined with measurements of the large-scale distribution of
galaxies, as mapped by the Sloan Digital Sky Survey (SDSS) and the 2dF Galaxy
Redshift survey, these data can be well described by models based on single field inflation.

Observations of galactic clusters continue to be of central importance in
understanding the dark matter problem.  Recent compelling evidence for the
existence of particle dark matter comes from the analysis of a unique cluster
merger event 1E0657-558 \cite{clowe06}.  Chandra observations reveal that the
distribution of the X-ray emitting plasma, the dominant component of the
visible baryonic matter, appears to be spatially segregated from the
gravitational mass (revealed by weak lensing data).  This result provides
strong evidence in favor of a weakly-interacting-particle dark matter, while
contradicting other explanations, such as modified gravity.

The primordial abundances of different particle species in the Universe are
determined by assuming that dark matter particles and all other particle
species are in thermal equilibrium until the expansion rate of the Universe
dilutes their individual reaction rates.  Under this assumption (which provides
stunningly accurate estimates of the abundance of light elements and
standard-model particles), particles that interact weakly fall out of
equilibrium sooner, escaping Boltzmann suppression as the temperature drops, and
hence have larger relic abundances in the current universe.
While a weakly-interacting thermal relic provides an appealing and
well-constrained candidate for the dark matter, nonthermal relics such as
axions or gravitinos, resulting from the decay of other relics, can also provide
contributions to the total matter density or even provide the dominant
component of the dark matter.  


Just as there is an unseen component of the universe required by astrophysical
observations, there are compelling theoretical arguments for the existence of
new particle degrees of freedom in the TeV to Planck scale energy range.  In
particle physics, a solution to the so-called hierarchy problem (the question
of why the expected mass of the Higgs particle is so low) requires new physics.
An example is provided by supersymmetry, a symmetry in nature between Fermions
and bosons, where the supersymmetric partners of standard model particles lead
to cancellations in the radiative corrections to the Higgs mass.  The hierarchy
problem in particle physics motivates the existence of new particle degrees of
freedom in the mass range of ~100~GeV to TeV scale.  It is a remarkable
coincidence that if dark matter is composed of a weakly interacting elementary
particle with an approximate mass of this order (i.e., on the scale of the weak
gauge bosons $\sim 100$ GeV), one could naturally produce the required
cosmological density through thermal decoupling of the DM component.
To make a viable candidate for the dark matter, one more ingredient is required;
the decay of such a particle must be forbidden by some
conserved quantity associated with an, as yet, undiscovered symmetry of Nature
so that the lifetime of the particle is longer than the Hubble time.


In supersymmetry, if one postulates a conserved quantity arising from some new
symmetry (R-parity), the lightest supersymmetric particle (LSP) is stable and
would provide a natural candidate for the dark matter. 
In fact, R-parity conservation is introduced into supersymmetry not to
solve the dark matter problem, but rather to ensure the stability of
the proton.
In many regions of supersymmetric parameter space, the LSP is the neutralino,
a Majorana particle (its own antiparticle) that is
the lightest super-symmetric partner to the electroweak and Higgs bosons.  

For a subset of the supersymmetric parameter space, these particles could be
within the reach of experimental testing at the Large Hadron Collider (LHC) (if
the rest mass is below about 500~GeV) \cite{baltz06}  or current or future
direct detection experiments XENON-I,II \cite{aprile05}, GENIUS
\cite{klapdor02,klapdor05} ZEPLIN-II,III,IV \cite{bisset07},
SuperCDMS\cite{akerib05}, and EDELWEISS-I,II\cite{sanglard07} (if the nuclear
recoil cross-section is sufficiently large).  While it is possible that the LHC
will provide evidence for supersymmetry, or that future direct detection
experiments will detect a clear signature of nuclear-recoil events produced by
dark matter in the local halo, {\emph{gamma-ray observations provide the only
avenue for measuring the dark matter halo profiles and illuminating the role of
dark matter in structure formation.}} 

Neutralinos could also be observed through other indirect astrophysical
experiments searching for by-products of the annihilation of the lightest
supersymmetric particle, such as positrons, low-energy antiprotons, and
high-energy neutrinos.  Since positrons and antiprotons are charged particles,
their propagation in the galaxy suffers scattering off of the irregular
inter-stellar magnetic field and hides their origin.  Electrons with energy
above $\sim$10~GeV suffer severe energy losses due to synchrotron and
inverse-Compton radiation, limiting their range to much less than the distance
between Earth and the galactic center.  However, cosmic-ray observations could
provide evidence for local galactic substructure through characteristic
distortions in the energy spectra of these particles. 
Detection of electrons from dark matter annihilation thus depend critically
on large uncertainties in the clumpiness of the local halo. 
Neutrinos would not suffer these difficulties and, like photons, would point
back to their sources.  But given the very low detection cross section compared
with gamma-rays, the effective area for a $\sim$km$^3$ neutrino experiment is
many orders of magnitude smaller than for a typical ground-based gamma-ray
experiment.  While detection of neutrinos directly from discrete sources (e.g.,
the Galactic center) would be difficult for the current generation of neutrino
detectors there is a reasonable prospect for detection of neutrinos from WIMPs
s in the local halo that are captured by interactions with the earth or sun
where they might have sufficient density to give an observable neutrino signal.
{Compared with all other detection techniques (direct and indirect),
$\gamma$-ray measurements
of dark-matter are unique in going beyond a detection of the local
halo to providing a measurement of the actual distribution of dark matter on
the sky.  Such measurements are needed  to understand the nature of the
dominant gravitational component of our own Galaxy, and the role of dark matter
in the formation of structure in the Universe.}


In other regions of supersymmetric parameter space, the dark matter particle
could be in the form of a heavy scalar like the sneutrino, or Rarita-Schwinger
particles like the gravitino.  In general, for gravitino models, R-parity
need not be conserved
and gravitinos could decay very slowly (with a lifetime on the
order of the age of the universe) but could still be visible in gamma-rays
\cite{ibarratran08}.
Supersymmetry
is not the only extension to the standard model of particle
physics that provides a dark matter candidate, and there is no guarantee that
even if supersymmetry is discovered it will provide a new particle that
solves the dark matter problem. 
Other extensions of the standard model involving TeV-scale extra dimensions,
include new particles in the form of
Kaluza-Klein partners of ordinary standard-model particles.  The
lightest Kaluza-Klein particle (LKP) 
could be stable and hence provide a candidate for the dark matter
if one invokes an absolute symmetry (KK parity
conservation) resulting from momentum conservation along the extra dimension.
The mass of the lightest Kaluza-Klein particle (e.g, the $B^{(1)}$ particle
corresponding to the first excitation of the weak hypercharge boson) is related
to the physical length scale of the extra dimension and could be on the
TeV-scale (but not much smaller) and provide a viable CDM candidate.  
The $B^{(1)}$
is expected to annihilate mainly to quarks or charged leptons
accompanied by an internal bremsstrahlung photon by the process
$B^{(1)}+B^{(1)}\rightarrow l^+ + l^- + \gamma$
\cite{bergstrom06}. 
The high energy of the LKP ($\gsim$1 TeV), and very-hard spectrum gamma-ray
production make ground-based gamma-ray and high-energy cosmic-ray electron
measurements promising avenues for discovery. 

As an interesting aside, TeV-scale extra dimensions may also manifest
themselves in a dispersion in the propagation velocity of light in
extragalactic space \cite{amelino98}.  Observations of the shortest flares, at
the highest energies from the most distant objects can place tight constraints
on theories with large extra dimensions.  Such constraints have already been
produced by TeV measurements \cite{biller99} and could be dramatically improved
with a future higher-sensitivity gamma-ray instrument, capable of detecting
shorter flares from distant AGNs and GRBs.  {Thus, ground-based TeV
gamma-ray astronomy probes TeV-scale particle physics both by providing a
possible avenue for detection of a Kaluza-Klein particle and by constraining
the 
the TeV$^{-1}$-scale
 structure of space-time from gamma-ray propagation effects.}

A new class of theories (the so-called ``little Higgs'' or LH models)
has been proposed to extend the standard model to the TeV scale
and offer an explanation for the lightness of the Higgs.
The LH models predict a light (possibly composite) Higgs boson as well
as other TeV-scale particles that could provide candidates
for the dark matter in the $\sim$100~GeV or $\gsim$500~GeV mass range 
\cite{birkedal06}.
However, only a small subset of the LH models have weak-scale masses
and interactions together with a symmetry principle
that protects the stability of the particle on a lifetime comparable
to the age of the universe.  In fact, for the composite
Higgs, the particles (like their analog, the neutral pion) could decay with
relatively short lifetimes.  Still, this class of models (like other new
physics at the TeV scale) could provide a viable dark matter candidate
with an observable gamma-ray signature.

\begin{figure}[tb]
\label{fig:bringmann}
\includegraphics[width=0.48\textwidth]{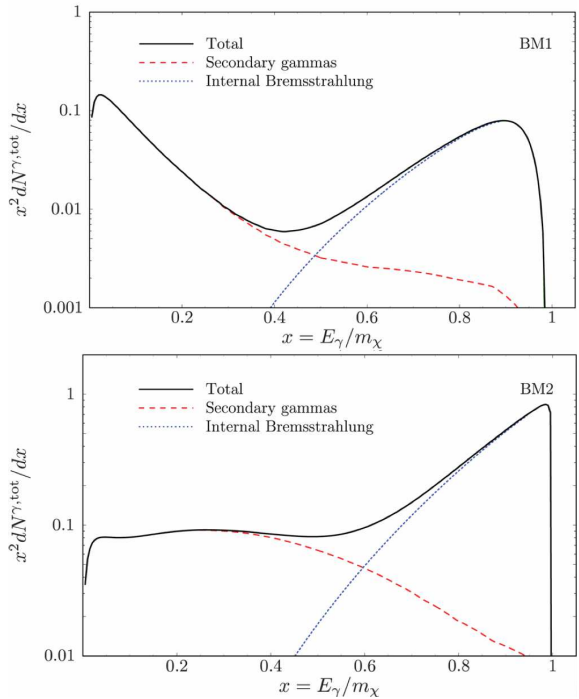}
\caption{Continuum emission from neutralino annihilation from mSUGRA models.}

\end{figure}

%
\begin{figure}[tb]
\includegraphics[width=0.48\textwidth]{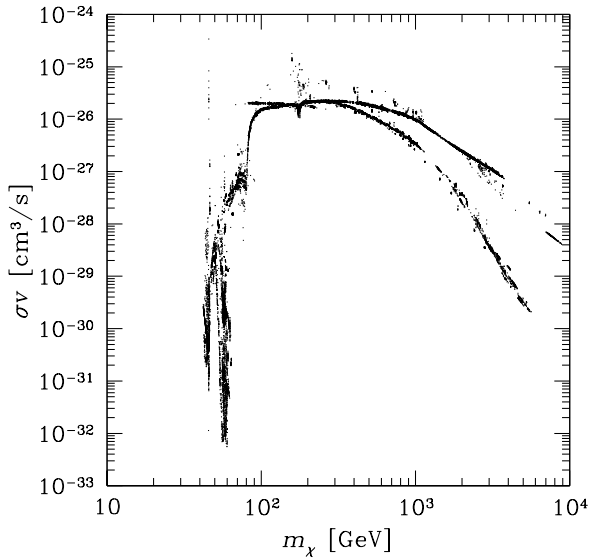}
\caption{
Scatter plot of neutralino annihilation cross section versus neutralino
mass for supersymmetric models that satisfy accelerator and WMAP constraints.
A typical cross-section (assumed in our estimates) is $\sigma v \approx
2\times 10^{-26}{\rm cm}^3 {\rm s}^{-1}$.
\label{fig:gondolosigma}}
\end{figure}
The recent discoveries of neutrino mass from measurements of atmospheric and
solar neutrinos may also have a bearing on the prospects for gamma-ray
detection of dark-matter.  While the primordial density of light standard-model
(SM) neutrinos $\nu_e$, $\nu_\mu$ and $\nu_\tau$ will provide a very small
hot-dark-matter contribution to the energy budget of the universe, they are
ruled out as candidates for the CDM component needed to explain structure
formation.  However, a new heavy neutrino (or the superpartner thereof)
may provide a viable candidate for
the CDM.  Krauss, Nasri and Trodden \cite{kraussnasri} proposed that a
right-handed neutrino with TeV mass could play a role in giving masses to
otherwise massless standard model neutrinos through high-order loop
corrections.  This model is a version of the Zee model \cite{zee80} that has
been successfully applied to results on solar and atmospheric neutrino
observations to explain the observed parameters of the mass and mixing matrix.
A discrete $Z_2$ symmetry, and the fact that the right-handed Majorana neutrino
$N_R$ is typically lighter than the charged scalars in the theory, make the
massive neutrino stable, and a natural dark matter candidate
\cite{baltzbergstrom}.  Direct annihilation to a gamma-ray line $N_R
N_R\rightarrow \gamma\gamma$ with a cross-section $\langle\sigma_{N_R
N_R\rightarrow \gamma\gamma} v\rangle \approx 10^{-29}{\rm cm}^{3}{\rm s}^{-1}$
is at the limit of detectability and direct annihilation to charged leptons is
also expected to give a very small cross-section.  However, \cite{baltzbergstrom} have
shown that internal bremsstrahlung can give rise to an observable gamma-ray
continuum from decays to two leptons and a gamma-ray $N_R N_R\rightarrow l^+
l^- \gamma$.  The three-body final state gives rise to a very hard spectrum
that peaks near the $N_R$ mass, then drops precipitously.  Unlike direct
annihilation to leptons, this non-helicity-suppressed process can have a large
cross-section, with an annihilation rate a factor of $\alpha/\pi$ (where $\alpha$ is the fine structure constant) times the
annihilation rate at freeze-out (with cross section $\langle \sigma v\rangle
\approx 3\times 10^{-26}{\rm cm}^3{\rm s}^{-1}$), and orders of magnitude lager
than the helicity-suppressed two-body $N_R N_R\rightarrow l^+ + l^-$ rate
typically considered in the past \cite{baltzbergstrom}.

Recently, Bringmann, Bergstr\"om and Edsj\"o \cite{bringmann07} have pointed out
that internal-bremsstrahlung process 
could also play a role in neutralino annihilation, and in some cases result in
a large enhancement in the continuum gamma-ray signal for certain model
parameters.  Fig.~\ref{fig:bringmann}
shows the continuum emission from neutralino annihilation from mSUGRA models
with particularly pronounced IB features, that could be observed in the
gamma-ray spectrum. 
{There are a number of different particle physics and
astrophysical scenarios that can lead to the production of an observable
gamma-ray signal with a spectral form
that contains distinct features that can
be connected, with high accuracy, to the underlying particle physics}.

In what follows, we focus on predictions for the neutralino.  While we show
detailed results for the specific case of SUSY models and the neutralino, for
any theory with a new weakly interacting thermal relic (e.g., the LKP) the
model parameter space is tightly constrained by the observed relic abundance
and hence the results for the overall gamma-ray signal level are fairly generic
for any WIMP candidate.  In the case of neutralino dark matter, the
cross-sections for annihilation have been studied in detail by a number of
groups.  Fig.~\ref{fig:gondolosigma} shows the cross-section calculated for a
range of parameters in supersymmetric parameter space as a function of mass.
Only points that satisfy accelerator constraints and are compatible with a
relic abundance matching the WMAP CMB measurements are shown.  At high
energies, the neutralino is either almost purely a Higgsino (for mSUGRA) or
Wino (for anomaly-mediated SUSY breaking) resulting in the relatively narrow
bands.  {Thus, the annihilation cross-section predictions for gamma-ray
production from  higher energy ($\sim$100~GeV--TeV) candidates are well
constrained, with the particle-physics uncertainty contributing $\sim$ one
order of magnitude to the range of the predicted gamma-ray fluxes.} 

We elaborate further on the potential of $\gamma$-ray experiments to play a
pivotal role in identifying the dark matter particle and in particular, how a
next-generation $\gamma$-ray experiment can in fact provide information on the
actual formation of structure in the Universe.

\subsection{Dark Matter Annihilation into $\gamma$-rays, and the uncertainties in 
the predicted flux}


For any of the scenarios that have been considered, the dark-matter particle 
must be neutral and does not couple directly to photons, however 
most annihilation channels ultimately lead to the production of photons through
a number of indirect processes.  While the total cross-section for gamma-ray
production is constrained by the measured relic abundance of dark matter, the
shape of the gamma-ray spectrum is sensitive to the details of the specific
particle-physics scenario.
Summarizing the 
previous discussion,
dark matter annihilation may yield photons in three ways: (1) by the direct
annihilation into a two-photon final state (or a $Z^0 \gamma$ or $H\gamma$
final state) giving a nearly monoenergetic line,  (2) through the annihilation
into an intermediate state (e.g. a quark-antiquark pair), that subsequently
decays and hadronizes, yielding photons through the decay of neutral pions and
giving rise to a broad featureless continuum spectrum or (3) through
internal-bremsstrahlung into a three-particle state, e.g. $\chi\chi\rightarrow
W^+W^-\gamma$ yielding gamma-rays with a very hard spectrum and sharp cutoff.
The cross section for the direct annihilation into two photons, or a photon and
$Z^0$ are loop-suppressed and can be at least 2 orders of magnitude less than the processes that lead
to the continuum emission.
However, for some cases of interest (e.g., a massive
Higgsino) the annihilation line can be substantially enhanced.
Also, in the next-to-minimal supersymmetric standard model (NMSSM) with an
extended Higgs sector, one-loop amplitudes for NMSSM neutralino pair
annihilation to two photons and two gluons, extra diagrams with a light CP-odd
Higgs boson exchange can strongly enhance the cross-section for the
annihilation line.  Such models have the added feature of providing a mechanism
for electroweak baryogenesis \cite{ferrer06}.  {{By combining Fermi
measurements of the continuum, with higher energy constraints from ground-based
ACT measurements, one can obtain constraints on the line to continuum ratio
that could provide an important means of discriminating between different
extensions to minimal supersymmetry or other
dark matter scenarios}}

In general, the flux of $\gamma$-rays from a high-density annihilating
region can be written as 
\begin{equation} 
\frac{dN_\gamma}{dAdt} =L  {\cal P} 
\end{equation}
where,
\begin{equation} 
L  = \frac{1}{4 \pi} \int_{\rm LOS} \rho^2(r) dl 
\label{eq:ldef}
\end{equation}
contains the dependence to the
distribution of dark matter, and
\begin{equation} 
{ \cal P} = \int_{{E_{th}}}^{{M_\chi}}\sum_i \frac{
\langle \sigma v \rangle_i }{M_\chi^2} \frac{dN_{\gamma, i}}{dE} \, dE
\end{equation}
is the particle physics function that contains the detailed physical
properties of the dark matter particle. The sum over
the index $i$ represents the sum over the different photon production
mechanisms.  (In Eq.~\ref{eq:ldef}, $M_\chi$ is the neutralino mass,
$l$ is the line-of-sight distance
while $r$ is the radial distance from the center of the halo distribution.
Note that this definition of $L$ is similar to the definition of the $J$-factor
used elsewhere in the literature (e.g., \cite{bub98})  

Given the fact that supersymmetry has not been detected yet, the uncertainty in
the value of ${\cal P}$ is rather large. Sampling of the available
supersymmetric parameter space reveals that the uncertainty in cross sections
can be as large as 5 orders of magnitude if one covers the entire mass range
down that extends over several orders of magnitude (see
Fig.~\ref{fig:gondolosigma}), but collapses considerably for $M_\chi \gsim$100
GeV.  For supersymmetric dark matter, ${\cal P}$ can take a {\it maximum} value
of approximately ${\cal P} \approx 10^{-28} \, {\rm cm}^3 {\rm s}^{-1} {\rm
GeV^{-2}}$ when $M_\chi \approx 46 \, {\rm GeV}$, $\sigma v = 5 \times 10^{-26}
\, {\rm cm^3} {\rm s^{-1}}$ and $E_{\rm th}=5 \, {\rm GeV}$ (with a more
typical value of $\approx 2\times 10^{-26} \, {\rm cm}^3{\rm s}^{-1}$ at
energies between 100 GeV and 1 TeV . On the other hand, for a threshold energy
of $E_{\rm th}=50 \, {\rm GeV}$ and a particle mass of $M_\chi \approx 200
\,{\rm GeV}$, the value is ${\cal P} \approx 10^{-31} \, {\rm cm}^3 {\rm
s}^{-1} {\rm GeV^{-2}}$. 

It is important to emphasize that even though the actual value of ${\cal P}$
from supersymmetry can be 
orders of magnitude smaller, in theories
with universal extra dimensions, both the cross section into a photon final
state and the mass of the particle can actually be higher than this value.  

The quantity $L$, on the other hand,
contains all the information about the spatial distribution of
dark matter. Specifically, $L$ is proportional to the
line of sight (LOS) integration of the square of the dark matter density.
Dissipationless N-body simulations suggest the density profiles of dark
matter halos can be described by the functional form
\begin{equation}
\rho (\tilde{r}) = \frac{ \rho_s }{\tilde{r}^\gamma 
( 1 + \tilde{r} ) ^{\delta-\gamma}}
\label{eq:densityfunction}
\end{equation}
where $\tilde{r} = r / r_s$ (e.g., \cite{nfw97,moore99}). 

The quantities $\rho_s$ and $r_s$ are the characteristic density and radius
respectively, while $\gamma$ sets the inner, and $\delta$ the outer slope  of
the distribution. Recent simulations suggest that $\delta \approx 3 $, while
the value  of $\gamma$ has a range of values, roughly $0.7 \le \gamma \le 1.2$
down to $\sim 0.1 \%$ of the virial radius of the halo \cite{Netal,DZMSC}. A
change in the value of the inner slope $\gamma$ between the values of 0.7 and
1.2 for a fixed halo mass results in a change in the value of $L$ that is
roughly 6 times smaller or higher respectively \cite{SKBK06}.  The values of
$\rho_s$ and $r_s$ for a dark matter halo of a
given mass are obtained if one specifies the virial mass and concentration
parameter. In general $\rho_s$ (or the concentration parameter)
depends solely on the redshift of collapse, while $r_s$ depends on both the
mass of the object as well as the redshift of collapse.  In many previous
studies the ``fiducial'' halo profile is that of Navaro, Frenck and White
(NFW; \cite{nfw97}) 
derived from an empirical fit to the halo profile 
determined by N-body simulations and
corresponding to
Eq.~\ref{eq:densityfunction} with $\delta=3$ and $\gamma=1$.

The main difficulty in estimating the value of $L$ for a dark matter halo is
due to the unknown density profiles in the regions from which the majority of
the annihilation flux is emitted.  Experimental data on the inner kiloparsec of
our Galactic (or extragalactic) halos is sparse and theoretical understanding
of these density profiles is limited by our lack of knowledge about the initial
violent relaxation in dark matter halos, and the complicated physics behind the
evolutionary compression of DM during the condensation of baryons in galactic
cores. Both processes still lack a complete theoretical understanding. The
uncertainty in the first is due to the unknown spectrum of density fluctuations
at small spatial scales and difficulties of predicting their evolution in high
resolution numerical simulations.  The uncertainty in the second is due to the
complexity of the gravitational interaction of the dark matter with the
dissipative baryonic matter on small scales and in regions of high density.
Experimentally, measurements of rotation curves and stellar velocity dispersion
are limited by finite angular resolution and geometric projection effects.
While progress is being made on both theoretical and experimental fronts, large
uncertainties remain.

\subsection{Targets for Gamma-Ray Detection}

The Galactic center has been considered the most promising target for the
detection of dark matter annihilation, with a flux more than an order of
magnitude larger than any potential galactic source (e.g., \cite{bub98}).
The detection of $\gamma$-rays from the region of Galactic Center by the
Whipple and H.E.S.S. collaborations~\cite{kosack04, GCHESS2004} can, in
principle, include a contribution from 
annihilating dark matter~\cite{Horns2005}.
While the flux and spectra of
the Whipple and HESS detections are in agreement, the Cangaroo-II group  
reported the detection of high-energy gamma-ray emission from the GC region
\cite{tsuchiya04},
with a considerably softer spectrum that now 
appears to be a transient effect (due to a variable source, or spurious
detection) in view of the latest, detailed HESS results. 
\begin{figure}[tb]
\includegraphics[width=0.48\textwidth]{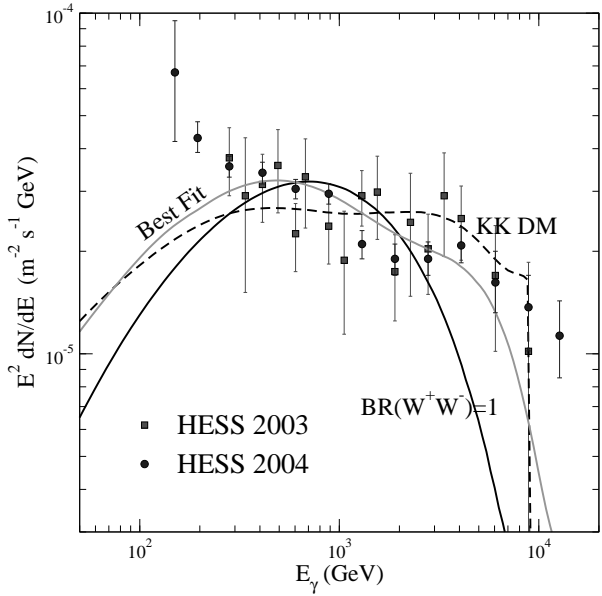}
\caption{
The HESS 2003 (grey squares) and HESS 2004 (filled circles) data on the flux of
GR from the GC, and the best fit to those data with a KK $B^{(1)}$
pair-annihilating lightest KK particle (dashed line), with a WIMP annihilating
into a $W^+W^-$ pair (black solid line), and with the best WIMP spectral
function fit (light grey line).}
\label{fig:dnde}
\end{figure}

\begin{figure}[tb]
\includegraphics[width=0.5\textwidth]{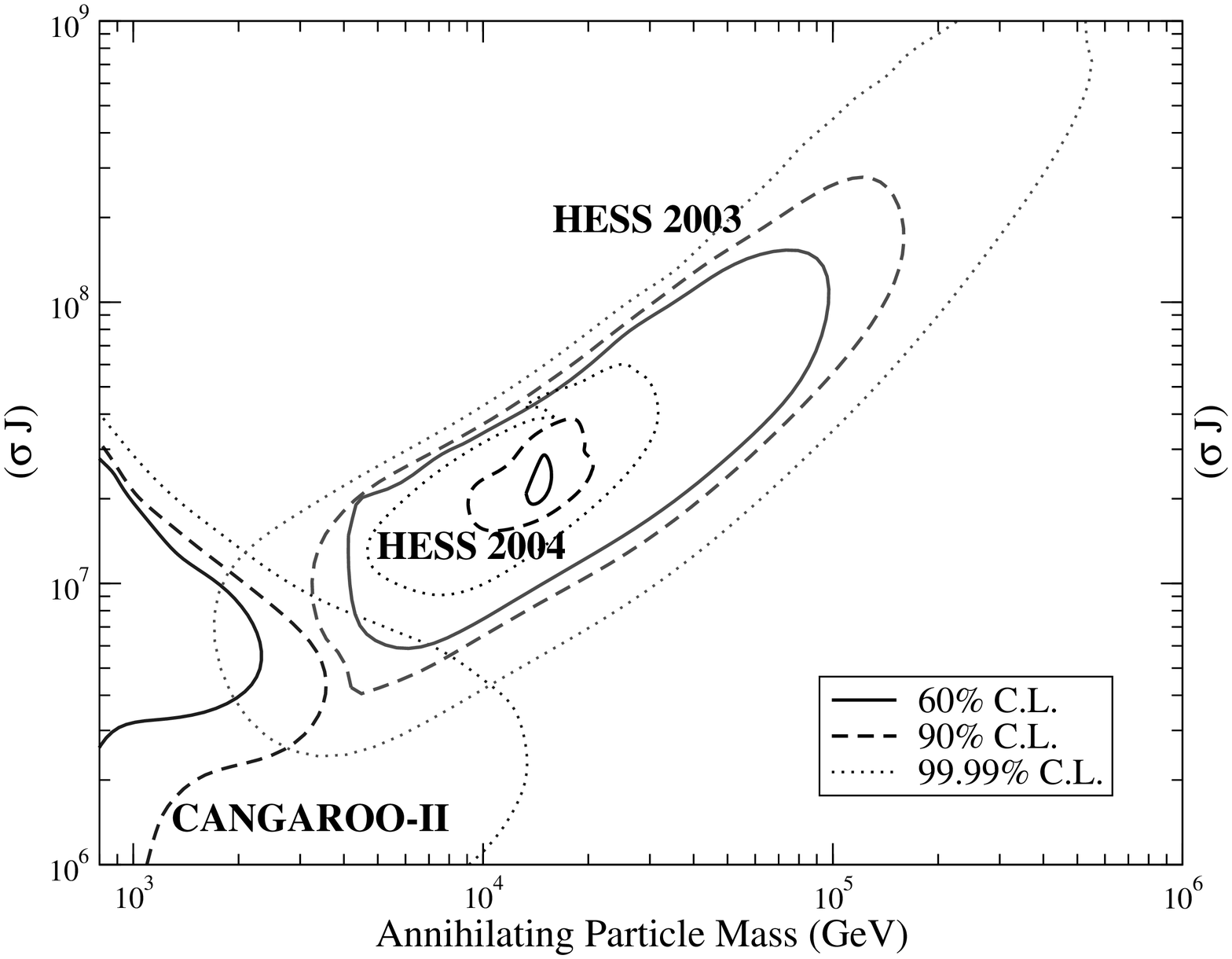}
\caption{
Iso-confidence-level contours of ``{\em best spectral functions}'' fits to the
Cangaroo-II and to the 2003 and 2004 HESS data, in the plane defined by the
annihilating particle mass and by the quantity $(\sigma J)$.
}
\label{fig:chi2_cntr}
\end{figure}

In Ref.~\cite{Profumo:2005xd} the possibility of interpreting the GR data from
the GC in terms of WIMP pair annihilations was analyzed in full generality.
Examples of fits to the HESS data with a Kaluza-Klein (KK) $B^{(1)}$ DM
particle, with WIMPs annihilating into $W^+W^-$ in 100\% of the cases and with
the best possible combination of final states, namely $\sim30$\% into $b\bar b$
and $\sim70$\% into $\tau^+\tau^-$ are shown in fig.~\ref{fig:dnde}. Those
options give a $\chi^2$ per degree of freedom of around 1.8, 2.7 and 1: only
the best-fit model is found to be statistically viable.

Using the Galactic-center data and assuming that the observed gamma-ray
emission arises from dark-matter annihilation,
Profumo \cite{Profumo:2005xd} derived 
confidence intervals for the product of the total annihilation cross-section
$\sigma$ and the $J$-factor (characterizing the astrophysical uncertainty 
from the halo density profile) versus the neutralino mass $m_\chi$.
Iso-confidence-level contours in the $(m_{\chi},(\sigma {J}))$ plane 
are shown in fig.~\ref{fig:chi2_cntr}. From the
figure, it is clear that a dark-matter origin for the emission requires
a DM mass range between 10-20 TeV. Further,
a value of $(\sigma{J})\approx 10^7$ implies either a very large
astrophysical boost factor 
($\approx 10^3$ larger than what expected for a NFW DM profile), or a similar
enhancement in the CDM relic abundance compared with the expectations for
thermal freeze-out

Ref.~\cite{Profumo:2005xd} showed that some supersymmetric models can
accommodate large enough pair annihilation cross sections and masses to both
give a good fit to the HESS data and thermally produce the right DM abundance
even though, from a particle physics point of view, these are not the most
natural models.
An example is a minimal anomaly-mediated SUSY breaking scenario with
non-universal Higgs masses.
For some choices of model parameters, such a dark matter particle
could even be directly detected at ton-sized direct detection
experiments, even though the lightest neutralino mass is in the several TeV
range \cite{Profumo:2005xd}.


However, the interpretation is particularly complicated since the center of our
own Milky Way galaxy has a relatively low mass-to-light ratio and is dominated
by matter in the form of a central massive black hole and a number of
other young massive stars, supernova remnants and compact stellar remnants.
Moreover, the lack of any feature in the power-law spectrum measured b HESS, and
the extent of this spectrum up to energies above 10 TeV makes a dark-matter
interpretation difficult.

A way of dealing with this background is to exclude the galactic center source
seen by HESS, and instead look at an annulus about the Galactic center position
\cite{Stoehr:2003hf, serpico08}.  Even though the background grows in proportion to the
solid angle of the annular region (and the sensitivity degrades as the
square-root of this solid angle) for a sufficiently shallow halo profile, the
signal-to-noise ratio for detection continues to grow out to large angles.
Moreover, any component of diffuse contaminating background falls off more
steeply as a function of latitude than the annihilation of the smooth component
of the dark matter halo. This result may even be enhanced by the presence of
other bound high density structures within the inner parts of the Milky Way
\cite{Diemand:2006ik}. 

\begin{figure*}[tb]\begin{center}
\includegraphics[height=10cm]{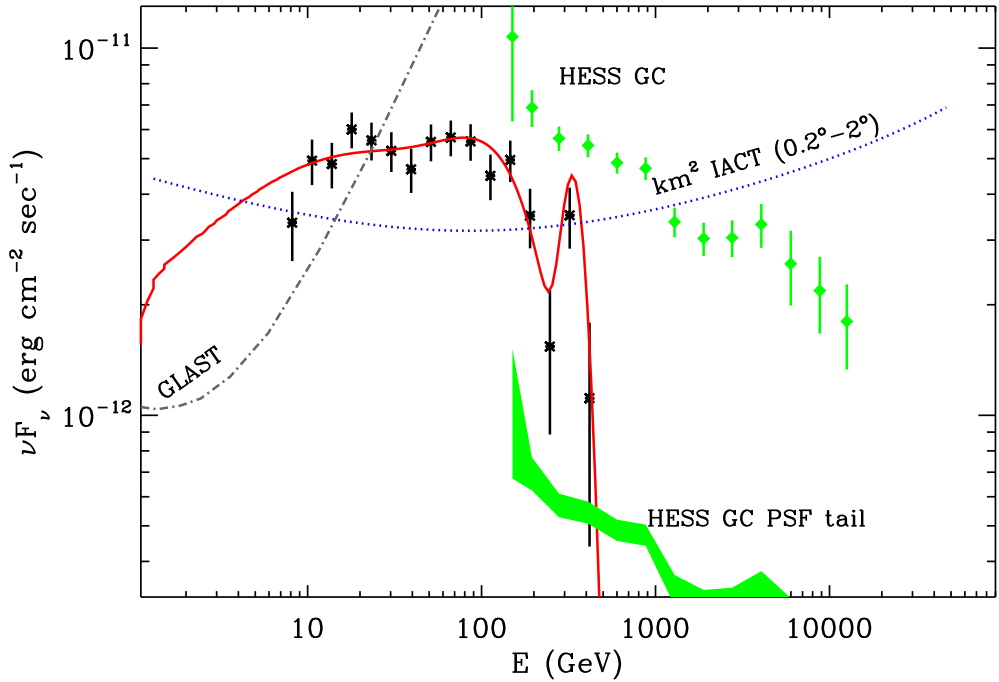}\end{center}
\caption{
Gamma-ray spectrum from dark matter annihilation in an annulus between 0.2$^\circ$
and 2$^\circ$ about the Galactic center assuming an NFW halo with a
central density of $\rho_s = 5.4\times 10^6\, M_\odot/{\rm kpc}^3$ and
a scale radius of $r_s = 21.7\,{\rm kpc}$.   We show the HESS spectrum
of the point source near the GC, and 10\% of this value assumed to bleed
into the annulus from the tails of the gamma-ray point-spread-function.
Here we assume a 200~hour exposure of a
a km$^2$ IACT instrument.  The reduced sensitivity, compared with that for
a point source, comes from integrating the hadronic, electron, and diffuse
gamma-ray background over the relatively large solid angle of the annulus.
}
\label{fig:gcannulus}
\end{figure*}

We make a conservative estimate of the signal from an
annulus centered on the galactic center.  For this calculation, we assume that
the Milky Way halo has a profile as given by Navaro, Frenck and White
\cite{nfw97} (NFW profile) with a scale radius of $r_s = 21.7$~kpc and a
central density of $\rho_s=5.38\, M_\odot\, {\rm kpc}^{-3}$ from Fornengo et
al. \cite{fornengo04}.  To be somewhat more conservative, in light of more
recent N-body simulations that show a flattening of the inner halo profile, we
assume a 10~pc constant density core.  The minimum angle for the annular
region is set by the assumed PSF for a future instrument.  We assume that the
flux from the point source at the GC (or from the diluted contribution from the
galactic ridge emission) will fall below 10\% of the GC value, 0.2 deg from the
position of Sgr~A*.  The optimum angular radius for the outer bound on the
annulus is 12~deg (see \cite{serpico08} for details), somewhat beyond the largest field of view envisioned for a
future imaging ACT (with a more realistic value of 6-8~deg).  As shown in 
Fig.~\ref{fig:gcannulus},
Fermi might also have adequate sensitivity and angular resolution to detect
the continuum emission and separate this from the other point sources.
If the
neutralino mass is large enough (above several TeV) and one chooses favorable
parameters for the annihilation cross-section and density, 
EAS detectors have the large field-of-view required to observe such 
extended sources as well as other regions of emission along the galactic
plane.  However, these detectors lack the good angular and energy resolution
to separate this emission from other point sources and would require follow-up
observations by more sensitive instruments such as imaging ACT arrays.
For the IACT sensitivity, we assume that we have an instrument with effective
area of 1~km$^2$, an exposure of 200~hrs, and that the background comes from
cosmic-ray electrons, cosmic-ray atmospheric showers, and diffuse gamma-rays
following the method given in Ref.~\cite{bub98}.  For the diffuse gamma-ray
spectrum, we take the EGRET diffuse flux, and assume that it continues with a
relatively hard $\sim E^{-2.5}$ spectrum up to TeV energies.  We also assume
that the largest practical angular radius of the annular region is 2~deg, a
reasonable value for a moderately wide-field-of view future instrument.  The
simulated spectrum is calculated for a typical annihilation cross section of
$\langle \sigma v\rangle = 2\times 10^{-26}{\rm cm}^3{\rm s}^{-1}$ and for an
arbitrary set of branching ratios corresponding to 50\% $\tau\bar{\tau}$, 50\%
$b\bar{b}$ and a line-to-continuum ratio of $6\times 10^{-3}$.  Assuming a 15\%
energy resolution, we obtain the simulated spectrum shown in
Fig.~\ref{fig:gcannulus}.  This demonstrates that a future instrument could
observe a spectral signature of dark matter annihilation in the region around
the GC, above the residual astrophysical backgrounds.  To search for gamma-ray
emission from dark-matter annihilation in the Galactic center region, the
requirements for the future instrument include: a large effective area
($\sim$1~km$^2$), a moderately large field of view ($\gsim 7^\circ$ diameter),
a good energy resolution ($\lsim 15$\%), a low energy threshold ($\lsim
50$~GeV), excellent angular resolution to exclude contributions from
astrophysical point-sources ($<\sim 0.1^\circ$) and a location at low
geographic latitude (preferably in the southern hemisphere) for
small-zenith-angle low-threshold measurements of the GC region.

However, given the large backgrounds in our own galaxy,
the observation of a wider class of astrophysical targets
is desirable.  A future km$^2$ ACT array should, 
for the first time, have the sensitivity required to detect extragalactic
sources such as Dwarf galaxies, without resorting to very optimistic 
assumptions about the halo distribution.
The VERITAS collaboration previously undertook such an observing
program with the Whipple 10m telescope and reported upper limits for several
extragalactic targets (M33, Ursa Minor \& Draco dwarf galaxies,
M15)~\cite{Vassiliev2003,LeBohec2003,woodetal07}.   The HESS group published
limits on the Sagittarius dwarf galaxy and the resulting constraints on the halo
models \cite{moulin07}.
However, more sensitivity is required to detect a more generic annihilation
flux from such sources. 

\begin{figure*}[tb]\begin{center}
\includegraphics[height=10cm]{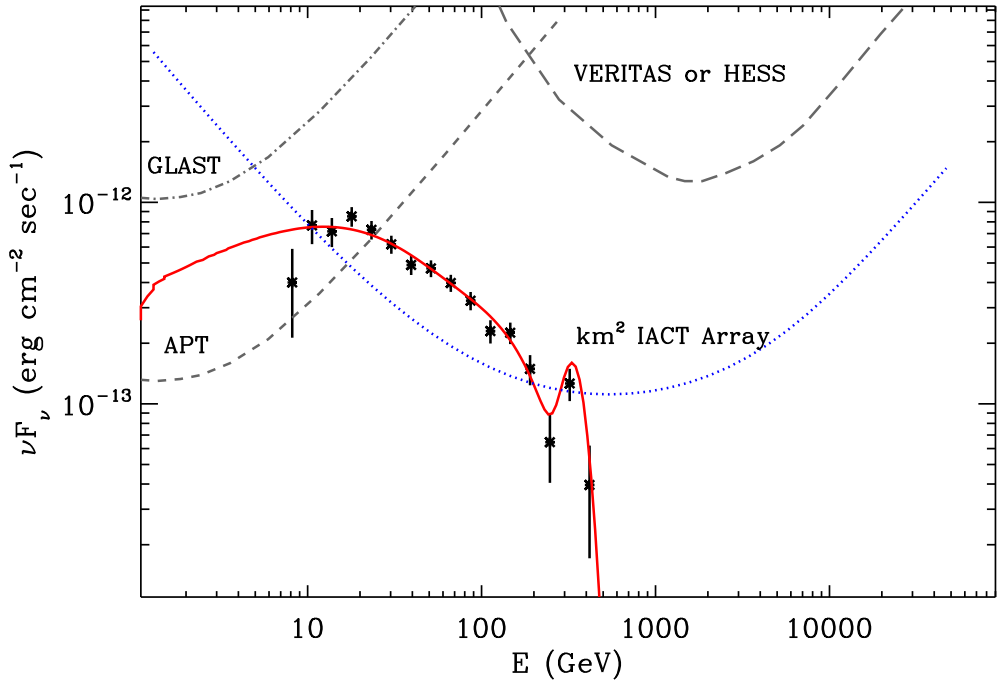} \end{center}
\caption{
Predicted gamma-ray signal from the dwarf spheroidal galaxy
Ursa Minor for
 neutralino mass of 330 GeV, branching into $\tau^+\tau-$ 20\% of the
time, and into $b\bar{b}$ 80\% of the time and with a line to continuum
ratio of 2$\times 10^{-3}$.  We assume
a typical annihilation cross-section of $2\times 10^{-26}{\rm cm}^3{\rm s}^{-1}$
the halo values
from Strigari et al. \cite{Strigari:2007at} with $r_s = 0.86\,{\rm kpc}$
and central density $\rho_s = 7.9\times 10^7\, M_\odot/{\rm kpc}^3$.
We also assume a modest boost factor of
$b=3$ from halo substructure.  We assume an ideal instrument with an
effective area of 1~km$^2$ and sensitivity limited only by the electron
background, diffuse gamma-ray background (assuming an $\sim E^{-2.5}$ spectrum
connecting to the EGRET points) and cosmic-ray background (10 times lower
than current instruments).  For this idealized IACT array, we do not
include the effect of a threshold due to night-sky-background, and assume
an energy resolution of 15\%.  The data points are simulated given the
signal-to-noise expected for the theoretical model compared with our anticipated instrument sensitivity.
}
\label{fig:ursaminor}
\end{figure*}

\subsubsection{Dwarf Spheroidals} Dwarf spheroidal (dSph) systems are 
ideal dark matter laboratories because astrophysical  backgrounds and
baryon-dark matter interactions are expected not to play a major role in the
distribution of dark matter. Furthermore, the mass--to--light ratio in dSphs 
can be very large, up to a few hundred,
showing that they are largely dark-matter dominated systems.
Numerous theoretical studies point to the potential for
detecting dark matter annihilation in dwarf
spheroidal galaxies or galaxies in the local group based on rough assumptions
of the distribution of dark matter \cite{Profumo:2005xd,Betal02,Tyler02,
BH06,PB04}.  However, with the advent of more data on the stellar content
of dSphs, it has recently been possible to perform a likelihood analysis on the
potential dark matter profiles that these systems could posses.  Under the
assumption that dSphs are in equilibrium, the radial component of the stellar
velocity dispersion is linked to the gravitational potential of the system
through the Jeans equation. This approach (utilized in
\cite{EFS04, SKBK06,Strigari:2007at}) has the significant advantage 
that observational data dictate the distribution of dark
matter with a minimum number of theoretical assumptions.
The main results of these studies are that dSphs are
very good 
systems for the search for dark matter annihilation, because most of the
uncertainties in the distribution of dark matter can be well quantified and
understood.  In addition, dSphs 
are expected to be
relatively free
of intrinsic
$\gamma$-ray emission from other astrophysical sources, thus eliminating contaminating background that may
hinder 
the interpretation of
any observation.  Assuming a scenario for supersymmetric dark
matter where $M_\chi = 200$\,GeV, $E_{\rm th}=50$\,GeV and ${\cal P} \approx 10^{-31} \, {\rm cm}^3 {\rm s}^{-1} {\rm
GeV^{-2}}$, the maximum expected fluxes from 9 dSphs studied in
\cite{SKBK06,Strigari:2007at} can be as large as $10^{-12}$ photons cm$^{-2}$
s$^{-1}$ (for Willman 1).  Observing $\gamma$-rays from dark matter
annihilation in dwarf spheroidals is of fundamental importance for 2 reasons:
First and foremost, these observations can lead to
an identification of the dark matter, especially if
line emission or other distinct features in the continuum
are detected and second, they will provide information on the actual
spatial distribution of dark matter halos
in these important objects.
If there is a weakly interacting thermal relic, then
$\gamma$-ray telescopes can 
tell us something about non-linear structure formation, a task unattainable by
any other experimental methods.

Fig.~\ref{fig:ursaminor} shows an example of one possible spectrum that might
be measured for Ursa Minor given conservative assumptions including: a typical
annihilation cross-section, a halo distribution constrained by stellar velocity
measurements ( from Strigari et al. \cite{Strigari:2007at}) and a modest boost
factor of $b=3$ at the low end of the expected range for such halos.  This
prediction demonstrates that detection from Dwarf galaxies is most likely out
of reach of the current generation of IACT experiments (HESS and VERITAS) or
proposed EAS experiments, but may be within reach of a future km$^2$IACT
instrument, if the point-source sensitivity is improved by an order of
magnitude, the energy resolution is good enough to resolve the spectral
features (better than 15\%) and the energy threshold can be pushed well below
100~GeV.    

With the advent of the Sloan Digital Sky Survey (SDSS), the number of known
dSph satellites of the local group has roughly doubled during the last decade
\cite{belokurov}.  Since the survey is concentrated around the north Galactic
pole, it is quite likely that there are many more dSph satellites waiting to be
discovered. For an isotropic distribution, and assuming that SDDS has found all
the satellites in its field of view, we would expect $\sim$ 50 dwarfs in all.
Since simulation data suggests that dwarf satellites lie preferentially along
the major axis of the host galaxy, the number of Milky-Way dwarf satellites
could be well above this estimate.  With more dwarf galaxies, and increasingly
detailed studies of stellar velocities in these objects, this class of sources
holds great promise for constraints on dark matter halos and indirect detection
of dark matter.  Since many of these discoveries are very new, detailed
astronomical measurements are still required to resolve the role of dark 
matter in individual sources.  For example, for the new object
Willman I, some have argued that this is a globular cluster while others
have made the case that despite
it's relatively small mass, this is a dark-matter
dominated object and not a globular cluster \cite{martin07}.  Other studies
challenge the inferences about the dark matter dominance in dSphs attributing
the rise in rotation velocities in the outer parts of dSphs to tidal effects
rather than the gravitational potential 
\cite{metz}.
Future progress in this blossoming area of astronomy could provide important
additional guidance for a more focused survey on the most promising sources
using pointed observations with very deep exposures. 


\begin{figure}[tb]
\includegraphics[height=8cm]{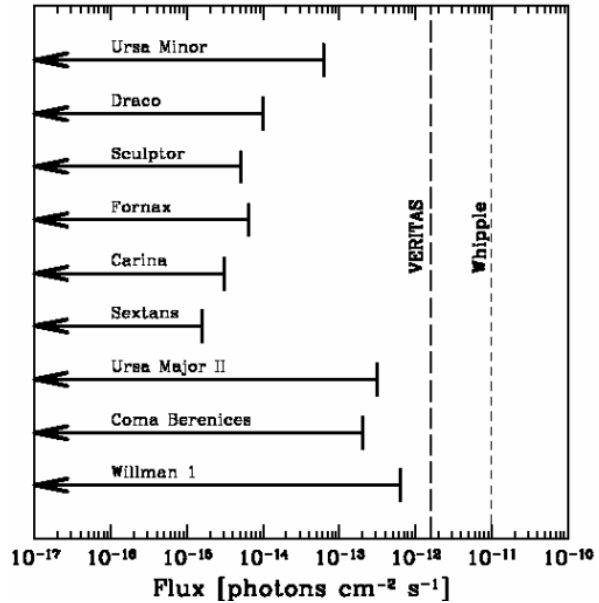}
\caption{Prospects for detecting the most prominent Dwarf-galaxy 
targets for dark matter
annihilation. Upper-limit bars show the range of theoretical predictions \cite{koushiappas_santafe06} with 
fluxes dropping below the level of detectability as one traverses the
full range of parameter space including the neutralino mass, cross-section and halo distribution. The plot includes dark-matter dominated 
dwarf spheroidal systems in the Milky Way halo, including promising sources 
located at high galactic latitude and with virtually no known
intrinsic $\gamma$ ray emission from astrophysical sources. The thin-dashed
line represents the sensitivity of Whipple, while the long-dashed line depicts
the sensitivity of VERITAS.}
\label{fig:dSphs}
\end{figure}

\subsubsection{Local group galaxies} Local group galaxies offer attractive
targets for the search of $\gamma$-rays form dark matter annihilation for many
of the
same reasons dSph galaxies do: they are relatively small systems, with
relatively high mass-to-light ratios (except M31). Relative to dSphs, the
influence of baryons in the central regions is higher, especially if a black
hole is present (such as M32). Nevertheless, their relative proximity and size
make them viable targets that should be explored. Recently, Wood et al. (2007)
\cite{woodetal07} used the Whipple 10m telescope and placed bounds on the
annihilation cross section of neutralinos assuming a distribution of dark
matter in the halos of M32 and M33 that resembles dark matter halos seen in
N-body simulations.  
While these observations with Whipple and now with VERITAS and HESS provide
interesting limits on some of the more extreme astrophysical or particle
physics scenario, more sensitive observations are needed if one makes more
conservative estimates.  Even with an order of magnitude increase in
sensitivity over the current generation experiments, it is still possible that
Dwarf or local-group galaxies will evade detection with the next generation
detector without some enhancement in the central halo (e.g. a cusp steepened by
the stellar population or a large boost factor).  Given this uncertainty
{the best strategy for detecting dark matter from Dwarf galaxies,
or local group galaxies is to observe an ensemble of sources,
taking advantage of the source-to-source variance in the halo profile
until better constraints are available from new astronomical measurements
(e.g., stellar velocity dispersion or rotation curves).}

\subsubsection{Detecting the Milky Way\\Substructure}

A generic prediction of the hierarchical structure formation scenario in cold
dark matter (CDM) cosmologies is the presence of rich substructure; bound dark
matter halos within larger, host halos.  Small dark matter halos form earlier,
and therefore have higher characteristic densities.  This makes some
of these subhalos able to withstand
tidal disruption as they sink in the potential well of their host halo
due to dynamical friction. Unfortunately, even though this is a natural outcome
of CDM, there is no clear explanation as to why the Milky Way 
appears to contain a
factor of 10-100 {\it fewer} subhalos than it should, based on CDM predictions
\cite{Klypin:1999uc,Moore:1999wf}.  Several solutions to this problem have been
suggested, such as changing the properties of the dark matter particle (e.g.,
\cite{1992ApJ...398...43C,Spergel:1999mh,Kaplinghat:2000vt}), modifying the
spectrum of density fluctuations that seed structure growth (e.g.,
\cite{Kamionkowski:1999vp, Zentner:2002xt}), or invoking astrophysical feedback
processes that prevent baryonic infall and cooling (e.g.,
\cite{1986ApJ...303...39D, 1999ApJ...523...54B,2001ApJ...548...33B}).  
The most
direct 
experimental
way to probe the presence of otherwise dark substructure in the Milky
Way is through $\gamma$-ray observations. 
Theoretical studies
\cite{Koushiappas:2003bn}, as well as numerical simulations of a Milky Way-size
halo \cite{Diemand:2006ik}, predict that given the probability of an
otherwise completely
dark subhalo nearby, the expected flux in $\gamma$-rays can be as large as
$\sim 10^{-13}$ cm$^{-2}$ s$^{-1}$. 

\subsubsection{Detecting Microhalos}

The {\it smallest} dark matter halos formed are set by the RMS dark matter
particle velocities at kinetic decoupling, the energy scale at which
momentum--changing  interactions cease to be effective
\cite{Setal99,Hetal01,Chenetal01,Berezinskyetal03,Green04,
Green05,LZ05,Bertch}. For supersymmetric dark matter this cutoff scale fives a
mass range for {\it microhalos} of around $10^{-13} \le [M/M_\odot] \le
10^{-2}$, depending on the value of the  kinetic decoupling temperature which
is set by the supersymmetric parameters.  While the survival of microhalos in
the Solar neighborhood is still under debate, there are indications that some
fraction ($\sim 20\%$) may still be present. In this case, microhalos could
even be detected via the proper motion of their $\gamma$-ray  signal
\cite{Mooreetal06,K06}. Microhalos that exhibit proper motion must be close
enough that their proper motion is above a detection threshold set by the
angular resolution and length of time over which the source can be
monitored(given by the lifetime of the observatory).  Microhalos must be
abundant enough so that at least one is within the volume set by this proper
motion requirement. The expected flux from a microhalo that may exhibit
detectable proper motion \cite{K06} is $\sim 10^{-15}$ cm$^{-2}$ s$^{-1}$. 
Such objects are most likely to be detected by very wide-field instruments 
like Fermi.  Follow-up measurements with IACT arrays would be required to
determine the characteristics of the spectrum and angular extent of these
sources at higher energies.

\subsubsection{Spikes around Supermassive and Intermediate-Mass Black Holes} 
There are other potential dark matter sources in our own Galaxy that may
be formed by a gravitational interplay of dark halos and baryonic matter.
In particular, it is possible that a number of intermediate-mass black holes (IMBHs)
with cuspy halos, might exist in our own
galaxy.  
The effect of the formation of a central object on the surrounding distribution
of matter has been investigated in
Refs.~\cite{peebles:1972,young:1980,Ipser:1987ru,Quinlan:1995} and for the
first time in the framework of DM annihilations in Ref.~\cite{Gondolo:1999ef}.
It was shown that the {\it adiabatic} growth of a massive object at the center
of a power-law distribution of DM, with index $\gamma$, induces a
redistribution of matter into a new power-law (dubbed ``spike'') with index
\begin{equation} 
\gamma_{sp} = (9-2\gamma)/(4-\gamma) \;\; .
\end{equation}
This formula is valid over a region of size $R_{sp} \approx 0.2 \, r_{BH}$,
where $r_{BH}$ is the radius of gravitational influence of the black hole,
defined implicitly as $M(<r_{BH})=M_{BH}$, where $M(<r)$ denotes the mass of
the DM distribution within a sphere of radius $r$, and where $M_{BH}$ is the
mass of the Black Hole~\cite{Merritt:2003qc}.
The process of adiabatic growth is, in particular, valid for the SMBH at the
galactic center.
A critical assessment of the formation {\it and survival} of the central spike,
over cosmological timescales, is presented in
Refs.~\cite{Bertone:2005hw,Bertone:2005xv} and references therein.
Adiabatic spikes are rather fragile structures, that require fine-tuned
conditions to form at the center of galactic halos~\cite{Ullio:2001fb}, and
that can be easily destroyed by dynamical processes such as major
mergers~\cite{Merritt:2002vj} and gravitational scattering off
stars~\cite{Merritt:2003eu,Bertone:2005hw}.

\begin{figure}
\label{fig:imbhs}

\includegraphics[height=8.3cm]{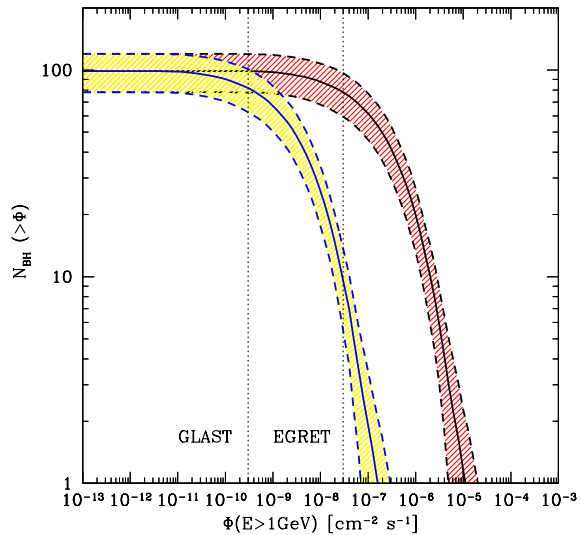}

\caption{IMBHs integrated luminosity function,
i.e. number of IMBHs that can be detected
from experiments with point source sensitivity $\Phi$ (above 1 GeV),
as a function of $\Phi$. We show for comparison the 5$\sigma$ point
source sensitivity above $1$~GeV of EGRET and Fermi (GLAST) in 1 year. From
Ref.~\cite{Bertone:2005xz}.}
\end{figure}
However Intermediate Mass BHs, with mass $10^{2} < M/{\rm M_{\odot}} <
10^{6}$, are not affected by these destructive processes.  Scenarios that seek
to explain the observed population and evolutionary history 
of supermassive-black-holes
actually result in the prediction of a large population of wandering
IMBHs, with a number in our own Galaxy. They may form in rare, overdense regions at high
redshift, $z \sim 20$, as remnants of Population III stars, and have a
characteristic mass-scale of a few $10^{2} \, {\rm M_{\odot}}$
\cite{Madau:2001,Bertone:2005xz, Zhao:2005zr,islamc:2004,islamb:2004}.
Alternatively, IMBHs may form directly out of cold gas in early-forming halos
and are typified by a larger mass scale of order $10^{5} \, {\rm
M_{\odot}}$~\cite{Koushiappas:2003zn}. We show in Fig.~\ref{fig:imbhs} the
number of objects that can be detected as a function of the detector
sensitivity.
{The spiky halos around galactic intermediate-mass black holes could
provide a large enhancement in the gamma-ray signal that could be effectively
detected by all-sky low-threshold instruments such as Fermi 
then followed-up by ground-based measurements.}  Over
most of the allowed parameter space, Fermi would detect the onset of the
continuum spectrum but would lack the sensitivity to measure the detailed
spectral shape above hundreds of GeV.  Ground-based measurements with good
point-source sensitivity, and good energy resolution (10-15\%) would be
necessary to follow-up these detections to measure the spectral cutoff and
other features of the annihilation spectrum needed to clearly identify a
dark-matter origin for the gamma-ray signal. 

High energy gamma-ray astronomy can also indirectly provide information about
the formation history of IMBHs through a very different avenue, i.e.,
infrared absorption measurements of gamma-rays from distant AGN.
For example, the early population-III stars that may seed the growth of IMBHs
are likely to be massive (100 $M_\odot$) stars that form in dark matter clumps
of mass $\sim 10^6 M_\odot$.  These short lived stars would result in a large
contribution to the total amount of visible and UV light in the early
(large-redshift) universe, that contribute to the present-day diffuse infrared
background.  Present observations by Whipple, HEGRA, MAGIC and HESS already
provide constraints on the contribution from population-III stars.  
{Gamma-ray astronomy has the unique potential to
provide important constraints on the history of structure formation in the
universe through observations of the annihilation signal from dark-matter halos on a 
range of mass scales (including IMBH halos)
in addition to probing the history of star formation through measurements of
the diffuse infrared background radiation.} 


\subsubsection{Globular clusters} Globular clusters are relatively low
mass-to-light ratio bound systems in the Milky Way that are dominated by a
dense stellar core. The presence of dark matter in the core of a collapsed
globular cluster is questionable because it is expected that 2-body stellar
interactions will deplete dark matter from the region. On the other hand, if
there is any dark matter left-over from
the core-collapse relaxation process, it is
possible that the dense stellar core would adiabatically steepen the
distribution of dark matter, thus making some dense globular clusters potential
targets for dark matter detection.  Wood et al. (2007)  \cite{woodetal07}
observed the relatively close M15 globular cluster with the Whipple 10m
telescope, and placed upper bounds on the cross section for dark matter
annihilation. 

\subsection{Complementarity of $\gamma$-Ray Searches with Other Methods for Dark Matter Searches}

Both Fermi and the LHC are expected to become operational in 2008.   What
guidance will these instruments provide for a future ground-based experiment?
The ATLAS and CMS experiments at the Large Hadron Collider (LHC) are designed
to directly discover new supersymmetric particles in the range of a few $\sim 100$ GeV/c$^{2}$ and
will start collecting data in the very near future.  The LHC alone will not, under even the
most optimistic circumstances, provide all of the answers about the nature of
dark matter. 
{In general, a combination of laboratory (LHC, ILC)
detection and astrophysical observations or direct detection experiments will
be required to pin down all of the supersymmetric parameters and to make the
complete case that a new particle observed in the laboratory really constitutes
the dark matter.}
Due to the fact that the continuum gamma-ray signal depends
directly on the total annihilation cross-section, there are relatively tight
constraints on the gamma-ray production cross-section from the cosmological
constraints on the relic abundance.  For direct detection, on the other hand,
the nuclear recoil cross-section is only indirectly related to the total
annihilation cross-section and thus there are a number of perfectly viable
model parameters that fall many orders of magnitude below any direct detection
experiment that may be built in the foreseeable future.
{Thus gamma-ray astronomy is unique in that the detection
cross-section is closely related to the total annihilation cross section
that determines the relic abundance}.  
A given theoretical scenario of SUSY
breaking at low energies, e.g. mSUGRA, SplitSUSY, non-universal SUGRA, MSSM-25,
AMSB, etc., reduces the available parameter phase space. Therefore, it is
natural to expect that, for some set of the parameters, the neutralino might be
detected by all experimental techniques, while in other cases only a single
method has sufficient sensitivity to make a detection~\cite{JE2004}. Only a
combination of accelerator, direct, and indirect searches would cover the
supersymmetric parameter space~\cite{BG2002}.  For example, the mass range of
neutralinos  
in the MSSM is
currently constrained by accelerator searches
to be above a few GeV~\cite{Bottino2003,Bottino2004} and by the
unitarity limit on the thermal relic to be below $\sim 100$ TeV~\cite{GK1990}
(a narrower region would result if specific theoretical assumptions are made,
e.g. mSUGRA).

For the LHC to see the lightest stable SUSY particle, it must first produce a
gluino from which the neutralino is produced.  This limits
the reach of the LHC up to neutralino masses of $m_\chi \approx 300$GeV, well
below the upper end of the allowed mass range.  Direct detection of
WIMP-nucleon recoil is most sensitive in the $60$ to $600$ GeV regime. Indirect
observations of self-annihilating neutralinos through $\gamma -$rays with
energies lower than $\sim 100$ GeV will best be accomplished by Fermi, while
VERITAS and the other ground-based $\gamma -$ ray observatories will play
critical role in searches for neutralinos with mass larger than $\sim 100$ GeV.

While direct detection and accelerator searches have an exciting discovery
potential, it should be emphasized that there is a large region of parameter
space for which gamma-ray instruments could provide the only detection for 
cases where the nuclear recoil cross-section falls below the threshold of
any planned direct detection experiment, or the mass is out of range of the
LHC or even the ILC.  {Any comprehensive scientific roadmap that
puts the discovery of dark matter as its priority must include support for
a future, high-sensitivity ground-based gamma-ray experiment in addition to
accelerator and direct searches}

But the next 5-10 years of DM research may provide us with a large amount of
experimental results coming from LHC, direct DM searches 
\cite{aprile05,klapdor02,klapdor05,bisset07,akerib05,sanglard07}
 and indirect
observations of astrophysical $\gamma$-rays.  Current gamma-ray experiments
such as AGILE, Fermi, VERITAS, HESS and MAGIC will continue making observations
of astrophysical sources that may support very high density dark matter spikes
and may, with luck, provide a first detection of dark matter.
The wide field-of-view Fermi instrument could provide serendipitous detections
of otherwise dark, dark matter halos, and search for the unique dark matter
annihilation signal in the isotropic cosmological background.  EAS experiments
will provide evidence about the diffuse galactic background at the highest
energies, helping to understand backgrounds for dark matter searches and even
offering the potential for discovery of some unforeseen
very high mass, nonthermal relic
that form the dark matter.  All of these
results will guide the dark matter research which can be conducted by a future
ground-based observatory needed to study the dark matter halos, and would
affect strongly the design parameters of such an observatory.

To briefly summarize the interplay between the LHC, Fermi and a future
ground-based gamma-ray instrument, it is necessary to consider several
different regimes for the mass of a putative dark matter particle:

\begin{itemize}

\item {\sl Case I:} If $m_\chi\sim  100\, {\rm GeV}$ and the LHC sees the LSP, Fermi will
probably provide the most sensitive measurements of the continuum radiation and
will be needed to demonstrate that a supersymmetric  particle constitutes the
dark matter \cite{koushiappas_santafe06}.  Ground-based measurements will be
needed to constrain the line-to-continuum ratio to better determine the
supersymmetric parameters or to obtain adequate photon statistics
(limited by the $\sim$m$^2$ effective area of Fermi) to obtain
the smoking gun signature
of annihilation by observing line emission. 

\item {\sl Case II:} If $100\, {\rm GeV} <  m_\chi < 300\,  {\rm GeV}$,
 the LHC could still see
the neutralino, but both the line {and continuum emission} could be better
detected with a 
a low-threshold (i.e., 20-40~GeV threshold)
ground-based experiment than with Fermi, if the source location
is known.  Again these gamma-ray measurements are still required to demonstrate
that a supersymmetric particle constitutes astrophysical halos, and to further
measure supersymmetric parameters \cite{baltz_p5_06}.

\item {\sl Case III:} If $m_\chi > 300\, {\rm GeV}$ future direct-detection experiments
and ground-based gamma-ray experiments may be able to detect
the neutralinos.  Only ground-based instruments will be able to determine the
halo parameters, and will provide additional constraints on SUSY parameter
space somewhat orthogonal to the constraints provided by the determination of
the direct detection cross-sections. For a sizeable fraction of parameter space, nuclear recoil cross-sections may be too small for direct detection but the
total annihilation cross section could still be large enough for a gamma-ray
detection.
Detection at very high energies would be particularly important for 
non-SUSY dark matter candidates such as the lightest Kaluza-Klein partner,
where current constraints put the likely mass range above the TeV scale.
Since TeV-scale neutralinos are likely to be either pure Higgsino or pure
Wino particles, particle-physics uncertainties are expected to be smaller
in this VHE energy regime.
\end{itemize}

\subsection{Conclusions}

A next-generation $\gamma$-ray telescope has the unique ability to make the
connection from particles detected in the laboratory to the dark matter that 
dominates the density of matter in the universe, and to provide important 
constraints that help to identify the nature of the dark matter particle. 
The main findings of our study about
the potential impact of gamma-ray measurements on the dark-matter problem
and the requirements for a future instrument are summarized below:

\begin{itemize}

\item Compared with all other detection techniques (direct and indirect),
$\gamma$-ray measurements of dark-matter are unique in going beyond a detection
of the local halo to providing a measurement of the actual distribution of dark
matter on the sky.  Such measurements are needed  to understand the nature of
the dominant gravitational component of our own Galaxy, and the role of dark
matter in the formation of structure in the Universe.

\item
There are a number of different particle physics and astrophysical
scenarios that can lead to the production of a gamma-ray signal with large
variations in the total flux and spectral shape. The spectral form of the
gamma-ray emission will be universal, and contains distinct features that can
be connected, with high accuracy, to the underlying particle physics.

\item The annihilation cross-section for gamma-ray production from
higher energy (TeV) candidates are well constrained by measurements of
the relic abundance of dark matter, with the particle-physics
uncertainty contributing $\sim$ one order of magnitude to the range of the
predicted gamma-ray fluxes. 

\item The Galactic center is predicted to be the strongest source of gamma-rays
from dark matter annihilation but contains large astrophysical backgrounds.
To search for gamma-ray emission from dark-matter annihilation in the
Galactic center region, the requirements for the future instrument include:
extremely good angular resolution to reject background from other point sources,
a moderately large field of view
($\gsim 7^\circ$ diameter), a good energy resolution ($\lsim 15$\%), a
low energy threshold $\lsim 50$~GeV, and location at a southern hemisphere
site.

\item Observations of local-group dwarf galaxies may provide the cleanest
laboratory for dark-matter searches, since these dark-matter dominated objects
are expected to lack other astrophysical backgrounds.  For these observations,
a very large effective area and
excellent point-source sensitivity down to $\lsim$50 GeV is required.
Energy resolution better than 15-20\% is required to determine the spectral
shape.  Currently, the best strategy for
detecting dark matter from dwarf galaxies, globular clusters or local group
galaxies is to observe an ensemble of sources, taking advantage of the
source-to-source variance in the halo profile that may lead to large
enhancements in the signal from some sources, although improvements in
constraints on the dark-matter density profile from future detailed astronomical measurements (e.g., from stellar velocity dispersion) will allow for a 
refinement of the list of most promising targets.

\item Observations of halo-substructure could provide important new constraints
on CDM structure formation, providing information on the mass of the first
building blocks of structure, and on the kinetic decoupling temperature.  The
most direct experimental way to probe the presence of otherwise dark halo
substructure in the Milky Way is through $\gamma$-ray observations.
Space-based low-threshold all-sky measurements will be most effective for
identifying candidate objects, but ground-based measurements will be required
to determine the detailed spectral shape (cutoff, line-to-continuum ratio)
needed to identify the dark matter candidate.

\item {The spiky halos around galactic intermediate-mass black holes could
provide a large enhancement in the gamma-ray signal that could be effectively
detected by all-sky low-threshold instruments such as Fermi or a future
space-based instrument, then followed-up by ground-based measurements.  Over
most of the allowed parameter space, Fermi would detect the onset of the
continuum spectrum but would lack the sensitivity to measure the detailed
spectral shape above hundreds of GeV.  Ground-based measurements with good
point-source sensitivity, and good energy resolution (10-15\%) would be
necessary to follow-up these detections to measure the spectral cutoff and
other features of the annihilation spectrum needed to clearly identify a
dark-matter origin of the gamma-ray signal.} 

\item {While a space-based instrument or future
IACT arrays are probably the only means of providing the large
effective area, low threshold, energy and angular resolution for detailed
measurements of gamma-rays from dark matter annihilation, future EAS
experiments like HAWC can also play a useful role.  Future EAS experiments,
with their wide field of view and long exposure time, also have the potential
for serendipitous discovery of some corners of parameter space, in particular
for nonthermal relics and mass close to the unitarity limit.   The good
sensitivity of EAS experiments can provide important measurements of diffuse,
hard-spectra galactic backgrounds.}  

\item {Gamma-ray astronomy has the unique potential to
provide important constraints on the history of structure formation in the
universe through dark-matter observations of dark-matter halos on a 
range of mass scales (including IMBH halos)
in addition to probing the history of star formation through measurements of
the diffuse infrared background radiation.} 

\item {In general, a combination of laboratory (LHC, ILC)
detection and astrophysical observations or direct detection experiments will
be required to pin down all of the supersymmetric parameters and to make the
complete case that a new particle observed in the laboratory really constitutes
the dark matter.}

\item {Gamma-ray astronomy is unique in that the detection
cross-section is closely related to the total annihilation cross section
that determines the relic abundance.}

In closing, we reiterate that a comprehensive plan for uncovering the nature
of dark matter must include gamma-ray measurements.  With an order of magnitude improvement in sensitivity and reduction in energy threshold, a future IACT
array should have adequate sensitivity to probe much of the most generic
parameter space for a number of sources including Galactic substructure,
Dwarf galaxies and other extragalactic objects.

\end{itemize}




\clearpage

\section{Extragalactic VHE astrophysics}
\label{EGS-subsec}
Group membership:\\ \\
A.Atoyan, M. Beilicke, M. B\"ottcher, A. Carraminana, P. Coppi, C. Dermer,
B. Dingus, E. Dwek, A. Falcone, J. Finley, S. Funk, M. Georganopoulos,
J. Holder, D. Horan, T. Jones, I. Jung, P. Kaaret, J. Katz,
H. Krawczynski, F. Krennrich, S. LeBohec, J. McEnery,
R. Mukherjee, R. Ong, E. Perlman, M. Pohl, S. Ritz, J. Ryan,
G. Sinnis, M. Urry, V. Vassiliev, T. Weekes, D. A. Williams

\subsection{Introduction}

A next-generation gamma-ray experiment will make extragalactic 
discoveries of profound importance. Topics to which gamma-ray observations can make unique
contributions are the following:
(i) the environment and growth of Supermassive Black Holes; (ii) the acceleration
of cosmic rays in other galaxies; (iii) the largest particle accelerators in
the Universe, including radio galaxies, galaxy clusters, 
and large scale structure formation shocks; (iv) study of the 
integrated electromagnetic luminosity of the Universe and
intergalactic magnetic field strengths through processes including
pair creation of TeV gamma rays interacting with infrared photons
from the Extragalactic Background Light (EBL).

The following sections will describe the science opportunities in these
four areas. Gamma-ray bursts and extragalactic searches for dark matter
annihilation gamma rays are discussed in separate sections.
\subsection{Gamma-ray observations of supermassive black holes}
\begin{figure*}[t!h]
\includegraphics[angle=270,width=6.5in]{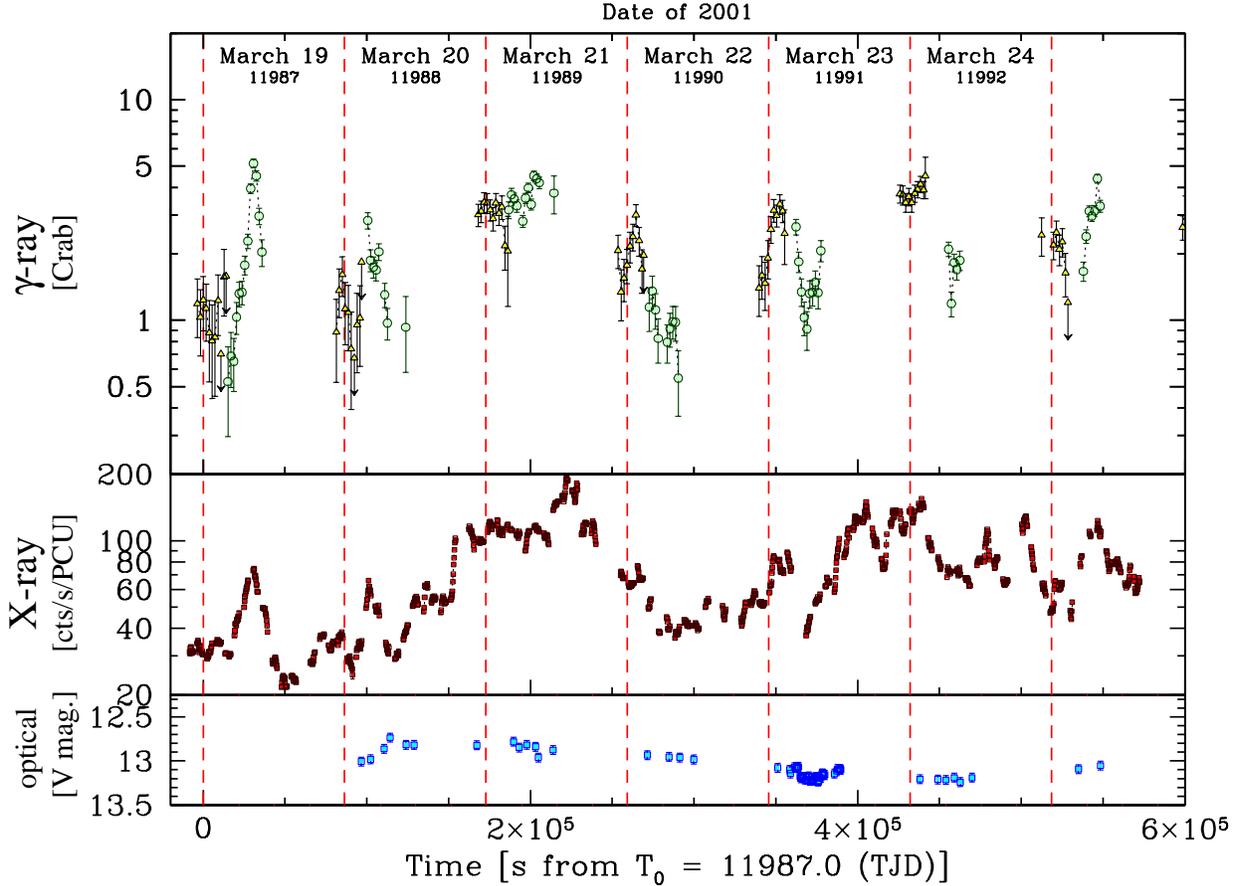} 
\caption{\label{agn} Results from 2001 Rossi X-ray Timing Explorer (RXTE) 2-4 keV X-ray and
Whipple (full symbols) and HEGRA (open symbols) gamma-ray observations 
of Mrk 421 in the year 2001 \cite{Foss:08}.
The X-ray/gamma-ray fluxes seem to be correlated. However, the
interpretation of the data is hampered by the sparse coverage at
TeV gamma rays.}
\end{figure*}

Supermassive black holes (SMBH) have masses between a million and several
billion solar masses and exist at the centers of galaxies. 
Some SMBHs, called Active Galactic Nuclei (AGN) are strong emitters
of electromagnetic radiation. 
Observations with the {\it EGRET Energetic Gamma-Ray Experiment Telescope}
on board of the Compton Gamma-Ray Observatory (CGRO) revealed that
a certain class of AGN known as blazars are powerful and variable 
emitters, not just at radio through optical wavelengths, but also 
at $\ge$100 MeV gamma-ray energies \cite{Hartman1999}. 
EGRET largely detected quasars, the most powerful blazars in the Universe.
Observations with ground-based Cherenkov telescopes showed that blazars 
emit even at TeV energies \cite{Punch1992}. In the meantime, more than
twenty blazars have now been identified as sources of $>$200~GeV gamma rays 
with redshifts ranging from 0.031 (Mrk 421) \cite{Punch1992} 
to 0.536 (3C~279) \cite{2007arXiv0709.1475T} \footnote{Up-to-date 
lists of TeV $\gamma$-ray sources can be found at the web-sites:
http$://$tevcat.uchicago.edu and 
http$://$www.mpp.mpg.de$/\sim$rwagner$/$sources$/$.}. 
Most TeV bright sources are BL Lac type objects, 
the low power counterparts of the quasars detected by EGRET. 
The MeV to TeV gamma-ray emission from blazars is commonly thought to 
originate from highly relativistic collimated outflows (jets) from mass 
accreting SMBHs that point at the observer \cite{Tavecchio2005,Krawczynski2006}.
The only gamma-ray emitting AGN detected to date that are not blazars are the 
radio galaxies Centaurus A \cite{Sreekumar1999} and M87 \cite{Aharonian2007M87}.
The observation of blazars in the gamma-ray band has had a major
impact on our understanding of these sources. The observation of rapid
flux variability on time scales of minutes together with 
high gamma-ray and optical fluxes
\cite{1996Natur.383..319G,2007ApJ...664L..71A} implies that the accreting 
black hole gives rise to an extremely relativistic jet-outflow with a bulk 
Lorentz factor exceeding 10, most likely even in the range between 10 and 50 
\cite{2001ApJ...559..187K,2008MNRAS.384L..19B}.
Gamma-ray observations thus enable us to study plasma which moves with
$\ge$99.98\% of the speed of light.
Simultaneous broadband multiwavelength observations of blazars have
revealed a pronounced correlation of the X-ray and TeV gamma-ray
fluxes \cite{1996ApJ...472L...9B,1996ApJ...470L..89T,2000A&A...353...97K,Foss:08}. 
The X-ray/TeV flux correlation (see Fig. \ref{agn}) suggests that the emitting particles
are electrons radiating synchrotron emission in the radio to X-ray
band and inverse Compton emission in the gamma-ray band. 

Blazars are expected to be the most copious extragalactic sources
detected by ground-based IACT arrays like VERITAS and by the satellite
borne gamma-ray telescope Fermi. For extremely strong
sources, IACT arrays will be able to track GeV/TeV fluxes on 
time scales of seconds and GeV/TeV energy spectra on time scales 
of a few minutes.
Resolving the spectral variability during individual strong flares
in the X-ray and gamma-ray bands should lead to the 
unambiguous identification of the emission mechanism. 
The present generation of IACTs will be able to track spectral
variations only for a very small number of sources and only during
extreme flares. The next-generation gamma-ray experiments will be able
to do such studies for a large number of sources on a routine basis.
Sampling the temporal variation of broadband energy spectra from a few
tens of GeV to several TeV will allow us to use blazars as precision
laboratories to study particle acceleration and turbulence in
astrophysical plasmas, and to determine the physical parameters
describing a range of different AGN.
The observations of blazars hold the promise to reveal details about the inner
workings of AGN jets. Obtaining realistic estimates of the power in
the jet, and the jet medium will furthermore constrain the origin of
the jet and the nature of the accretion flow.

Recently, spectracular results have been obtained by combining monitoring VLBA, 
X-ray and TeV $\gamma$-ray observations. This combination has the potential to pinpoint 
the origin of the high energy emission based on the high resolution radio images, 
and thus to directly confirm or to refute models of jet formation.
For example, radio VLBA, optical polarimetry, X-ray and TeV $\gamma$-ray observations 
of the source BL Lac seem to indicate that a plasma blob first detected with the 
VLBA subsequently produces an X-ray, an optical and a $\gamma$-ray flare \cite{2008Natur.452..966M}.
A swing of the optical polarization seems to bolster the case for a helical magnetic field
as predicted by magnetic models of jet formation and acceleration.
Presently such observations are extremely difficult as the current instruments 
can detect sources like M 87, BL Lac, W Com only in long observations or during extreme flares.
Next-generation $\gamma$-ray instruments will allow us to study the correlation 
of fast TeV flares and radio features on a routine basis.

In addition to ground-based radio to optical coverage, 
several new opportunities might open up within the next decade.
The Space Interferometry Mission (SIM) will be able to image emerging
plasma blobs with sub milli-arcsec angular resolution \cite{Unwi:02}. 
The center may be located with an accuracy of 
a few  micro-arcsec. For a nearby blazar at z=0.03, 1 milli-arcsec 
corresponds to a projected distance of 0.6 pc. The SIM observations 
could thus image the blobs that give rise to the flares detected in 
the gamma-ray regime. Joint X-ray/radio interferometry observations already give some tentative evidence for the emergence of radio blobs 
correlated with X-ray flares. 
If a Black Hole Finder Probe like the Energetic X-ray Imaging Space 
Telescope (EXIST) \cite{Grin:05} will be launched, it would provide reliable
all-sky, broad-bandwidth (0.5-600 keV), and high-sensitivity X-ray coverage 
for all blazars in the sky. EXIST's full-sky sensitivity would be 
2 $\times$ 10$^{-12}$ ergs cm$^{-2}$ s$^{-1}$ for 1 month of integration.
For bright sources, EXIST could measure not only flux variations but also
the polarization of hard X-rays. 
Opportunities arising from neutrino coverage will be described below.

At the time of writing this white paper, the Fermi gamma-ray telescope is in the process
of detecting a few thousand blazars. The source sample will make it possible to
study the redshift dependent luminosity function of blazars, although
the identification of sources with optical counterparts may be
difficult for the weaker sources of the sample, owing to Fermi's
limited angular resolution. Another important task for the next-generation 
instrument will be to improve on the Fermi localization accuracies, and thus 
to identify a large number of the weaker Fermi sources.

Independent constraints on the jet power, kinematics, and emission 
processes can be derived from GeV-TeV observations of the large scale 
(up to hundreds of kpc) jets recently detected by Chandra. Although such 
large scale jets will not be  spatially resolved, the fact that the 
gamma-ray emission from the quasar core is highly variable 
permits  us to set upper 
limits to the steady GeV-TeV large scale jet emission \cite{2006ApJ...653L...5G}.
In the case of the relatively nearby 3C 273, for example, the electrons that 
produce the large scale jet IR emission will also produce a flat GeV component. 
The fact that this emission is weaker than the EGRET upper limit constrains 
the Doppler factor of the large scale jets to less than 12, a value that can 
be pushed down to 5 with Fermi observations.  Such low values of delta have 
implications on the nature of the large scale jet X-ray emission observed 
by Chandra. In particular, they disfavor models in which the X-ray emission is 
inverse Compton scattering of the cosmic microwave background (CMB), because 
the jet power required increases beyond the so-called Eddington luminosity, 
thought by many to be the maximum luminosity that can be channeled continuously 
in a jet.  A synchrotron interpretation for the X-ray emission, requiring 
significantly less jet power, postulates a population of multi-TeV electrons 
that will unavoidably  up-scatter the CMB to TeV energies. The existing 3C~273 
shallow HESS upper limit constrains the synchrotron interpretation to Doppler 
factors less than 10. Combining deep TeV observations with a next-generation
experiment with Fermi observations holds the promise of  confirming 
or refuting the synchrotron interpretation and constraining the jet power. 

Whereas the X-ray/gamma-ray correlation favors leptonic models with electrons
as the emitters of the observed gamma-ray emission, hadronic models are not ruled out.
In the latter case, the high-energy component is synchrotron emission, either 
from extremely high-energy (EHE) protons \cite{Aharonian2000,Muecke2001,Muecke2003}, 
or from secondary $e^+/e^-$ resulting from synchrotron and pair-creation cascades 
initiated by EHE protons \cite{Mannheim1993} or high-energy electrons or photons 
\cite{Lovelace1979,Burns1982,Blandford1995,Levinson1995}. 
If blazars indeed accelerate UHE protons, it might even be possible to correlate their
TeV gamma-ray emission with their flux of high-energy neutrinos detected by
the IceCube detector \cite{Halz:05}. The high sensitivity of a next-generation
ground-based experiment would be ideally suited to perform such multi-messenger
studies.

Although most observations can be explained with the emission of high-energy particles that
are accelerated in the jets of AGN , the observations do not exclude that the emitting particles 
are accelerated closer to the black hole. If the magnetic field in the black hole magnetosphere has a poloidal net component on the 
order of $B_{\rm 100}\,=$ 100~G, both the spinning black hole \cite{Blandford1977} and 
the accretion disk \cite{Lovelace1976,Blandford1976} will produce strong electric fields that could 
accelerate particles to energies of $2\times 10^{19}$ $B_{\rm 100}$ eV.
High-energy protons could emit TeV photons as curvature radiation \cite{Levinson2000}, and high-energy
electrons as Inverse Compton emission \cite{2007ApJ...659.1063K}.
Such models could be vindicated by the detection of energy spectra, which are inconsistent
with originating from shock accelerated particles. An example for the latter would be 
very hard energy spectra which require high minimum Lorentz factors of accelerated particles.

The improved data from next-generation gamma-ray experiments can be compared 
with improved numerical results. The latter have recently made very substantial progress.
General Relativistic Magnetohydrodyamic codes are now able to test magnetic models of jet formation and
acceleration (see the review by \cite{2008arXiv0804.3096S}). The Relativistic-Particle-in-Cell technique 
opens up the possibility of greatly improving our understanding a wide range of issues 
including jet bulk acceleration, electromagnetic energy transport in jets, and particle
acceleration in shocks and in magnetic reconnection while
incorporating the different radiation processes \cite{Nogu:07,Nish:07,Love:05,Chan:07}.

Blazar observations would benefit from an increased sensitivity in the 
100 GeV to 10 TeV energy range to discover weaker sources and to sample the energy
spectra of strong sources on short time scales. A low energy threshold in the
10-40 GeV range would be beneficial to avoid the effect of intergalactic 
absorption that will be described further below. 
Increased sensitivity at high energies would be useful for measuring the energy spectra
of a few nearby sources like M 87, Mrk 421, and Mrk 501 at energies $\gg$10 TeV
and to constrain the effect of intergalactic absorption in the wavelength range above
10 microns. The interpretation of blazar data would benefit from dense temporal sampling
of the light curves. Such sampling could be achieved with gamma-ray experiments
located at different longitudes around the globe.

\subsection[Cosmic rays from star-forming galaxies]{Cosmic rays from star-forming galaxies}
\label{CR-subsect}
More than 60\% of the photons detected by EGRET
during its lifetime were produced as a result of interactions
between cosmic rays (CRs) and galactic interstellar gas and dust. 
This diffuse radiation represents 
approximately $90\%$ of the MeV-GeV gamma-ray
luminosity of the Milky Way~\cite{SMR2000}. Recently H.E.S.S.
reported the detection of diffuse radiation at TeV energies
from the region of dense molecular clouds in the innermost 200\,pc around
the Galactic Center~\cite{DiffGC2006}, 
confirming the theoretical expectation
that hadronic CRs could produce VHE radiation in their interaction with
atomic or molecular targets, 
through the secondary decay of $\pi^\circ$'s. 
Only one extragalactic source of diffuse GeV radiation was
found by EGRET: the Large Magellanic Cloud, 
located at the distance of $\sim 55$\,kpc~\cite{Sreekumar1992}. 
A simple re-scaling argument suggests that a putative galaxy
with Milky-Way-like gamma-ray luminosity, located at the distance of 
1\,Mpc, would have a flux of 
approximately $2.5\times 10^{-8}$ cm$^{-2}$ s$^{-1}$ ($>100$MeV), 
well below the detection limit of EGRET and $\sim 2\times
10^{-4}$ of the Crab Nebula flux ($>1$ TeV), well below the sensitivity of
VERITAS and H.E.S.S. Thus, a next-generation gamma-ray observatory 
with sensitivity at least an order of magnitude better than VERITAS would 
allow the mapping of GeV-PeV cosmic rays in normal local 
group galaxies, such as M31, and study diffuse radiation from more distant 
extragalactic objects if their gamma-ray luminosity is enhanced by a 
factor of ten or more over that of the Milky Way.

Nearby starburst galaxies (SBG's), such as NGC253, M82, IC342, M51 exhibit
regions of strongly enhanced star formation and supernova (SN) explosions,
associated with gas clouds which are a factor of $10^{2}-10^{5}$ more dense
than the average Milky Way gas density of $\sim 1$ proton per cm$^{3}$. 
This creates nearly ideal conditions for the emission of intense, diffuse VHE
radiation, assuming that efficient hadronic CR production takes place in the
sites of the SNR's (i.e. that the galactic CR origin paradigm is valid) and in
colliding OB stellar winds~\cite{Volk1996}. In addition, leptonic 
gamma-ray production through inverse-Compton scattering of 
high density photons produced by OB associations may become effective 
in star forming regions~\cite{Pohl1994}. Multiple attempts to detect SBGs
have been undertaken by the first generation ground-based gamma-ray 
observatories. At TeV energies, M82, IC342, M81, and NGC3079 were observed 
by the Whipple 10\,m telescope~\cite{Nagai2005}, while M82 and NGC253 were 
observed by HEGRA. However, none of these objects were 
detected. A controversial detection of NGC 253 by the CANGAROO collaboration 
in 2002~\cite{Itoh2002} was ruled out by H.E.S.S. observations~\cite{HESS253}. 
The theoretical predictions of TeV radiation from starburst galaxies 
have not yet been confirmed by observations and these objects will be 
intensively studied by the current generation instruments during the next 
several years. The optimistic theoretical considerations suggest that a 
few SBG's located at distances less than $\sim 10$ Mpc may be discovered. 
Should this prediction be confirmed, a next-generation gamma-ray 
observatory with sensitivity at least an order of magnitude better than VERITAS 
will potentially discover thousands of such objects within the $\sim 100$ 
Mpc visibility range. This will enable the use of SBG's as 
laboratories for the detailed study of the SNR CR acceleration paradigm
and VHE phenomena associated with star formation, including quenching effects 
due to evacuation of the gas from star forming regions by SNR shocks 
and UV pressure from OB stars. 

If accelerated CR's are confined in the regions of high gas or photon density
long enough that the escape time due to diffusion through the magnetic field
exceeds the interaction time, then the diffuse gamma-ray flux cannot be
further enhanced by an increased density of target material, 
and instead an increased SN rate is needed. Ultra Luminous InfraRed 
Galaxies (ULIRGs), which have SN rates on the scale of a few per year 
(compared to the Milky Way  rate of $\sim 1$ per century) 
and which also have very large amounts of molecular material,
are candidates for VHE emission~\cite{Torres2004}. Although
located at distances between ten and a hundred times farther than the most
promising SBG's, the ULIRG's Arp220, IRAS17208, and NGC6240 may be within the
range of being detected by Fermi, VERITAS and 
H.E.S.S.~\cite{Torres2004a}. Next-generation gamma-ray instruments 
might be able to detect the most luminous objects of this type even if they 
are located at $\sim 1$ Gpc distances. Initial studies of the population of ULIRGs 
indicate that these objects underwent significant evolution through the
history of the Universe and that at the moderate redshift ($z<1$) the abundance 
of ULIRGs increases. Any estimate of the number of ULIRGs that may be detected is 
subject to large uncertainties due to both the unknown typical gamma-ray 
luminosity of these objects and their luminosity evolution. However, if theoretical 
predictions for Arp220 are representative for objects of this type, 
then simple extrapolation suggests $>10^{2}$ may be detectable. 

The scientific drivers to study ULIRG's are similar to those of SBGs and may include research of 
galaxy gamma-ray emissivity as a function of target gas density, 
supernova rate, confining magnetic field, etc. In addition, research of ULIRGs 
may offer a unique possibility to observe 
VHE characteristics of star formation in the context of the recent history of 
the Universe ($z<1$) since ULIRGs might be detectable to much further distances.
Other, more speculative, avenues of research may also be available. A 
growing amount of evidence suggests that AGN feedback mechanism connects episodes 
of intense starbursts in the galaxies with the accretion activity of central 
black holes. One can wonder then if a new insight into this phenomena can be 
offered by observation of VHE counterparts of these processes detected from 
dozens of ULIRGs in the range from 0.1-1 Gpc.

\subsection[The largest particle accelerators in the universe: radio galaxies,
galaxy clusters, and large scale structure formation shocks]{The largest particle  accelerators in the Universe: radio galaxies,
galaxy clusters, and large scale structure formation shocks}
\label{CR-clusters}
\begin{figure*}[t!h]
\begin{minipage}[!ht]{11.2cm}
\includegraphics[width=11cm]{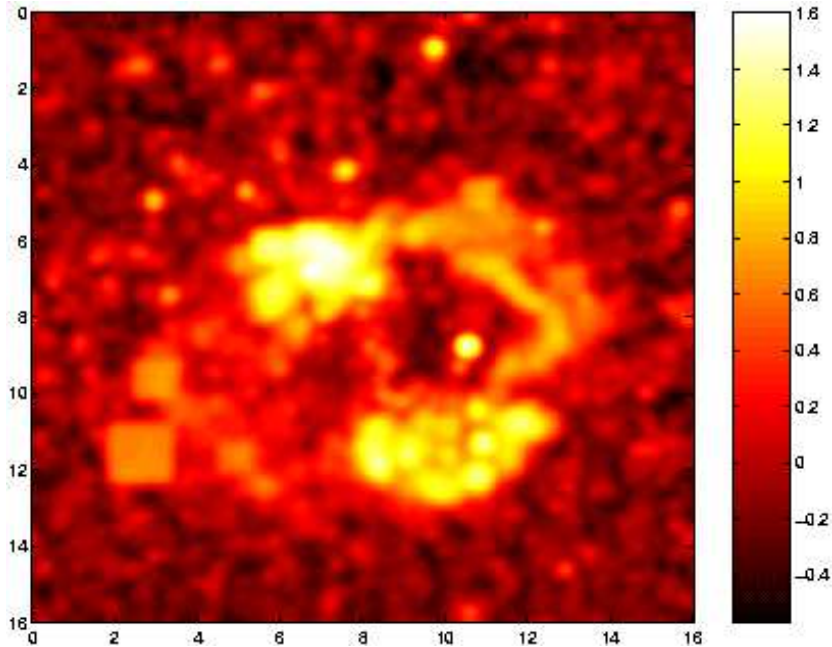}
\end{minipage}
\begin{minipage}[ht]{5cm}
\caption{\label{keshet} Results from a cosmological simulation showing how the 
$>10$ GeV gamma-ray emission from a nearby rhich galaxy cluster could look like when mapped
with a gamma-ray telescope with 0.2$^{\circ}$ angular resolution. The image covers a 
16$^{\circ}\times 16^{\circ}$ region (color scale: log($J/\bar{J}$) for an average 
$>$10 GeV flux of $\bar{J}\,=$ 8.2$\times 10^{-9}$ cm$^{-2}$ sec$^{-1}$ sr$^{-1}$) (from \cite{Keshet2002})}
\end{minipage}
\end{figure*}

%

The possibility of observing diffuse GeV and TeV radiation from even
more distant, rich galaxy clusters (GCs) has widely been discussed in the literature.
As the Universe evolves, and structure forms on increasingly larger scales, 
the gravitational energy of matter is converted into random kinetic 
energy of cosmic gas. In galaxy clusters, collisionless structure formation 
shocks, triggered by accretion of matter or mergers, are thought to be the 
main agents responsible for heating the inter-cluster medium (ICM) to 
temperatures of $\sim 10$ keV. Through these processes a fraction of 
gravitational energy is converted into the kinetic energy of 
non-thermal particles: protons and electrons. Galactic 
winds~\cite{VoelkAtoyan1999} and re-acceleration of mildly relativistic 
particles injected into the ICM by powerful cluster 
members~\cite{EnsslinBiermann1998} may accelerate 
additional particles to non-thermal energies. Cosmic ray protons can 
escape clusters diffusively only on time scales much longer than the 
Hubble time. Therefore, they accumulate over the entire formation 
history~\cite{VoelkAtoyan1999} and interact with the intercluster thermal 
plasma to produce VHE gamma radiation. Theoretical predictions for 
the detection of such systems in gamma rays by VERITAS and H.E.S.S. 
include clusters in the range from $z=0.01$ to 
$z=0.25$ (see Fig.\ \ref{keshet}) \cite{Volk1996,Keshet2002,GB2003}. 
Objects of this category were observed with Whipple \cite{Perkins2006} and H.E.S.S.
\cite{HessCluster} but were not detected. Multiple attempts to find gamma-ray 
signals from GCs in EGRET data also failed. Nevertheless, a large theoretical 
interest~\cite{BD2004,R2004,Rf2004} motivates 
further observations of the particularly promising candidates, such 
as the Coma and Perseus clusters by VERITAS and H.E.S.S.. If nearby 
representatives of the GC class are detected, a next-generation gamma-ray 
observatory with sensitivity increased by a factor of $10$   
would be able to obtain spatially resolved energy spectra from 
the close, high-mass systems, and should be able to obtain flux 
estimates and energy spectra of several dozen additional clusters. 
The detection of gamma-ray emission from galaxy clusters would make 
it possible to study acceleration mechanisms on large scales 
($>10$ kpc). It would permit measurement of the energy density of 
non-thermal particles and investigation of whether they affect the process of star 
formation in GCs, since their equation of state and cooling behavior 
differs from that of the thermal medium. If cosmic ray protons indeed 
contribute noticeably to the pressure of the ICM, measurements 
of their energy density would allow for improved estimates 
of the cluster mass based on X-ray data, and thus improve estimates 
of the universal baryon fraction. Based on population studies
of the gamma-ray fluxes from GCs, one could explore 
the correlation of gamma-ray luminosity and spectrum with 
cluster mass, temperature, and redshift. If such correlations are 
found, one could imagine using GCs as steady 
\textquotedblleft standard candles\textquotedblright \ to measure 
the diffuse infrared and visible radiation of the Universe
through pair-production attenuation of gamma rays. From a theoretical 
point of view the spectral properties of gamma-ray fluxes from GCs  
might be better understood than the intrinsic properties of blazars.  

The anticipated discovery of extragalactic sources by VERITAS and H.E.S.S. will put theoretical predictions 
discussed here on firmer ground, at least for the number of sources that the next 
generation ground-based observatory may detect. Over the next five years, 
Fermi will make major contributions to this area of studies. If the origin of 
gamma radiation in these sources is hadronic, Fermi should be able 
to detect most of the SBGs, ULIRGs, and GCs, which could potentially be 
detected by VERITAS and H.E.S.S. Under some scenarios, in which gamma rays 
are produced via leptonic mechanisms, a fraction of sources may escape Fermi 
detection (M82 might be such example), yet may still be detectable 
with VERITAS and H.E.S.S. Future theoretical effort will be required to guide
observations of these objects. In general, benefiting from the full 
sky coverage of Fermi, a program to identify the Fermi 
sources using the narrow field of view ACT observatories of the present day will
be possible, and it is likely that diffuse gamma-ray extragalactic 
sources will be discovered. Fermi will measure the galactic and extragalactic 
gamma-ray backgrounds with unprecedented accuracy and will likely resolve
the main contributing populations of sources in the energy domain 
below a few GeV. The task of determining the contribution from the diffuse 
gamma-ray sources to the extragalactic background in the range above 
a few GeV to $\sim 100$ GeV will be best accomplished by the next 
generation ground-based instrument, capable of detecting a large number of sources 
rather than a few. Most of these sources are anticipated to be weak, so they will
require deep observations. 

Large scale structure formation shocks could accelerate protons and
high-energy electrons out of the intergalactic plasma. Especially in the 
relatively strong shocks expected on the outskirts of clusters and on the 
perimeters of filaments, PeV electrons may be accelerated in substantial numbers. CMB
photons Compton scattered by electrons of those energies extend into the
TeV gamma-ray spectrum. The energy carried by the scattered photons cools
the electrons rapidly enough that their range is limited to regions close
to the accelerating shocks. However, simulations have predicted that the
flux of TeV gamma rays from these shocks can be close to detection limits
by the current generation of ground-based gamma-ray telescopes \cite{2003MNRAS.342.1009M}.
If true, this will be one of the very few ways in which these shocks can be 
identified, since very low thermal gas densities make their X-ray detection
virtually impossible. Since, despite the low gas densities involved, these 
shocks are thought to be a dominant means of heating cluster gas, their study 
is vital to testing current models of cosmic structure formation.

The origin of ultra-high-energy cosmic rays (UHECRs, $E^{>}_{\sim} 10^{16}\,\rm eV$) 
is one of the major unsolved problems in contemporary astrophysics. 
Recently, the Auger collaboration reported tentative evidence for a correlation of
the arrival directions of UHECRs with the positions of nearby Active Galactic Nuclei. 
Gamma-ray observations may be ideally suited to study the acceleration process, as the UHECRs
must produce gamma rays through various processes. The UHECRs may be accelerated far away from 
the black hole where the kpc jet is slowed down and dissipates energy.
If they are accelerated very close to the black hole at $\sim$pc distances, the high-energy particle 
beam is expected to convert into a neutron beam through photohadronic interactions 
\cite{AtoyanDermer03}. 
On a length scale $l \sim 100\, (E_n/10^{19}\,\rm eV)\,$kpc the neutron beam would
convert back into a proton beam through beta decays. 

\begin{figure}[tb]
{\epsfig{file=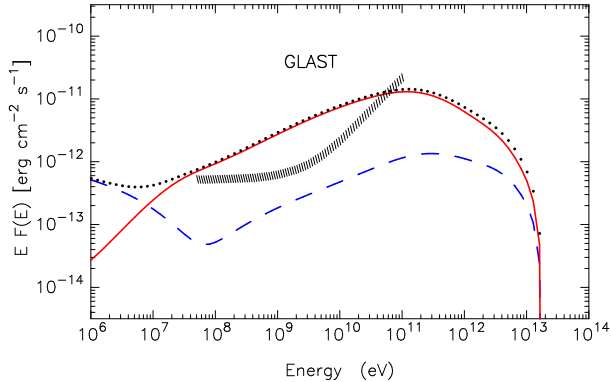,width=8cm}}
\caption{\label{cyg} Fluxes from the electromagnetic cascade initiated in  
Cyg A by UHECRs assuming the total 
injection power of secondary UHE electrons and gamma rays injected at
$\leq 1\,$Mpc distances about 
$10^{45}\, \rm erg/s$. The solid and dashed lines show the synchrotron and
Compton fluxes, respectively.}
\end{figure}

The interaction of UHECR with photons from the Cosmic Microwave Background (CMB)
creates secondary gamma rays and electrons/positrons. 
Depending on the strength of the intergalactic magnetic field ($B_{\rm IGMF}$), 
a next-generation ground-based gamma-ray experiment could detect GeV/TeV 
gamma rays from synchrotron emission of first generation electrons/positrons 
($B_{\rm IGMF}\ge 10^{-9}$~G), or inverse Compton radiation from an 
electromagnetic cascade ($B_{\rm IGMF}\le 10^{-9}$~G) \cite{2005PhRvL..95y1102G}. 
Figure \ref{cyg} shows gamma-ray fluxes expected from the electromagnetic cascade 
initiated in the CMBR and $B=3\, \mu G$ environment 
of Cyg A by injecting $10^{45}\,\rm erg/s$ of secondary electrons and/or gamma rays from 
GZK protons. For the distance to Cyg A of $\simeq 240\,\rm Mpc$
the assumed radial size of the cluster $R< \sim 1\, Mpc$ corresponds to an
extended source, or halo, of angular size $< \sim 14$ arcmin.  
Although the absorption in EBL at TeV energy is significant, the source should 
be detectable with a next-generation experiment because the source spectrum 
is very hard owing to synchrotron emission of UHE electrons. 
The detection of such emission could give information about the $\gg$TeV luminosity 
of these sources, about the intensity and spectrum of the EBL, and about the strength of the IGMF. A few aspects will be discussed further below.

A next-generation experiment might also be able to detect gamma-ray haloes with diameters 
of a few Mpc around superclusters of galaxies. Such haloes could be powered by all the sources 
in the supercluster that accelerate UHECRs. The size of the halo in these cases will be 
defined by the combination of gyroradius of the UHE electrons and their cooling path 
(synchrotron and Compton in Klein-Nishina regime). The spectral and spatial distibutions 
of such halos will contain crucial information about the EBL and 
intergalactic magnetic fields.
\subsection{Extragalactic radiation fields and extragalactic magnetic fields}
%
%
Very high-energy gamma-ray beams traveling over extragalactic distances
are a unique laboratory for studying properties of photons, to constrain 
theories that describe spacetime at the Planck scale and for testing radiation 
fields of cosmological origin.
The potential for probing the cosmic infrared background with TeV photons
was first pointed out by Gould and Schr\'eder \cite{Goul:67} and was revived 
by Stecker, de Jager \& Salamon \cite{Stec:92}, inspired by the detection of extragalactic TeV gamma-ray
sources in the nineties. 
High-energy gamma rays traveling cosmological  distances are attenuated 
en route to Earth by  $\gamma+\gamma \rightarrow e^+ + e^-$ interactions 
with photons from the extragalactic background light. 
While the  Universe is transparent for gamma-ray astronomy with energies below 10 GeV, 
photons with higher energy are absorbed by diffuse soft photons of wavelengths short 
enough for pair production. Photons from the EBL
in the 0.1 to 20 micron wavelength range render the Universe opaque in TeV gamma rays,  
similarly to the cosmic microwave background that constitutes a barrier for  100~TeV photons.  
The transition region from an observational window  turning opaque with 
increasing gamma-ray energy provides the opportunity for deriving observational 
constraints to the intervening radiation field.
Whereas the cosmic microwave background is accessible via direct measurements,
the cosmic infrared background (CIB) has been elusive and remains extremely 
difficult to discern by direct measurements. 
Energy spectra of extragalactic gamma-ray emitters between 10 GeV to 100 TeV allow us to 
extract information about the diffuse radiative background using spectroscopic 
measurements. Non-thermal gamma-ray emission spectra often extend over several orders of 
magnitude in energy and the high-energy absorption features expected from pair production can be 
adequately resolved with the typical energy resolution of 10\% to 20\% achievable with atmospheric
Cherenkov telescopes.

The EBL, spanning the UV to far-infrared wavelength region, consists of the cumulative 
energy releases in the Universe since the epoch of recombination  (see \cite{Haus:01} for a review). 
The EBL spectrum comprises of two distinct components. 
The first, peaking at optical to near-infrared wavelengths (0.5-2~$\mu$m), consists 
of primary redshifted stellar radiation that escaped the galactic environment either 
directly or after scattering by dust. 
In a dust-free Universe, the SED of this component can be simply determined from 
knowledge of the spectrum of the emitting sources and the cosmic history of their 
energy release. In a dusty Universe, the total EBL intensity is preserved, but the 
energy is redistributed over a broader spectrum, generating a second component 
consisting of primary stellar radiation that was absorbed and reradiated by dust 
at infrared (IR) wavelengths. This thermal emission component peaks at wavelengths 
around 100 to 140~$\mu$m. The EBL spectrum exhibits a minimum at 
mid-IR wavelengths (10 - 30 $\mu$m), reflecting the decreasing intensity of the 
stellar contribution at the Rayleigh-Jeans part of the spectrum, 
and the paucity of very hot dust that can radiate at these wavelengths. 

All energy or particle releases associated with the birth, evolution, and death 
of stars can ultimately be related to or constrained by the intensity or spectral 
energy distribution (SED) of the EBL.
The energy output from AGN represent a major non-nuclear contribution to the radiative 
energy budget of the EBL. Most of the radiative output of the AGN emerges at X-ray, UV, 
and optical wavelengths. However, a significant fraction of the AGN output can be 
absorbed by dust in the torus surrounding the accreting black hole, and reradiated at IR wavelengths. 
In addition to the radiative output from star forming galaxies and AGN, 
the EBL may also harbor the radiative imprint of a variety of ''exotic'' 
objects including Population III stars, decaying particles, and primordial
massive objects. EBL measurements can be used to constrain 
the contributions of such exotic components.\\[2ex]
%
%
Direct detection and measurements of the EBL are hindered by the fact that it has no 
distinctive spectral signature, by the presence of strong foreground emission from 
the interplanetary (zodiacal) dust cloud, and from the stars and interstellar medium of the Galaxy. 
Results obtained from TeV gamma-ray observations will complement the results 
from a number of NASA missions, i.e. Spitzer, Herschel, 
the Wide-Field Infrared Survey Explorer (WISE), 
and the James Webb Space Telescope (JWST).
In order to derive the EBL density and spectrum via gamma-ray absorption, 
ideally one would use an astrophysical standard candle of gamma rays to 
measure  the absorption component imprinted onto
the observed spectrum. In contrast, extragalactic TeV gamma-ray sources 
detected to date are highly variable AGN. Their 
gamma-ray  emission models are not unanimously agreed upon, making it 
impossible to predict the intrinsic source spectrum.
Therefore, complementary methods are required for a convincing detection of EBL attenuation.
Various approaches have been explored to constrain/measure the EBL 
intensity \cite{Stec:92,Bill:95,Vass:99,Dwek:05,Ahar:06,Mazi:07}, ranging from searching for 
cutoffs, the assumption of plausible theoretical source models, 
the possibility of using contemporaneous X-ray to TeV measurements combined 
with emission models and the concept of simultaneous constraints from direct
IR measurements/limits combined with TeV data via exclusion of unphysical gamma-ray spectra.
All of these techniques are useful; however, none has so far provided an unequivocal
result independent of assumed source spectra.

The next-generation gamma-ray experiments will allow us to use 
the flux and spectral variability of blazars \cite{Copp:99,Kren:02,Ahar:02} 
to separate variable source phenomena 
from external persistent spectral features associated with  absorption of the 
gamma-ray beam by the EBL.  
Redshift dependent studies are required to distinguish 
possible absorption by radiation fields nearby the source from extragalactic absorption. The most prominent feature of blazars is their occasional brightness (sometimes 
$>$ 10~Crab) yielding a wealth of photon statistics.  Those flares are to date
the most promising tests of the EBL density based on absorption. To constrain the EBL between 
the UV/optical all the way to the far IR a statistical sample of gamma-ray sources, and a broader 
energy coverage with properly matched sensitivity are required.

Since the cross-section for the absorption of a given gamma-ray energy is maximized
at a specific target photon wavelength (e.g., a 1 TeV gamma-ray encounters a 0.7 eV
soft photon with maximum cross-section), there is a natural division of EBL studies with 
gamma rays into three regions:  the UV to optical light, the
 near- to mid-IR and the mid- to far-IR portion of the EBL are the most effective
absorbers for $\approx$ 10 - 100 GeV, the $\approx$ 0.1 TeV to 10 TeV and the $\approx$ 
10 - 100 TeV regime, correspondingly.

In the search for evidence of EBL absorption in blazar spectra it is important 
to give consideration to the shape of the EBL spectrum showing a near IR peak, a mid IR valley 
and a far IR peak; absorption  could imprint different features onto the observed blazar spectra.
For example, a cutoff from the rapid increase of the opacity with gamma-ray energy and redshift 
is expected to be most pronounced in an energy spectral regime that corresponds to a 
rising EBL density; e.g., as is found between 0.1 - 2 micron.     This corresponds to gamma 
ray energies of 10 GeV - 100 GeV.  A survey with an instrument with sensitivity in the 10 GeV
to several 100s of GeV could measure a cutoff over a wide range of redshifts and constrain
the UV/optical IR part of the EBL.  Fermi, together with existing ground-based telescopes, is
promising in yielding first indications or maybe first conclusive results for a 
detection of the EBL absorption feature.    
However, an instrument 
with a large collection area over the given energy range by using the ground-based gamma-ray 
detection technique would allow stringent tests via spectral variability measurements. 

Similarly, a substantial rise in the opacity with gamma-ray energy is expected
in the energy regime above 20 TeV, stemming from the far IR peak.   
A corresponding cutoff should occur in the 20-50 TeV regime.   Prospective 
candidate objects are Mrk~421, Mrk~501 or 1ES1959+650, as they provide episodes of
high gamma-ray fluxes, allowing a search for a cutoff with ground-based instruments 
that have substantially enlarged collection areas in 10 - 100 TeV regime.  Sensitivity for 
detection of a cutoff in this energy regime requires IACTs with a collection area in excess 
of $\rm 1 km^{^2}$.  

Finally, a promising and important regime for ground-based telescopes to contribute
to EBL constraints lies in the  near and the mid IR (0.5 - 5 micron).   The peak in the
near IR and the slope of decline in the mid IR could lead to unique spectral imprints 
onto blazar spectra  around 1-2 TeV, assuming sufficient instrumental sensitivity.   A steep decline 
could lead to a decrease in opacity, whereas a minimal decline could result in steepening 
of the slope of the source  spectrum. If  this feature is sufficiently 
pronounced and/or the sensitivity of the instrument is sufficient, it could be a powerful method 
in unambiguously deriving the level of absorption and discerning the relative near to mid IR density.
  The location of the near IR peak and, consequently, the corresponding change in absorption, is 
expected to occur around 1.5 TeV, which requires excellent sensitivity between 100 GeV and 10 TeV.
The discovery of a signature for EBL absorption at a characteristic energy would be extremely 
valuable in establishing the level of absorption in the near to mid IR regime.  The origin of any 
signature could be tested using spectral variations in blazar spectra and discerning a 
stable component. 

A powerful tool for studying the redshift dependence of the
EBL intensity are pair haloes \cite{Ahar:94}.  For suitable IGMF strengths, such 
haloes will form around powerful emitters of $>$100 TeV gamma rays or UHECRs, 
e.g. AGN and galaxy clusters. 
If the intergalactic magnetic field (IGMF) is not too strong, the high-energy radiation 
will initiate intergalactic electromagnetic pair production and inverse Compton cascades. 
For an intergalactic magnetic field (IGMF) in the range between $10^{-12}$~G and $10^{-9}$~G
the electrons and positrons can isotropize and can result in a spherical halo glowing 
predominantly in the 100 GeV -- 1 TeV energy range. These haloes should have large
extent with radial sizes $>$ 1 Mpc. The size of a 100 GeV halo 
surrounding an extragalactic source at a distance of 1 Gpc could be 
less than 3$^{\circ}$ and be detectable with a next-generation IACT
experiment. The measurement of the angular diameter of such a halo
gives a direct estimate of the local EBL intensity at the redshift of
the pair halo. Detection of several haloes would thus allow us to
obtain unique information about the total amount of IR light produced
by the galaxy populations at different redshifts.

For a rather weak IGMF between $\sim\,10^{-16}$~G and $\sim\,10^{-24}$~G,
pair creation/inverse Compton cascades may create a GeV/TeV ''echo'' of a 
TeV GRB or AGN flare \cite{Plaga95}. The IGMF may be dominated by a primordial 
component from quantum fluctuations during the inflationary epoch of the Universe, 
or from later contributions by Population III stars, AGN, or normal galaxies. The time delay 
between the prompt and delayed emission depends on the deflection of the 
electrons by the IGMF, and afford the unique 
possibility to measure the IGMF in the above mentioned interval of 
field strengths.

\clearpage


\section{Gamma-ray bursts}
\label{grb-subsec}
Group membership:\\ \\ A. D. Falcone, D. A. Williams, M. G. Baring, R. Blandford, J. Buckley,
V. Connaughton, P. Coppi, C. Dermer, B. Dingus, C. Fryer, N. Gehrels,
J. Granot, D. Horan, J. I. Katz, K. Kuehn, P. M\'esz\'aros, J. Norris,
P. Saz Parkinson, A. Pe'er, E. Ramirez-Ruiz, S. Razzaque, X.-Y. Wang, and B. Zhang

\subsection{Introduction}


High energy astrophysics is a young and relatively undeveloped
field, which owns much of the unexplored ``discovery space'' in
contemporary astronomy. The edge of this discovery space has recently been illuminated by the current generation of very high energy (VHE) telescopes, which have discovered a diverse catalog of more than seventy VHE sources. At this time, gamma ray bursts (GRBs) have eluded attempts to detect them with VHE telescopes (although some tentative, low-significance detections have been reported). However, theoretical predictions place them 
near the sensitivity limits of current instruments.
The time is therefore at hand to increase VHE telescope sensitivity, thus facilitating the detection of these extreme and mysterious objects. 


Much has been learned since the discovery of GRBs in the late 1960s.
There are at least two classes of GRB, most conveniently referred to as ``long'' and ``short,''
based on the duration and spectral hardness of their prompt sub-MeV emission.   The
distribution of the types and star formation rates of the host galaxies suggests different 
progenitors for these two classes.
The exact nature of the progenitors 
nevertheless 
remains unknown, although it is widely believed that long GRBs come from the deaths of massive rotating stars and short GRBs result from compact object mergers. The unambiguous solution to this mystery is critical to astrophysics since it has fundamental importance to several topics, including stellar formation history and ultra high energy cosmic ray acceleration. A detection of VHE emission from GRBs would severely constrain the physical parameters surrounding the particle acceleration from GRBs and the energy injected into the particle acceleration sites, and would therefore constrain the properties of the GRB progenitors themselves. These same observations would constrain models for cosmic ray acceleration.

One of the big questions regarding GRBs is 
whether the jets are dominated by ultrarelativistic protons, that interact with either the radiation field or the
background plasma, or 
are dominated by e$^+$e$^-$ pairs.  
The combination of Fermi and
current generation VHE 
telescopes 
such as 
HESS,
MAGIC  
and VERITAS
will 
contribute to progress on these questions in the near term, but more sensitive observations will likely be needed.

The same shocks which are thought to accelerate electrons
responsible for non-thermal $\gamma$-rays in GRBs should also
accelerate protons. Both the internal and the external reverse shocks
are expected to be mildly relativistic, and are expected to lead to relativistic
protons. The maximum proton energies achievable in GRB shocks are estimated to be 
$\sim$10$^{20}$ eV, comparable to the highest energies of the mysterious ultra high energy cosmic rays measured with large ground arrays. 
The accelerated protons can interact with the fireball photons,
leading to 
pions, followed by high-energy gamma rays, 
muons, and neutrinos. Photopion production
is enhanced in conditions of high internal photon target density, 
whereas 
if the density of (higher-energy) photons is too large,
the fireball is optically thick to gamma-rays, 
even in a purely leptonic outflow.
High-energy gamma-ray studies of GRBs 
provide a direct probe of the shock
proton acceleration as well as of the photon density.

\subsubsection{Status of theory on emission\\models}

Gamma-ray burst $\nu F_\nu$ spectra have a peak at photon energies ranging from a few keV to several MeV, and the spectra are nonthermal. From EGRET data, it is clear that the 
spectra extend to at least several GeV \cite{Hurleyetal94,Dingus95,Dingusetal98,Gonzalezetal03}, and there is a possible detection in the TeV range by Milagrito \cite{Atkinsetal00,Atkinsetal03}.
These non-thermal spectra imply
that a significant fraction of the explosion energy is first converted into another form of energy before being dissipated and converted to nonthermal radiation. The most widely accepted interpretation is the conversion of the explosion energy into kinetic energy of a
relativistic flow \cite{Paczynski86,Goodman86,Paczynski90}. At a second stage,
the kinetic energy is converted into radiation via internal collisions 
(internal shock model) resulting from variability
in the ejection from the progenitor \cite{PaczynskiXu94,ReesMeszaros94} or an
external collision (external shock model) with the surrounding medium 
\cite{ReesMeszaros92,DermerMitman99,Dermer07}. The collisions
produce shock waves, which enhance and are believed to create magnetic fields, as well as to accelerate electrons to high energies \cite{Kazimuraetal98,Silvaetal03,Frederiksenetal04,%
Nishikawaetal05,Spitkovsky08}. 
In the standard theoretical model, the initial burst of emission described above (prompt emission) is followed by 
afterglow emission, discussed below, from an external shock that moves through the circumburst environment.

Flux variability in GRBs is seen on timescales as short as milliseconds and can occur at late times.
This rapid variability 
can be easily explained in the internal shock model, which
makes it the most widely used model. It can also be explained in the context of the external
shock model 
either 
if one assumes variations in the strength of the magnetic field or in the energy
transfer to the non-thermal electrons \cite{PanaitescuMeszaros98}, or 
by collisions of the outflow with  
small, high density clouds in the surrounding medium \cite{DermerMitman99,Dermer07}.

An alternative way of producing the emission involves conversion of the explosion energy
into magnetic energy \cite{Thompson94,Usov94,MeszarosRees97}, which
produces a flow that is Poynting-flux dominated. The emission is produced following dissipation
of the magnetic energy via reconnection of the magnetic field lines \cite{Drenkhahn02,%
DrenkhahnSpruit02,LyutikovBlandford03,GianniosSpruit05}. 
An apparent advantage of this model over the internal or external shock model is that the conversion of
energy to radiation is much more efficient (see \cite{Kumar99,Panaitescuetal99} on the
efficiency problem in the internal shocks model). The microphysics of the
reconnection process in this model, like the microphysics determining the 
fraction of energy in relativistic electrons and in the magnetic field
in the internal and external 
shock scenarios,
is not yet fully understood.

VHE observations 
probe the extremes of the efficiency of energy conversion for each of these models 
and 
simultaneously probe the environment where the emission originated. 


The dissipation of kinetic and/or magnetic energy leads to the emission of radiation.
The leading emission mechanism employed to interpret the GRB prompt emission in the keV-MeV region of the spectral energy distribution is nonthermal synchrotron radiation \cite{Meszarosetal93,MeszarosRees93b,Katz94b}.
An order of magnitude estimate of the maximum observed energy of photons produced
by synchrotron emission was derived in \cite{PeerWaxman04b}: Assuming that the
electrons are Fermi accelerated in the shock waves, the maximum Lorentz factor of the accelerated
electrons $\gamma_{\max}$ is found by equating the particle acceleration time
and
the synchrotron cooling time,
yielding $\gamma_{\max} = 10^5 / \sqrt{B /10^6}$, where $B$ is the magnetic
field strength in gauss.
For relativistic motion with bulk Lorentz factor $\Gamma$ 
at redshift $z$, synchrotron emission from electrons with $\gamma_{\max}$ 
peaks in the observer's frame at
energy  $70 \, (\Gamma/315) 
(1+z)^{-1}$ GeV,  
which is independent of the magnetic field.
Thus, synchrotron emission can 
produce photons 
with energies up to, and possibly exceeding, $\sim$100 GeV.

Many of the observed GRB spectra were found to be consistent with the synchrotron
emission interpretation \cite{Tavani96a,Tavani96b,Cohenetal97}. However, a significant fraction
of the observed spectra were found to be too hard (
spectral photon index harder than $2/3$ at low energies) to be accounted for by this model \cite{Crideretal97,Preeceetal98,Fronteraetal00,Preeceetal02,Ghirlandaetal03}.
This motivated studies of magnetic field tangling on very short
spatial scales \cite{Medvedev00}, anisotropies in the electron 
pitch angle distributions \cite{LloydPetrosian00,LloydRonningPetrosian02}, 
reprocessing of radiation by an optically thick cloud heated by the impinging gamma rays \cite{DermerBoettcher00} or by synchrotron self absorption \cite{Granotetal00}, 
and
the contribution of a photospheric (thermal) component \cite{MeszarosRees00,DaigneMochkovitch02,Meszarosetal02}. 
A
thermal component that accompanies the first stages of the overall non-thermal emission and decays after a few seconds was consistent with some observations \cite{Ryde04,Ryde05}. Besides explaining the hard spectra observed in some of the 
GRBs seen by the Burst and Transient Source Experiment (BATSE),
the thermal component provides seed photons that can be Compton scattered by relativistic electrons, resulting in a potential VHE gamma ray emission signature that can be tested.



A natural emission mechanism that can contribute to emission at high energies 
($\gtrsim$MeV) is inverse-Compton (IC) scattering. The seed photons for the scattering can be synchrotron photons emitted
by the same electrons, namely synchrotron self-Compton (SSC) emission 
\cite{MeszarosRees94,Meszarosetal94,PapathanassiouMeszaros96,%
Liangetal97,SariPiran97,PillaLoeb98,ChiangDermer99,PanaitescuMeszaros00},
although in some situations this generates MeV-band peaks broader than those
observed \cite{BaringBraby04}.
The seed photons can also be  thermal emission originating from the photosphere 
\cite{ReesMeszaros05,Peeretal06}, 
an accretion disk 
\cite{ShavivDar95}, 
an accompanying supernova 
remnant \cite{Lazzatietal00,Broderick05}, or supernova emissions in two-step collapse scenarios \cite{Inoueetal03}.
Compton scattering of photons can produce emission up to observed energies
$15 \, (\gamma_{\max}/10^5)\,(\Gamma/315) 
(1+z)^{-1}$ TeV, 
well into the VHE regime.

The shapes of the Comptonized emission spectra in GRBs depend on the spectra of the seed photons and the energy and pitch-angle distributions of the electrons. A thermal population of electrons can inverse-Compton scatter seed thermal photons \cite{Liang97} or photons at energies below the synchrotron self-absorption frequency to produce the observed peak at sub-MeV energies \cite{GhiselliniCelotti99}. Since the electrons cool by the IC process, a variety of spectra can be obtained \cite{PeerWaxman04b,Peeretal06}.
Comptonization can produce a dominant high-energy component \cite{GuettaGranot03a}
that can explain hard high-energy spectral components, such as that observed in
GRB 941017 \cite{Gonzalezetal03,GranotGuetta03a,PeerWaxman04a}. 
Prolonged higher energy emission could potentially be observed with a sensitive VHE gamma ray instrument.


The maximum observed photon energy
from GRBs 
is limited by the annihilation of  gamma rays with target photons, both extragalactic IR background and photons local to the GRB, to produce electron-positron pairs. This limit is sensitive to the uncertain value of the bulk motion Lorentz factor as well as to the spectrum at low energies, and is typically in the sub-TeV regime.
Generally, escape of high-energy photons requires large Lorentz factors.
In fact, observations of GeV photons have been used to constrain the minimum Lorentz factor of the bulk motion of the flow \cite{KrolikPier91,FenimoreEpsteinHo93,%
WoodsLoeb95,BaringHarding97,LithwickSari01,Razzaqueetal04}, 
and spectral 
coverage up to TeV energies could 
further constrain
the Lorentz factor \cite{KobayashiZhang03,Zhangetal03,Peeretal07}.
If the Lorentz factor can be determined independently, e.g.\ from afterglow modeling,
then the annihilation signature can be used to diagnose the gamma-ray emission 
region \cite{GuptaZhang08}.
The evidence for acceleration of leptons in GRB blast waves is based on fitting lepton synchrotron spectra models to GRB spectra. This consistency of leptonic models with observed spectra still allows the possibility of hadronic components in these bursts, and perhaps more importantly, GRBs with higher energy 
emission
have not been explored for such hadronic components due to the lack of sensitive instruments in the GeV/TeV energy range. The crucially important high-energy emission components, represented by only 5 EGRET spark chamber bursts, a handful of BATSE and EGRET/TASC GRBs, and a marginal significance Milagrito TeV detection, were statistically inadequate to look for 
correlations between high-energy and keV/MeV emission
that can be attributed to a particular 
process. Indeed, the 
prolonged
high-energy components in GRB 940217 and the ``superbowl'' burst, GRB 930131, and the anomalous gamma-ray emission component in GRB 941017, behave quite differently than the measured low-energy gamma-ray light curves. Therefore, it is quite plausible that hadronic emission components are found in the high energy spectra of GRBs.

Several theoretical mechanisms exist for hadronic VHE emission components. 
Accelerated protons can emit synchrotron radiation in the GeV--TeV energy band \cite{Vietri97,BottcherDermer98,Totani98}. The power emitted by a particle is 
$\propto \gamma^2/m^2$, where $\gamma$ is the Lorentz factor of the particle and $m$ is its mass. 
Given the larger mass of the proton, 
to achieve the same output luminosity, 
the protons have 
$\sim$1836 times 
higher mean Lorentz factor, 
the acceleration mechanism must convert 
$\sim$ 3 million times 
more energy to protons than electrons 
and the peak of the proton emission would be at 
$\gtrsim$ 2000 times higher  
energy 
than the peak energy of photons emitted by the electrons. 
Alternatively, 
high-energy baryons can produce energetic pions, via photomeson interactions with the low energy photons, creating 
high-energy photons and neutrinos following the pion decay \cite{BottcherDermer98,WaxmanBahcall97,WaxmanBahcall00,DermerAtoyan03,GuettaGranot03b,GranotGuetta03b}. This process could be the primary source of ultra high energy (UHE)
neutrinos. 
Correlations between gamma-ray opacity, bulk Lorentz factor, and neutrino production will test whether GRBs are UHE cosmic ray sources \cite{Dermeretal07}. 
If the neutrino production is too weak to be detected, then the former two measurements can be obtained independently with sensitive GeV-TeV $\gamma$-ray telescopes and combined to test for UHE cosmic ray production. 
Finally, 
proton-proton or proton-neutron collisions may also be 
a
source of 
pions 
\cite{PaczynskiXu94,Katz94a,DePaolisetal00,Derishevetal99,BahcallMeszaros00}, and in addition, 
if there are neutrons in the flow, then the neutron $\beta$-decay has a drag effect on the protons, which may produce another source of radiation \cite{Rossietal06}. Each of 
these cases has a VHE 
spectral shape and intensity 
that can be studied coupled with the 
emission measured at lower energies and with neutrino measurements.

Afterglow emission is explained in synchrotron-shock models by the same processes that occur during the prompt phase. The key difference is that the afterglow emission originates from large radii,
$\gtrsim 10^{17}$~cm, as opposed to the much smaller radius of the flow during the prompt emission
phase,  $\simeq 10^{12} - 10^{14}$~cm for internal shocks, and $\simeq 10^{14} - 10^{16}$~cm for external shocks. As a result, the density of the blast-wave shell material is smaller during the afterglow emission phase than in the prompt phase, and some of the radiative mechanisms, e.g. thermal collision processes, may become less important.


Breaks in the observed lightcurves,
abrupt changes in the power law slope, 
are attributed to a variety of phenomena, such as refreshed shocks originating from late time central engine activity \cite{ReesMeszaros98,Granotetal03}, aspherical variations in the energy \cite{KumarPiran00}, or variations in the external density \cite{WangLoeb00,Nakaretal03}. Blast wave energy escaping in the form of UHE neutrals and cosmic rays can also produce a rapid decay in the X-ray light curve \cite{Dermer07}. In addition, interaction of the blast wave with the wind termination shock 
of the progenitor 
may be the source of a jump in the lightcurve \cite{Wijers01,RamirezRuizetal05,PeerWijers06}, 
although this bump may not be present at a significant level \cite{NakarGranot07}. High-energy gamma-ray observations may show whether new photohadronic emission mechanisms are required, or if the breaks do not require new radiation mechanisms for explanation (see, e.g., \cite{Genetetal07,UhmBeloborodov07}).

\subsubsection{GRB Progenitors}

We still do not know the exact progenitors of GRBs, and it is therefore difficult, if not impossible, to understand the cause of these cosmic explosions. These GRB sources involve emission of energies that can
exceed $10^{50}$ ergs. The seat of this activity is
extraordinarily compact, as indicated by rapid variability of the
radiation flux on time scales as short as milliseconds. It is unlikely
that mass can be converted into energy with better than a few (up to
ten) percent efficiency; therefore, the more powerful short GRB sources
must ``process'' upwards of $10^{-3}M_\odot$ through a region which is
not much larger than the size of a neutron star (NS) or a stellar mass
black hole (BH). No other entity can convert mass to energy with such
a high efficiency, or within such a small volume.  
The leading contender for the production of the longer class of GRBs
---  supported by observations of supernovae associated with several bursts --- 
is the catastrophic collapse of massive, rapidly rotating stars.
The current preferred model for short bursts, the merger of binary systems of
compact objects, such as double neutron star systems (e.g. Hulse-Taylor pulsar systems)
is less well established. 
A fundamental problem posed by GRB sources is how to generate
over $ 10^{50}$ erg in the burst nucleus and channel it into
collimated relativistic plasma jets.

The progenitors of GRBs are essentially masked by the resulting fireball, which reveals little more than
the basic energetics and microphysical parameters of relativistic shocks.  
Although long and short bursts most likely have different progenitors, the observed radiation
is very similar.
Progress in understanding the progenitors can come from determining the burst environment, the kinetic energy and Lorentz factor of the ejecta, the duration of the central engine activity, and the redshift distribution. VHE gamma-ray observations can play a supporting role in this work.  To the extent that we understand GRB emission across the electromagnetic spectrum, we can look for the imprint of the burst environment or absorption by the extragalactic background light on the spectrum as an indirect probe of the environment and distance, respectively. VHE 
emission
may also prove to be crucial to the energy budget of many bursts, thus constraining the progenitor.

\subsection{High-energy observations of gamma-ray bursts}

Some of the most significant advances in GRB research have come from
GRB correlative observations at longer wavelengths. Data on
correlative observations at shorter wavelengths are sparse but
tantalizing and inherently very important. One definitive observation
of the prompt or afterglow emission could significantly influence our
understanding of the processes at work in GRB emission and its
aftermath. Although many authors have predicted its existence, the predictions
are near or below the sensitivity of current instruments, and there
has been no definitive detection of VHE emission from a GRB either
during the prompt phase or at any time during the multi-component
afterglow.

For the observation of photons of energies above 300\,GeV, only
ground-based telescopes are available. These ground-based
telescopes fall into two broad categories, air shower arrays and
imaging atmospheric Cherenkov telescopes (IACTs). The air shower arrays,
which have wide fields of view that are suitable 
for GRB searches, are relatively insensitive. There are several reports
from these instruments of possible TeV emission: 
emission $>$16 TeV from GRB\,920925c \cite{Padilla:98:Airobicc},
an indication of 10\,TeV emission in a stacked analysis of 57 bursts
\cite{Amenomori:01:TibetGRBs},
and
an excess gamma-ray signal during the prompt phase of
GRB\,970417a 
\cite{Atkinsetal00}. In all of
these cases however, the statistical significance of the detection is
not high enough to be conclusive. 
In addition to searching the Milagro
data for VHE counterparts for over 100 satellite-triggered GRBs since 2000
\cite{Atkins:05:MilagroGRBCounterparts,SazParkinson:07:MilagroGRBCounterparts,Abdo:07:MilagroGRBCounterparts},
the Milagro Collaboration
conducted a search for VHE transients of 40 seconds to 3 hours
duration in the northern sky \cite{Atkins:04:MilagroGRB}; 
no evidence
for VHE emission was found from 
these searches.

IACTs have better flux sensitivity and energy resolution 
than air shower arrays,
but are limited by their
small fields of view (3--5$^\circ$) and low duty cycle ($\sim$10\%).
In the BATSE \cite{Meegan:92} era (1991--2000), attempts at GRB monitoring were
limited by slew times and uncertainty in the GRB source position
\cite{Connaughton:97}. 
More recently, VHE upper
limits from 20\% to 62\% of the Crab flux at late
times ($\gtrsim$4 hours) were obtained with Whipple Telescope
for seven GRBs in 2002-2004 \cite{Horan:07}. 
The MAGIC Collaboration took
observations of GRB\,050713a beginning 40 seconds after the prompt
emission but saw no evidence for VHE emission
\cite{Albert:06:40sGRB}. Follow-up GRB observations have been made on
many more GRBs by the MAGIC Collaboration
\cite{Bastieri:07:ICRCStatus} but no detections have been
made \cite{Albert:2006:GRBs,Bastieri:7:ICRCGRBObs}.  Upper limits of 2--7\% of the 
Crab flux on the
VHE emission following three GRBs have also been obtained with VERITAS \cite{Horan:07ICRC}.

One of the main obstacles for VHE observations of GRBs is the distance
scale. Pair production interactions of gamma rays with the infrared
photons of the extragalactic background light attenuate the gamma-ray
signal,
limiting the distance over which VHE gamma rays can
propagate. 
The MAGIC
Collaboration 
has reported the detection of 
3C279, 
at 
redshift 
of 0.536
\cite{Teshima:07}. This represents a large increase in distance
to the 
furthest detected VHE source, revealing more of the
universe to be visible to VHE astronomers than was previously thought.






\subsection{High Energy Emission Predictions for Long Bursts}

As described earlier, long duration GRBs are generally believed to be associated with core
collapses of massive rotating stars \cite{Woosley93,MacFadyenWoosley99}, which lead to particle acceleration by relativistic internal shocks in jets. The 
isotropic-equivalent 
gamma-ray luminosity can vary
from $10^{47}~{\rm erg~s^{-1}}$ all the way to $10^{53}~{\rm erg~s^{-1}}$. 
They are distributed in a wide redshift range (from 
0.0085 for GRB 980425
\cite{Bloometal99} to 
6.29 for GRB 050904 \cite{Haislipetal06}, 
with a mean redshift of 
2.3--2.7
for Swift bursts, 
e.g. \cite{Bergeretal05a,Jakobssen06}). The low redshift long GRBs ($z \lesssim 0.1$, e.g. GRB 060218, $z=0.033$ \cite{Campanaetal06}) are typically sub-luminous with 
luminosities of 
$10^{47}-10^{49} ~{\rm erg~s^{-1}}$
and spectral peaks at lower energies,
so they are 
less likely 
detected at high energy.
However, one nearby, ``normal'' long GRB has been detected (GRB 030329, 
$z=0.168$), which has large fluences in both its prompt gamma-ray emission 
and afterglow.

\subsubsection{Prompt emission}
\begin{figure*}[!ht]
\centering
\includegraphics*[height=2in]{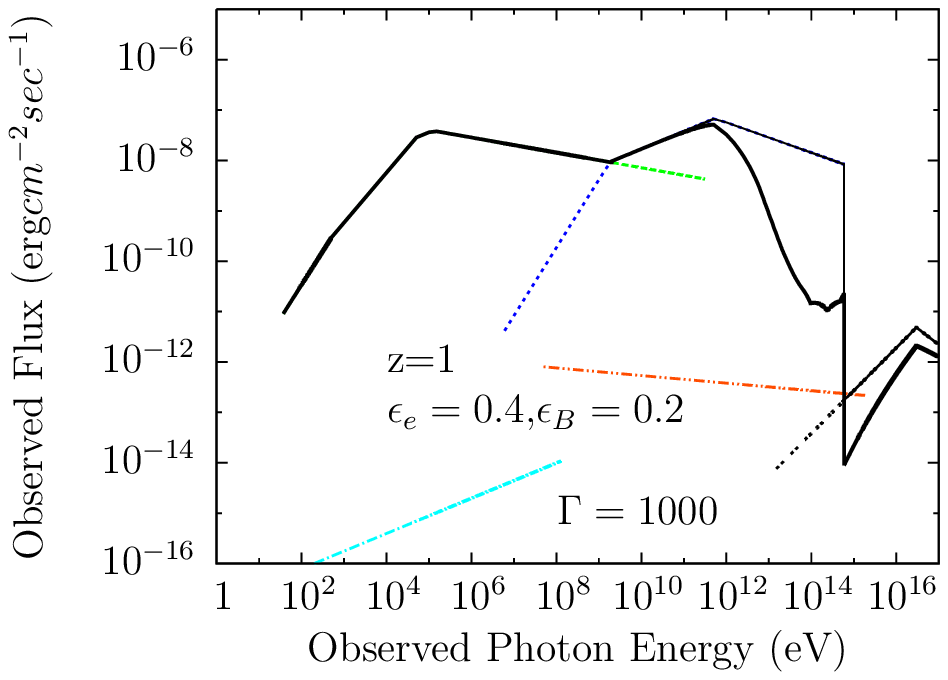}\quad
\includegraphics*[height=2in]{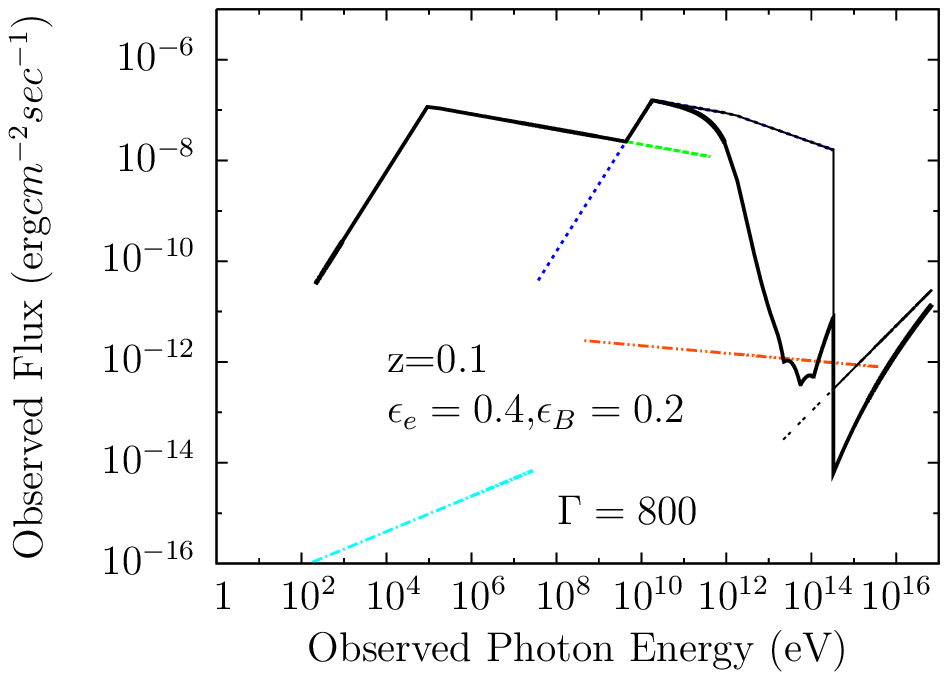}
\caption{Broad-band spectrum of the GRB prompt emission within the internal 
shock model 
(from \cite{GuptaZhang07}). 
(a) 
A 
long GRB with the observed sub-MeV luminosity of $\sim 10^{51}~{\rm
erg~s^{-1}}$,
is modeled 
for parameters as given in the figure.
The solid black lines represent the final spectrum before (thin line)
and after (thick line) including the effect of internal optical depths. 
The long dashed green line (mostly hidden) is the electron synchrotron component; the short-dashed blue line is the electron IC
component; the double short-dashed black curve on the right side is the $\pi^0$ decay 
component; the triple short-dashed dashed line represents the synchrotron radiation produced by 
e$^\pm$ from $\pi^\pm$ decays; the dash-dotted (light blue) line represents 
the proton synchrotron component. 
(b) The analogous spectrum of a bright short GRB with 
10$^{51}$ erg isotropic-equivalent energy release.} 
\label{low-compac}
\label{short}
\end{figure*}
The leading model of the GRB prompt emission is the internal shock model
\cite{ReesMeszaros94},
and we begin by discussing prompt emission in that context.
The relative importance of the leptonic vs. hadronic components for high
energy photon emission depends on the unknown shock equipartition parameters,
usually denoted as $\epsilon_e$, $\epsilon_B$ and $\epsilon_p$ for the
energy fractions carried by electrons, magnetic fields, and protons,
respectively. Since electrons are much more efficient emitters than protons,
the leptonic emission components usually dominate unless $\epsilon_e$ is 
very small. 
Figure~\ref{low-compac}a 
displays the broadband spectrum of a 
long GRB 
within the internal shock model for 
a particular choice
of parameters 
\cite{GuptaZhang07}. 
Since the phenomenological shock microphysics
is poorly known, modelers usually introduce 
$\epsilon_e$, $\epsilon_B$, $\epsilon_p$
as free parameters.
For 
$\epsilon_e$'s not too small (
$\gtrsim$10$^{-3}$), the high energy spectrum is dominated by the electron IC 
component, 
as 
in Fig.~\ref{low-compac}a.
For smaller $\epsilon_e$'s (e.g. $\epsilon_e =10^{-3}$), 
on the 
other hand, the hadronic components become at least comparable to the 
leptonic component above $\sim$100 GeV, and the $\pi^0$-decay component 
dominates the spectrum above $\sim$10 TeV. 

A bright GRB, 080319B, with a plethora of multiwavelength observations has recently allowed very detailed spectral modelling as a function of time, and it has shown that an additional high energy component may play an important role. For GRB 080319B, the bright optical flash suggests a synchroton origin for the optical emission and SSC production of the $\sim$500 keV gamma-rays \cite{Racusin08}.  The intensity of these gamma rays would be sufficient to produce a second-order IC peak around 10--100 GeV.

Due to the high photon number density in the emission region of GRBs, 
high energy photons 
have an optical depth for photon-photon pair production
greater than unity above a critical energy, producing a sharp spectral cutoff, which depends on the unknown bulk Lorentz 
factor of the fireball and the variability time scale of the central 
engine, which sets the size of the emission region.
Of course, the shape of time-integrated spectra will also be modified (probably to power laws rolling over to steeper power laws) due to averaging of evolving instantaneous spectra \cite{Granotetal08}. 
For the nominal bulk Lorentz factor $\Gamma=400$ (as suggested by recent 
afterglow observations, e.g. \cite{Molinarietal07}) and for a typical variability time 
scale 
$t_v = 0.01$ s,
the cut off energy is about several tens 
of GeV. Below 10 GeV, the spectrum is mostly dominated by the electron synchrotron
emission, so that with the observed high energy spectrum alone, usually there
is no clean 
differentiation of 
the leptonic vs. 
hadronic origin of the high energy gamma-rays. Such an issue may however be
addressed by collecting both prompt and afterglow data. Since a small $\epsilon_e$
is needed for a hadronic-component-dominated high energy emission, these fireballs
must have a very low efficiency 
for radiation,
$\lesssim\epsilon_e$,
and most
of the energy will be carried by the afterglow. As a result, a moderate-to-high 
radiative efficiency would suggest a leptonic origin of high energy photons, 
while a GRB with an extremely low radiative efficiency but an extended high 
energy emission component would be consistent with (but not a proof for)
the hadronic origin. If the fireball has a much larger Lorentz factor 
($\gtrsim 800$), 
the spectral cutoff energy 
is higher, as in Fig.~\ref{low-compac}. 
This would allow a larger spectral space to diagnose the origin of the GRB high energy emission and would place the cutoff energies in the spectral region that can only be addressed by VHE telescopes. 
At even higher energies, the fireball again becomes transparent to 
gamma rays 
\cite{Razzaqueetal04}, so that under ideal conditions, 
the $\sim$ PeV component due to $\pi^0$ decay
can escape the fireball. Emission above one TeV escaping from GRBs would suffer additional external 
attenuation by 
the cosmic infrared background (CIB) 
and the cosmic microwave background (CMB), thus limiting VHE observations of GRBs to lower redshifts (e.g. z$\lesssim$0.5--1).
The external shock origin of prompt emission is less favored by the Swift observations, which show a rapidly falling light curve 
following the prompt emission before the emergence of a more slowly
decaying component attributed to the external shock.
A small fraction of bursts lack the initial steep component,
in which case the prompt emission may result from an external
shock.  
Photons up to TeV energies are expected in the external
shock scenario \cite{Dermeretal00}, and
the 
internal pair cut-off energy
should be very high, 
more favorable to detection at VHE energies,
because of the less compact emission region.

The ``cannonball model'' of GRBs \cite{DardeRujula04}, in which the prompt GRB emission is produced by IC scattering from blobs of relativistic material (``cannonballs''), can also be used to explain the keV/MeV prompt emission, but it does not predict significant VHE emission during the prompt phase. Sensitive VHE observations would provide a strong constraint to differentiate between these models. The cannonball model could still produce delayed VHE emission during the deceleration phase, in much the  same way as the fireball model: as a consequence of IC scattering from relativistic electrons accelerated by the ejecta associated with the burst \cite{DadoDar05}.

\subsubsection{Deceleration phase}
\begin{figure*}[!ht]
\centering
\includegraphics[height=2.25in]{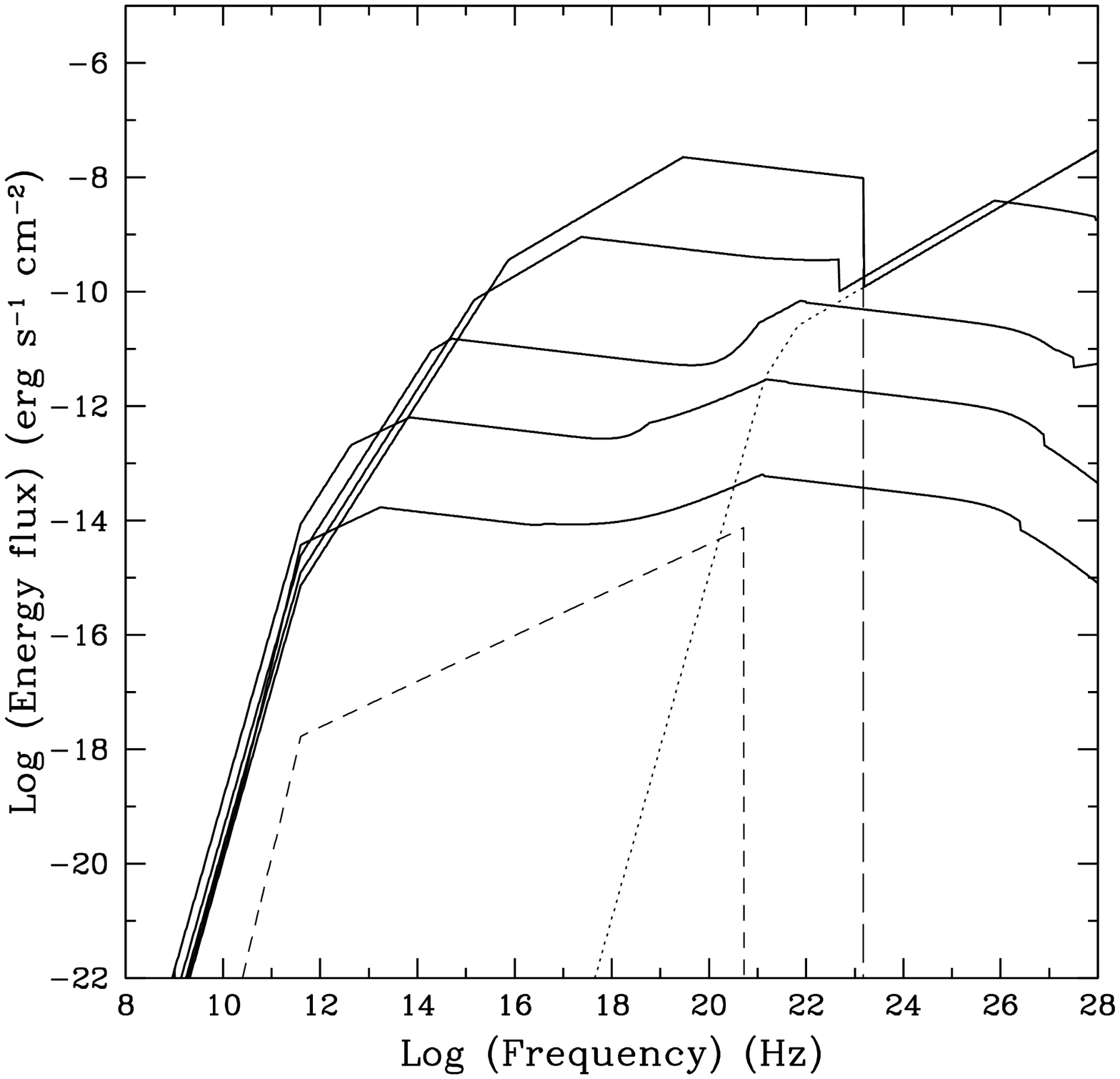} \quad
\includegraphics[height=2.25in]{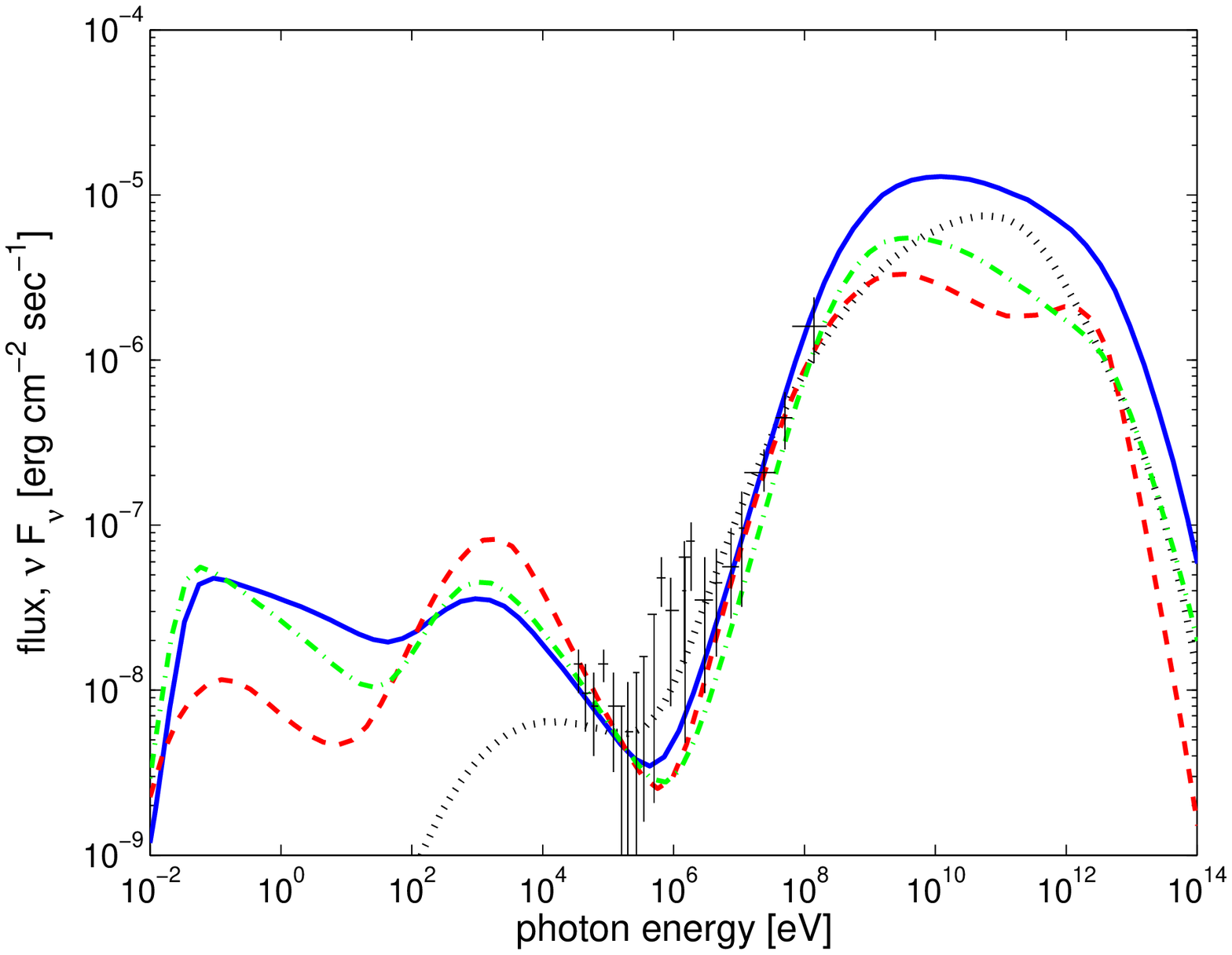}
\caption{
(a) The SSC emission from the forward shock region in the deceleration phase. Temporal evolution of
the theoretical models for synchrotron and SSC components for $\epsilon_e=0.5$, $\epsilon_B=0.01$; solid curves from top to bottom are at onset, 1 min, 1 hour, 1 day, 1 month. The contributions to
the emission at onset are shown as long-dashed (electron-synchrotron),
short-dashed (proton-synchrotron) and dotted (electron IC) curves \cite{ZhangMeszaros01}. (b) Fit to the prompt emission data of GRB 941017 using
the 
IC model of Ref.~\cite{PeerWaxman04a}.}
\label{SSC}
\end{figure*}
A GRB fireball would be significantly decelerated by the circumburst medium starting 
from a distance of $10^{16}-10^{17}$ cm from the central 
engine, at which point
a pair of shocks propagate into the circumburst
medium and the ejecta, respectively. 
Both shocks contain a similar amount of energy. Electrons from 
either shock region
would Compton scatter the soft seed synchrotron
photons from both regions 
to
produce high energy photons
\cite{MeszarosRees94,GranotGuetta03a,PeerWaxman04a,ZhangMeszaros01,Dermeretal00,Wangetal01a,Wangetal01b}.
Compared with the internal shock radius, the deceleration radius corresponds to 
a low ``compactness'' so that high energy
photons 
more readily
escape from the source. 
Figure~\ref{SSC}(a) presents the theoretical forward shock high energy emission components as a function of time for the regime of IC dominance (from \cite{ZhangMeszaros01}).
It is evident that during the first several minutes of the deceleration time, the high energy emission could extend to beyond $\sim$ 10 TeV.
Detection of this emission by ground-based VHE detectors, for sources close enough to have little
absorption by the IR background, would be an important test of this paradigm.


Various IC processes have been considered to interpret the distinct high energy
component detected in GRB 941017 \cite{Gonzalezetal03,GranotGuetta03a}. 
For preferable parameters, the IC emission of forward shock electrons off the
self-absorbed reverse shock emission can interpret the observed spectrum 
(Fig.~\ref{SSC}b, \cite{PeerWaxman04a}). 

\subsubsection{Steep decay}

Swift observations revealed new features of the GRB afterglow. A canonical X-ray
lightcurve generally consists of five components \cite{Zhangetal06,Nouseketal06}: a steep decay component (with decay index $\sim -3$ or 
steeper), a shallow decay component (with decay index $\sim -0.5$ but
with a wide variation), a normal decay component (with decay index 
$\sim -1.2$), a putative post-jet-break component seen in a small group
of GRBs at later times, and multiple X-ray flares with sharp rise and 
decay occurring 
in nearly half GRBs. Not all five components appear in
every GRB, and the detailed afterglow measurements of GRB 080319B
\cite{Racusin08}  present some challenges to the standard picture we describe here.
The steep decay component \cite{Tagliaferrietal05} is 
generally interpreted as the tail of the prompt gamma-ray emission 
\cite{Zhangetal06,Nouseketal06,KumarPanaitescu00}. 
Within this interpretation,
the steep decay phase corresponds to significant reduction of high energy
flux as well. On the other hand, Ref.~\cite{Dermer07} suggests that the steep decay
is the phase when the blastwave undergoes a strong discharge of its hadronic
energy. Within such a scenario, strong high energy emission 
of hadronic origin is expected. Detection/non-detection of strong high energy
emission during the X-ray steep decay phase would greatly constrain the origin
of the steep decay phase.

\subsubsection{Shallow decay}

The shallow decay phase following the steep decay phase is still not
well understood \cite{Meszaros06,Granot06,Zhang07}. The standard 
interpretation is that the external forward
shock 
is continuously refreshed by late energy injection, either
from a long-term central engine, or from slower shells ejected in the prompt
phase \cite{Zhangetal06,Nouseketal06,Panaitescuetal06a,GranotKumar06}.
Other options include delay of transfer of the fireball energy to the
medium \cite{KobayashiZhang07}, a line of sight outside the region of prominent
afterglow emission \cite{EichlerGranot06}, a two-component jet model
\cite{Granotetal06}, and time varying shock micro-physics
parameters \cite{Granotetal06,Iokaetal06,Panaitescuetal06b}.

Since the pre-Swift knowledge of the 
afterglow kinetic energy comes from the late afterglow observations, the
existence of the shallow decay phase suggests that the previously estimated
external SSC emission strength is over-estimated during the early afterglow. A modified SSC model 
including the energy injection effect indeed gives less significant SSC
flux 
\cite{GouMeszaros07,Fanetal07}. 
The SSC component nonetheless is still 
detectable by Fermi and higher energy detectors 
for some choices of parameters.  Hence, detections or limits from VHE 
observations constrain those parameters.
If, however, the shallow decay
phase is not the result of a smaller energy in the afterglow shock at
early times, compared to later times, but instead due to a lower
efficiency in producing the X-ray luminosity, the luminosity at higher
photon energies could still be high, and perhaps comparable to (or even
in excess of) pre-Swift expectations. Furthermore, the different
explanations for the flat decay phase predict different high-energy
emission, so 
the latter could help distinguish between the various
models. For example, in the energy injection scenario, the reverse shock
is highly relativistic for a continuous long-lived relativistic wind from
the central source, but only mildly relativistic for an 
outflow that was
ejected during the prompt GRB with a wide range of Lorentz factors and that
gradually catches up with the afterglow shock. 
The
different expectations for the high-energy emission 
in these two cases
may be tested against
future observations.

\subsubsection{High-energy photons associated with X-ray flares}

X-ray flares have been detected during the early afterglows in a
significant fraction of gamma-ray bursts (e.g. 
\cite{Burrowsetal05,Chincarinietal07,Falconeetal07}). The amplitude of an X-ray flare with 
respect to the background afterglow flux can be up to
a factor of $\sim$500 and the fluence can approximately equal 
the 
prompt emission fluence (e.g. GRB 050502B \cite{Burrowsetal05,Falconeetal06}). The rapid rise
and decay behavior of some flares suggests that they are caused by
internal dissipation of energy due to late central engine
activity \cite{Zhangetal06,Burrowsetal05,Falconeetal06,Liangetal06}. 
There are two likely processes that can produce very
high energy (VHE) photons. One process is that the inner flare
photons, when passing through the forward shocks, would
interact with the shocked electrons and get boosted to higher
energies. Another process is the SSC 
scattering within the X-ray flare region \cite{Fanetal07,Wangetal06}.

\begin{figure*}[!th]
\centering
\includegraphics[width=5.9in]{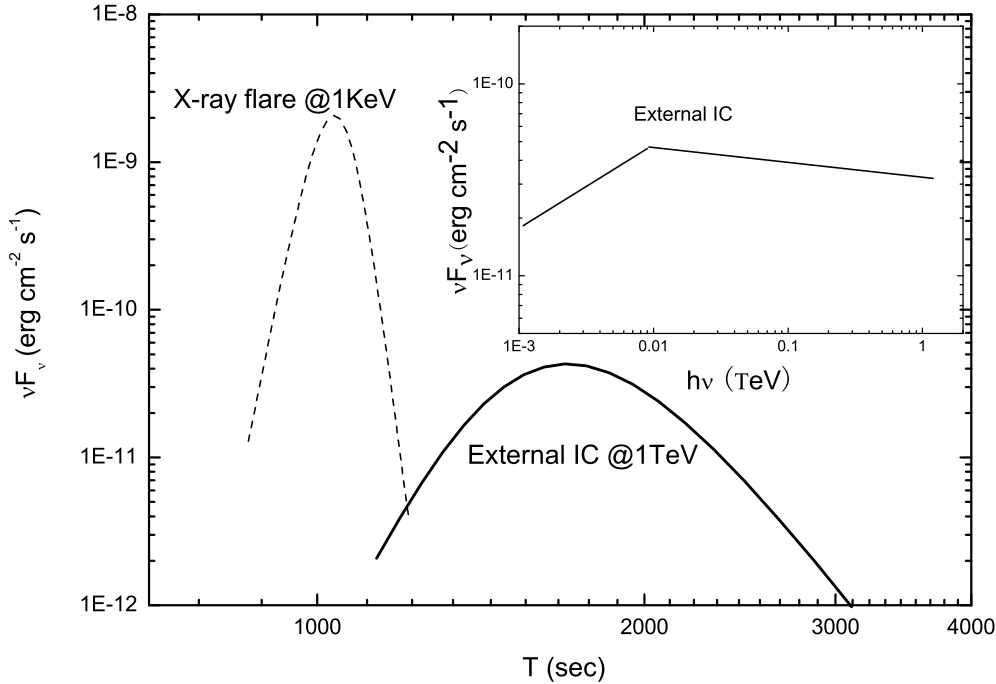}
\caption{The expected light
curves (main figure) and spectral energy distribution (insert
figure) of 
IC scattering of 
X-ray
flare photons by 
forward shock electrons. The flux is
calculated according to 
\cite{Wangetal06}, based
on the following parameters: 
$10^{53}$ erg blast wave energy, 
electron energy distribution index $2.2$, 
electron equipartition factor $\epsilon_e=0.1$, 
1 keV peak energy of the X-ray flare,
$10^{28.5}$ cm source distance
and that the flare has $\delta t/t_f=0.3$. }
\label{flare}
\end{figure*}

Figure~\ref{flare} shows an example of IC scattering of flare 
photons by the afterglow electrons for a flare of duration
$\delta t$ superimposed upon an underlying power law X-ray
afterglow around time $t_f=1000 \,{\rm s}$ after the burst, as
observed in GRB 050502B. 
The duration of the IC emission 
is lengthened by the angular spreading effect
and the anisotropic scattering effect as well 
\cite{Fanetal07,Wangetal06}.   
Using the calculation of \cite{Wangetal06},
for typical parameters as given in the caption,
$\nu F_\nu$ at 1 TeV reaches about $4\times10^{-11}$ erg cm$^{-2}$ s$^{-1}$, with a total duration of about 2000 s.

The
peak energy of the 
SSC scattering within the X-ray flare region
lies at tens of
MeV \cite{Wangetal06} to a few hundreds of MeV \cite{Fanetal07}.
The 
flares may come from internal dissipation processes
similar to the prompt emission, so their dissipation radius may be
much smaller than that of the afterglow external shock. A smaller
dissipation radius causes strong internal absorption to very high
energy photons. For a flare with luminosity $L_x\sim10^{48}$ 
erg s$^{-1}$ and duration $\delta t=100$ s, 
the VHE photons can escape 
only if the dissipation radius is
larger than $\sim$10$^{16}$ cm.  
So in
general, even for a strong X-ray flare occurring at small
dissipation radius, the SSC 
emission at TeV
energies should be lower than the 
IC component above.

\subsubsection{High-energy photons from\\external reprocessing}

Very high energy photons above 100 GeV produced by GRBs at
cosmological distances are subject to photon-photon attenuation 
by the 
CIB 
(
e.g.\ \cite{Primacketal99,Steckeretal06}) and 
CMB. 
The attenuation of $E$ 
TeV photons by the CIB would produce
secondary electron-positron pairs with a Lorentz factor of
$\gamma_e\simeq 10^6 E$, 
which in turn IC scatter off CMB photons to
produce  MeV--GeV emission 
\cite{Razzaqueetal04,ChengCheng96,DaiLu02,Wangetal04}. This emission is 
delayed relative to the primary
photons 
by two
mechanisms: one is the 
opening angles of
the scattering processes, producing a 
deviation from the direction of the original TeV photons by
an angle $1/\gamma_e$;  the other 
is 
the deflection of 
the secondary pairs in
the intergalactic magnetic field \cite{Plaga95}. Only if the intergalactic
magnetic field is 
less than $\sim$10$^{-16}$ G would the
delayed secondary gamma-rays still be beamed from the same direction
as the GRB.  

\subsection{High Energy Emission Predictions for Short Bursts}

Recent observational breakthroughs 
\cite{Gehrelsetal05,Foxetal05,Barthelmyetal05,Bergeretal05b,Bloometal06} suggest that at least some short GRBs 
are nearby low-luminosity GRBs that are associated with old stellar populations and 
likely to be 
compact star mergers. The X-ray afterglows of short duration
GRBs are typically much fainter than those of long GRBs, which is consistent with
having a smaller total energy budget and a lower density 
environment 
as expected from
the compact star merger scenarios. Observations suggest
that except being fainter, the afterglows of short GRBs are not distinctly
different from those of long GRBs. 
The long duration
GRB 060614 has a short, hard emission episode followed by extended softer 
emission. It is 
a nearby GRB, but 
has no supernova association, suggesting 
that 060614-like GRBs are 
more energetic versions of short
GRBs \cite{Gehrelsetal06,Zhangetal07}. 

The radiation physics of short GRBs is believed to be similar to that of
long GRBs. As a result, all the processes discussed
above for long GRBs are relevant to short GRBs as well. 
The predicted prompt emission spectrum of a bright short GRB is presented in
Fig.~\ref{short}b 
\cite{GuptaZhang07}. 
Figure~\ref{short}b
is calculated for 
a comparatively bright, 1-second burst at redshift 0.1
with isotropic-equivalent luminosity 10$^{51}$ erg s$^{-1}$. 
Fig.~\ref{short}b 
suggests that 
the high energy component of such a burst is barely detectable by Fermi. Due to 
internal optical depth, the spectrum is cut off beyond about 100 GeV.  VHE observations can constrain the bulk Lorentz factor, since VHE 
emission can be 
achievable if the bulk
Lorentz factor is even larger (e.g. 1000 or above).


No evidence of strong reverse shock emission from short GRBs exists.
For the forward shock, the flux is typically nearly 100 times fainter than
that of long GRBs. This is a combination of low isotropic energy and presumably
a low ambient density. The SSC component in the forward shock region still
leads to GeV-TeV emission, but the flux 
is scaled down by the same factor
as the low energy afterglows. 
Multiple late-time X-ray flares have been detected
for some short GRBs (e.g. GRB 070724 and GRB 050724),
with at least some properties similar to the flares in long GRBs,
so that the emission mechanisms discussed above for long GRB flares
may 
also apply, 
scaled down accordingly. 
In general, short GRBs
may be
less prominent emitters of high energy photons than long GRBs, mainly due
to their low fluence observed in both prompt emission and afterglows. A potential
higher bulk Lorentz factor on the other hand facilitates the escaping of 100 GeV
or even TeV photons from the internal emission region. Furthermore, a few short GRBs are detected at redshifts lower than 0.3, 
and the average short GRB redshift is much lower than that of long GRBs. This is favorable for TeV detection since
the CIB 
absorption 
is greatly reduced at these
redshifts.



\subsection{Supernova-associated gamma-ray bursts}

Nearby GRBs have been 
associated with
spectroscopically identified supernovae, {\it e.g.}, 
GRB 980425/SN 1998bw, GRB 031203/SN 2003lw, GRB 060218/SN 1006aj, and
GRB 030329/SN 2003lw. 
The processes discussed in the section on high-energy emission from
long GRBs can 
all apply in 
these 
bursts, and
with 
their  close distances, 
VHE emission from 
these sources
would not be significantly attenuated by the CIB{}. 
These bursts
have low luminosity, but the internal absorption by 
soft prompt
emission photons may therefore be lower, so that VHE photons originating from the
internal shock are more likely to escape without significant
absorption, compensating for the overall low flux.
In addition, if there is a highly relativistic jet component
associated with the supernovae, 
supernova
shock breakout photons would be scattered to high energies by the
shock-accelerated electrons in the forward shocks \cite{WangMeszaros06}. 
The strong thermal X-ray emission
from GRB 060218 
may be such 
a relativistic supernova shock 
breakout \cite{Campanaetal06,Waxmanetal07}. 
It has been shown \cite{WangMeszaros06} 
that if the wind mass loss rate from
the progenitor star is 
low, the $\gamma\gamma$ absorption
cutoff energy at early times can be larger than $\sim$100 
GeV, so VHE emission could be detected from these nearby
SN-GRBs.

\subsection{Ultra High Energy Cosmic Rays and GRBs}

The origin of the UHE cosmic rays (UHECR) is an important
unsolved problem. The idea that they originate from 
long duration GRBs
is argued for a number of reasons. 
%
First, 
the power required for the cosmic rays above the ``ankle'' ($\sim
10^{19}$~eV) is within one or two orders of magnitude equal to the
hard X-ray/$\gamma$-ray power of BATSE GRBs, assumed to be at average
redshift unity \cite{Waxman95,Vietri95,Dermer02}. 
Second, GRBs form powerful relativistic flows, providing extreme sites for
particle acceleration consistent with the known physical limitations, e.g. size,
required to achieve ultra high energy. 
Third, GRBs are expected to be associated with star-forming galaxies, so
numerous UHECR sources would be found within the $\sim 100$~Mpc GZK
radius, thus avoiding the situation that there is no persistent
powerful source within this radius.  
And, finally,  
various features in the medium- and high-energy $\gamma$-ray spectra of
GRBs may 
be attributed to hadronic emission processes.

The required Lorentz factors of UHECRs, $\gtrsim 10^{10}$, exceed by
orders of magnitude the baryon-loading parameter $\eta \gtrsim 100$
thought typical of GRB outflows. Thus the UHECRs must be accelerated
by processes in the relativistic flows. The best-studied mechanism is
Fermi acceleration at shocks, including external shocks when the GRB
blast wave interacts with the surrounding medium, and internal shocks
formed in an intermittent relativistic wind. 

Protons and ions with nuclear charge $Z$ are expected to be
accelerated at shocks, just like electrons.  
The maximum energy in the internal shock model~\cite{Waxman95} 
or in the case of an external shock in a uniform density medium~\cite{Vietri98,DermerHumi01} are both of order a few $Z$ 10$^{20}$ eV for
typical expected burst parameters. 
Thus GRBs can 
accelerate UHECRs.  The ultrarelativstic
protons/ions in the GRB jet and blast wave can interact with ambient
soft photons 
if the corresponding opacity
is of the order of unity or higher, to form escaping neutral
radiations (neutrons, $\gamma$-rays, and neutrinos).  They may also
interact with other baryons via inelastic nuclear production
processes, again producing neutrals.  
So VHE gamma rays are a natural consequence of UHECR acceleration in 
GRBs.
While leptonic models explain keV--MeV data as synchrotron or
Compton radiation from accelerated primary electrons, and GeV--TeV
emission from inverse-Compton scattering, a hadronic emission component
at GeV--TeV energies can also be present.



Neutrons are coupled to the jet protons by elastic
$p$-$n$ nuclear scattering and, depending on injection conditions in the
GRB, can decouple from the protons during the expansion phase.  As a
result, the neutrons and protons travel with different speeds and will
undergo inelastic $p$-$n$ collisions, leading to $\pi$-decay radiation,
resulting in tens of GeV photons~\cite{Derishevetal99,BahcallMeszaros00}.  The decoupling
leads to subsequent interactions of the proton and neutron-decay
shells, which may reduce the shell Lorentz factor by heating~\cite{Rossietal06}.
The $n$-$p$ decoupling occurs in short GRBs for values of the
baryon-loading parameter $\eta \sim 300$ \cite{RazzaqueMeszaros06a}.  The relative
Lorentz factor between the proton and neutron components may be larger
than in long duration GRBs, leading to energetic ($\sim$50 GeV)
photon emission. 
Applying this model to several short GRBs in the field of view of 
Milagro~\cite{Abdoetal07} gives fluxes of a few 10$^{-7}$ cm$^{-2}$ s$^{-1}$ for
typical bursts, suggesting that a detector of large
effective area, $\gtrsim 10^7$~cm$^{2}$, at low threshold energy is
needed to detect these photons.  For the possibly nearby ($z=0.001$) GRB 051103,
the flux could be as large as $\sim$10$^{-3}$ cm$^{-2}$ s$^{-1}$.

%


Nuclei accelerated in the GRB jet and blast wave to
ultra high energies can make $\gamma$-rays through the synchrotron
process; 
photopair production, which converts the target photon
into an electron-positron pair with about the same Lorentz factor
as the ultrarelativistic 
nucleus; and 
photopion production, which makes pions that decay into 
electrons and positrons, photons, and neutrinos.
The target photons for the 
latter two
processes are usually considered
to be the ambient synchrotron and synchrotron self-Compton photons
formed by leptons accelerated at the forward and reverse shocks of
internal and external shocks.  If the pion-decay muons decay before
radiating much energy~\cite{RachenMeszaros98}, the secondary leptons,
$\gamma$-rays, and neutrinos each carry about 5\% of the primary
energy.

About one-half of the time, neutrons are formed in a photopion
reaction.  If the neutron does not undergo another photopion reaction
before escaping the blast wave, it becomes free to travel until it
decays. Neutrons in the neutral beam \cite{AtoyanDermer03}, collimated by the
bulk relativistic motion of the GRB blast wave shell, travel $\approx
(E_n/10^{20}$ eV) Mpc before decaying. A neutron decays into neutrinos
and electrons with $\approx 0.1$\% of the energy of the primary.
Ultrarelativistic neutrons can also form secondary pions after
interacting with other soft photons in the GRB enviroment. The
resulting decay electrons form a hyper-relativistic synchrotron
spectrum, which is proposed as the explanation for the anomalous
$\gamma$-ray emission signatures seen in GRB 941017 \cite{DermerAtoyan04}.

The electromagnetic secondaries generate an electromagnetic cascade
when the optical depth 
is sufficiently large. The photon number
index of the escaping $\gamma$-rays formed by multiple generations of
Compton and synchrotron radiation is generally between $-3/2$ and $-2$
below an exponential cutoff energy, which could reach to GeV or,
depending on parameter choices, TeV energies 
\cite{DermerAtoyan03,AtoyanDermer03}.


Gamma-ray observations of GRBs will help distinguish between leptonic
and hadronic emissions.  
VHE 
$\gamma$-ray emission
from GRBs can be modeled by synchro-Compton processes of
shock-accelerated electrons 
\cite{PeerWaxman04b,Meszarosetal94,PapathanassiouMeszaros96,ChiangDermer99,Dermeretal00}, or by
photohadronic interactions of UHECRs and subsequent cascade emission
\cite{BottcherDermer98,Totani99,Fragileetal04}, or by a combined leptonic/hadronic model.
The clear distinction between the two models from $\gamma$-ray
observation will not be easy. The fact that the VHE $\gamma$-rays are
attenuated both at their production sites and in the 
CIB
restricts measurements to energies below  150 GeV ($z \sim 1$) or 5
TeV ($z \lesssim 0.2$).  Distinctive features of 
hadronic models are:
\begin{itemize}
\item 
Photohadronic interactions and subsequent electromagnetic $\gamma$-ray
producing cascades develop over a long time scale due to slower energy
loss-rate by protons than electrons. The GeV-TeV light curves arising
from hadronic mechanisms then would be longer than those expected from
purely leptonic processes \cite{BottcherDermer98},
facilitating detection with pointed instruments. 
\item
Cascade $\gamma$ rays will be harder than a $-2$ spectrum below an
exponential cutoff energy, and photohadronic processes can make hard,
$\sim-1$ spectra from anisotropic photohadronic-induced cascades,
used to explain GRB 941017 \cite{Gonzalezetal03,DermerAtoyan04}. A ``two zone'' leptonic
synchro-Compton mechanism can, however, also explain the same
observations \cite{GranotGuetta03a,PeerWaxman04a,ZhangMeszaros01,Wangetal05}, with low energy emission
from the prompt phase and high energy emission from a very early
afterglow.
\item
Another temporal signature of hadronic models is delayed emission from 
UHECR cascades in the CIB/CMB \cite{WaxmanCoppi96} or
$\gtrsim$PeV energy $\gamma$-rays, from $\pi^0$ decay, 
which may escape the GRB fireball \cite{Razzaqueetal04}. 
However, $\gtrsim$TeV
photons created by leptonic synchro-Compton mechanism in external
forward shocks may imitate the same time delay by cascading in the
background fields \cite{Wangetal04}.
\item 
Quasi-monoenergetic $\pi^0$ decay $\gamma$-rays from $n$-$p$
decoupling, which are emitted from the jet photosphere prior to the
GRB, is a promising hadronic signature \cite{Derishevetal99,BahcallMeszaros00,RazzaqueMeszaros06a}, though
it requires that the GRB jet should contain abundant free neutrons as
well as a large baryon load.
\end{itemize}
Detection of high-energy neutrino emissions would conclusively
demonstrate 
cosmic ray 
acceleration in GRBs,
but non-detection would not conclusively rule out GRBs as a source
of UHECRs, since the $\nu$ production level even for optimistic parameters
is small.
\subsection{Tests of Lorentz Invariance with Bursts}

Due to quantum gravity effects, it is possible that the speed of light is energy dependent and that ${\Delta}c/c$ scales either linearly or quadratically with ${\Delta}E/E_{QG}$, where $E_{QG}$ could be assumed to be at or below the Planck energy, $E_{P}$  
\cite{ellis1992,amelino1997,gambini1999}. Recent detections of flaring from the blazar Mrk 501, using the MAGIC IACT, have used this effect to constrain the quantum gravity scale for linear variations to $\gtrsim 0.1E_{P}$ \cite{albert2007}. This same technique could be applied to GRBs, which have fast variability, if they were detected in the TeV range and if the intrinsic chromatic variations were known.
However, there may be intrinsic limitations 
to some approaches \cite{Scargleetal07}.
By improving the sensitivity and the energy range with a future telescope array, the current limit could be more tightly constrained, particularly if it were combined with an instrument such as Fermi at lower energies, thus increasing the energy lever arm.

\begin{figure*}[!th]
\centering
\includegraphics[width=5in]{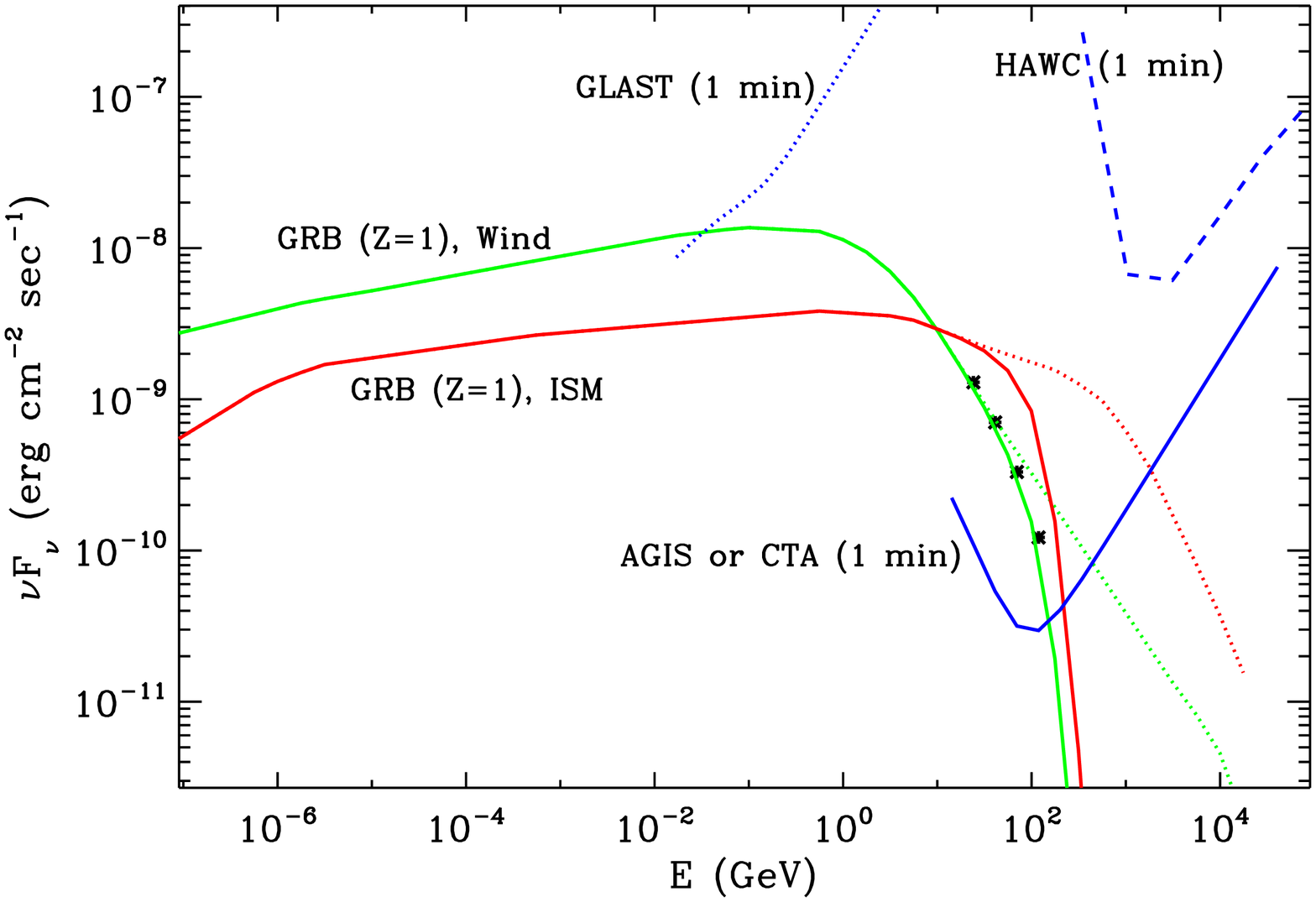}
\caption{A plot of the predicted gamma-ray spectrum from a GRB at a redshift of z=1 adapted from Pe'er and Waxman 
\cite{PeerWaxman05b},
reduced by a factor of 10 to illustrate the sensitivity even to 
weaker bursts.
The green and red curves show the calculation for a wind environment and an ISM-like environment. The dotted curves give the source spectrum, while the solid curves include the effects of intergalactic absorption using a model from Franceschini et al.\  \cite{Franceschinietal2008}.
The blue curves show the differential sensitivity curves for Fermi (GLAST; dotted), a km$^2$ IACT array like AGIS or CTA (solid) and 
the HAWC air shower array (dashed). 
For the AGIS/CTA curve we show the differential sensitivity for 
0.25 decade bins, while for the HAWC instrument we assume 
0.5 decade bins.
The sensitivity curve is based on a 5 sigma detection and at least 25 detected photons. Black points and error bars (not visible) are simulated independent spectral points that could be obtained with AGIS/CTA.}
\label{grb_sedsim}
\end{figure*}

\subsection{Detection Strategies for VHE Gamma-Ray Burst Emission}



Ground-based observations of TeV emission from gamma ray bursts are difficult. The fraction of GRBs close enough to 
elude attenuation 
at TeV energies by the CIB 
is small. Only $\sim$10\% of long bursts are within z$<$0.5, 
the redshift of the most distant detected VHE 
source, 3C 279 \cite{Teshima:07}. Short bursts are more nearby with over 50\% detected within z$<$0.5, but the prompt emission has ended prior to satellite notifications of the burst location.

Therefore, wide field of view detectors with high duty cycle operations 
would be ideal to observe the prompt emission from 
gamma-ray bursts. Imaging atmospheric Cherenkov telescopes (IACT) can be made 
to cover large sections of the sky by either having many 
mirrors each pointing in a separate direction or by employing secondary optics to expand the field of view of each mirror. 
However, the duty factor is still $\sim$10\% 
due to solar, lunar, and weather constraints. IACTs could also be made with fast slewing mounts to allow them to slew to most GRBs within $\sim$20 seconds, thus allowing them to observe some GRBs before the end of the prompt phase. 
Alternatively, extensive air shower 
detectors intrinsically have a field of view of $\sim$2 sr and operate with $\sim$95\% 
duty factor. These observatories, especially if located at very high altitudes, can detect gamma rays down to 100 GeV, but at these low energies they lack good energy resolution and have a point spread function of $\sim$1 degree. 
The traditionally less sensitive extensive air shower detectors may have difficulty achieving the required prompt emission sensitivity on short timescales ($>5\sigma$ detection of 10$^{-9}$ erg s$^{-1}$ cm$^{-2}$ in $\lesssim20$ sec integration). The combined observations of both of these types of detectors would yield the most complete picture of the prompt high energy emission. 
The expected performance of the two techniques relative to 
a particular 
prompt GRB emission 
model 
is shown in Figure~\ref{grb_sedsim}. 

The detector strategy for extended emission associated with traditional afterglows or with late-time flares from GRBs is far simpler than the strategy for early prompt emission. The high sensitivity and low energy threshold of an IACT array are the best way to capture photons from this emission at times greater than $\sim$1 min, particularly if fast slewing is included in the design.
\subsection{Synergy with other instruments}
While GRB triggers are possible from wide angle VHE instruments, a space-based GRB 
detector will be needed. Swift, Fermi, or future wide field of view X-ray monitors such as EXIST or JANUS must provide lower energy observations. GRBs with observations by both Fermi and VHE 
telescopes will be particularly exciting and may probe high Lorentz factors. Neutrino 
telescopes such as IceCube, UHECR telescopes such as Auger, and next generation VHE observatories 
can supplement one another in the search for UHECRs from GRBs, since neutrinos are 
expected along with VHE gamma rays. 
Detection of gravitational waves from GRB progenitors with instruments such as LIGO have the potential to reveal the engine powering the GRB fireball.  Correlated observations between gravitational wave observatories and VHE gamma-ray instruments will then 
be important for understanding which type(s) of engine can power
VHE emission.

Correlated observations 
between TeV gamma-ray detectors and neutrino detectors 
have 
the potential for significant reduction in background for the
participants. 
If TeV gamma sources are observed, observers will know where and when to look for neutrinos (and vice versa 
\cite{KowalskiMohr07}, 
though the advantage in that direction is less significant).  
For example, searches for GRB neutrinos have used the known time and
location to reduce the background
by a factor of nearly 10$^5$ compared to an annual all-sky diffuse search \cite{Achterbergetal07}.
Beyond decreasing background, correlated observations also have the potential to increase the expected signal rate.  
If the spectrum of high-energy gamma rays is known, then constraints on the expected neutrino spectrum can also be introduced, allowing the signal-to-noise ratio of neutrino searches to be 
significantly improved \cite{StamatikosBand06}.
In the case of the AMANDA GRB neutrino search, which is based on a specific 
theoretical neutrino spectrum, the expected signal collection efficiency is nearly
20 times higher than the less constrained search for diffuse UHE
neutrinos.
With combined photon and neutrino observational efforts, there is a much better chance of eventual neutrino detection of sources such as GRBs (and AGN).



\subsection{Conclusions}




Gamma-ray bursts undoubtedly involve a population of high-energy 
particles responsible for the emission detected
from all bursts (by definition) at energies up to of order 1 MeV,
and for a few bursts so far observable by EGRET, up to a few GeV.  Gamma-ray 
bursts may in fact be the source of the highest energy particles in the universe.
In virtually all models, this high-energy population can also produce
VHE gamma-rays, although in many cases the burst environment would be 
optically thick to their escape.  The search for and study of VHE emission from GRB therefore tests theories about the nature of these
high energy particles (Are they electrons or protons? What is their
spectrum?) and their environment (What are the density and bulk
Lorentz factors of the material? What are the radiation fields?
What is the distance of the emission site from the central source?).
In addition, sensitive VHE measurements would aid in assessing the 
the total calorimetric radiation output from bursts.  
Knowledge of the VHE gamma-ray properties of bursts will therefore
help complete the picture of these most powerful known accelerators.


An example of the insight that can be gleaned from VHE data
is that leptonic synchrotron/SSC
models can be tested, and model parameters extracted, by 
correlating the peak energy of
X-ray/soft $\gamma$-ray emission
with GeV--TeV data.
For long lived GRBs, the spectral properties of late-time flaring in the X-ray band can be compared to the measurements in the VHE band, where associated emission is expected.
Of clear interest is whether there are distinctly evolving high-energy $\gamma$-ray  
spectral components, whether at MeV, GeV or TeV energies, unaccompanied
by the 
associated lower-energy component 
expected in leptonic
synchro-Compton models. 
Emission of this sort is most easily explained in models involving
proton acceleration.
As a final example, the escape of VHE photons from the burst fireball provides a tracer of the minimum Doppler boost and bulk Lorentz motion of the emission region 
along the line of site, since the inferred opacity of the emission region declines with increasing boost.   






There are observational challenges for detecting VHE emission during the initial
prompt phase of the burst.  The short duration of
emission leaves little time (tens of seconds) for repointing 
an instrument, and the opacity of the compact fireball is at its 
highest.  For the majority of bursts having redshift $\gtrsim$0.5, the 
absorption of gamma rays during all phases of the burst by 
collisions with the extragalactic background light reduces the 
detectable emission, more severely with increasing gamma-ray energy.
With sufficient sensitivity, an all-sky
instrument is the most desirable for studying the prompt phase,
in order to measure the largest sample of bursts and to catch 
them at the earliest times.
As discussed in the report of the Technology Working Group,
the techniques used to implement all-sky compared to 
pointed VHE instruments result in a trade-off of energy threshold and
instantaneous sensitivity for field of view.  More than an order of magnitude improvement in sensitivity to GRBs is envisioned
for the next-generation instruments of both types, giving both 
approaches a role in future studies of GRB prompt emission.

The detection of VHE afterglow emission, delayed prompt emission from large radii, and/or late X-ray flare-associated emission simply requires a sensitive instrument 
with only moderate slew speed. It is likely that an instrument with significant sensitivity improvements over the current generation of IACTs will detect GRB-related VHE emission from one or all of these mechanisms which do not suffer from high internal absorption, thus making great strides towards understanding the extreme nature and environments of GRBs and their ability to accelerate particles.

In conclusion, large steps in understanding GRBs have frequently resulted from particular new characteristics measured for the first time in a single burst. 
New instruments improving sensitivity to very-high-energy gamma-rays
by an order or magnitude or more compared to existing observations
have the promise to make just
such a breakthrough in the VHE band.

\clearpage

\section{Technology working group}
\label{twg-subsec}
Group membership:\\ \\
K. Byrum, J. Buckley, S. Bugayov, B. Dingus, S. Fegan, S. Funk, E. Hays, J. Holder, A. Konopelko,  H. Krawczynski, F. Krennrich, S. Lebohec, A. Smith, G. Sinnis, V. Vassiliev, S. Wakely

\subsection[Introduction and overview]{Introduction and Overview}

High-energy gamma rays can be observed from the ground by detecting 
secondary particles of the atmospheric cascades initiated by the interaction 
of the gamma-ray with the atmosphere. Imaging atmospheric Cherenkov 
telescopes (IACTs) detect broadband spectrum Cherenkov photons ($\lambda > 
300$ nm), which are produced by electrons and positrons of the cascade and 
reach the ground level without significant attenuation.    
The technique utilizes large mirrors to focus Cherenkov photons onto a finely 
pixelated camera operating with an exposure of a few nanoseconds, and 
provides low energy threshold and excellent calorimetric capabilities. The 
IACTs can only operate during clear moonless and, more recently, partially-moonlit nights. Alternatively, the 
extended air shower (EAS) arrays, which directly detect particles of the 
atmospheric cascade (electrons, photons, muons, etc.) can be operated 
continuously but require considerably larger energy of the gamma rays 
necessary for extensive air showers to reach the ground level.

The field of TeV gamma-ray astronomy was born in the years 1986 to 1988 
with the first indisputable detection of a cosmic source of TeV gamma
rays with the Whipple $10$~m IACT, the Crab Nebula 
\cite{1989ApJ...342..379W}. Modern IACT observatories such as VERITAS 
\cite{Week:02,Maie:07}, MAGIC \cite{2004NewAR..48..339L,Goeb:07}, and 
H.E.S.S. \cite{2004NewAR..48..331H,Horn:07} can detect point sources with a 
flux sensitivity of $1\%$ of the Crab Nebula corresponding to a limiting $\nu 
$F$_{\nu }$-flux of $\sim 5\times 10^{-13}$ ergs cm$^{-2}$ s$^{-1}$ at 1 TeV. 
The improvement of sensitivity by two orders of magnitude during the last two 
decades has been made possible due to critical advances in IACT technology 
and significantly increased funding for ground-based gamma-ray astronomy. 
The high point-source flux sensitivity of IACT observatories is a result of  
their large gamma-ray collecting area ($\sim 10^{5}$ m$^{2}$), relatively 
high angular resolution ($\sim 5$ arcminutes), wide energy coverage (from 
$<100$ GeV to $>10$ TeV), and unique means to reject cosmic ray background 
($> 99.999\%$ at 1 TeV). The limitations of the IACT technique are the small 
duty cycle ($\sim 10\%$), and narrow field of view ($\sim 4\deg $; $3.8\times 
10^{-3}$ sr for present-day IACTs).

Large EAS arrays provide complementary technology for observations of very 
high-energy gamma rays. Whereas their instantaneous sensitivity is 
currently a factor $\sim 150$ less sensitive than that of IACT observatories, their 
large field of view ($\sim 90\deg $; $1.8$ sr) and nearly $100\%$ duty cycle 
makes these observatories particularly suited to conduct all-sky surveys and 
detect emission from extended astrophysical sources (larger than 
$\sim 1\deg $, e.g. plane of the Galaxy). Milagro \cite{Smit:05}, the first 
ground-based gamma-ray observatory which utilized EAS technology to 
discover extended sources \cite{Abdo:07}, has surveyed $2\pi $~sr of 
the sky at $20$~TeV for point sources to a sensitivity of $3\times 10^{-12}$ 
ergs cm$^{-2}$ s$^{-1}$. Due to the wide field of view 
coverage of the sky and uninterrupted operation, the EAS technique also has 
the potential for detection of Very High Energy (VHE) transient phenomena. 
The current limitations of EAS technique are high-energy threshold ($\sim 10$ 
TeV), low angular resolution ($\sim 30$ arcminutes), and limited capability to reject cosmic-ray background and measure energy.

The primary technical goal for the construction of the next generation of observatories is to 
achieve an improvement of sensitivity by a factor of $\alpha $ at the cost 
increase less than a factor of $\alpha ^{2}$, the increase that would be required if the observatory were constructed by simply cloning present day instrumentation~\footnote{Background dominated regime of observatory 
operation is assumed}. The history of ground-based gamma-ray astronomy 
over the last two decades has shown twice an improvement in the sensitivity of the observatories by a factor of ten while the cost has increased each time only by a factor of ten  
\cite{2007ebhe.conf..282W}. 

The construction of a large array of IACTs covering an area of $\sim 1$ km$^2$ will enable ground-based $\gamma$-ray astronomy to achieve another order of magnitude improvement in sensitivity. This next step will be facilitated by several technology improvements. First, large arrays of IACTs should have the capability to operate over a broad energy range with significantly improved 
angular resolution and background rejection as compared to the present day 
small arrays of telescopes, such as VERITAS or H.E.S.S.. Second, the capability of using subarrays to fine tune the energy range to smaller intervals will allow for considerable reduction of aperture of 
individual telescopes and overall cost of the array while maintaining the 
collecting area at lower energies equal to the smaller array of very large 
aperture IACTs. Finally, the cost per telescope can be significantly reduced 
due to the advancements in technology, particularly the development of low cost 
electronics, novel telescope optics designs, replication methods for 
fabrication of mirrors, and high efficiency photo-detectors, and due to the 
distribution of initial significant non-recurring costs over a larger number of 
telescopes. 

In the case of EAS arrays, the breakthrough characterized by the 
improvement of sensitivity faster than the inverse square root of the array 
footprint area is possible due to mainly two factors. First, next generation 
EAS array must be constructed at a high elevation ($>4000$ m) to increase the number of particles in a shower by being closer to the altitude where the shower has the maximum number of particles. Thus, a lower energy threshold is possible and energy resolution is improved.  Second, the size of the EAS array needs to be increased in order to more fully contain the lateral distribution of the EAS.  A larger array improves the angular resolution of the gamma-ray showers and also dramatically improves the cosmic ray background rejections.  The lateral distribution of muons in a cosmic ray shower is very broad, and identification of a muon outside the shower core is key to rejecting the cosmic ray background.

The science motivations for the next generation ground-based gamma-ray 
observatories are outlined in this document.  There are clear cost, reliability, maintenance, engineering, 
and management challenges associated with construction and operation of a future 
ground-based astronomical facility of the order $\sim $100M dollar scale. 
Detailed technical implementation of a future observatory will benefit from current and future R\&D efforts that will provide better understanding 
of the uncertainties in evaluation of the cost impact of 
improved and novel photon detector technologies and from the current incomplete  
simulation design studies of the large optimization space of parameters of 
the observatory. 
In the remainder of this section, we outline a broadly defined technical roadmap for 
the design and construction of future instrumentation which could be realized within the next decade.  We start with a status of the field, 
identify the key future observatory design decisions, technical drivers, 
describe the current state of the art technologies, and finally outline a plan for 
defining the full technology approach.

\subsection{Status of ground-based gamma-ray observatories}{Status of Ground-Based Gamma-ray\\Observatories}
\begin{figure*}[t]
\begin{center}

\includegraphics[angle=0,width=6.0in]{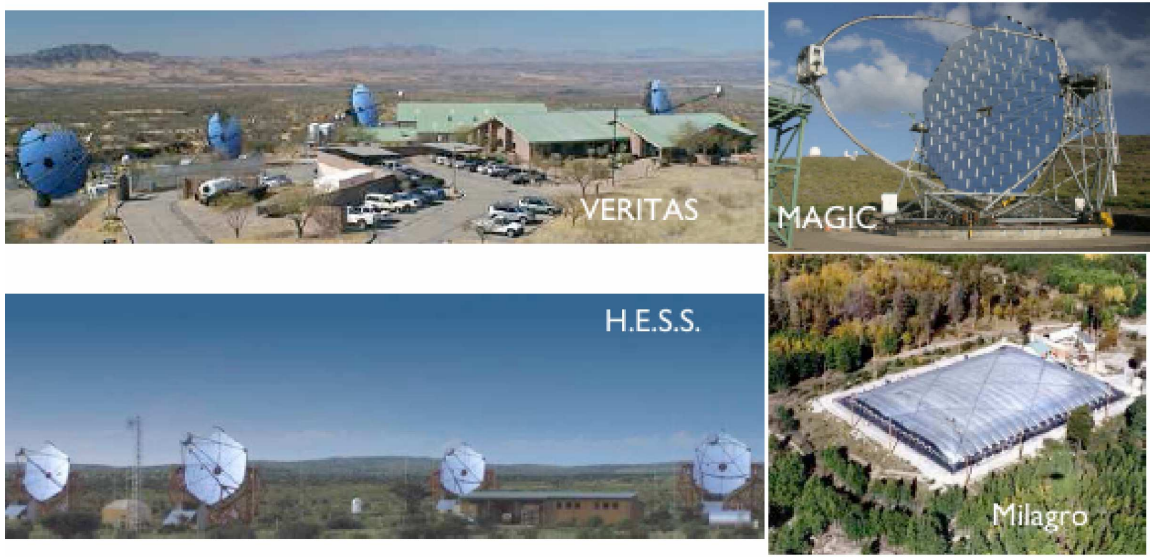}

\caption{\label{fig:exp} The images show four major ground-based 
gamma-ray observatories currently in operation: VERITAS, MAGIC, H.E.S.S.\ 
, and MILAGRO. A future ground-based gamma-ray project can build on the 
success of these instruments.}
\end{center}

\end{figure*}

At present, there are four major IACT and three EAS observatories worldwide conducting routine astronomical observations, four of which are shown in Fig \ \ref{fig:exp}.  Main parameters of these instruments are the following:

\paragraph{VERITAS} is a four-telescope array of IACTs located at the Fred Lawrence Whipple Observatory in Southern Arizona (1268 m a.s.l.). Each telescope is a 12 m diameter Davies-Cotton (DC) reflector (f/1.0) and a high resolution 3.5$\deg$ field of view camera assembled from 499 individual photo multiplier tubes (PMTs) with an angular size of 0.15 deg.  The telescope spacing varies from 35~m to 109~m.  VERITAS was commissioned to scientific operation in April 2007.

\paragraph{The H.E.S.S.\  array} consists of four 13 m DC IACTs (f/1.2) in the Khomas Highlands of Namibia (1800 m a.s.l.). The 5 deg field of view cameras of the telescopes contain 960 PMTs, each subtending 0.16deg angle.  The current telescopes are arranged on the corners of a square with 120m sides. H.E.S.S.\  has been operational since December 2003. The collaboration is currently in the process of upgrading the experiment (H.E.S.S.\ -II) by adding a central large (28 m diameter) telescope to the array to lower the trigger threshold for a subset of the events to 20 GeV and will also improve the sensitivity of the array above 100 GeV. 

\paragraph{MAGIC} is a single 17 m diameter parabolic reflector (f/1.0) located in the Canary Island La Palma (2200 m a.s.l.). It has been in operation since the end of 2003. The 3.5 deg non-homogenous camera of the telescope is made of 576 PMTs of two angular sizes 0.1deg (396 pixels) and 0.2deg (180 pixels). The MAGIC observatory is currently being upgraded to MAGIC-II with a second 17-m reflector being constructed 85 m from the first telescope. The addition of this second telescope will improve background rejection and increase energy resolution. 

\paragraph{CANGAROO-III} consists of an array of four 10 m IACTs (f/0.8) located in Woomera, South Australia (160 m a.s.l.) \cite{Mori:07}.  The telescope camera is equipped with an array of 552 PMTs subtending an angle of 0.2deg each.  The telescopes are arranged on the corners of a diamond with sides of 100 m.

\paragraph{Milagro} is an EAS water Cherenkov detector located near Los Alamos, New Mexico (2650 m a.s.l.). Milagro consists of a central pond detector with an area of 60 x 80m$^2$ at the surface and has sloping sides that lead to a 30 x 50 m$^2$ bottom at a depth of 8 m. It is filled with 5 million gallons of purified water and is covered by a light-tight high-density polypropylene line.  Milagro consists of two layers of upward pointing 8'' PMTs.  The tank is surrounded with an array of water tanks. The central pond detector has been operational since 2000. The array of water tanks was completed in 2004.

\paragraph{The AS-\large{$\gamma$} and ARGO arrays} are located at the YangBaJing high-altitude laboratory in Tibet, China.  AS-$\gamma$, an array of plastic scintillator detectors, has been operational since the mid 1990s. ARGO consists of a large continuous array of Resistive Plate Counters (RPCs) and will become operational in 2007 \cite{Zao:05}.

\bigskip

The current generation of ground based instruments has been joined in mid-2008 by the space-borne \textbf{Fermi Gamma-ray Space Telescope} (formerly GLAST). Fermi comprises two instruments, the Large Area Telescope (LAT) \cite{McEn:07} and the Fermi Gamma-ray Burst Monitor (GBM) \cite{Lich:07}. The LAT covers the gamma-ray 
energy band of 20 MeV - 300 GeV with some spectral overlap with IACTs.  The present generation of IACTs 
match the $\nu F_{\nu}$-sensitivity of Fermi. Next-generation ground-based observatories with one order of 
magnitude higher sensitivity and significantly improved angular resolution would be ideally suited to 
conduct detailed studies of the Fermi sources.

\begin{table*}[!ht]
\begin{center}
\caption{\label{regimes} Gamma-ray energy regimes, scientific highlights and 
technical challenges.}

{\footnotesize

\begin{tabular}{p{0.6in} p{0.6in} p{1.9in} p{2.6in}}

\hline\hline

Regime & Energy Range & Primary Science Drivers & Requirements/Limitations \\ 
\hline

{\bf multi-GeV}: & $\leq$50~GeV &
extragalactic sources (AGN, GRBs) at cosmological distances ($z>1$),
Microquasars, Pulsars  &
very large aperture or dense arrays of IACTs, preferably high altitude operation \& 
high quantum efficiency detectors
required; 
 angular resolution and energy resolution will be limited by shower 
fluctuations, cosmic-ray background rejection utilizing currently available technologies is inefficient.

\\

{\bf sub-TeV}: & 50~GeV -- 200~GeV & extragalactic sources at intermediate redshifts($z < 1$), search for 
dark matter, Galaxy Clusters, Pair Halos, 
Fermi sources &  very-large-aperture telescopes or dense arrays of mid-size telescopes and high light detection efficiency required;
limited but improving with energy cosmic-ray background rejection based on imaging analysis. For gamma-ray bursts, high altitude EAS array.

\\ 

{\bf TeV}: & 200~GeV -- 10~TeV &
nearby galaxies (dwarf, starburst), nearby AGN, detailed 
morphology of extended galactic sources (SNRs, GMCs, PWNe) 
& large arrays of IACTs: best energy flux sensitivity, best angular and energy resolutions, best cosmic-ray hadron background rejection, new backgrounds from cosmic-ray electrons may ultimately limit sensitivity in some regions of the energy interval.  At the highest energy end, an irreducible background may be due to single-pion sub-showers. EAS arrays for mapping Galactic diffuse emission, AGN flares, and sensitivity to extended sources.

\\

{\bf sub-PeV}:  & $\geq$10~TeV & Cosmic Ray PeVatrons (SNRs, PWNe, GC, ...), 
origin
of galactic cosmic rays & 
requires very large (10 km$^2$ scale) detection areas; large arrays of IACTs equipped with very wide ($\ge 6^\circ$) FoV cameras and separated with distance of several hundred meters may provide adequate technology.  Background rejection is excellent and sensitivity is $\gamma$-ray count limited.  Single-pion sub-showers is ultimate background limiting sensitivity for very deep observations.  Regime of best performance of present EAS arrays; large EAS arrays ($\ge 10^{5}m^{2}$).

\\

\hline

\end{tabular}
}

\end{center}

\end{table*}

\subsection[Design considerations for a next-generation gamma-ray detector]{Design Considerations for a Next-Generation Gamma-Ray Detector}

At the core of the design of a large scale ground-based gamma-ray
 observatory is the requirement to improve the integral flux sensitivity by an order of magnitude over instruments employed today in 
the 50 GeV-20~TeV regime where the techniques are proven to give excellent 
performance. At lower energies (below 50 GeV) and at much higher energies 
(50-200 TeV) there is great discovery potential, but new technical approaches 
must be explored and the scientific benefit is in some cases less certain. 
For particle-detector (EAS) arrays, it is possible to simultaneously improve 
energy threshold and effective area by increasing the elevation, and the 
technical road-map is relatively well-defined.  In considering the design 
of future IACT arrays, the development path allows for complementary branches to more fully maximize the greatest sensitivity for a broad energy 
range from 10~GeV up to 100~TeV. 
Table \ref{regimes} summarizes specific issues of the detection technique and 
scientific objectives for four broad energy regimes (adapted from 
\cite{AharT:05,AharT:08}).
\subsection[Future IACT arrays]{Future IACT Arrays}

\begin{figure*}[t]

\begin{center}

\includegraphics[angle=0,width=3.in]{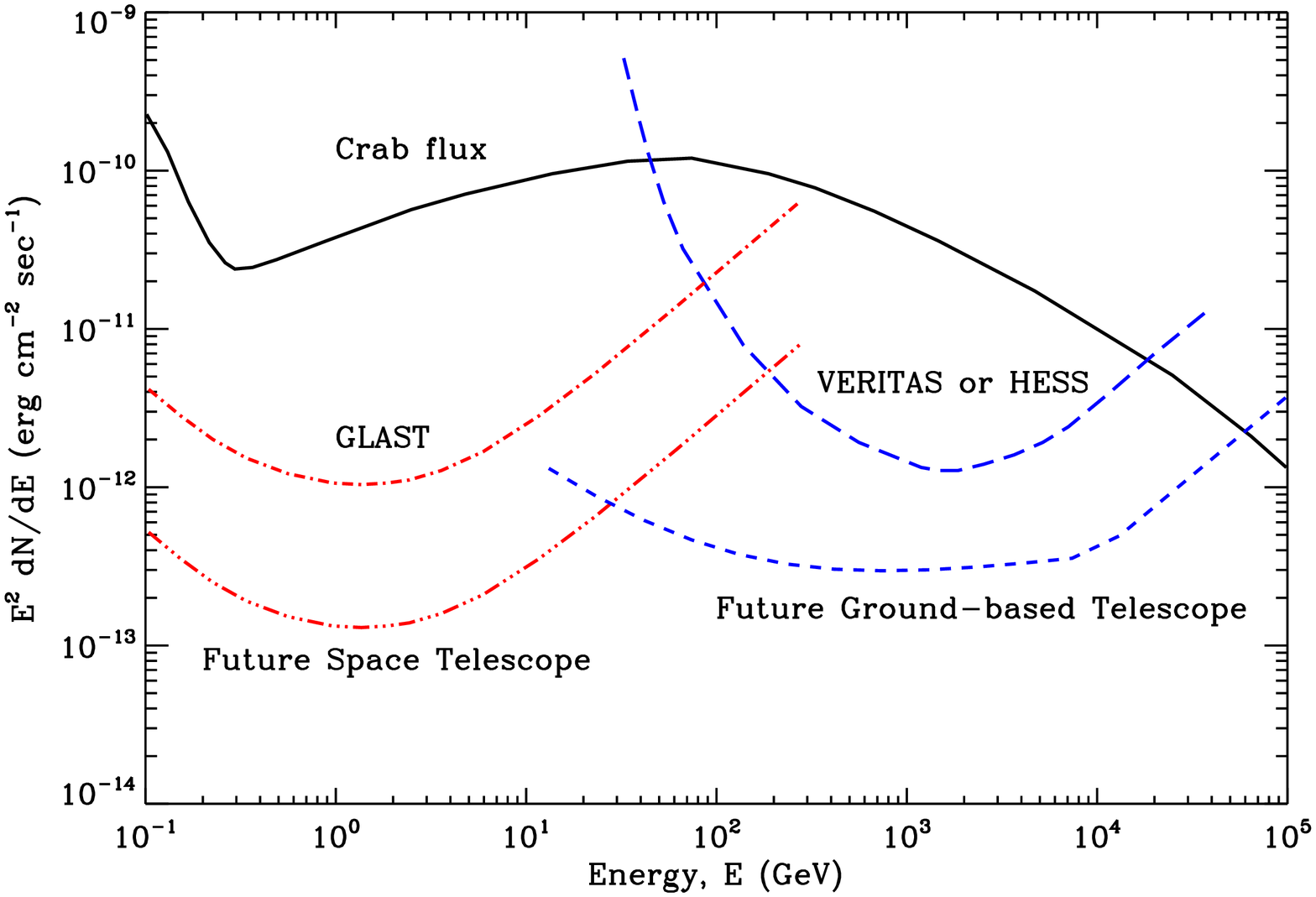}
\includegraphics[angle=0,width=3.in]{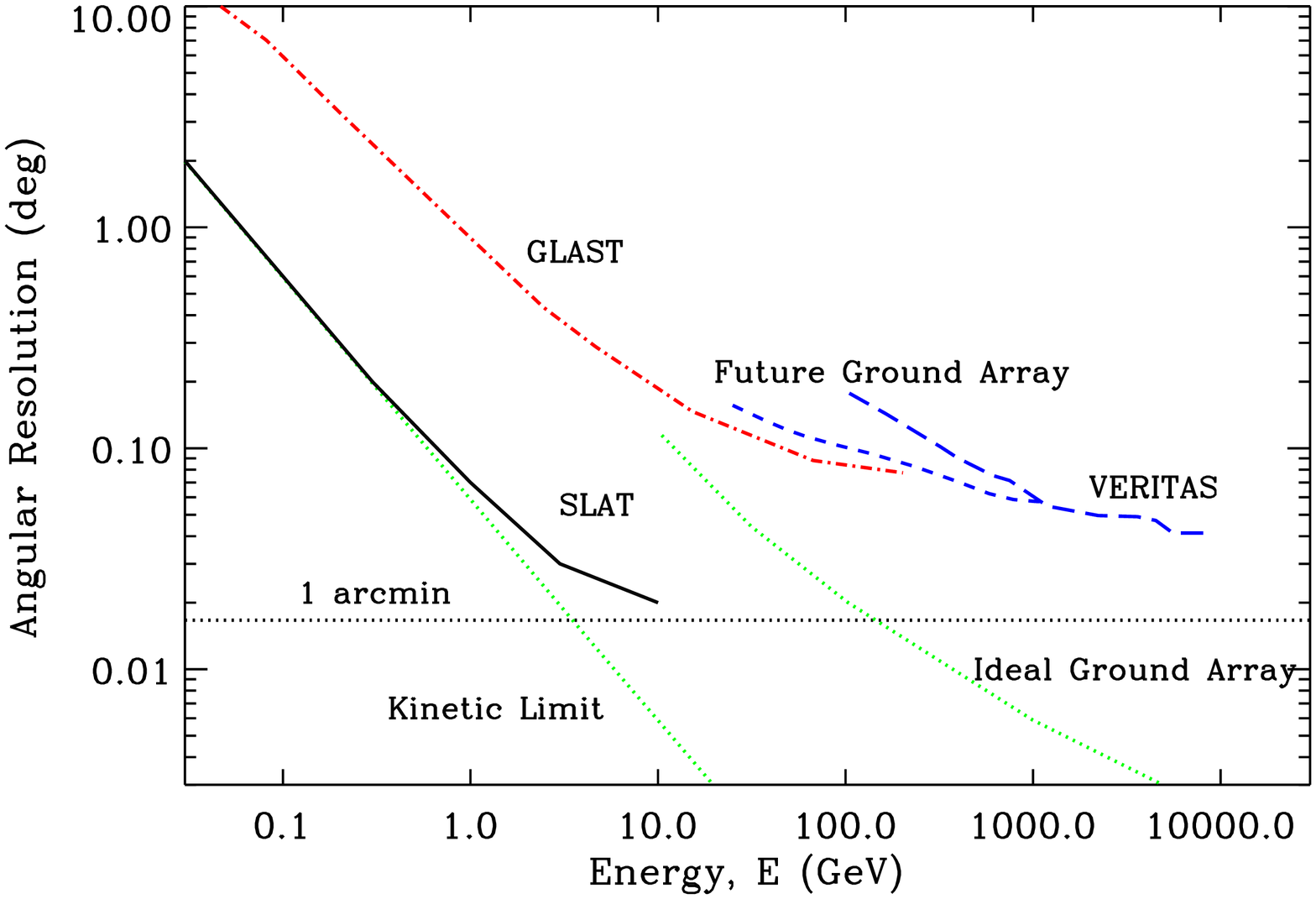}

\caption{ \textit{Left:} Differential sensitivities calculated for present and future gamma-ray experiments. For the future IACT array, an area of $\sim$1~km$^2$, no night-sky-background, a perfect point spread function \cite{Bugaev:07}, and an order of magnitude improvement in cosmic-ray rejection compared with current instruments has been assumed.  All sensitivities are 5 sigma detections in quarter decade energy intervals (chosen to be larger than the expected full-width energy resolution). \textit{Right} Angular resolution for Fermi (GLAST)
\cite{GLAST}, VERITAS \cite{Krawcz:06} and for ideal future space-borne and ground based \cite{Hofmann2005} gamma-ray detectors.}

\label{fig:sensang}

\end{center}

\end{figure*}

The scientific goals to be addressed with a future IACT array require 
a flux sensitivity at least a factor of ten better than  present-day observatories, 
and an operational energy range which extends preferably into the sub-100 GeV domain 
in order to open up the $\gamma$-ray horizon to observations of cosmologically distant 
sources. These requirements can be achieved by an array with a collecting area of 
$\sim 1$~km$^2$ (see Fig 1).  

The intrinsic properties of a $\sim 1$~km$^2$ IACT array could 
bring a major breakthrough for VHE gamma-ray astronomy since it combines several key
 advantages over existing 4-telescope arrays: 

\begin{itemize}

\item A  collection area that is 20 times larger than that of
existing arrays. Comparison of the collection area of a  $\sim 1$ km$^2$ array 
with the characteristic  size of the Cherenkov light pool ($\sim 5 \times 10^4$ m$^2$) 
suggests that the array should be populated
with  50-100 IACTs.

\item Fully contained events for which the shower core falls well within the geometrical
dimensions of the array,  thus giving better  angular reconstruction and much improved 
background rejection.   The performance of a typical IACT array in the energy regime below a few TeV  
is limited by the cosmic-ray background. The sensitivity of a future  
observatory could be further enhanced through improvements of its  angular  
resolution and  background rejection capabilities.  It is known that the  
 angular resolution of the present-day arrays of IACTs, which typically have  
four telescopes, is not limited by the physics of atmospheric cascades, but  
by the pixelation of their cameras and by the number of telescopes  
simultaneously observing a $\gamma$-ray event 
\cite{VF2005,Hofmann2005,FV2007}.

\item  Low energy threshold compared to existing small arrays, since contained events
           provide  sampling  of the inner light pool  where the
           Cherenkov light density is highest.  
Lower energy thresholds  (below 100~GeV) generally require larger  aperture 
($>15$ m) telescopes; however,  a  $\sim 1$ km$^2$ IACT has an intrinsic advantage
to  lower the energy  threshold due to the detection of fully contained events.

\item A wider field of view and the ability to operate the array as a survey instrument.
         
 \end{itemize}

 In order to maximize the scientific capabilities of
a $\sim 1$ km$^2$ array with respect to angular resolution, background suppression, 
energy threshold and field of view, it is necessary to study a range 
of options including the design of the individual telescopes and the array footprint.
Furthermore, it is necessary to determine the most cost effective/appropriate technology 
available. The  reliability of the individual telescopes is also a key consideration to 
minimize operating costs.

The history of the development of instrumentation for ground-based 
$\gamma$-ray astronomy has shown that a significant investment into the design  
and construction of new instruments ($\sim 10$ times the cost of previously  
existing ACTs) has yielded significant increases in sensitivity. For example,  
the construction of high resolution cameras in the 1980s assembled from  
hundreds of individual PMTs and fast electronics made the ``imaging''  
technique possible. This advancement improved the sensitivity of the observatories by  
a factor of 10 through the striking increase of angular resolution and  
cosmic-ray background rejection, and ultimately led to a detection of the  
first TeV source \cite{1989ApJ...342..379W}. Another factor of ten investment into the development of  
small arrays of mid-sized IACTs ($12$~m) demonstrated the benefits of  
``stereoscopic'' imaging and made possible the H.E.S.S. and VERITAS  
observatories. The sensitivity of these instruments improved by a factor of  
10 due to the increase of angular resolution and CR background  
discrimination, despite their only relatively modest increase in the  
$\gamma$-ray collecting area compared to the previous-generation Whipple  
$10$~m telescope. 

The next logical step in the evolution of the IACT technique is the $\sim 1$ km$^2$ array concept. 
Technological developments such as novel multi-pixel high-quantum-efficiency 
photo-detectors (MAPMTs, SiPMs, APDs, CMOS sensors, etc.) or PMTs with 
significantly improved QE, new telescope optical design(s), 
and modular low-cost electronics based on ASICs (Application-Specific Integrated Circuits)  
and intelligent trigger systems based on FPGAs (Field Programmable Gate Arrays)
hold the promise to (i) significantly reduce the price per telescope, and 
(ii) considerably improve the reliability and versatility  of IACTs. 

The improvement in sensitivity with a  $\sim 1$ km$^2$ array is in part achieved
by increasing the number of telescopes. Simple scaling suggests that a 
factor of $10^1$ improvement in  sensitivity requires a factor of $10^2$ increase
 in the number of telescopes and observatory cost.   However, this is not the case
for the $\sim 1$ km$^2$ IACT array concept, since the $\sim 1$ km$^2$ concept inherently
 provides a  better event reconstruction so that the  sensitivity improves far beyond 
 simple scaling arguments.
For the current generation of small arrays, the shower core mostly falls outside the 
physical array dimensions. 
 A $\sim 1$ km$^2$ array could, for the first time, fully constrain the air shower 
based on  many view points from the ground. This leads to several  substantial 
improvements  and can be understood by considering the
Cherenkov light density distribution at the ground. 
 
The Cherenkov light pool from an atmospheric cascade consists of three  
distinct regions: an inner region ($r<120$~m) in which the photon density is  
roughly constant, an intermediate region where density of the Cherenkov  
photons declines as a power law ($120$~m $<r<$ $300$ m) and an outer region  
where the density declines exponentially. 
A small array (VERITAS, HESS) samples the  majority of cascades in the intermediate and outer 
regions of the light pool.  A $\sim 1$ km$^2$ array samples for its mostly contained events,
 the inner, intermediate and outer region of the light pool and allows
 a much larger number of telescopes to participate in the event reconstruction with several
important consequences:

\begin{itemize}

\item First of all, at the trigger level this results in a lower energy threshold
since there are always telescopes that 
fall into the inner region where the light density is highest. For example, the 
$12$~m reflectors of the VERITAS array sample a majority of $100$ GeV 
$\gamma$ rays at distances of $\sim 160$~m
 and collect $\sim 105$ PEs per event. The same median number of photons  
would be collected by $9.3$ m reflectors, if the atmospheric cascades were  
sampled within a  distance of $~\sim 120$ m.  A $\sim 1$ km$^2$ array
 of IACTs with fully contained events could operate  effectively at energies below 
100~GeV despite having a telescope aperture  smaller than that 
of  VERITAS~\cite{VF2005,JKBF2005}.
Reducing the telescope size translates into  a reduction 
of cost per telescope and total cost for a future observatory.

\item The second factor which significantly affects the sensitivity and cost of  
future IACT arrays is the angular resolution for $\gamma$-rays. Due to the  
small footprint of the VERITAS and H.E.S.S. observatories, the majority  
of events above $\sim 100$~GeV are sampled outside the boundaries of the  
array, limiting the accuracy to which the core of atmospheric cascade can be  
triangulated.   Even higher resolution pixels will not help to improve the
angular resolution below $\sim 9$ arc-minutes ~\cite{Bugaev:07} for small
arrays.  However,  contained events in a  $\sim 1$ km$^2$ array
 of IACTs provide a nearly ideal  reconstruction based on simultaneous observations of the 
shower from all directions while sampling multiple core distances. 
Simulations of idealized  (infinite) large arrays of IACTs equipped with cameras composed  
from pixels of different angular sizes suggest that the angular resolution  
of the reconstructed arrival direction of $\gamma$-rays improves with finer  
pixelation up to the point at which the typical angular scale,  
determined by the transverse size of the shower core is  
reached~\cite{FV2007}.  Figure~\ref{fig:sensang} shows the angular  
resolution that can be achieved (few minutes of arc) with an ideal ``infinite'' array of IACTs  
when instrumental effects are neglected \cite{Hofmann2005}.

\item The third factor improving the sensitivity of  $\sim 1$ km$^2$ arrays of IACTs  
comes through enhanced background discrimination. For atmospheric cascades  
contained within the array footprint, it is possible to determine  
both the depth of the shower maximum and the cascade  
energy relatively accurately, thereby enabling better separation of hadronic and electromagnetic  
cascades. Multiple viewpoints from the ground at different core distances 
also allow the detection of fluctuations in light density and further improve background rejection.
Additional  improvements extending to energies below 200~GeV may be possible by 
picking up muons from hadronic cascades, a technique that is used in air shower 
arrays.  A ``muon veto'' signal  present in the images  obtained of a large
array could improve the technique even further. Another method to reject cosmic-ray background
at the lowest energies and low light levels \cite{FK:1995} is based  on the parallactic displacement
of images.
The images viewed from multiple viewpoints at the ground show significant fluctuations in
lateral displacements for hadronic showers and simulations indicate appreciable $\gamma$/hadron 
separation capabilities in a regime where faint Cherenkov light images can no longer be resolved
for the calculation of standard image parameters.  This technique could
become effective close to the trigger threshold of large arrays.

\end{itemize}

In summary, the concept of ``large IACT arrays'' provides strongly  
improved sensitivity at mid-energies, $\sim 1$ TeV, not only due to  
increased collecting area, but also due to enhanced angular  
resolution and CR background rejection. It also presents a 
cost-effective solution for increasing the collecting area of the  
observatory at lower energies.

For energies above $>10$~TeV, the collecting area of the $\sim 
1$ km$^2$ IACT array will be approximately two times larger than its geometrical area 
due to events impacting beyond the perimeter of  the array. 
It must be noted that in this energy regime the observatory is no  
longer background limited and therefore its sensitivity scales inversely  
proportional to the collecting area and exposure. 

Clearly, versatility is another virtue of a ``large IACT array''.  If the  
astrophysics goal is to only measure the high-energy part of the spectrum  
($>10$~TeV) of a given source, e.g. the Crab Nebulae or Galactic Center, only  
$1/10^{\mathrm{th}}$ of the observatory telescopes, spaced on the grid of  
$\sim 300$~m, would be required to participate in the study to gain a  
required sensitivity, while at the same time other observation programs 
could be conducted.   The  flexibility  of a large array also allows
operation in a sky survey mode to detect transient galactic or  
extragalactic sources~\cite{VF2005}. In this mode of operation a large field  
of view would be synthesized by partially overlapping the fields of view of  
individual telescopes. Survey observations, in which collecting  
area has been traded for wide solid-angle coverage, could then be followed up  
 by more sensitive ``narrow-field'' of view for detailed source studies.

Although the design considerations outlined above are relevant for any 
``large IACT array'',  realistic implementations of this concept could vary.
An alternative approach to the array, consisting of identical telescopes, is being developed,  
based on an extrapolation from small arrays, H.E.S.S. and  
VERITAS, and is known as the hybrid array concept.
In this approach the limitation  of the cost of the future observatory is addressed 
through a design with multiple types of IACTs, each addressing a different energy range.
For  example, a central core composed of a few very large aperture  
telescopes ($\sim 20$~m) equipped with fine pixel cameras (or very high  
spatial density mid-size reflectors~\cite{JKBF2005} ), provides for the low  
energy response of the array. A significantly larger, $\sim 1$~km$^2$, ring  
area around the array core is populated with VERITAS class telescopes  
($>12$~m) to ensure improved collecting area and performance at mid-energies,  
$\sim 1$ TeV. Finally, a third ring surrounds the 1~km$^2$ array with a very  
spread-out array of inexpensive, small ($2$~m aperture), wide-field IACTs  
outfitted with coarsely pixelated cameras ($0.25^{\circ}$), which would cover  
areas up to $10$~km$^2$. On the order of $100$ telescopes with $300$~m  
spacing might be required to gain the desired response at the highest  
energies ($> 10$ TeV)~\cite{stamatescu07}. 
 
The hybrid array concept with a central region of several large  
aperture telescopes is motivated  by  significant changes in  
the distribution of Cherenkov photons at energies considerably smaller than  
$\sim 100$ GeV.  At very low energies, $\sim  
10$ GeV, the Cherenkov light is distributed over a relatively large area, but  
with lower overall density. Therefore, large aperture telescopes arranged in  
an array with significant separation between them may provide a cost  
effective solution to improve the low energy response.

Independently from exact implementation of the IACT array layout, the  
sensitivity of future ground-based observatories could be improved through the  
increase of both camera pixelation  and the number of  
telescopes. The low energy sensitivity will also be  
affected by the telescope aperture. Therefore, a trade-off optimization of  
these factors should also be performed under a constraint of constant cost of  
the observatory. For example, if the camera dominates the overall cost of the  
IACT significantly, then a reduction of camera pixelation and increase of the  
number of telescopes is suggested for optimizing cost.
 If the telescope optical and positioning systems dominate the cost,  
then reducing the number of telescopes and improving their angular resolution  
is preferential for achieving the highest sensitivity. The cost per pixel and 
of the indivisual  telescopes of a given apearture are the most critical 
parameters required for future observatory design decisions.
 
Through the design and construction of H.E.S.S., VERITAS, and MAGIC, 
considerable experience has been gained in  
understanding the cost and technical challenges of constructing prime  
focus, Davies-Cotton (DC) and parabolic reflectors and assembling cameras  
from hundreds of individual PMTs. 
The relatively inexpensive, DC telescope design has been used in ground-based $\gamma$-ray  
astronomy for almost fifty years successfully and provides an excellent baseline option  
for a future observatory.  For example, the HESS 13~m aperture telescopes have an optical
pointspread function of better than 0.05 deg. FWHM  over a 4 degree field of view
and pixel size of 0.15~deg., demonstrating that this telescope design could in principle
accommodate a few arc minute camera resolution.
  To reach significantly better angular resolution in conjunction with wider field of view 
systems,  alternative designs are being considered.

An alternative telescope design that could be  
used in future IACT array is based on the Schwarzschild-Couder (SC) optical  
system (see Fig. \ref{fig:vass_fig2.ps})~\cite{Vass:07}, which consists of  
two mirrors configured to correct spherical and coma aberrations, and  
minimize astigmatism. For a given light-collecting area, the SC optical system  
has considerably shorter focal length than the DC optical system, and is  
compatible with small-sized, integrated photo-sensors, such as Multi Anode  
PMTs (MAPMTs) and possibly Silicon PMs (SiPMs). Although the SC telescope  
optical system, based on aspheric mirrors, is more expensive than that of a  
DC design of similar aperture and angular resolution, it offers a  
reduction in the costs of focal plane instrumentation using pixels that are physically
substantially smaller.
In addition, the SC telescope offers a wide, unvignetted, 6 degree field-of-view, 
unprecedented  for ACTs, which can be further extended up to 12 degrees, if necessary, when  
a modest degradation of imaging and loss of light-collecting area can be  
tolerated. Unlike a DC telescope, the two-mirror aplanatic SC design does not  
introduce wavefront distortions, allowing the use of fast $>$~GHz electronics to  
exploit the very short intrinsic time scale of Cherenkov light pulses ($<$3  
nsec). 
The Schwarzschild telescope design was proposed in 
1905~\cite{Schwarzschild1905}, but the construction of an SC telescope only  
became technologically possible recently due to fundamental advances in the  
process of fabricating aspheric mirrors utilizing replication processes such  
as glass slumping, electroforming, etc.  It is evident that the SC design
requires novel  technologies and  is scientifically attractive.  Prototyping
and a demonstration of its performance and cost are required to fully explore its
potential and scientific capabilities.

To summarize, ``large'' IACT array concept provides the means to achieve  
the required factor of 10 sensitivity improvement over existing instruments.
 Significant simulations and design studies are required to make an informed 
decision on the exact array implementation, such as deciding between uniform 
or graded arrays. Two  
telescope designs, DC \& SC, offer a possibility for the largest collecting  
area, largest aperture, and highest angular resolution IACT array options.  
Studies of the tradeoff of performance costs and robustness of operation are  
necessary for design conclusions.

\begin{figure*}[t] 
 
\begin{center} 
 
\includegraphics[angle=0,width=6.1cm]{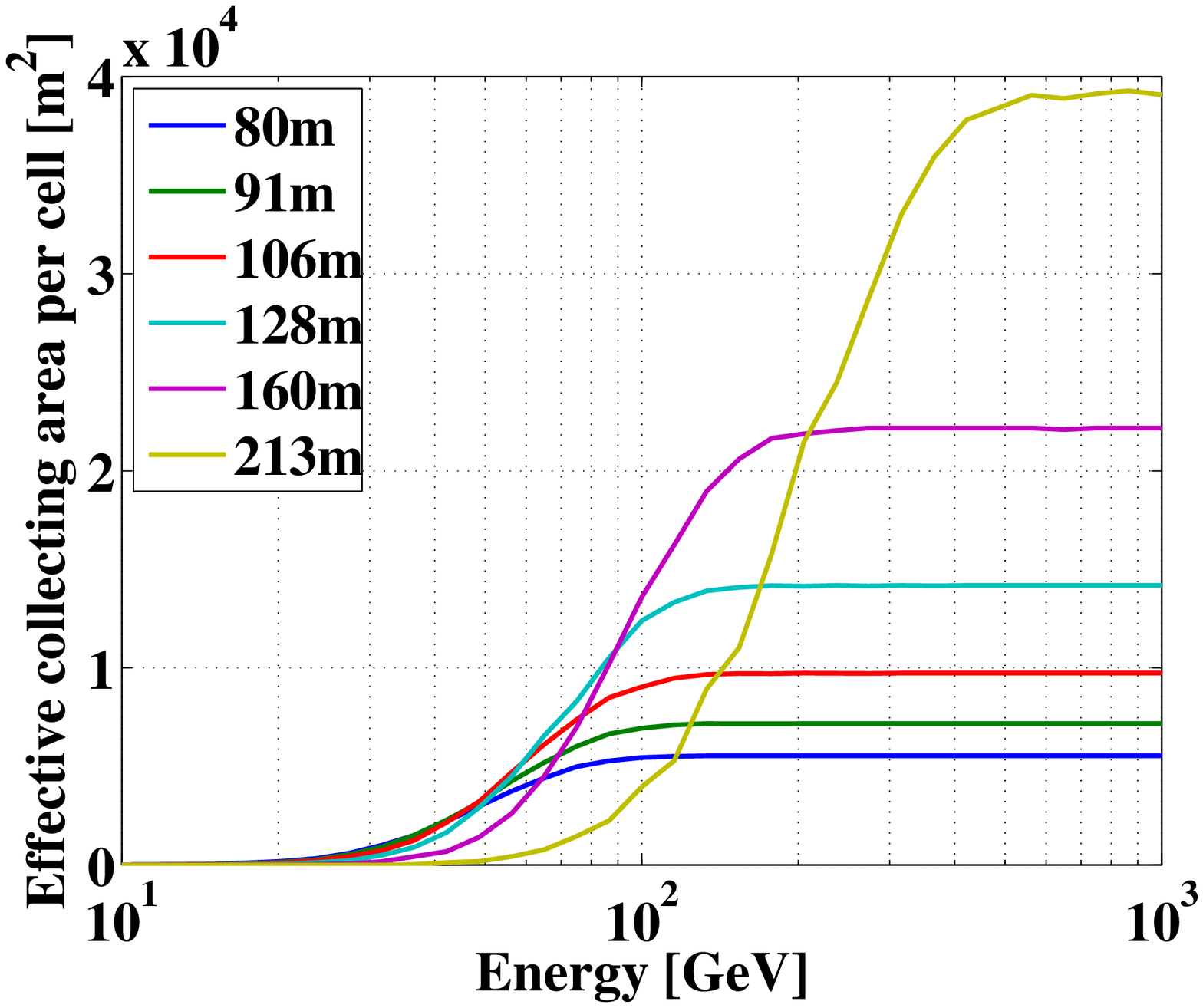} 
\includegraphics[angle=0,width=8.8cm]{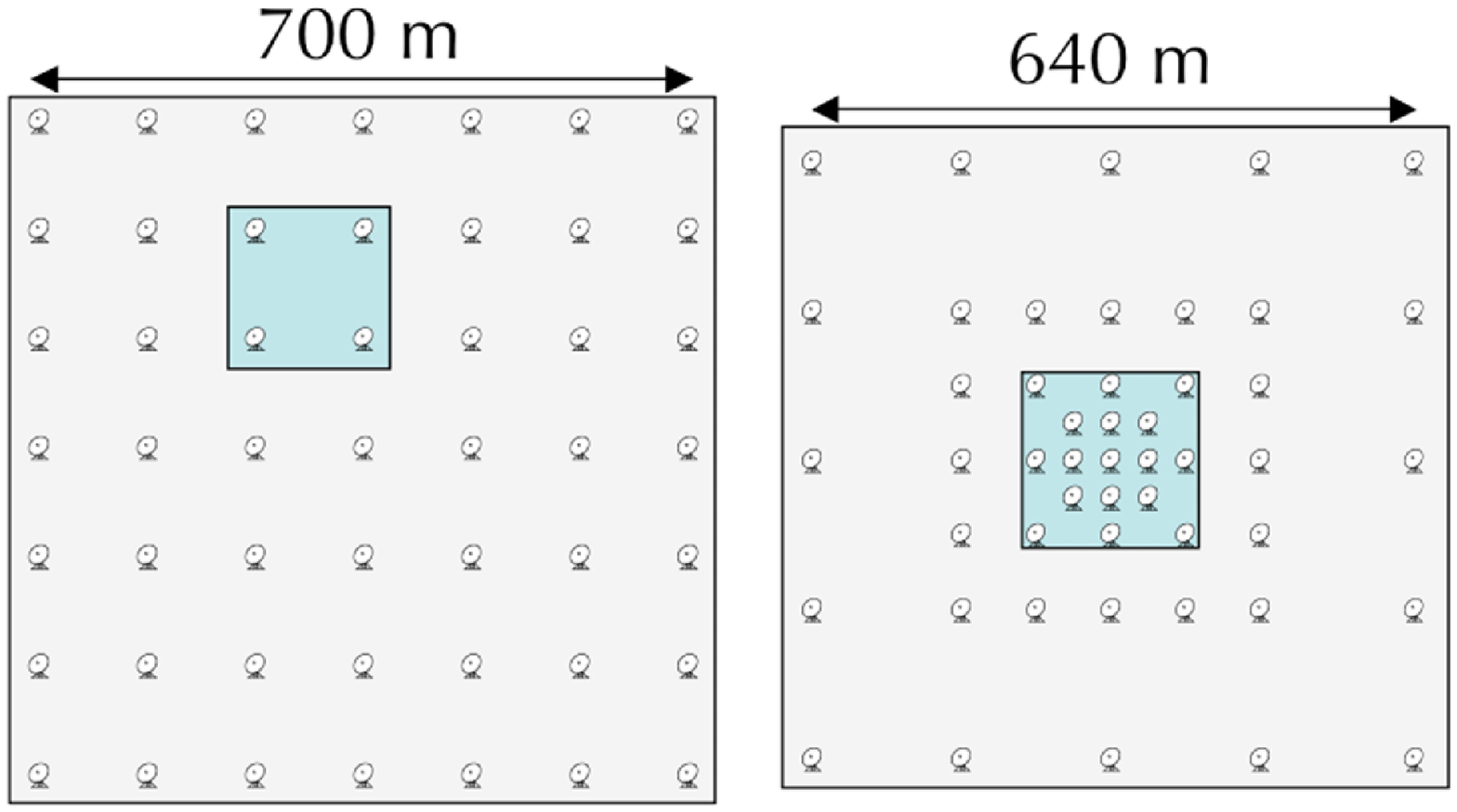} 
 
\caption{ \textit{Left:} Effective area vs. energy for a single cell for different telescope spacings; for a very large array with a fixed number of telescopes, the total effective area will be proportional to this number. \textit{Center,Right:} Two possible array configurations showing a uniform array and one where the central cluster of telescopes is more densely packed to achieve a balance between the desires for low threshold and large effective are at higher energies.}
 
\label{fig:array1}

\end{center} 
 
\end{figure*}

\begin{figure*}[t] 
 
\begin{center} 
 
 
\includegraphics[angle=0,width=4.0in]{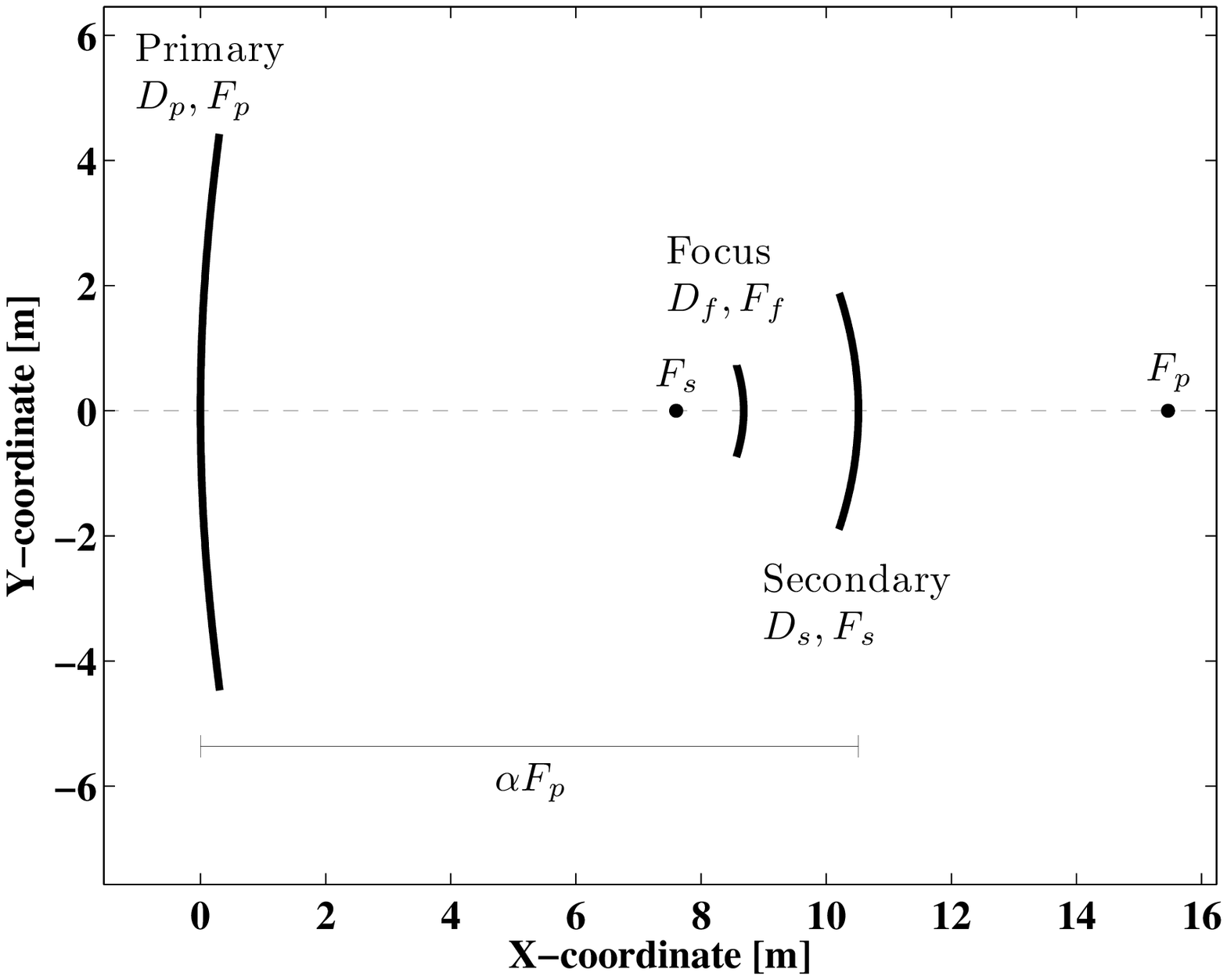} 
 
\caption{\label{fig:optics} A future Cherenkov telescope array may use  
conventional Davies-Cotton or parabolic optical reflectors 
similar to the ones used by VERITAS, MAGIC, and H.E.S.S., or may use novel  
Schwarzschild-Couder optical designs that  
combine wide field of views 
with excellent point spread functions and a reduction of the plate-scale, and  
thus of the camera size, weight, and costs. 
The image shows the cross-section of an exemplary Schwarzschild-Couder design  
(from \cite{Vass:07}).} 
\label{fig:vass_fig2.ps} 
\end{center} 

\end{figure*}

\subsection[Future EAS observatory]{Future EAS Observatory} 

The success of EAS observatories in gamma-ray astronomy is relatively recent, with the
first detection of new sources within the last couple of years \cite{Abdo:07}, as compared to the
over 20 year history of successes with IACTs.  However, EAS observatories have
unique and complementary capabilities to the IACTs. 
The strengths of the technique lie in the ability to perform unbiased all-sky surveys (not
simply of limited regions such as the Galactic plane), to measure spectra
up to the highest energies, to detect extended sources
and very extended regions of diffuse emission such as the Galactic plane, and
to monitor the sky for the brightest transient emission
from active galaxies and gamma-ray bursts and search for
unknown transient phenomena.

The instantaneous field of view of an EAS detector is $\approx$2 sr and is
limited by the increasing depth of the atmosphere that must be traversed by
the extensive air shower at larger zenith angles. 
However, for higher energy gamma rays, the showers are closer to
shower maximum and have more particles; thus the resolution improves. 
As the Earth
rotates, all sources that pass within $\approx$45 degrees of the detector's zenith
are observed for up to 6 hours.  For a source with a Crab-like spectrum, the flux sensitivity
of an EAS detector varies by less than 30\% for all sources located within $\approx$2$\pi$ sr.

The angular resolution, energy resolution, and $\gamma$-hadron separation  
capabilities of EAS technique are limited by the fact that the detectors 
sample the particles in the tail of the shower development well past the shower
maximum. The angular resolution improves  
at higher energies ($>$ 10 TeV), and the best single-photon angular  
resolution achieved to date is 0.35$^{\circ}$ which was achieved with the highest energy observations of Milagro.
Placing an extensive shower detector at a higher elevation will allow the particles
to be detected closer to the shower maximum.  For example,  an observatory at 4100m
above sea level detects 5-6 times as many particles for the same energy primary
as an observatory at 2650m (the elevation of Milagro).

Also, increasing the size of a detector will increase the collection area and thus the 
sensitivity. As both signal and background are increased, the relative sensitivity would
 scale proportional to Area$^{0.5}$ if there were no other improvements. However, the effectiveness 
of the gamma-hadron cuts improves drastically with detector size, because the lateral 
shower distribution is more thoroughly sampled. The background hadron induced showers 
can be efficiently rejected through the identification of muons, hadrons and secondary 
electromagnetic cores. But the large transverse momentum of hadronic interactions spreads 
the shower secondaries over a much larger area on the ground than the gamma-ray initiated 
showers.  Detailed simulations using Corsika to simulate the air showers and GEANT4 to simulate
a water Cherenkov observatory show that most background hadronic showers can be rejected by 
identifying large energy deposits separated from the shower core\cite{Smith_GLAST:07}. 
Simulations of larger versions of such a detector demonstrate that sensitivity scales as 
Area$^{0.8}$ at least up to 300m x 300m.

The high-energy sensitivity of all gamma-ray detectors is limited by the total exposure
because the flux of gamma rays decreases with energy.  An EAS detector has a very large
exposure from observing every source every day.  For example, a detector of area 2 $\times$ 10$^4$m$^2$ 
after 5 years will have over 1 km$^2 \times$ 100 hours of exposure. And as the energy increases, EAS
observatories become background free because the lateral distribution of muons, hadrons and secondary
cores in hadronic showers is better sampled.

The low energy response of EAS detectors is very different from IACTs, again because only the
tail of the longitudinal distribution of the shower is observed.  Past shower maximum, the 
number of particles in the shower decreases with each radiation length.  However, the 
probability of a primary penetrating several radiation lengths prior to first interaction
in the atmosphere decreases exponentially with radiation length.  These two facts, as well as
the number of particles at shower maximum is proportional to the primary energy,
imply the effective area increases with energy E as E$^{2.6}$ until a threshold energy 
where the shower can be detected if the primary interacts within the first radiation
length in the atmosphere. Therefore, EAS detectors can have an effective area up to 100 m$^2$ at 
the low energies of $\sim$ 100 GeV.  This area is considerably larger than Fermi's of $\sim$ 1
m$^2$, and is sufficient to observe bright, extragalactic sources such as active
galactic nuclei and possibly gamma-ray bursts.  The wide field of view of EAS observatories
is required to obtain long term monitoring of these transient sources and EAS observatories
search their data in real time for these transient events to send notifications within a few
seconds to IACTs and observers at other wavelengths.

The HAWC (High Altitude Water Cherenkov) observatory is a next logical step in the development
of EAS observatories\cite{dingus:07}.  It will be located in Mexico at Sierra Negra at an altitude of 4100 m and
will have 10-15 times the sensitivity of Milagro. The (HAWC) observatory will re-use the 
existing photomultiplier tubes from Milagro in an approximately square array of 900 large water tanks.  The tanks
will be made of plastic similar to the Auger tanks, but will be larger, with a diameter of 5 m and
4.3 m tall. An 8" diameter PMT would be placed at the bottom of each tank and look up
into the water volume under $\approx$4 m of water.
The array would enclose 22,500 m$^2$ with $\approx$75\% active area.
Thus, unlike Milagro, the same layer of
PMTs would be used to both reconstruct the direction of the primary gamma ray
and to discriminate against the cosmic-ray background.   
The optical isolation of each PMT in a separate tank allows a single layer to accomplish
both objectives. A single tank has been tested in conjunction with Milagro and its 
performance agrees with Monte Carlo simulation predictions.  The optical isolation also
improves the background discrimination (especially at the trigger level), and the 
angular and energy resolution of the detector.

\begin{figure*}[ht] 
\begin{center}
\includegraphics[height=5.2in]{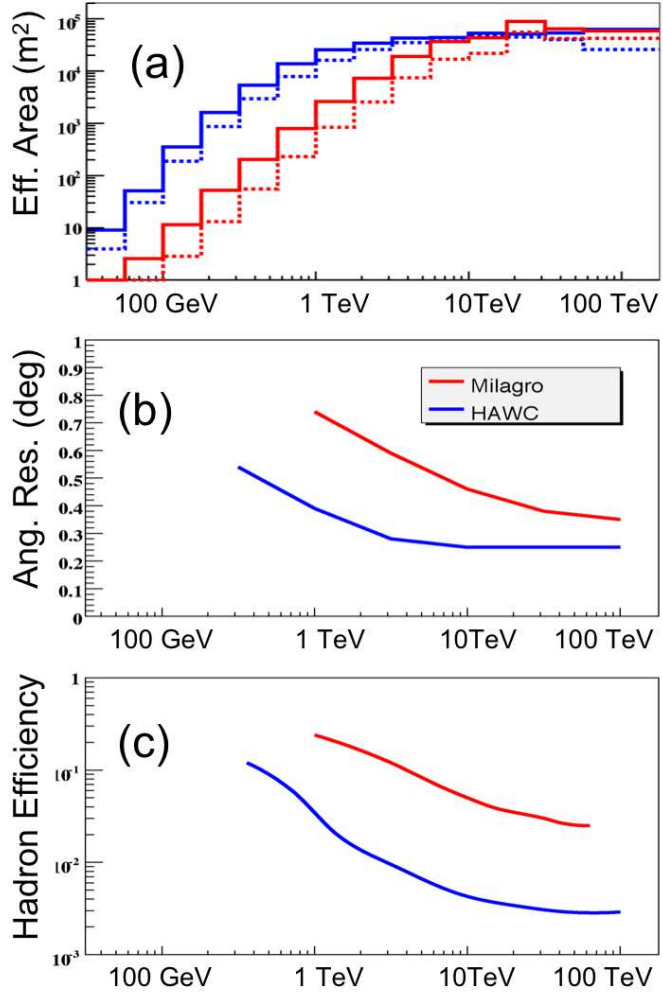}  
\caption{\label{fig:hawc}: The sensitivity of HAWC and Milagro versus primary gamma-ray energy. 
Panel (a) shows the effective area, (b) the angular resolution, and (c) the efficiency with the hadronic background showers are rejected when half of the gamma-ray events are accepted. }
\end{center}  
\end{figure*}

The performance of HAWC is shown in Figure \ref{fig:hawc} and is compared to Milagro.  These detailed
calculations use the same Monte Carlo simulations that accurateley predict the performance of Milagro.
The top panel shows the large increase in the effective area at lower energies as expected from the 
increase in altitude from 2600m to 4100m.  At higher energies the geometric area of HAWC is similar
to the geometric area of Milagro with its outrigger tanks.  However, the improved sampling of the
showers over this area with the continuous array of HAWC tanks results in improved angular resolution
and a major increase in background rejection efficiency.  Therefore, the combined sensitivity improvement 
for a Crab-like source is a factor of 10-15 times better than Milagro.  This implies that the Crab 
can be detected in one day as compared to three months with Milagro.   

The water Cherenkov EAS detector  can be extrapolated to enclose even larger areas and
the sensitivity of such a detector is relatively straight forward to calculate.
Earlier work in this area discussed an array enclosing 100,000 m$^2$, with
two layers of PMTs \cite{Sinnis2004, Sinnis2005}.  Recent work
indicates that a single deep layer (as in the HAWC design)
will perform as well as the previous two-layer design. 
For example, a detector with an active detection
area 100,000 m$^2$ (HAWC100), located at 5200 m above sea level, would have an effective
area at 100 GeV of $\sim$10,000 m$^2$ for showers from zenith.
The low-energy response allows for the detection of gamma-ray bursts at larger
redshifts than current instruments ($z\sim$1 for HAWC compared to $z\sim$0.3
for Milagro if, at the source, the TeV fluence is equal to the keV fluence).
While current instruments, such as Milagro, indicate that the
typical TeV fluence from a GRB is less than the keV fluence, instruments such as
HAWC100 and HAWC would be sensitive to a TeV fluence 2-3 orders of magnitude
smaller than the keV fluence of the brightest gamma-ray bursts.

\subsection[Technology roadmap]{Technology Roadmap}
\label{sec:RoadMap}
The recent successes of TeV $\gamma$-ray astronomy both in terms of  
scientific  
accomplishments and in terms of instrument performance have generated 
considerable interest in next-generation instruments. Part of the excitement  
originates from the fact that an order of magnitude sensitivity improvement  
seems to be in reach and at acceptable costs for making use of existing technologies.  
New technologies could result in even better sensitivity improvements.
A roadmap for IACT instruments over 
the next 3 years should focus on design studies to understand the trade-
offs between performance, costs, reliability of operation of IACT arrays, 
and on carrying out prototyping and the required research and development. 
It is anticipated that, at the end of this R\&D phase, a full proposal for construction of an observatory would be submitted.  
A next generation instrument could be built on a time scale of $\sim$5 years  
to then be operated for between 5 years (experiment-style operation) and  
several decades (observatory-style operation). 
For IACT instruments, the following R\&D should be performed: 
\begin{itemize}

\item Monte Carlo simulations of performance of large IACT arrays to optimize array configuration parameters such as array type (hybrid or homogeneous), array layout, aperture(s) of the telescope(s), and pixilation of the cameras, with a fixed cost constraint.  Effects of these parameters on energy threshold, angular resolution, and sensitivity of the observatory should be fully understood, together with associated cost implications.

\item The conservative Davies-Cotton telescope design with $f - \frac{F}{D} \sim 1$ should be considered as 
a baseline option for the future observatory.  However, limitations of this design and benefits and cost impact of alternative options should be investigated.  These alternatives include large focal length Davies-Cotton or parabolic prime-focus reflectors with $f\sim 2$ and aplanatic two-mirror optical systems, such as Schwarzschild-Couder and Ritchey-Chr\'{e}tien telescopes.  The latter designs have the potential to combine significantly improved off-axis point spread functions, large field-of-views, and isochronicity with reduced plate scales and consequently reduced costs of focal plane instrumentation.  Prototyping of elements of the optical system of SC or RC telescopes is required to assess cost, reliability and performance improvement.  Mechanical engineering feasibility studies of large focal length prime focus telescopes and two-mirror telescopes should be conducted.

\item The development and evaluation of different camera options should be  
continued. Of particular interest  
are alternative photo-detectors (photomultiplier tubes with ultra high  
quantum efficiency,  
multi-anode photomultipliers, multi channel plates, Si photomultipliers,  
Geiger mode Si detectors,  
and hybrid photodetectors with semiconductor photocathodes such as GaAsP or  
InGaN) and a modular design  
of the camera which reduces the assembly and maintenance costs. Compatibility of these options with different telescope designs and reliability of operation and cost impact should be evaluated.

\item The development of ASIC-based front-end-electronics should be continued  
to further minimize the power  
and price of the readout per pixel. 

\item A next-generation experiment should offer the flexibility to operate in  
different configurations, so that specific 
telescope combinations can be used to achieve certain science objectives.  
Such a system requires the development of 
a flexible trigger system.  Furthermore, the R\&D should explore the possibility of combining the trigger signals of  
closely spaced telescopes to synthesize a single telescope of larger  
aperture. A smart trigger could be used  
to reduce various backgrounds based on parallactic displacements of Cherenkov light images \cite{FK:1995}. 

\item The telescope design has to be optimized to allow for mass production  
and to minimize the maintenance costs. 

\item The telescopes should largely run in robotic operation mode to enable a  
small crew to operate the entire system.  The reliability of operation of large IACT arrays should be specifically researched, including tests of instrumentation failure rates and weathering to evaluate required maintenance costs.

\end{itemize}

A roadmap for EAS array over the next 5 years (HAWC) is well defined by the benefits of moving the experiment to high altitudes and   
enlarging the detection area.  The cost of this path is  $<$ \$10M USD.  A site in Mexico has been identified and is a few km from the Large Millimeter Telescope; it is a 2 hour drive from the international airport in Puebla, and has existing infrastructure of roads, electricity, and internet.  The HAWC project will be a joint US and Mexican collaboration with scientists from Milagro, Auger, and other astronomical and high-energy physics projects.

The R\&D for IACT could be finalized on a time scale of  
between 3 (IACTs).  
The R\&D should go hand in hand with the establishment of a suitable  
experimental site and the build-up of basic infrastructure. 
Ideally, the site should offer an easily accessible area exceeding 1 km$^2$.  
For an IACT array, an altitude between 2 km and 3.5 km will give the best  
tradeoff between low energy thresholds,  
excellent high-energy sensitivity, and ease of construction and operation.

The U.S.\ teams have pioneered the field of ground based $\gamma$-ray  
astronomy during the last 50 years. The U.S. community has formed the  
``AGIS'' collaboration  
(Advanced Gamma ray Imaging System) to  optimize the design of a future  
$\gamma$-ray detector.  
A similar effort is currently under consideration in Europe by the CTA  
(Cherenkov Telescope Array) group, and the 
Japanese/Australian groups building CANGAROO are also exploring avenues for  
future progress. 
Given the scope of a next-generation experiment, the close collaboration of  
the US teams with the European and  
Japanese/Australian groups should be continued and intensified. If funded  
appropriately, the US teams are in  
an excellent position to lead the field to new heights.

\clearpage

\appendix
\appendixpage
\addappheadtotoc
\section{Glossary}
\label{glossary-appendix}
\subsection[Astronomical and physics terms]{Astronomical and Physics Terms} 
\textbf{30 Dor C} - 30 Dor C is a {\it superbubble} in the {\it Large Magellanic Cloud} coinciding with an {\it OB association}.
                    The detection of non-thermal radio emission indicates the presence of relativistic electrons.

\medskip 

\textbf{AGN} - An Active galactic nucleus is a compact region at the center of a galaxy that
has a much higher than normal luminosity over most of the electromagnetic spectrum 
ranging from the radio, infrared, optical, ultra-violet, X-ray and high energy to very high
energy  {\it VHE} gamma-ray energies. Active galactic nuclei (AGN) often show a pair of {\it relativistic jets} that 
are powered by accretion  onto  a {\it supermassive black hole} the size of our solar system.  
These large scale jets are prospective sites for particle acceleration since they often reveal relativistic phenomena 
(see also {\it blazars}).

\medskip 

\textbf{Blazar} - Blazars are {\it AGN} that have their {\it jet} axis closely aligned with the observer's line of
                  sight.  Consequently, {\it relativistic Doppler boosting} of emission regions moving along
                  the {\it jet} axis  causes blazars to appear  
                  extremely bright and to exhibit  rapid flux variations.               

\medskip 

\textbf{Bremsstrahlung} - The radiation that is emitted by the deceleration of an electron in the
                         electric field of an atomic nucleus. 

\medskip 

\textbf{Cosmic rays} - Cosmic rays are energetic particles, mostly protons and helium nuclei that impinge on the
                       Earth's atmosphere. Although cosmic rays were discovered in 1912 by Victor Hess,  their origin
                       is still unknown.
                       Their vast energy range ($10^9$- $10^{20}$~eV)  suggests  that cosmic rays
                       are produced in  astrophysical environments of vastly different size scales and magnetic field strengths.
                      Currently favored sites are parsec (pc) scale {\it supernova remnants} with the potential 
                      to accelerate atomic nuclei to a few PeV and kpc scale
                       {\it jets} associated with {\it AGN} that might achieve
                       energies in the  $10^{20}$~eV ({\it EHE}) regime. 

\medskip 

\textbf{Crab Nebula} - The Crab Nebula is a {\it supernova remnant} with a {\it pulsar wind nebula} that dates back to a {\it supernova}
                     that occurred
                       in 1054 AD. Due to its strong and steady emission of {\it synchrotron radiation} from the nebula, the Crab is used as 
                       a standard candle in X-ray and gamma-ray astronomy and source fluxes are often given in units of 1~Crab.

\medskip 

\textbf{Cygnus region} - The Cygnus region is a prominent bright feature in the galactic sky map across many wavelengths with
                          gamma-ray emission extending up to several TeV. The presence of {\it supernova remnants}, {\it OB associations} 
                        and {\it Wolf-Rayet stars} makes it a very promising site for relativistic particle acceleration in our
                       galaxy. 

\medskip 

\textbf{Dark matter} - Approximately  80\% of the matter in the universe is made from Dark matter  which is a hypothetical 
                   form of matter of unknown composition that does not emit or interact with light at a level 
                        that is directly observable.  Only less than 20\% of the  matter in the universe appears luminous
                       through the emission of light and is made from baryonic matter such as protons and  nuclei. 
                       The existence of dark matter is concluded from its gravitational effects on light 
                        and the dynamics of individual galaxies and {\it galaxy clusters}.

\medskip 

\textbf{Dwarf galaxies} - A dwarf galaxy is a small galaxy composed of up to several billion stars, a small number 
                          compared to our own {\it Milky Way} galaxy with  200-400 billion stars.

\medskip 
\textbf{EBL} - Extragalactic background light describes the diffuse background radiation with wavelengths between 0.1~micron and 1000~micron.
               The EBL is produced by  all cumulative radiative  energy releases after the epoch of recombination and constitutes 
               the second most important radiative energy density permeating the universe after the {\it cosmic microwave background}.

\medskip 
\textbf{EHE} - Extremely high energy stands for $\rm 10^{18}$~eV and is used for describing the energy scale 
               of the highest energy {\it cosmic rays}.

\medskip 
\textbf{Galactic Center} - The Galactic Center refers to the center of the Milky Way galaxy that 
                            contains a {\it supermassive black hole}
                            accreting mass from stars, dust and gas.

\medskip 
\textbf{Galaxy cluster} - Galaxies occur in groups, the larger groups with 50 to 1000 galaxies are called galaxy clusters,
                          the space in between galaxies is filled with a hot gas, the {\it intracluster medium} (ICM).

\medskip 
\textbf{Gamma-ray burst} - Gamma-ray bursts (GRBs) are flashes of gamma-rays originating in random places in space
                            and are the most luminous events since the big bang.

\medskip 
\textbf{Heliosphere} - The heliosphere is a bubble in the {\it interstellar medium } produced by the solar wind 
                       and surrounds the sun out to a 
                       radial distance of $\approx$~100  astronomical units  (1 AU = distance between  Earth and Sun).

\medskip 

\textbf{Intracluster medium} - The intracluster medium (ICM) permeates the space between galaxies in a {\it galaxy cluster }
                               and typically has a temperature of $\rm 10^{7} - 10^8$~K and is detected in  X-rays
                                from thermal {\it bremsstrahlung} and atomic line emission.

\medskip 

\textbf{Inverse Compton scattering} - Inverse Compton scattering is the process in which a photon gains energy 
                                     from the interaction with a relativistic electron. 

\medskip 

\textbf{Interstellar medium} - The interstellar medium (ISM) consists of a dilute mixture of mostly hydrogen 
                 and helium gas (together about 99\% by mass), 1\% dust 
               grains, cosmic rays and a magnetic field that permeates the interstellar space
               between stars in a galaxy.

\medskip
 
\textbf{LMC} - The Large Magellanic Cloud (LMC) is a nearby satellite galaxy of the Milky Way and resides at a distance of 50~kpc.
              It contains just about 1/10 of the Milky Way's mass.

\medskip
 
\textbf{Microquasar} - Microquasars share important similarities with {\it quasars}: strong and variable radio
                       emission, a pair of radio {\it jets}  and a compact region surrounded by an accretion disk.
                       Microquasars are found in stellar binary systems and may be powered by  accretion of mass 
                       from a companion star  onto a 
                      compact object, either a neutron star or a black hole.
                     Microquasars play an important role in the study of {\it relativistic jets} since 
                      they are  miniature versions  of {\it quasars}.

\medskip
 
\textbf{Milky Way Galaxy} - Our solar system belongs to the Milky Way Galaxy that consists of over 200-400~billion
                            stars, mostly distributed in a disk of a few hundred light years thick and a radial
                             extension of 100,000 light years across. 

\medskip

\textbf{Millisecond pulsars} - Millisecond pulsars are extremely rapid spinning neutron stars with rotation periods of 1 - 10 milliseconds.
                               These are extreme rotation periods even for {\it pulsars}, and could arise from angular momentum
                               transfer via accretion.  Millisecond pulsars are found in X-ray binary star systems with a massive 
                                companion star consistent with 
                              the idea that they have evolved from regular {\it pulsars} and were spun up by accretion.

\medskip 

\textbf{Molecular cloud} - A molecular cloud is an interstellar cloud  containing hydrogen  gas with a density and temperature
                           cool enough to allow the formation of $\rm H_2$ molecules.  Since $\rm H_2$ is difficult to
                           detect, the presence of carbon monoxide (CO) is often used as a proxy for tracing $\rm H_2$ since
                           the ratio between CO luminosity and the mass in $\rm H_2$  is thought to be constant.

\medskip

\textbf{Neutrino} - A subatomic particle with little mass that interacts only weakly making it difficult to detect.
                    Neutrinos are generated in reactors, inside the Sun, in a {\it supernova} and  in {\it cosmic-ray} interactions 
                    therefore probing a variety of astrophysical phenomena.  Their study is particularly 
                    useful for studying  processes in dense and hidden environments of astrophysical sources.

\medskip 

\textbf{OB associations} - The term OB association describes star forming regions comprised of 10 to 100 massive stars,
                           mostly O and B (very hot) stars.

 \medskip 

\textbf{Parsec} - Distance unit often used in astronomy where 1 parsec corresponds to 3.26 light years. 

\medskip 

\textbf{Periastron} - Periastron describes the closest approach of a star orbiting the center of attraction in a binary system.

\medskip 

\textbf{Pions} - In particle physics, pion is the collective name for three subatomic particles: a neutral, a negatively and a
                 positively charged particle. Pions are the lightest mesons and play an important role in explaining 
                 low-energy properties of the strong nuclear force.

\medskip 

\textbf{PeV} - 1 petaelectronvolt corresponds to an energy of $\rm 10^{15}$~eV.

\medskip 

\textbf{Pulsar} - Pulsars are highly magnetized rapidly rotating neutron stars which emit a beam of detectable electromagnetic 
                  radiation in the form of radio waves.  The beam can only be observed when it is directed towards the 
                  observer's line of sight.

\medskip 

\textbf{PWN} - A Pulsar wind nebula is a {\it synchrotron radiation} emitting nebula powered by the relativistic wind 
                of an energetic pulsar. The most prominent example is the {\it Crab nebula}.

\medskip 

\textbf{Quasar} - Quasars are extremely bright radio sources located at the centers of very distant active galaxies. Historically, quasars
                  were detected as radio sources initially without optical counterpart, hence their name (QUASi-stellAR radio source).
                   It is generally agreed upon that a quasar is a halo of matter surrounding a supermassive black hole at the center
                   of an active galaxy.    Quasars have been detected out a redshift of 6.43, corresponding to a  distance of 8.5~Gpc.

\medskip 

\textbf{Relativistic Doppler boosting } -  Relativistic Doppler boosting refers to the increase in luminosity for a light
                    source moving at relativistic speed towards the observer,  whereas a reduction in luminosity  
                    is seen for a light source moving in opposite direction.  {\it Blazars} are therefore extremely 
                    bright  as one of  their jets is directed towards the observer.

\medskip 

\textbf{Relativistic jets} - Relativistic jets are narrow, pencil-beam structures 
of plasma that move at relativistic speeds from the centers of active galactic nuclei ({\it AGN}).
Mildly relativistic jets are also observed in galactic objects such as {\it microquasars}.
               The large scale jets in  {\it AGN} often reach several thousand {\it parsec} in scale and  are
               prime candidates for producing the highest energy cosmic rays.

\medskip 

\textbf{SS 433} - SS 433 is an X-ray binary system (13.1 day orbital period) with either 
                  a neutron star a black hole as the compact
                  object. Strong evidence for two mildly (0.26 c) {\it relativistic
                  jets} is given by varying Doppler-shifted emission lines.

\medskip 

\textbf{Starburst galaxies} - Starburst galaxies exhibit star formation rates hundreds of times larger 
                               than in our own galaxy.   Such high star formation rates are attributed
                               to a collision with another galaxy.

\medskip 

\textbf{Supermassive black holes} - Supermassive black holes have masses ranging from a hundred thousand up to  several tens of billions of solar masses.
                                    Most if not all galaxies appear to harbor a supermassive black hole at their center.

\medskip 

\textbf{Superbubble} - Superbubbles are regions in interstellar space that contain hot gas of $\rm 10^6$~Kelvin most likely
                       produced by multiple {\it supernovae} and stellar winds.

\medskip 

\textbf{Supernova} - A star that can no longer support its own weight collapses.  This  occurs in stars that
                     accrete matter from a companion star or in stars that run out of fuels for nuclear fusion.
                      As a result it throws off its outer layer causing a bright burst of electromagnetic radiation that 
                      can outshine an entire galaxy.

\medskip 

\textbf{Supernova remnants (SNR)} - A supernova remnant is the result of the gigantic explosion of a star, 
                      a {\it supernova}. The remnant is shaped by an expanding shock wave that consists of ejected material 
                       sweeping up  interstellar gas leading to the acceleration of charged particles
                       along the way. 

\medskip 

\textbf{Synchrotron radiation} - Synchrotron radiation is electromagnetic radiation, similar to cyclotron radiation, 
                                 but generated by the acceleration of ultrarelativistic 
                                 (i.e., moving near the speed of light) charged particles through magnetic fields.

\medskip 
\textbf{UHE} - Ultra high energy refers to {\it cosmic ray} particle energies between $\rm 10^{14}$~eV to  $\rm 10^{20}$~eV.

\medskip 
\textbf{Ultraluminous infrared galaxies} - Ultraluminous infrared galaxies (ULIRG) are galaxies that emit most of their
                                           energy output at far-infrared  wavelengths with luminosities exceeding
                                    more than 
                         $\rm 10^{12}$ solar luminosities.   This indicates that they contain a large amount of dust.

\medskip
 
\textbf{X-ray binary} - X-ray binary star systems are  powerful X-ray sources.  The X-ray emission  
                        results from accretion from a companion star onto a compact object.

\medskip 

\textbf{VHE} - Very high energy refers to the gamma-ray energy region of $\rm 10^{10}$~eV to $\rm 10^{14}$~eV
               and is generally covered by ground based gamma-ray observatories.

\medskip 

\textbf{Wolf-Rayet stars} - Wolf-Rayet stars are evolved, massive stars (over 20 solar masses), and are losing their 
                           mass rapidly by means of a very strong stellar wind, with speeds up to 2000 km/s. 

\subsection[Abbreviations and acronyms]{Abbreviations \&  Acronyms} 
\textbf{AGIS} - Advanced Gamma-ray Imaging System is a concept of a large future array of imaging atmospheric Cherenkov telescopes
               {\it (IACTs)} with a collection area of 1~$\rm km^2$ and is currently pursued by U.S. scientists. A similar effort 
               ({\it CTA}) is being studied by groups in Europe.

\medskip

\textbf{ANITA} - The ANtarctic Impulsive Transient Antenna is a balloon-borne neutrino detector circling Antartica, looking for radio
                  evidence of particle showers generated from extremely high-energy neutrinos interacting with the Antartic ice sheet.

\medskip 

\textbf{CGRO} - The Compton Gamma-Ray Observatory was launched in 1990 and is the second of NASA's four great 
               observatories. This observatory
                consisted of four instruments (EGRET, Comptel, OSSE and BATSE) and provided sensitivity in the
                gamma-ray regime between  20~keV and 30 GeV.

\medskip 

\textbf{CTA} - Cherenkov Telescope Array (CTA) is pursued by European scientists, is similar to {\it AGIS}
                   and aims to build a 1~$\rm km^2$ array of 
                imaging atmospheric Cherenkov telescopes.

\medskip 

\textbf{Chandra} - X-ray observatory that was launched in 1999 and is one of NASA's great observatories.

\medskip 

\textbf{EXIST} - Energetic X-ray Imaging Space Telescope would provide an all-sky survey at energies  between 
                 0.5-600~keV.

\medskip

\textbf{EGRET} - The Energetic Gamma-Ray Experiment Telescope was operated on the Compton Gamma-Ray Observatory satellite
                 in the 1990s and was sensitive to 20~MeV - 30~GeV.

\medskip 

\textbf{Fermi/GLAST} - Gamma-ray Large Area Space Telescope, a 20~MeV - 300~GeV pair conversion telescope to be launched
                in early 2008. GLAST will also have a burst monitor on board that is sensitive between 8~keV and 25~MeV.

\medskip 

\textbf{HAWC} - The High Altitude Water Cherenkov  observatory is a proposed wide field of view gamma-ray experiment over an
                energy range of $\rm approx$ 700~GeV - 50~TeV.

\medskip 

\textbf{H.E.S.S.} - High Energy Stereoscopic System, an array of four imaging atmospheric Cherenkov telescope 
                (see also {\it IACT}) systems in Namibia and has been operating since 2003.

\medskip 

\textbf{IACT} - Imaging Atmospheric Cherenkov Telescopes use  a technique to detect gamma-rays with ground based telescopes
                which was pioneered by the Whipple collaboration.  The technique is based on measurements of the secondary
                particle cascade (air shower) from a gamma-ray primary in the atmosphere.   The air shower is detected by recording
                Cherenkov light images of the shower.

 \medskip 

\textbf{IceCube} - The high energy neutrino telescope using a 1~km$^3$  of the ice  at the south pole.

\medskip 

\textbf{LHC} - Large Hadron Collider experiment at CERN.
 
\medskip 

\textbf{LIGO} - The Laser Interferometer Gravitational-wave Observatory, consisting of two 4-km laser interferometers
                to detect gravitational waves.

\medskip

\textbf{LISA} - The Laser Interferometer Space Antenna.

\medskip 

\textbf{LOFAR} - The LOw Frequency Array, a joint Dutch-U.S. initiative to study radio wavelengths longer than
                 2~m.

\medskip 

\textbf{LSST} - The Large-aperture Synoptic Survey Telescope, a 6.5~m class optical telescope.

\medskip 

\textbf{MAGIC} - Major Atmospheric Gamma-ray Imaging Cerenkov telescope,  a 17~m diameter imaging atmospheric Cherenkov telescope on the island of La Palma to detect sub-TeV - 10~TeV gamma radiation.

\medskip 

\textbf{Milagro} - A water Cherenkov detector for the detection of very high energy gamma-rays.

\medskip 

\textbf{Milagrito} - A  prototype water Cherenkov detector for the detection of very high energy gamma-rays that ultimately 
                    became the Milagro  observatory.

\medskip 

\textbf{PAMELA} - The Payload for Antimatter Matter Exploration and Light-nuclei Astrophysics, is the first satellite dedicated to detecting 
      cosmic rays and also antimatter from space, in the form of positrons and antiprotons. PAMELA pursues a long-term monitoring of the solar 
     modulation of cosmic rays and measurement of energetic particles from the Sun. 

\medskip 

\textbf{SIM} - The Space Interferometry Mission to be launched in 2011 will provide accurate distance measurements
               using parallax out to distances of 250~kpc.

\medskip

\textbf{Swift} - The Swift gamma-ray burst ({\it GRB}) mission is a multi-wavelength observatory dedicated to the study of {\it GRBs}.
                 Its three instruments can observe a {\it GRB} at gamma-ray, X-ray, ultraviolet and optical wavebands. It also provides 
                  burst notification within a few seconds time 
                 allowing both ground-based and other space-based telescopes around the world the opportunity to observe the burst's afterglow. 

\medskip

\textbf{VERITAS} - The Very Energetic Radiation Imaging Telescope Array System is an array of four
                    imaging atmospheric Cherenkov telescopes ({\it IACTs})
                   located in southern Arizona and has been operating since spring of 2007.
\medskip

\textbf{WCA} - Water Cherenkov Array, a technique to detect gamma-rays from ground, was pioneered by the 
              Milagro collaboration. 

\medskip 

\textbf{Whipple 10~m gamma-ray telescope} - The Whipple 10~m gamma-ray telescope is the pioneering instrument that 
                      was used to develop the imaging atmospheric Cherenkov technique and is located in southern
                     Arizona at Whipple Observatory. 

\newpage

\onecolumn
\section{Charge from APS}
\label{charge-appendix}
\begin{figure*}[!hb]
\includegraphics[height=8in]{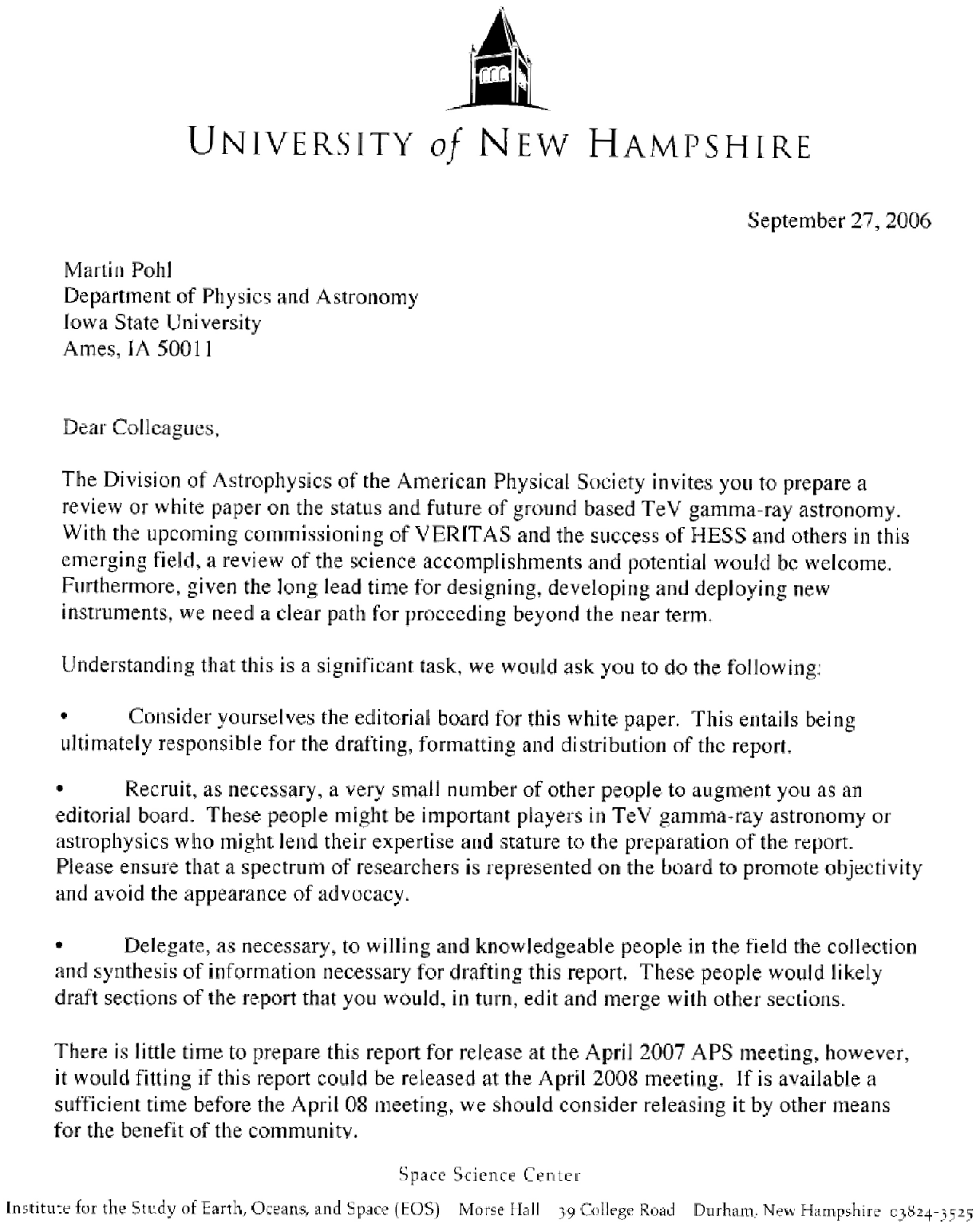}
\end{figure*}
\newpage
\begin{figure*}[!ht]
\includegraphics[width=6in]{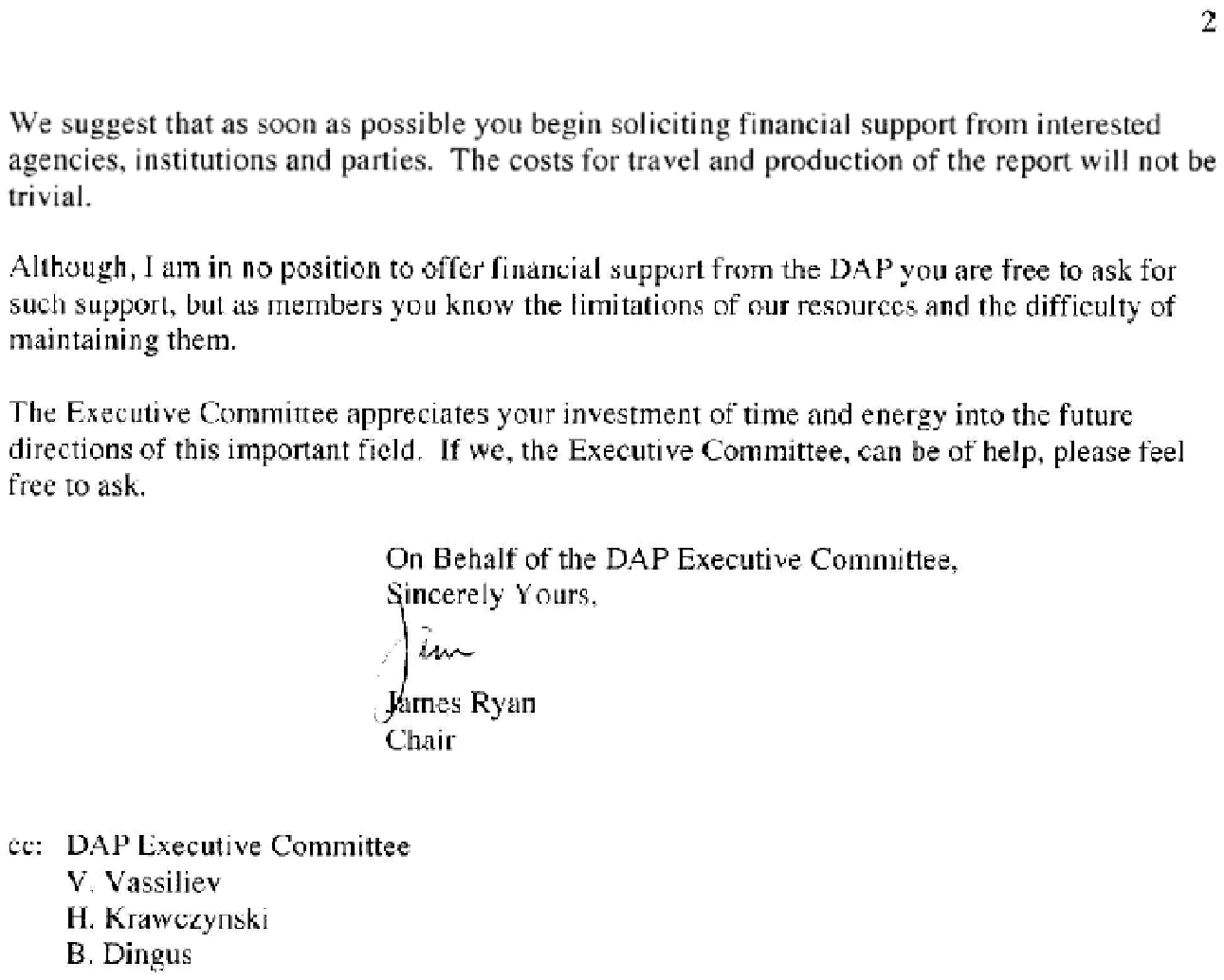}
\end{figure*}
\newpage

\section{Letter to community (via e-mail)}
\label{letter-appendix}
Dear Colleague,\\ \\

\noindent
In recent years, ground-based gamma-ray observatories have made a number of important
astrophysical discoveries which have attracted the attention of the wider scientific
community. The high discovery rate is expected to increase during the forthcoming years,
as the VERITAS observatory and the upgraded MAGIC and HESS observatories
commence scientific observations and the space-based gamma-ray telescope, GLAST, is
launched. The continuation of these achievements into the next decade will require a new
generation of observatories. In view of the long lead time for developing and installing
new instruments, the Division of Astrophysics of the American Physical Society has
requested the preparation of a White Paper (WP) on the status and future of ground-based
gamma-ray astronomy to define the science goals of the future observatory, to determine
the performance specifications, and to identify the areas of necessary technology
development. The prime focus of the WP will be on the astrophysical problems which
can be addressed at energies above 10 GeV. No particular experiment or technology will
be endorsed by the WP; instead it will enumerate available space and ground-based
technological alternatives. On behalf of the working groups, we would like to invite both
US and international scientists from the entire spectrum of astrophysics to contribute to
the concepts and ideas presented in the White Paper.\\ \\

\noindent
The work on the WP is organized through six science working groups (SWGs) and a
technology working group (TWG). We would like to invite you to contribute your
expertise and time to our efforts on (fill in specific working group). We anticipate that in
the final version, the WP will consist of a brief Executive summary designed for non-
scientists, an extended summary written for physicists, and it will also include detailed
"Appendices" written by each working group for specialists in the appropriate fields of
astrophysics or technology. While the Executive and extended summaries will be
compiled by the WP editorial board, the Appendices will be written by the members of
the SWGs and TWG. All contributing scientists will be authors of the WP, and the paper
will be endorsed by them.\\ \\

\noindent
The approximate timeline for writing the WP is as follows. A first brief public meeting to
discuss the White Paper will be held on Feb 8, 2007, the last day of the GLAST
symposium at Stanford University, California. Information about this splinter meeting is
available at: \verb!http://cherenkov.physics.iastate.edu/wp/glast.html!. We invite contributions
from all interested scientists. The format of the presentations can be further discussed
with the organizers of the WP SWGs and TWG, which are listed below. We expect that
the first drafts of the Appendices will be completed by the working groups in late March,
2007. A special session devoted to the White Paper will be organized at the meeting of
the American Physical Society, April 14-17, 2007 in Jacksonville, FL. The revised
versions of the working group reports (the Appendices) are expected to be produced in
early May, 2007. We plan to have a designated one day meeting on the WP during the
workshop, "Ground Based Gamma Ray Astronomy: Towards the Future," May 13-14,
2007 in Chicago, IL \verb!http://www.hep.anl.gov/byrum/next-iact/index.html!. The final
version of the Appendices is expected in early July. We plan to complete WP in late fall
2007. \\ \\

\noindent
Further information about the White Paper, the science and technology working groups,
and the associated meetings can be found at the web-site:\\

\noindent
\verb!http://cherenkov.physics.iastate.edu/wp/!\\

\noindent
To directly contact a current working group organizer, send an e-mail to:\\

\noindent
Henric Krawczynski \emph{krawcz@wuphys.wustl.edu} (Extragalactic Astrophysics),\\
Eric Perlman \emph{perlman@jca.umbc.edu} (10 GeV Sky Survey sub-group),\\
Phil Kaaret \emph{philip-kaaret@uiowa.edu} (Galactic compact objects),\\
Martin Pohl \emph{mkp@iastate.edu} (SNR and cosmic rays),\\
Jim Buckley \emph{buckley@wuphys.wustl.edu} (Dark matter),\\
Abe Falcone \emph{afalcone@astro.psu.edu} or\\
David Williams \emph{daw@scipp.ucsc.edu} (Gamma-ray bursts),\\
Karen Byrum \emph{byrum@hep.anl.gov} (Technology).\\ 

\noindent
For additional information please contact a member of the editorial board:\\ 

\noindent
Brenda Dingus \emph{dingus@lanl.gov}\\
Francis Halzen \emph{halzen@pheno.physics.wisc.edu}\\
Werner Hofmann \emph{Werner.Hofmann@mpi-hd.mpg.de}\\
Henric Krawczynski \emph{krawcz@wuphys.wustl.edu}\\
Martin Pohl \emph{mkp@iastate.edu}\\
Steven Ritz \emph{Steven.M.Ritz@nasa.gov}\\
Vladimir Vassiliev \emph{vvv@astro.ucla.edu}\\
Trevor Weekes \emph{weekes@egret.sao.arizona.edu}\\ 

\noindent
We understand that you have many commitments and appreciate any investment of time
and expertise you could provide to this endeavor. We would be grateful for your reply to
this invitation during the next week. \\ \\

\noindent
Sincerely,

\newpage

\section[Agenda for Malibu meeting]{Agenda for Malibu meeting\footnote{\url{http://gamma1.astro.ucla.edu/future_cherenkov}}}
\label{malibu-appendix}
\begin{center}\textbf{October 20, Thursday: Science (Mays' landing)}\end{center}

\noindent Please arrive to Mays' Landing no earlier than 8:30AM. Parking is strictly limited, carpooling (at least two participants per car) is a must. Meeting begins at 9:00AM. Coffee will be available from 8:45 to 9:00\\
Chairperson: Vladimir Vassiliev\\

\noindent
\begin{enumerate}
\item Recent progress in ground-based gamma-ray astronomy and the goals of this meeting.
(Simon Swordy) 10 min
\item Overview of Particle Astrophysics in the U.S. (roadmaps, future missions, funding).
(Rene Ong) 20 min
\item High energy astrophysics after the GLAST mission (5 years of operation)
(Julie McEnery) 25 min
\begin{center}
                                    \textbf{Coffee Break (30 min)}\\
\end{center}
\item Development of ideas in ground based gamma-ray astronomy, status of the field, and
scientific expectations from HESS, VERITAS, MAGIC, CANGAROO.
(Trevor Weekes) 20 min
\item Scientific expectations from MILAGRO and motivations for a 2 pi sr, 95
detector (Brenda Dingus) 20 min
\item Discussion: Politics, etc. (Rene Ong) 40 min
\begin{center}
                                    \textbf{LUNCH}\\
\end{center}
\item Very high energy extragalactic transient sources: AGN\/EBL. (Henric Krawczynski) 20
min
\item Very high energy extragalactic transient sources: GRBs. (David Williams) 20 min
\item Galactic science in the 1 GeV -1 TeV domain (Martin Pohl) 20 min
\begin{center}
                                    \textbf{Coffee Break (30 min)}\\
\end{center}
\item Galactic science in the 1-100 TeV domain. (Stephan LeBohec) 10 min
\item Astroparticle physics: Dark Matter, Annihilation of cosmological defects, exotic
particles (James Buckley) 15 min
\item Astroparticle physics: Non-standard model, broken symmetries, PBH searches, etc.
(Frank Krennrich) 15 min
\item Brief outline of a few science ideas for VHE observations (Paolo Coppi) 15 min
\item Discussion: Scientific justification of future observatory and required performance
characteristics (solid angle, energy domain, etc.) (Gus Sinnis) 1 hour
\end{enumerate}
\begin{center}
                                    \textbf{Meeting ends at 4:45PM.}\\
\end{center}                                  
 DINNER - 6:30pm, The Sunset Restaurant \verb!(http://www.thesunsetrestaurant.com)! \\
\newpage

\begin{center}
\textbf{October 21, Friday: Technical design options for a future observatory\\
(Mays' landing)}\\\end{center}

\noindent Please arrive to Mays' Landing no earlier than 8:30AM. Parking is strictly limited, so carpooling is a must. Meeting begins at 9:00AM. Coffee will be available from 8:45 to 9:00.\\
Chairperson: Simon Swordy\\

\noindent
\begin{enumerate}
\item Concept(s) for very low energy observations (\textless10 GeV) (John Finley for Alexander
Konopelko) 20 min
\item High energy transient observatory (HE-ASTRO 1km$^{2}$ array) (Stephen Fegan) 20
min
\item Small Telescope Arrays (STAR) (Henric Krawczynski) 20 min
\begin{center}
                                    \textbf{Coffee Break (15 min)}\\
\end{center}
\item Optimization of High Altitude Water Cherenkov (HAWC) detector (Gus Sinnis) 20
min
\item Design, sensitivity, and cost of miniHAWC (Andrew Smith) 20 min
\item Strategies associated with observations at \textgreater10 TeV regime (Stephan LeBohec) 10 min
\item Increasing the collection area for Cherenkov telescopes at high energies (Jamie
Holder) 10 min
\item Effect of the FoV of IACTs on the sensitivity for point and extended sources (David
Kieda) 10 min
\item Discussion: Design evaluation criteria. (Simon Swordy) 50 min
\begin{center}                                           \textbf{LUNCH}\\ \medskip 

\noindent Chairperson: Karen Byrum \end{center}

\item Considerations of beyond-CANGAROO projects (Takanori Yoshikoshi) 10 min
\item Summary of thoughts from HESS \& MAGIC members (Vladimir Vassiliev) 10 min
\item Wide field of view Cherenkov Telescopes: Cassegrain Cherenkov telescopes
 (James Buckley) 15 min
\item Wide FoV: Initial design considerations (Vladimir Vassiliev) 15 min
\item High Elevation Sites in Mexico (Alberto Carraminana) 15 min
\item Site Considerations (Stephen Fegan) 15 min
\begin{center}
                                   \textbf{Coffee Break (15 min)}\\ \medskip
Technology development:\end{center}

\item Intelligent trigger concepts: Scientific motivations (Frank Krennrich) 15min
\item Photodetectors and electronic readout options (James Buckley) 15 min
\item Digital Asic (Gary Drake) 15 min
\item HE-ASTRO: focal plane instrument and trigger concept (Vladimir Vassiliev) 15 min
\item Triggering and data acquisition (high data rates regime) (Jim Linnemann) 15 min
\item Discussion: What systems should be pursued with further R\&D? (Karen Byrum) 1+
hour
\end{enumerate}
\begin{center}
\textbf{Meeting ends at 4:45PM.}\medskip
\end{center}
\medskip
\begin{center}\textbf{October 22, Saturday: Organizational issues, R\&D plans (UCLA)}\\
Meeting begins at 9:00AM\end{center}
Depending on the number of contributions, part of the agenda of this day may be moved
to the evening of the previous day.\\ 
\begin{enumerate}
\item Discussion (tasks, responsibilities, goals) \& Organization of working groups (science,
detectors, simulations, ...)
\item Upcoming funding opportunities (Potential proposals)
\item Future conferences
\end{enumerate}
\newpage

\section[Agenda for Santa Fe meeting]{Agenda for Santa Fe meeting\footnote{\url{http://www.lanl.gov/orgs/p/g_a_d/p-23/gammaworkshop/}}}
\label{santafe-appendix}
\small
\noindent
\begin{tabular*}{0.75\textwidth}{@{\extracolsep{\fill}} r c l }

\multicolumn{3}{c}{\textbf{Thursday, May 11, 2006}}\\ 

  &  \\

9:00 - 9:10  & Welcome       &                          Gus Sinnis\\ 

9:10 - 9:20 & Science Section of the White Paper    &  Brenda Dingus\\ 

9:20 - 9:50 & Extragalactic Sources Working Group   &  Henric Krawczynski\\ 

9:50 - 10:20 & Particle Acceleration in TeV Gamma-Ray Sources & Siming Liu\\ 

10:20 - 10:35  &          &                             Paolo Coppi\\ 

10:35 - 11:00 & Break\\ 

11:00 - 11:30 & Gamma Ray Bursts            &             Abe Falcone\\ 

11:30 - 12:00 & Milagro Detection of the Cygnus Region &  Aous Abdo\\ 

12:00 - 12:45 & Lunch\\ 

12:45 - 1:15 &  Galactic Diffuse Working Group      &     Martin Pohl\\ 

1:15 - 1:45  & Dark Matter \& Gamma-Ray All-Sky Surveys & Savvas Koushiappas\\ 

1:45 - 2:15  & Dark Matter and Other Particle Physics Working Group &  Jim Buckley\\ 
 
2:15 - 2:45 &   Break\\ 

2:45 - 3:15 & Galactic Sources             &                Phil Kaaret\\ 

3:15 - 3:45   & Physics Motivations      &                    Martin Pohl\\ 

3:45 - 4:15 &   Extending GLAST Science Simulations to TeV  & Julie McEnery\\ 

4:15 - 5:30 &  Science Working Groups Meet Separately\\ 

6:00  &   Banquet\\

 & \\

\multicolumn{3}{c}{\textbf{Friday, May 12, 2006}} \\

  &  \\

8:30 - 8:50  &  ACT Overview: Technology Drivers and Design Metrics &  Jim Buckley\\

8:50 - 9:05 &   Mirror Design   &                             John Finley\\

9:05 - 9:20 &   Digitizers       &                            Gary Drake\\

9:20 - 9:35 &   Photodetectors           &                    Gary Drake\\

9:35 - 9:50  &  Trigger Electronics        &                  Frank Krennrich\\

9:50 - 10:05   & DAQ Electronics        &                      Scott Wakely\\ 

10:05 - 10:20 &  Mobile ACTs        &                          Henric Krawczynski\\ 

10:20 - 10:35 &  LANL CMOS Camera Development      &           Kris Kwiatkowski\\ 

10:35 - 11:00 & Break\\ 

11:00 - 11:15 & VERITAS Update   &                           Trevor Weekes\\ 

11:15 - 11:30 & TRICE   &                                    Karen Byrum\\ 

11:30 - 12:00 & miniHAWC   &                                 Andy Smith\\ 

12:00 - 12:15 & Mexico Sites      &                          Alberto Carraminana\\ 

12:15 - 12:30 & Utah Sites       &                           Dave Kieda\\ 

12:30 - 1:30 & Lunch\\ 

1:30 - 2:00  & \textless100 GeV ACTs    &                           Alex Konopelko\\ 

2:00 - 2:10  & Experimental Approaches to the High Energy &  Stephan LeBohec\\ 
& End of $\gamma$-Ray Source Spectra & \\ 

2:10 - 2:20 & ACTs and Intensity Interferometry     &      Stephan LeBohec\\ 

2:20 - 2:35  & HE-ASTRO     &                               Stephan Fegan\\ 

    &   STAR          &                             Abe Falcone\\ 

2:35 - 3:05 &       HESS \& Beyond      &                        German Hermann\\ 

3:05 - 3:30 &  Break\\ 

3:30 - 4:00 &  Summary \& Comparison of Future Projects  &           Frank Krennrich\\ 

4:00 - 5:00 &   White Paper Discussion   &            Jim Ryan\\

  & \\ 

\multicolumn{3}{c}{\textbf{Saturday, May 13, 2006}} \\

\multicolumn{3}{c}{Tour of Milagro Site in the morning} \\
\end{tabular*}
\normalsize
\newpage

\section[Agenda for Chicago meeting]{Agenda for Chicago meeting\footnote{\url{http://www.hep.anl.gov/byrum/next-iact/agenda.html}}}
\label{chicago-appendix}
\small 
\begin{center} Future in Gamma-ray Astronomy Meeting\\May 13 \& 14, 2007 at the Wyndham Hotel in downtown Chicago\end{center}

\begin{tabular*}{0.75\textwidth}{@{\extracolsep{\fill}} r c l }
\multicolumn{3}{c}{\textbf{Sun. Morning Session: 9:00 am - 12:30 pm}}\\
08:15-09:00  &  Continental Breakfast \\

09:00-09:10  &  Welcome and Introduction        &  Scott Wakely\\

09:10-09:40 &  Current Status and Near Future  &  Trevor Weekes\\

09:40-10:00  &  White Paper Status and Timeline &  Vladimir Vassiliev\\

10:00-10:30  &  Coffee Break\\

10:30-10:50  &  Extragalactic Talk            &    Markos Georganopoulos\\

10:50-11:10  &  Extragalactic Talk            &    Charles Dermer\\

11:10-11:30  &  Extragalactic Discussion      &    Led by Henric Krawczynski\\

11:30-12:00  &  Gal. Compacts Talk            &   Roger Romani\\

12:00-12:20  &  Gal Compacts Discussion      &     Led by Phil Kaaret\\

12:20-12:30  &  Group Photo\\

12:30-14:00  &  Lunch (provided)\\

\multicolumn{3}{c}{\textbf{Sun. Afternoon Session: 14:00 pm - 18:30 pm}}\\

14:00-14:20  &       SNR/CR Talk              &              Patrick Slane\\

14:20-14:40   &      SNR/CR Talk              &           Igor Moskalenko\\

14:40-15:00   &   SNR/CR Discussion          &          Led by Martin Pohl\\

15:00-15:20   &     Dark Matter Talk         &             Tim Tait\\

15:20-15:40   &    Dark Matter Talk      &            Savvas Koushiappas\\

15:40-16:00  &  Dark Matter Discussion       &     Led by Jim Buckley\\

16:00-16:30  &       Coffee Break\\

16:30-16:50    &       GRB Talk             &        Shri Kulkarni\\

16:50-17:10    &       GRB Talk                &              Neil Gehrels\\

17:10-17:30  &      GRB Discussion  & Led by David Williams\\

17:30-18:30   &    Spare/Extra Time\\
\multicolumn{3}{c}{\textbf{Dinner: 7:00pm}}\\

\multicolumn{3}{c}{\textbf{Mon. Morning Session: 9:00 am - 12:30 pm}}\\
08:15-09:00    &           Continental Breakfast\\

09:00-09:30 & Summary \& Status of Tech. Section of the WP & Frank Krennrich\\

09:30-09:50 & Status of CTA focusing on the science & Agnieszka Jacholkowska\\

09:50-10:10      &    Status of CTA focusing on possible instrument studies & German Hermann\\

10:10-11:00       &   Coffee Break \& Poster Viewing \& Mingling \\

11:00-11:15     &      Interesting sites for Cherenkov telescopes in Argentina & Adrian Rovero\\

11:15-11:30 & Mexican proposal for hosting Cherenkov detectors  &      Alberto Carraminana\\

11:30-11:50 & The ILC detector R\&D Model      & Harry Weerts\\

11:50-12:05     &      Update on Decadel Survey       &     Brenda Dingus\\

12:05-13:30  &             Lunch (provided)  \\

\multicolumn{3}{c}{\textbf{Afternoon Session: 13:30 pm - 18:30 pm - Chair: Dave Kieda}}\\

13:30-14:00  & Future Directions: Steps towards the future     &       Martin Pohl\\

14:00-14:30    &                    R\&D Proposal  & Jim Buckley\\

14:30-15:00 &  Continuation of discussion of Next Steps\/    &         Dave Kieda\\
 & Establishing a new Collaboration\/Team & \\

15:00-15:30     &    Coffee Break\\

15:30-17:30 &    \multicolumn{2}{c}{Break into smaller working groups and identify tasks\/action items.}\\
\end{tabular*}
\normalsize 
\newpage

\section[Agenda for SLAC meeting]{Agenda for SLAC meeting\footnote{\url{http://www-conf.slac.stanford.edu/vhegra/agenda.htm}}}
\label{slac-appendix}
\begin{tabular*}{0.75\textwidth}{@{\extracolsep{\fill}} r l }
\multicolumn{2}{c}{\textbf{Day 1}}\\
     8:00   &  Breakfast\\
\multicolumn{2}{c}{\textbf{Welcome and overview session:}}\\
     8:30   &  Welcome note          (R. Blandford - KIPAC)\\
     8:40   &  Current status of the field and future directions (W. Hofmann - MPI-K Heidelberg) \\
     9:10   &  Summary of the White Paper (H. Krawczynski - Washington University) \\
     9:40   &  Discussion\\
    10:00  &   Coffee Break\\
\multicolumn{2}{c}{\textbf{Science Session:}}\\
    10:40   &  Galactic talk - Top 10 Science Questions (S. Funk, KIPAC) \\
    11:10   &  Extragalactic talk - Top 10 Science Questions (P. Coppi – Yale)\\ 
    11:40   &  New physics talk - Top 10 Science Questions (L. Bergstrom - Stockholm University)\\ 
    12:10   &  The connection to GLAST  (O. Reimer - Stanford University)\\ 
 12:40  &  Lunch break\\
\multicolumn{2}{c}{\textbf{Projects session:}}\\
 14:30   & HAWC       (B. Dingus - Los Alamos National Lab)\\ 
 15:00  &  AGIS     (J. Buckley - Washington University)\\ 
 15:30  &  CTA    (M.Martinez – IFAE Barcelona)\\ 
 16:00   & Status of the Japanese Gamma-ray Community (T. Tanimori - Kyoto University).\\ 
 16:30  &  Other ideas for Gamma-ray instruments (S. LeBohec - University of Utah)\\ 
 17:00  &  Coffee Break\\
 17:30  &  Technical challenges and parameters for a future design (S. Swordy - University of Chicago)\\ 
 18:00  &  Wrap up and social event\\
\multicolumn{2}{c}{\textbf{Day 2: Technical Session}}\\
  8:30   & Breakfast\\
  9:00   & Wide field of view instruments and secondary optics (V. Vassiliev - UCLA)\\ 
  9:25  &  Monte Carlo Studies for CTA    (K. Bernloehr – MPIK Heidelberg)\\ 
  9:50  &  Monte Carlo Studies for a future instrument (S. Fegan - UCLA)\\ 
 10:15  &  Coffee Break\\
 10:45  &  Survey instrument  (G. Sinnis - Los Alamos National Lab)\\ 
 11:10  &  Backend electronics and readout  (H. Tajima – SLAC)\\ 
 11:35  &  Triggering etc.        (F. Krenrich - Iowa State University)\\ 
 12:00  &  Current and future Photodetectors for AGIS (Bob Wagner – ANL)\\ 
 12:25  &  Photodetectors in gamma-astronomy (M. Teshima – MPIP Munich)\\
 12:50  &  Lunch Break\\
 14:00  &  Future of Space-based Gamma-Astronomy (N. Gehrels - NASA/GSFC)\\
& AGIS Session (3pm-6pm) devoted to AGIS collaboration and future R\&D proposals\\
 15:00  &  Opening comments               (V. Vassiliev)\\
 15:30  &  Short presentations (1 transparency) of visions for the future instrument\\
& (discussion moderated by J. Buckley)\\
 16:00  &  General discussion of AGIS collaboration issues: International collaboration, \\
& schedule for collaboration meetings, timescale for proposals, possible site (north versus south).\\
16:30 & Presentation on optimization of design parameters for an array of IACTs (S. Bugaev)\\
16:40 & The Low Energy Array of A Major Future VHE Gamma-Ray Experiment (A. Konopelko)\\ 
16:50 & Discussion of cost of different cameraapproaches (H. Tajima)\\ 
17:05 & Coffee Break\\
17:20 & Discussion of mechanical design and fabrication, schedule (B. Wagner and V. Gaurino)\\
17:40 & Presentation on the SPM site\\
\end{tabular*}
\newpage

\begin{figure*}[!ht]
\centering
\includegraphics[width=\textwidth]{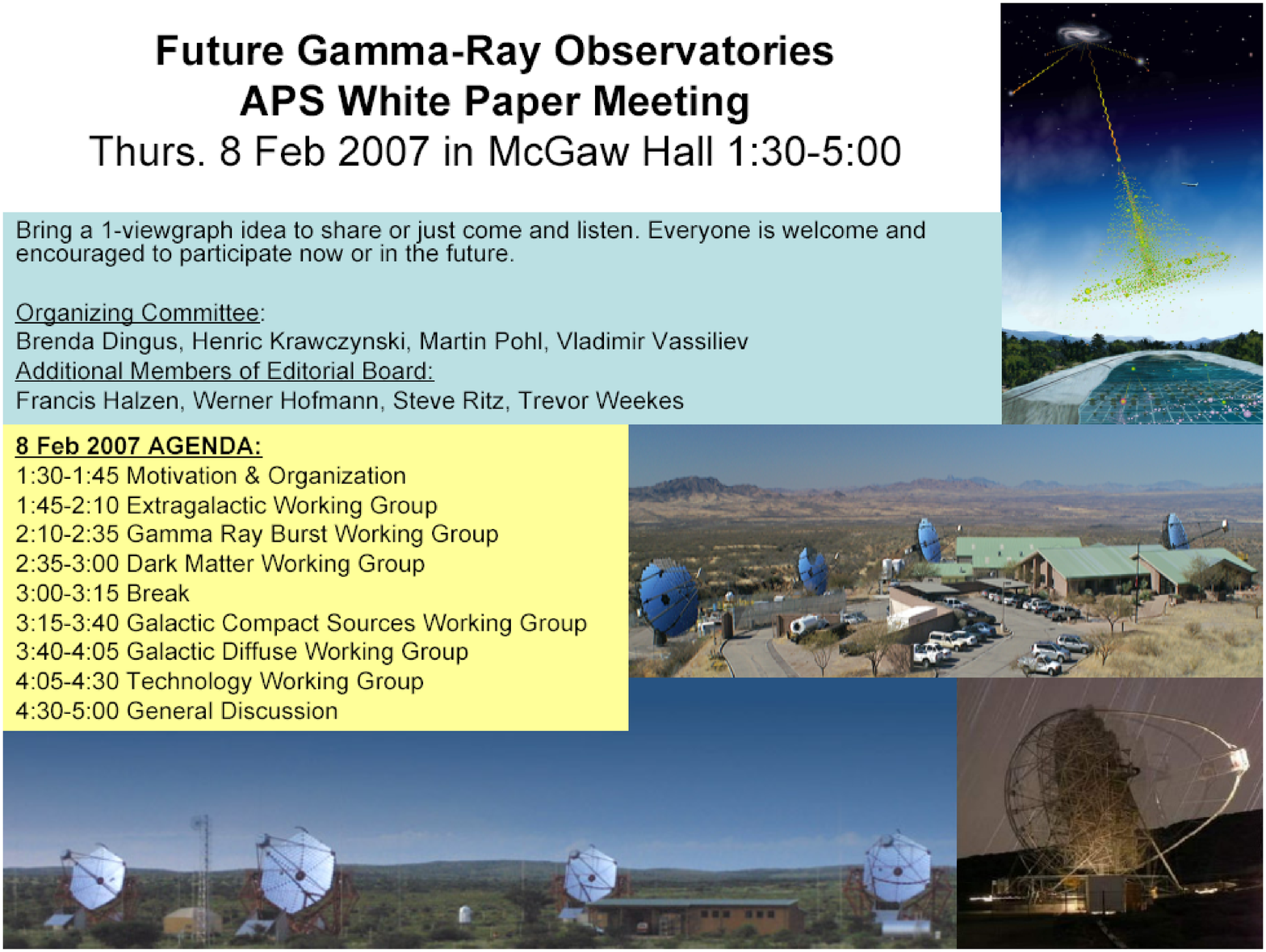}
\label{meeting-appendix}
\end{figure*}
\newpage

\section{Agenda and record of white paper teleconference meetings}
\label{conference-calls-appendix}
\noindent
Conference call on Jun 09, 2006 \\
       - Discussion of organizational structure\\
       - Nomination of Working Group chairs\\
       - Milestones and deadlines.\\

\noindent
Conference call on Jun 15, 2006\\
       - Web page discussion\\
       - Description of draft letter to the community describing the white paper\\
       - Preliminary organization for a fall whitepaper meeting in St. Louis\\

\noindent
Conference call on Aug 22, 2006\\
       - Discussion of external participation on organizing committee\\
       - Update from working groups.\\
       - Plans for one-day workshop at the GLAST symposium\\

\noindent
Conference call on Sep 13, 2006\\
       - Update from working groups; working group reports posted on our web site.\\
       - Action items discussed for GLAST symposium\\
       - Jim Ryan announces APS-DAP plans to officially invite us to produce a white paper.\\

\noindent
Conference call on Oct 12, 2006\\
       - Detailed update from working groups.\\
       - Formalize plans for whitepaper workshop at GLAST symposium.\\

\noindent
Conference call on Nov 01, 2006\\
       - Election of external organizing committee members\\
       - Update from working groups\\
       - Outline plans for reaching out to a broader community for support\\

\noindent
Conference call on Nov 08, 2006\\
       - Discussion on members for editorial board\\

\noindent
Conference call on Nov 15, 2006\\
       - Discussion of gamma-ray astrophysics session at April 2007 APS\\
       - Initial planning for a Chicago Spring meeting\\
       - Working group discussions\\

\noindent
Conference call on Nov 22, 2006\\
       - Expansion of the editorial board\\
       - Working group updates\\
       - Webpage update\\
       - Organization of GLAST symposium talks\\

\noindent
Conference call on Nov 29, 2006\\
       - Working group updates and planning for GLAST symposium\\

\noindent
Conference call on Dec 04, 2007\\
       - Working group update\\

\noindent
Conference call on Jan 11, 2007\\
       - Update on submissions of White Paper abstracts to APS\\
       - Discussion of template of invitation e-mail to broader community\\

\noindent
Conference call on Mar 19, 2007\\
       - Update on progress of working groups\\

\noindent
Conference call on Apr 23, 2007\\
       - Working group updates.\\
       - Discussion following APS meeting\\

\noindent
Conference call on Aug 28, 2007\\
       - Working group updates\\
       - Discussion of summary sections\\

\noindent
Conference call on Oct 29, 2007\\
       - Remaining White Paper contributions\\

\noindent
Conference call on Dec 04, 2007\\
       - Discussion of White Paper working group contributions.\\

\newpage

\section{Membership}
\label{member-appendix}
\oddsidemargin -0.5in
\evensidemargin -0.5in
\textwidth 7.5in

\begin{longtable}{@{\extracolsep{\fill}}p{1.65in} p{4in} p{1.45in}}

\textbf{Editors} & \textbf{Affiliation}\\
& \\
Brenda Dingus & Los Alamos National Laboratory \\
Henric Krawczynski & Washington University (St.Louis) \\
Martin Pohl & Iowa State University  \\
Vladimir Vassiliev & University of California Los Angeles \\

& \\
\textbf{Senior Advisors} & \textbf{Affiliation}\\
& \\

Francis Halzen & University of Wisconsin, Madison \\
Werner Hofmann & Max-Planck-Institut f\"ur Kernphysik (Heidelberg) \\
Steven Ritz & NASA Goddard Space Flight Center \\
Trevor Weekes & The Harvard Smithsonian Center for Astrophysics \\

& \\

\textbf{Group Chairs} & \textbf{Affiliation} & \textbf{Group}\\
& \\
Jim Buckley & Washington University (St.Louis)  & DM\\
Karen Byrum &  Argonne National Laboratory  & Tech\\
Abe Falcone & Penn State  & GRB\\
Phil Kaaret & The University of Iowa & GCO\\
Henric Krawczynski & Washington University (St.Louis)  & EG\\
Martin Pohl & Iowa State University &  SNR\\
David Williams & University of California, Santa Cruz  & GRB\\ 

& \\
\textbf{Secretarial staff} & \textbf{Affiliation}\\
& \\

Alan Hulsebus & Iowa State University \\
Iris D. Peper & Washington University (St.Louis) \\
& \\

\textbf{Group Members} & \textbf{Affiliation} & \textbf{Group}\\
& \\
Armen Atoyan & Montreal University & EG \\
Ahmad Abdo & Michigan State University & SNR, GCO\\
Jonathan Arons & University of California, Berkeley  & GCO\\
Armen Atoyan & Universite de Montreal  & EG, SNR\\
Edward Baltz & Kavli Institute for Particle Astrophysics and Cosmology & DM\\
Matthew Baring & Rice University & SNR, GCO, GRB\\
John Beacom & Ohio State University & SNR\\
Matthias Beilicke & Washington University (St. Louis)  & EG\\
Gianfranco Bertone & Institut d'Astrophysique de Paris  & DM\\
Roger Blandford & Kavli Institute for Particle Astrophysics and Cosmology, Stanford Linear Accelerator Center & EG, SNR, GRB\\
Markus B\"ottcher & Ohio University  & EG\\
Jim Buckley & Washington University (St. Louis)  & DM, Tech, GRB\\
Slava Bugayov & Washington University (St. Louis)  & Tech\\ 
Yousaf Butt & The Harvard Smithsonian Center for Astrophysics  & SNR\\
Andrei Bykov & Ioffe Physico-Technical Institute, St. Petersburg  & SNR\\
Karen Byrum & Argonne National Laboratory  & DM, Tech\\
Alberto Carraminana & Instituto Nacional de Astrof\'isica, \'Optica y Electr\'onica & EG\\
Valerie Connaughton &  National Space Science and Technology Center & GRB\\
Paolo Coppi & Yale University  & EG, GRB\\
Wei Cui & Purdue University &  GCO\\
Charles Dermer & U.S. Naval Research Laboratory &  EG, GRB\\
Brenda Dingus &  Los Alamos National Laboratory &  DM, EG, GCO, GRB, Tech\\
Gary Drake & Argonne National Laboratory &  Tech\\
Eli Dwek & Goddard Space Flight Center &  EG\\
Don Ellison & North Carolina State University &  SNR\\
Abe Falcone & Pennsylvania State University &  EG, GRB\\
Steve Fegan & University of California, Los Angeles &  DM, Tech\\
Francesc Ferrer & Case Western Reserve University & DM\\
John Finley & Purdue University &  EG, GCO,  Tech\\
Chris Fryer & Los Alamos National Laboratory & GRB\\
Stefan Funk & Kavli Institute for Particle Astrophysics and Cosmology, Stanford Linear Accelerator Center & EG, GCO, Tech\\
Bryan Gaensler & The University of Sydney, Austrailia &  GCO\\
Neil Gehrels & NASA Goddard Space Flight Center &  GRB\\
Markos Georganopoulos & NASA Goddard Space Flight Center &  EG\\
Paolo Gondolo & The University of Utah &  DM\\
Jonathan Granot & Stanford Linear Accelerator Center &  GRB\\
Jeter Hall & Fermi National Accelerator Laboratory &  DM\\
Francis Halzen & University of Wisconsin, Madison &  SNR\\
Alice Harding & NASA Goddard Space Flight Center &  GCO\\
Elizabeth Hays &  NASA Goddard Space Flight Center &  GCO, SNR, Tech\\
Sebastian Heinz & University of Wisconsin, Madison &  GCO\\
Jamie Holder & University of Delaware &  EG, GCO, Tech\\
Dan Hooper & Fermi National Accelerator Laboratory & DM\\
Deirdre Horan & Argonne National Laboratory &  DM, EG, GRB, Tech\\
Brian Humensky & The University of Chicago &  SNR\\
Tom Jones & University of Minnesota &  EG, SNR\\
Ira Jung & Washington University (St. Louis) &  EG\\
Philip Kaaret & The University of Iowa &  EG, GCO, SNR\\
Jonathan Katz & Washington University (St. Louis) &  EG, GRB\\
Dave Kieda & The University of Utah &  SNR, GCO\\
Alexander Konopelko & Purdue University & Tech\\
Savvas Koushiappas & Brown University &  DM\\
Henric Krawczynski & Washington University (St. Louis) &  DM, EG, Tech\\
Frank Krennrich  & Iowa State University &  EG, Tech\\
Kyler Kuehn & Center for Cosmology and Astroparticle Physics & GRB\\
Stephan LeBohec  & The University of Utah &  DM, EG, GCO, SNR, Tech\\
Amir Levinson  & Tel Aviv University, Israel &  GCO\\
Julie McEnery  & NASA Goddard Space Flight Center &  EG\\
Peter M\'esz\'aros  & Pennsylvania State University &  GRB, SNRB\\
Igor Moskalenko  & Stanford Linear Accelerator Center &  SNR\\
Reshmi Mukherjee  & Columbia University &  EG, GCO\\
Jay Norris & NASA Goddard Space Flight Center & GRB\\
Pablo Saz Parkinson & University of California, Santa Cruz & GRB\\
Rene Ong  & University of California, Los Angeles &  EG, GCO\\
Asaf Pe'er  &  University of Amsterdam \& Space Telescope Science Institute &  GRB\\
Eric Perlman  & Flordia Institute of Technology &  EG\\
Martin Pohl  & Iowa State University &  DM, EG, GCO, SNR\\
Stefano Profumo & University of California, Santa Cruz & DM\\
Ken Ragan  & McGill University &  GCO\\
Enrico Ramierez-Ruiz  & University of California, Santa Cruz &  GRB\\
Soeb Razzaque  & Pennsylvania State University &  GRB\\
Steven Ritz  & NASA Goddard Space Flight Center &  EG\\
James Ryan  & University of New Hampshire &  EG\\
Joe Silk  & Oxford University &  DM\\
Gus Sinnis  & Los Alamos National Laboratory &  EG, Tech\\
Patrick Slane  & Harvard-Smithsonian Center for Astrophysics &  SNR, GCO\\
Andrew Smith  & University of Maryland &  GCO, Tech\\
Andy Strong  & Max-Planck-Institut fuer extraterrestrische Physik &  SNR\\
Tim Tait  & Argonne National Laboratory \& Northwestern University &  DM\\
Diego Torres  & Instituci\'o Catalana de Recerca i Estudis Avançats (Barcelona) &  GCO\\
Meg Urry  & Yale University &  EG\\
Vladimir Vassiliev  & University of California, Los Angeles &  DM, EG, Tech\\
Robert Wagner  & Argonne National Laboratory &  DM, Tech\\
Scott Wakely  & The University of Chicago &  DM, SNR, Tech\\
Trevor Weekes  & Smithsonian Astrophysical Observatory &  EG, GRB\\
Matthew Wood  & University of California, Los Angeles &  DM, Tech\\
Xiang-Yu Wang  & Pennsylvania State University &  GRB\\
David A. Williams  & University of California, Santa Cruz &  EG, GRB\\
Gabrijela Zaharijas  & Argonne National Laboratory &  DM\\
Bing Zhang  & University of Nevada, Las Vegas &  GRB\\
\end{longtable}

\begin{center}
\begin{tabular}{r l}
\multicolumn{2}{c}{\textbf{Group Legend}}\\
& \\
DM & Dark Matter\\
EG & Extragalactic Very-High-Energy Astrophysics\\
GCO & Galactic Compact Objects\\
GRB & Gamma-Ray Bursts\\
SNR & Supernova Remnants\\
Tech & Technology\\
\end{tabular}
\end{center}
\newpage






\end{document}